%% file: thesis.tex
\pdfoutput=1

\documentclass[12pt,lot, lof, singlespace]{puthesis}
\newcommand{\printmode}{}

\title{Two--dimensional and novel quasi--two--dimensional quantum liquids}

\submitted{January 2013}  
\copyrightyear{2013}  
\author{Marco Nava}
\adviser{D. E. Galli}  
\department{Physics}

\usepackage{graphicx}
\usepackage{verbatim}

\usepackage{multirow}
\usepackage{longtable}

\usepackage{booktabs}

\usepackage{amsmath,amssymb,amsthm}

\ifdefined\printmode

\usepackage{url}

\else

\ifdefined\proquestmode

\usepackage{hyperref}
\hypersetup{bookmarksnumbered}
\def\onlinecite{\cite}

\makeatletter
\hypersetup{pdftitle=\@title,pdfauthor=\@author}
\makeatother

\else

\usepackage{hyperref}
\hypersetup{colorlinks,bookmarksnumbered}

\makeatletter
\hypersetup{pdftitle=\@title,pdfauthor=\@author}
\makeatother

\fi 
\fi 

\ifodd 0
\renewcommand{\maketitlepage}{}

\else

\acknowledgements{
I would like to thank and dedicate this work to friends and foes. To friends for the time shared 
together, in both bliss and woe. To foes for their opposition that gives me fuel to improve myself.
Well, actually I have no foes but rather friends with different views and friends with more different views; 
anyway, I like the opening effect of the previous sentences. Now, to acknowledgements: as almost everyone 
does in acknowledgements, I wish to thank my parents for the support that they have been giving till now. I hope to have 
the opportunity to pay back. Of course, it's not that everyone who writes acknowledgements 
says ``thank you" to {\it my} parents, but I'm sure a lot of phd students owe much to {\it theirs}. 
Particular thanks also to Natasha, my beloved girlfriend, and no, 
she's not threatening me with dagger and sword to write something 
cute for her. I'm really happy with her and I wish to thank her for her patience with me when I had to write 
this never ending document. Indeed, my struggles with blank pages that somehow had to be filled with something brought me 
to discover new original ways to launch curses and imprecations; in such moments, she has been very kind and brought me 
cats (I thank them to, for their {\it furriness}), tea, biscuits and her sweet eyes to calm me down. 

A lot of thanks to my collaborators, I would not have made it till now if not for the deep physical insight of Luciano, 
the everlasting ``holiness" of Maurizio (and His nemesis that can't be named), the hyperfine mathematic formalism of 
Ettore (the section about the mathematics of Markov chains has been possible only thanks to him). Thanks also to Filippo, 
he has been using the QMC++ library since September and has been very kind to report various minor bugs and also 
a ``thank you" to Mario; no, not because I don't want to exclude him. He's the new entry of the group but we have had chance 
to do some work together with Jellium and I enjoyed the time spent. But he should learn to be less shy and take freely the 
tea leaves that I leave in lab for everyone's tea time.

Finally, particular and sincere acknowledgements to Davide, he has been a superb advisor and I hope I'll have the opportunity 
to work again with him and learn even more. His guidance led me to work on very exciting projects and his wisdom and cunning
contributed a lot in the developments of the works presented here. In these years I've come to know him (at least 
for what regards academic life): his passion for research and his dedication to be a collaborator instead 
of a competitor should be taken as example. I'm very proud to know people like him. Indeed, I really mean it. 
I swear that it's not because he's gonna examine me !

Who else to thank? I'm surely forgetting an awful lot of people, but to excuse myself I write here that 
I offer a beer to anyone shows me that has been forgotten in this section; at only one condition: he has 
to answer correctly to a question of my choice that concerns this work. 
}

\fi  

\begin{document}

\makefrontmatter



\include{ch-intro/chapter-intro}

\include{ch-methods/chapter-methods}

\include{ch-polarization/chapter-polarization}

\include{ch-dynamics/chapter-dynamics}

\include{ch-fluorographene/chapter-fluorographene}

\include{ch-conclusion/chapter-conclusion}

\include{ch-details/chapter-details}

\appendix 
\include{ch-appendicies/estimators}

\include{ch-appendicies/higord}


\singlespacing
\bibliographystyle{plain}

\cleardoublepage
\ifdefined\phantomsection
  \phantomsection  
\else
\fi
\addcontentsline{toc}{chapter}{Bibliography}

\bibliography{thesis}

\end{document}

%% file: ch-intro/chapter-intro.tex
\def\onlinecite{\cite}

\chapter{Introduction} \label{ch:intro}

\subsection{Motivations}

Strongly interacting quantum many--body systems have been one of the main challenges of quantum physics and are still not well 
understood in many aspects; many novel intriguing phenomena may in fact be originated from 
the strong interactions among particles in these systems\cite{anderson}. A strongly interacting system can be described as a system where one can not define a 
small parameter on which a perturbative theory can be built. This complication inspired the development of  
numerical approaches based on the variational principle\cite{intro_refm5} and also quantum simulations\cite{intro_refm12,intro_refm14} 
that, in the case of bosonic systems, are in principle ``exact". 
In this work we have considered two systems that can be regarded as the archetype for neutral strongly interacting systems:
 $^4$He, which is a bosonic system, and its fermionic counterpart, $^3$He. More specifically, in this work we employed Quantum 
 Monte Carlo (QMC) techniques at zero and at finite temperature, respectively the Path Integral Ground State\cite{intro_refm12} (PIGS) and the 
 Path Integral Monte Carlo\cite{intro_refm14} (PIMC), to the study of a system of two dimensional $^3$He ($2d$-$^3$He) and 
 to the study of  $^4$He adsorbed on Graphene-Fluoride (GF, called also Fluorographene) and Graphane (GH), namely two corrugated substrates 
 that can be derived\cite{intro_ref5} from Graphene. Our main purpose in the case of adsorbed $^4$He was the research of new 
 physical phenomena, whereas in the case of 2$d$ $^3$He it was the application of novel methodologies\cite{intro_ref05} 
 for the study of static properties of Fermi systems and the extension of such methodologies for an {\it ab--initio} study of 
 the low energy excitations of a strongly interacting fermionic system.
 
  Apart from being both strongly interacting, the systems that we have considered are interesting also from a methodological point 
  of view, as they can be used to test the limits of the employed techniques. In the case of $2d$-$^3$He the main technical 
  difficulty relies in the well known {\it sign problem}\cite{intro_refm15pre}, which, on one side, poses a severe limit on the number of particles that can be 
  simulated by QMC and, on the other side, limits the study of imaginary--time dynamics to small values of imaginary--time. 
  For $^4$He on GF and GH the geometry of the confinement 
  gives rise to rare tunneling events that are relevant in both the static and dynamic properties of the system and must 
  thus be correctly described by the used QMC technique. 
  The relevance of these system is also increased by the fact that experiments are feasible on both systems, indeed 
  for $2d$-$^3$He there is already a number of experimental works in literature and comparison with experiments has been done wherever possible;
  our study of $^4$He on GF and GH instead is novel and up to now there is no experimental data with which we can compare our 
  predictions, however it has been shown in Ref.~\onlinecite{intro_ref5} that the substrates that were considered are available to the experimentalists and we hope that this work 
  will inspire some new experiments on this topic. 
  
In the remnant of this section we give a first introduction of the systems that have been considered.

  \subsubsection{Two dimensional $^3$He}
Two dimensional bulk $^3$He at zero temperature is a model well suited for the study of strongly correlated 
Fermi systems. This is because, as shown in Ref.~\onlinecite{intro_ref29}, the model is a good approximation for liquid 
$^3$He adsorbed on preplated graphite substrates. 
Indeed, much experimental work has been done on such systems, we mention heat capacity measurements in Ref.~\onlinecite{intro_ref06,intro_ref07} and more recently \onlinecite{intro_ref08}, 
the study of the thermodynamic behavior of the second layer of $^3$He has been done in Ref.~\onlinecite{intro_ref09}, the study of magnetic properties of liquid $^3$He films\cite{intro_ref010,intro_ref011} 
and the study of low energy excitations with neutron scattering experiments\cite{intro_ref012,intro_ref013}; another feature of such systems is also the possibility to realize small clusters 
with a controlled number of particles\cite{intro_ref014}; this is appealing because those systems can possibly be simulated with ``exact'' QMC techniques. 

Also from the theoretical side, 2$d$ $^3$He has been the subject of many works, we mention the thermodynamic study of 2$d$ Fermi liquid with and without external  magnetic
field\cite{intro_ref015,intro_ref016}, a many--body study of elementary excitations is reported in Ref.~\onlinecite{intro_ref017}, a QMC computation of the zero temperature equation state of pure 2$d$
$^3$He\cite{intro_ref018} and an estimation of its effective mass\cite{intro_ref019};

The experimental works in Ref.~\onlinecite{intro_ref29} revealed 
that quasi--two--dimensional $^3$He has a nearly perfect Fermi liquid behavior, in particular, they showed 
that the effective mass $m^*$ and the spin susceptibility $\chi/\chi_0$ increase with the density. 
This behavior, consistent with a divergence of $m^*$ near the freezing density, has been interpreted\cite{intro_ref01} as a signal of 
Mott transition to an insulating crystal. On the other hand, quasi--two--dimensional $^3$He has been studied by 
theoretical means\cite{intro_ref019} that suggested that the freezing and the divergence of $m^*$ may not have the same physical origin,
 in particular the freezing density is influenced by the preplated substrate. In this context, the study of the strictly 
 2$d$ $^3$He becomes valuable in order to isolate the effect of correlations on the system near freezing density. A further 
 advantage in the theoretical study of this system is that the properties of the liquid phase are largely independent on 
 the choice of the substrate and thus it is possible to make a comparison with experimental data\cite{intro_ref29}. 
 An even greater interest in 2$d$ $^3$He has been also inspired by the recent work in Ref.~\onlinecite{intro_ref017,intro_ref04b} in which, 
 {\it for the first time}, the collective zero--sound mode has been observed as a well defined excitation crossing and 
 possibly reemerging from the particle--hole continuum.  
 
 We have thus performed a Quantum Monte Carlo study of a two--dimensional bulk sample of $^3$He using the unbiased Fermionic 
 Correlations (FC) technique that has been successfully employed in the 2$d$ electron gas in Ref.~\onlinecite{intro_ref05}. 
 This technique is a formally exact method that makes use of bosonic imaginary--time correlation functions of operators 
 suitably chosen in order to extract fermionic energies. In this work 
 we computed the energy per particle as function of the polarization of the system at different fluid densities, from this 
 data we obtained a spin susceptibility that is in very good agreement with experiments. As a further study of the system, 
 we have extended the FC method to study dynamical properties; we computed an {\it ab--initio} low--energy 
 excitation spectrum of 2$d$ $^3$He obtaining a well defined zero--sound mode in remarkably good agreement with 
 Ref.~\onlinecite{intro_ref04b}.

  \subsubsection{$^4$He on Graphane and Graphene-Fluoride}
  Experiments on the adsorption of Helium on Graphite have been carried out in the seventies at 
  the University of Washington; those experiments revealed for the first time a behavior corresponding to a two--dimensional gas. Moreover, 
  the appearance of a peak in the specific heat of $^4$He near a critical temperature $T_c =3$ K showed evidence of a phase 
  transition from a high $T$ fluid to a low $T$ commensurate ($\sqrt 3 \times \sqrt 3$~ R30$^o$) phase, an ordered phase 
  in which the $^4$He atoms are localized on second--nearest neighbors hexagons. Following this discovery, a number of 
  experimental and theoretical works followed and now the Helium monolayer on Graphite is probably one of 
  the most studied adsorbed quantum systems.
  
  On the experimental side we mention specific heat 
  measurements in Ref.~\onlinecite{intro_ref13,intro_ref13b}, chemical potential measurements in Ref.~\onlinecite{intro_ref14} and 
  neutron scattering experiments in Ref.~\onlinecite{intro_ref14b}. The phase diagram of the first layer of $^4$He on Graphite 
  has been inspected in Ref.~\onlinecite{intro_ref15,intro_ref16,intro_ref17}. As for the second layer, we mention the 
  experimental work in Ref.~\onlinecite{intro_ref18}. Superfluid properties of Helium on Graphite were investigated in Ref.~\onlinecite{intro_ref19,intro_ref20,intro_ref21}.
  
  On the theoretical side we mention the work on the interaction potential of He on Graphite by Carlos 
  and Cole\cite{intro_ref22} and a study on the possible commensurate solid phases of the second layer presented in 
  Ref.~\onlinecite{intro_ref23,intro_ref23b}. There are also many simulations\cite{intro_ref24,intro_ref25,intro_ref26,intro_ref27,intro_ref27b,intro_ref27c} 
  on strictly 2$d$ $^4$He. As for Helium on Graphite, the role of corrugation has been studied with Path Integral in Ref.~\onlinecite{intro_ref28} 
  whereas the properties of the adsorbed layers have been studied with Monte Carlo simulations 
  in Ref.~\onlinecite{intro_ref29,intro_ref30,intro_ref31,intro_ref32} and more recently in Ref.~\onlinecite{intro_ref33}.
  There has also been works on Helium on Graphene, the phase diagram has been calculated in Ref.~\onlinecite{intro_ref34} 
  and superfluid properties in Ref.~\onlinecite{intro_ref35}. 
    
  The availability of Graphene and especially its derivatives like Graphane and Graphene-Fluoride makes possible the study 
  of new adsorbed systems. No special phenomenon is expected for Helium adsorbed on Graphene because the interaction is 
  geometrically similar to that on graphite, but in the case of GF and GH the adsorption potential is qualitatively different 
  from the case of Graphite and indeed we found a unique behavior of the adsorption system. 
  The difference of GF and GH from Graphite is due to their conformation; GF and GH are respectively Graphene sheets to which 
  are chemically bonded planes of either Fluorine or Hydrogen atoms; in the case of GH, for example, the substrate is made 
  of a Graphene sheet with Hydrogen atoms attached above and below the C atoms, in an alternating pattern. Such atomic 
  structure provides an Helium--substrate interaction potential which, compared with the Helium-Graphite potential, has 
  twice the number of adsorption minima located on an honeycomb lattice; compared with Graphite, the tunneling between the adsorption 
  sites of GF and GH is also enhanced along three spatial directions that cross saddle points of the potentials. These 
  properties of the GF(GH) adsorption potential, as shown in Sec.~\ref{sec:fluorographene}, not only confine Helium in a 
  multi--connected space but also destabilize the analogue of the $\sqrt 3 \times \sqrt 3$~ R30$^o$ on Graphite: we found that 
  the ground state at equilibrium density, for both GF and GH, is indeed a {\it modulated superfluid} that in GF has
  anisotropic rotons in the excitation spectrum. Also high coverages of $^4$He monolayer on GF and GH show novel properties 
  that have been described in Sec.~\ref{sec:fluorographene}; we found in fact a stable commensurate solid phase that is 
  the analogue of the theoretically predicted $4/7$ phase on Graphite, moreover we have preliminary evidence that this 
  solid phase possesses also a relevant superfluid fraction.

  \section{Implemented Methodologies}  
 Quantum Monte Carlo methods are largely employed in the study of strongly interacting quantum systems; the main reason for that 
  is because they can provide expectation values that can be in principle ``exact'' in the case of Bose systems. In the case of Fermi systems, 
  QMC methods are still an highly accurate tool. The word ``exact'' here means that the used approximations may be reduced 
 below the statistical error of the QMC method.
 To make a few examples of successful applications of QMC methods,
  we mention the quantitative evaluation\cite{intro_refm1} of the Bose--Einstein condensate fraction in liquid $^4$He at zero temperature, the 
  phase diagram of $^4$He adsorbed on Graphite\cite{intro_ref33} and, more recently, the low energy 
  excitation spectrum\cite{intro_refm3} of $^4$He at zero temperature and the computation of the normal--state equation of a Fermi ultra--cold gas at 
  unitary regime\cite{intro_refm4}. 
  
  The first QMC method that appeared was a variational technique named Variational Monte Carlo\cite{intro_refm5} (VMC). This technique expresses a zero temperature expectation 
  value on a given family of variational wave functions as a multi--dimensional integral and then compute the integral with the Metropolis algorithm\cite{intro_refm6}.
  Originally it was implemented with Jastrow wave functions\cite{intro_refm7}, but better classes of trial wave functions were introduced; it is worth to mention here 
  the Shadow Wave Functions (SWF) for Bosons\cite{intro_refm8} and for Fermions\cite{slek1}, that introduce many--bodies correlations in an implicit way and is able to describe a system in both the liquid and the solid phases, 
  without introducing explicitly any equilibrium lattice for the solid state. Beyond the variational level, the first introduced ``exact'' QMC technique was 
  the Diffusion Monte Carlo\cite{intro_refm9} (DMC) that solves the Schr\"odinger equation for the ground state of a many--body system taking advantage of its similarity with 
  the diffusion equation in imaginary time. Another exact technique valid at zero temperature that was developed soon after DMC is the Green's Function Monte Carlo\cite{intro_refm11} (GFMC); this method 
  exploits an integral formulation of the Schr\"odinger equation in order to express ground state quantum averages; on the same line, another very successful method is 
  the Path Integral Ground State\cite{intro_refm12} (PIGS) that expresses a ground state expectation value through Feynman's path integrals as a sufficiently long imaginary--time evolution 
  of a trial wave function; an improvement of DMC that had been introduced in the same years of PIGS is the Reptation Monte Carlo\cite{intro_refm10} (RMC).
  Like in the case of VMC, better trial wave functions have been constantly introduced in PIGS; one of the last advancements in zero temperature path integral simulations 
  on Bose systems is the Shadow Path Integral Ground State\cite{intro_refm13} (SPIGS) which makes use of SWF as trial wave functions. A very strong feature of PIGS and SPIGS is that they are formally similar 
  to the Path Integral Monte Carlo\cite{intro_refm14} (PIMC) method; PIMC, in fact, uses Feynman's path integrals in order to compute quantum thermal averages; apart from that, its remarkable formal similarity with PIGS 
  comes also from the similarity between the thermal density matrix and the quantum imaginary--time evolution operator. This feature has a practical value because the two methodologies can be implemented 
  within the same framework. 
  
  The mentioned ``exact'' methodologies, if applied to Fermi systems, suffer from the {\it sign problem}\cite{intro_refm15pre}. This problem occurs because the Fermi symmetry introduces a nodal surface in the ground state wave function 
  (or in the density matrix elements in the case of PIMC) that, as consequence, is no longer a probability density that can be sampled with Monte Carlo; the same problem is also present in Bosonic systems 
  if an excited state instead of the ground state is considered. There are workarounds but they result in a 
  signal to noise ratio that decreases exponentially with the number of particles; exact Fermi simulations, as well as the study of the excitations of Bosonic systems, are thus restricted to 
  system with small number of particles. Among the adaptations that allow the QMC computation on Fermi systems there is the Fixed Node\cite{intro_refm15} approximation (FN), a variational technique that approximates 
  the true nodal surface of the ground state with that of a trial wave function, and its evolution, the Released Node\cite{intro_refm16}, that has shown to be exact for small systems\cite{intro_refm16}; we also mention the 
  Restricted Path\cite{intro_refm18} method that extends PIMC to Fermi systems and a more recent evolution\cite{slek2} of the DMC method that gives exact results for small systems.  
  
  The mentioned techniques work in real coordinates space; another rather new and promising approach to the study of Fermi systems is the formulation of novel QMC techniques; we mention here the Auxiliary Fields Quantum Monte Carlo\cite{intro_refm19} (AFQMC) and the Bold Diagrammatic Monte Carlo\cite{intro_refm20} (BDMC).  
  
  In this work we have studied Fermi systems with SPIGS; for this purpose we have adopted another recently developed technique named Fermionic Correlations\cite{intro_ref05} (FC). FC can be defined as a ``cross--over'' technique because 
  its basic idea is to obtain informations on a Fermi system through the computation of an imaginary--time correlation function on a fictitious Bose system; with this approach, the sign problem is avoided and the simulation is in principle exact.
  However, to obtain the informations on the Fermi system from the imaginary--time correlation function one has to compute a numerical inversion of the Laplace transform in  
  ill--posed conditions, this is a difficult inverse problem that, again, results in severe limits on the number of particles that can be studied. If the number of particles is 
  small enough, however, the FC technique is an unbiased, {\it ab--initio} method that gives access to the energy (and possibly its derivatives) of strong interacting Fermi systems; 
  moreover, the FC technique has been extended in this work to study collective excitations of Fermi systems.

 \section{Thesis Outline}
 In this work we have made an ''unconventional" choice: instead of making a single chapter devoted 
 to the full theoretical introduction of the methodologies, we have introduced the essentials 
 in chapter \ref{ch:methods} and the technical details of the methodologies in the chapter {\it after} the 
 conclusions. With this choice, a reader that is not interested in technical details can safely ignore 
 anything written after the conclusions. 
 
 This document is organized as follows. 
 
 \begin{itemize}

 \item The present section provides a background on both the studied physical systems and the employed methodologies. 

\item In chapter \ref{ch:methods} we provide a basic description of the PIGS and PIMC techniques. In this chapter, 
a methodological work is also presented.
 We show that, on a realistic model potential for 
 $^4$He, the PIGS method does not suffer from any bias deriving from the choice of the trial wave function. 
 
 This work has been published on {\it J. Chem. Phys.}, {\bf 131}, 154108 (2009).
 
\item In chapter \ref{ch:polarization} is presented the study of 2$d$ $^3$He at zero temperature with the FC technique. 
The energy of the system for various densities and polarizations is reported as well as the resulting 
spin susceptibility as function of the density. 

The work includes also comparison with experimental data and Fixed Node simulations and has been published on {\it Phys. Rev. B}, {\bf 85}, 184401 (2012).

\item In chapter \ref{ch:dynamics} we adapted the FC technique to study the excitations of a Fermi system. The reader can find an {\it ab--initio} computation of the dynamic structure factor of 
2$d$ $^3$He at zero temperature compared with recent experimental data, the static response function and the approximate static structure factor.

These results are in preparation for submission to {\it Phys. Rev. B.}

\item In chapter \ref{sec:fluorographene} we present the study of Helium adsorbed on Graphene-Fluoride (GF) and Graphane (GH). The section will present one body properties, such as the ground state energy 
of one atom of $^3$He and $^4$He on GF and GH and the first energy band in the four cases; it will treat then many--body properties of the first layer of $^4$He, 
such as the stability of various commensurate phases, the equation of state at zero temperature, the condensate fraction in the liquid phases, the zero temperature low energy excitation spectrum at the equilibrium density 
and superfluid properties at both zero and finite temperature. We also present preliminary data on a possible 
supersolid phase present at high coverages on both GF and GH.

Many of these results have been published on:

{\it J. Phys.: Conference Series} {\bf 400}, 012010 (2012) - proceedings of the LT26 conference.

{\it J. Low. Temp. Phys.} - proceedings of the QFS2012 conference. DOI: 10.1007/s10909-012-0770-9

{\it Phys. Rev. B.} {\bf 86}, 174509 (2012).

\item In chapter \ref{ch:conclusions} we draw the conclusions of this work. 
 
\item in chapter \ref{ch:details} the computational details of the PIGS and PIMC methods are thoroughly described, from 
the mathematics of the Markov chain to the implementation of the Metropolis algorithm and the derivation of 
estimators that compute expectation values of various physical quantities.

 \end{itemize}

%% file: ch-methods/chapter-methods.tex
\def\onlinecite{\cite} 
\chapter{Path Integral Methods\label{ch:methods}}

In this Chapter the general basis of two Monte Carlo techniques will be described; the technical details will instead 
be discussed in Chapter \ref{ch:details}. 
The method used for zero temperature simulations is the 
 Path Integral Ground State\cite{pigs} (PIGS) whereas, that used for finite temperature simulations is 
 the Path Integral Monte Carlo\cite{m:pimc} (PIMC).
 The PIGS and the PIMC techniques are ``exact'' methods if the studied system has the Bose symmetry; 
 the word ``exact'' in the context of Quantum Monte Carlo (QMC) means that the systematic errors due to the used 
 approximations can be arbitrarily reduced below the Monte Carlo
 statistical uncertainty.
 The two techniques have also a similar 
 formalism. For this reason, they are easily implementable in a unified computer library.

 \section{Path Integral Ground State}\label{secpigs}
 In Sec. \ref{ch:details} we show that, using Monte Carlo techniques, it is indeed possible to sample an arbitrary 
 probability distribution and that with the resulting sampling it is possible to evaluate $N$--dimensional integrals. 
 We now specialize that methodology to the problem of calculating the expectation values of a bosonic $N$--particle system.
 
 Let's thus consider a system of $N$ atoms of mass $m$ at a temperature $T=0$ K, in a box of volume $V_b$ in 
 periodic boundaries conditions, with 
 an interatomic potential $V(r)$, the Hamiltonian operator is
 \begin{eqnarray}\label{hamiltonianop}
 \hat{H} = \hat{T} + \hat{V} 
 \end{eqnarray} 
 where the kinetic term is
 \begin{eqnarray}
 \hat{T} = -\frac{\hbar^{2}}{2m}\sum_{i=1}^{N}\nabla_{i}^{2}
 \end{eqnarray}
 and the potential term is
 \begin{eqnarray}
 \hat{V} = \sum_{i<j}v\left(\left|\vec{r}_{i}-\vec{r}_{j}\right|\right) 
 \end{eqnarray}  
 this Hamiltonian is used for ease of writing, but a more general Hamiltonian with anisotropic interactions and external 
 potential can be used as well. Defined $\Psi(R)$ as the ground state wave function, we want to compute the quantity 
\begin{eqnarray}\label{expavg1}
\langle\hat{O}\rangle=\int dR\:O(R)\Psi^2(R)
\end{eqnarray}
where $R=\left\lbrace\vec{r}_{i}\right\rbrace_{i=1}^{N}$ is a many--body variable  and $\vec{r}_i$ is the position of the 
$i$--th particle of the system and 
$\hat{O}$ is an operator that is diagonal in the coordinate representation.
 The square of the wave function, $\Psi^2(R)$, real and nodeless because we are considering a bosonic system, is 
 proportional to the 
the probability distribution to be sampled with the Metropolis algorithm (see Chapter \ref{ch:details}). The quantity $\Psi^2(R)$ is in 
general unknown but a workaround that has been very successful among $T$=0 K methods is to exploit the quantum evolution in 
imaginary time.

 Given an initial state $\left|\Psi\left(0\right)\right\rangle$, the quantum time--evolution is determined by the 
 Schr\"odinger's equation and
 \begin{eqnarray}\label{qevoop}
 \left|\Psi\left(t\right)\right\rangle = e^{-\frac{i}{\hbar}t\hat{H}}\left|\Psi\left(0\right)\right\rangle
 \end{eqnarray} 
 where the time evolution operator $\hat{U}(t)=e^{-\frac{i}{\hbar}t\hat{H}}$. If $\left|\Phi_i\right\rangle$ is an 
 eigenvector of $\hat{H}$, its overlap with the state
 $\left|\Psi(\tau)\right\rangle$ can be expressed as
 \begin{eqnarray}\label{qevo1}
 \left\langle\Phi_i|\Psi \left(\tau\right)\right\rangle = \sum_j \left\langle\Phi_i|e^{-\tau\hat{H}}|\Phi_j\right\rangle\left\langle\Phi_j|\Psi\left(0\right)\right\rangle 
 \end{eqnarray} 
where we have defined the quantum imaginary--time evolution operator $\hat{U}\left(\tau\right)=e^{-\tau\hat{H}}$ by 
substituting $\tau = \frac{i}{\hbar}t$. Eq. \eqref{qevo1} can be rewritten as
 $\left\langle\Phi_i|\Psi(\tau)\right\rangle = e^{-\tau E_i}\left\langle\Phi_i|\Psi\left(0\right)\right\rangle$. 
 For a sufficiently long $\tau$, if the initial state $\Psi\left(0\right)$ is not orthogonal to the ground 
 state, only the eigenstate corresponding to the lowest eigenvalue has a relevant overlap on the evolved trial wave 
 function $\left|\Psi(\tau)\right\rangle$. 
 The ground state wave function $\Psi_0$ in coordinate representation can be thus expressed 
 as the $\tau \rightarrow \infty$ limit of an imaginary time evolution of an arbitrary trial wave function
  $\Psi_T$ provided that $\left\langle\Psi_0|\Psi_T\right\rangle \neq  0$
  \begin{eqnarray}\label{conve}
  \Psi_0 = \lim_{\tau\rightarrow\infty} \frac{e^{-\tau\left(\hat{H}-E_0\right)}\Psi_T}{\left\langle\Psi_0|\Psi_T\right\rangle}
  \quad .
  \end{eqnarray}
The normalization factor is not involved in the Monte Carlo sampling; within the Green's function formalism, 
the ground state wave function can be approximated with $\tilde{\Psi}_{\tau}(R)$,

\begin{eqnarray} \label{pigs1}
\tilde{\Psi}_{\tau}(R) = \frac{1}{\mathcal{N}}\int dR'\:G\left(R,R',\tau\right)\Psi_T\left(R\right)
\end{eqnarray}

where $\mathcal{N}$ is the normalization constant and the term 
$G(R,R',\tau)=\langle R|e^{-\tau\hat{H}}|R'\rangle$ is the Green's function or density matrix. Here, the expectation 
value has just merely been rewritten in term of the Green's function, but the 
Green's function for a sufficiently large $\tau$ is still a generally unknown quantity. There are, however, known 
analytic approximations of the Green's function that are valid for small imaginary 
time $\delta\tau$ and the Path Integral formalism provides a way to express a large $\tau$ Green's function as a 
convolution of smaller imaginary--time Green's functions. 
This comes from an important property of the density matrix, 
 \begin{eqnarray}
 e^{-\tau\hat{H}} = \left(e^{-\delta\tau\hat{H}}\right)^M
 \end{eqnarray}
 where $\delta\tau = \frac{\tau}{M}$. In the coordinate representation, the product becomes a convolution
 \begin{eqnarray}\label{pintegral}
 G\left(R_1,R_{M+1},\tau\right) = \int ...\int\: dR_{2}...dR_{M}\prod_{j=1}^{M-1}G\left(R_{j},R_{j+1},\delta\tau\right)
\quad .
 \end{eqnarray}
 A density matrix at imaginary time $\tau$ can be represented as a convolution of $M$ density matrices at smaller 
 imaginary time $\tau/M$. This convolution is the Path Integral and, as the name 
 {\it Path Integral Ground State} may suggest, it is a fundamental element for
 the quantum simulation techniques that have been used throughout this work. 
 
Combining Eq. \eqref{expavg1} with \eqref{conve} and \eqref{pigs1}, a quantum average on the ground state thus becomes

\begin{eqnarray}\label{pigsunsimm}
 \left\langle\hat{O}\right\rangle = \frac{1}{\mathcal{N}}\int \left(\prod_{i=1}^{M}dR_{i}\right)\: \Psi_T(R_1)O\left(R_{M/2}\right)\times \nonumber \\
 \times\:\prod_{j=1}^{M-1}G\left(R_{j},R_{j+1},\delta\tau\right)\Psi_T(R_M)
 \quad .
 \end{eqnarray}
In the case that $\tau /2$ is sufficiently large to have a good approximation of Eq. \eqref{conve},
if the operator $\hat{O}$ commutes with $e^{-\tau\hat{H}}$, then it can be applied at any position $k$ of the 
path integral and it will give the ground state expectation value, if 
otherwise, $[\hat{O},\hat{H}] \ne 0$, then the operator applied at positions $k=1$ and $k=M$ will give mixed 
expectation values $\langle\Psi_T| \hat{O}| \Psi_0\rangle$ and 
for $k=2... M/2$ the expectation values $\langle\Psi_T|e^{-(k-1)\delta\tau\hat{H}}| \hat{O}| \Psi(0)\rangle$  will 
converge to the ground state value. 
  To go further and obtain an explicit definition of the quantum expectation value, an analytic approximation of the 
  small imaginary--time Green's function must be used. 
 The simplest one is the Primitive Approximation (PA); more advanced approximations are illustrated 
 in Appendix.~\ref{sechighorder}. The PA consists in neglecting the commutator between $\hat{T}$ and $\hat{V}$ when 
 factorizing the density matrix
 $e^{-\delta\tau\hat{H}}$, the error associated with this approximation is of the order $\delta\tau^2$.
 \begin{eqnarray}
 e^{-\delta\tau\hat{H}} \simeq e^{-\delta\tau\hat{T}}e^{-\delta\tau\hat{V}}
 \quad .
 \end{eqnarray}
 In this approximation the matrix elements of the two factors are easy to obtain,
 \begin{eqnarray}\label{papprox}
 \left\langle R_{i}\left|e^{-\delta\tau\hat{T}}\right|R_{i+1}\right\rangle = \frac{1}{\left(4\pi\lambda\delta\tau\right)^{\frac{dN}{2}}}e^{\left[-\frac{\left|R_{i}-R_{i+1}\right|^{2}}{4\lambda\tau}\right]} \\
  \left\langle R_{i}\left|e^{-\delta\tau\hat{V}}\right|R_{i+1}\right\rangle = e^{-\delta\tau V\left(R_{i}\right)}\delta\left(R_{i},R_{i+1}\right)
  \end{eqnarray}
  where we have defined $\lambda = \frac{\hbar^2}{2m}$, $\left|R_{i}-R_{i+1}\right|^2 = \sum_{j=1}^{N}\left|\vec{r}_{j}^{\:i}-\vec{r}_{j}^{\:i+1}\right|^2$ and $V\left(R_{m}\right) = \sum_{i<j}^{N}v\left(r_{ij}^{\:m}\right)$. We use the shorthand notation $r_{ij}^{\:m} =
  \left|\vec{r}_{i}^{\:m} - \vec{r}_{j}^{\:m}\right|$.
    
  The ground state expectation value of an operator $\hat{O}$ that is diagonal in the coordinate representation becomes
  \begin{eqnarray} \label{zeavg}
  \left\langle\hat{O}\right\rangle \simeq \frac{1}{\mathcal{N}}\int \prod_{i=1}^{M-1}dR_{i}\: \Psi_T(R_1)e^{-\frac{\delta\tau}{2}V\left(R_{i}\right)}
  e^{-\frac{\left(R_{i}-R_{i+1}\right)^{2}}{4\lambda\delta\tau}}e^{-\frac{\tau}{2}V\left(R_{i+1}\right)}O\left(R_{M/2}\right)\Psi_T(R_M)
  \end{eqnarray}
 where the primitive approximation has been written in a symmetric form, namely
 \begin{eqnarray} \label{pasimm}
 G_{PA}\left(R_{i},R_{i+1},\delta\tau\right) = e^{-\frac{\delta\tau}{2}V\left(R_{i}\right)}e^{-\frac{\left(R_{i}-R_{i+1}\right)^{2}}{4\lambda\delta\tau}}e^{-\frac{\delta\tau}{2}V\left(R_{i+1}\right)}
\quad .
 \end{eqnarray}
 In the limit $M\rightarrow\infty$, Eq. \eqref{zeavg} becomes exact due to the Trotter formula :
 \begin{eqnarray}
 e^{-\tau\hat{H}} = \lim_{M\rightarrow\infty}\left[e^{\left(-\delta\tau\hat{T}\right)}e^{\left(-\delta\tau\hat{V}\right)}\right]^{M}
 \quad .
 \end{eqnarray}
 Chosen a sufficiently high $M$, then, the error in Eq. \eqref{zeavg} can be arbitrarily reduced, and, chosen a 
 sufficiently high $\tau$, an arbitrary precise description of the ground state
  can be obtained.  
   The integral can be evaluated with Monte Carlo and the multi--dimensional probability distribution to sample with 
   the Metropolis algorithm is
 \begin{eqnarray}\label{probdist}
p\left(\lbrace R_n\rbrace\right) = \Psi_T(R_1)\prod_{j=1}^{M-1}G\left(R_{j},R_{j+1},\delta\tau\right)G\left(R_{M},R,\delta\tau\right)\Psi_T(R_M)
 \quad .
 \end{eqnarray}

 The value of $\tau$ that is sufficiently high to have convergence depends by a good degree on the choice of the trial 
 wave function $\Psi_T$. A trial wave function that 
  has an high overlap on the ground state could, in fact, enhance the convergence of Eq. \eqref{conve}. 
  We emphasize that this, however, is not necessary to obtain unbiased results: 
  a very strong feature of PIGS, as we have shown in Ref. \onlinecite{patatelesse},
  is that, indeed, the results of a PIGS calculation do not depend on the choice of the trial wave function.
   This is what we are going to show in Sec.~\ref{secpatatelesse}, 
  the practical implementation of the Metropolis algorithm will be instead described in Sec.~\ref{metromoves}.

 \subsection{Quantum--Classical Isomorphism} 
\begin{figure}[h]
 \begin{center}
 \includegraphics*[width=11cm]{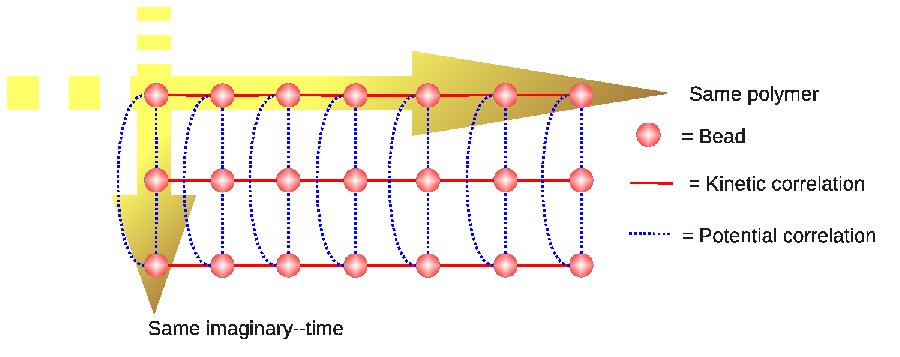}
 \caption{\label{figpol} 
 Schematic representation of the polymers in PIGS and their correlations.
 }
\end{center}
\end{figure}

Although path integrals and quantum evolution in imaginary time are very abstract topics, 
there is a simple interpretation of the probability distribution 
\eqref{probdist} that allows for an easy visualization of the Metropolis sampling; 
besides the practical advantages, it is also an interesting example in which 
many aspects of both the mathematics of Markov chains and the physics of the system take life in a fictitious 
classical system made of beads that have very special interactions between each other;
 this is why this correspondence is called {\it quantum--classical isomorphism}. More specifically, 
 Eq. \eqref{probdist} is the partition function of a {\it classical} system of $N$ {\it polymers} 
 composed by $M$ {\it beads} that have special interactions. The kinetic term of the Hamiltonian represents the 
 interaction between adjacent beads of the same polymer whereas the
   potential term of the Hamiltonian maps onto the interactions between beads of different polymers. 
   A polymer is essentially a set of beads corresponding to some integration variables in Eq. 
   \eqref{probdist}, namely the $i$--th polymer is 
   $\left\lbrace \vec{r}_{i}^{\:j}\right\rbrace_{j=1}^{M}$. The length of the polymer is $\tau$; due to the analogy 
   with the quantum evolution operator $e^{-i\tau\hat{H}}$,
   this length is an imaginary time. The index $j$ represents the position of the bead in the polymer, this position 
   corresponds to a discrete imaginary time $\tau_{j} = j\delta\tau = j\frac{\tau}{M}$. 
   The discretization of the polymer in the imaginary time is named {\it time--step}. The mean square displacement of 
   the beads in a polymer represents the indetermination of the position of the 
   corresponding particle in the quantum system. 
   A configuration of these special interacting polymers is thus defined by a set of coordinates 
   $\left\lbrace\vec{r}_{i}^{\:j}\right\rbrace$, where $i$ represents the polymer and $j$ represents the bead in the 
   $i$--th polymer; Figure \ref{figpol} shows a schematic representation of the polymers in PIGS and their correlations. 
   A quantum observable is mapped to an operator that acts on such configurations that in this context shall be 
   referred as {\it estimator}. 

   Here on, we will focus on Bose systems. In Sec.~\ref{metromoves} the practical implementation of the Metropolis algorithm 
   will be described; in that section, two different Monte Carlo algorithms used to sample the space of permutations will also be shown.

 \subsubsection{Path Integral Ground State in action}\label{secpatatelesse}
 In my work for the Master degree I developed a library that can run PIGS simulations as well as Path Integral 
 simulations at finite temperature (Path Integral Monte Carlo, PIMC, see Sec.~\ref{sec:pimc}). 
 The work presented in this section is one of the early employments of the library and it is a benchmark for both 
 the library itself and the PIGS technique. 
 A benchmark for the library because the library came through extensive testing during this work, 
 a benchmark for the PIGS technique because this work shows that PIGS is really unbiased, in other words,
 the choice of the trial wave function does not affect the final results, provided that the projection time 
 is large enough and the time--step $\delta\tau$ is sufficiently small. 
 This is a very strong feature of PIGS and is a necessary condition 
 for a truly {\it ab--initio} method because it allows the study of a quantum system even if we don't know 
 {\it anything} about its many--body ground state wave function.
 \paragraph{Test systems}\label{subsec:syst}
 
We have considered two bulk phases of a
many--body strongly interacting Boson system: liquid and solid $^4$He. 
Dealing with low temperature properties, $^4$He atoms are described as structureless zero--spin 
bosons, interacting through a realistic two--body potential, that we assume to be the HFDHE2 
Aziz potential~\cite{Aziz}; we remark here that our results are thus valid on this interaction potential 
and have not general validity.

For the liquid phase, we have considered a cubic box with periodic boundary conditions, 
containing $N=64$ atoms at the equilibrium density $\rho_l=0.0218$\AA$^{-3}$.
For the solid phase we have considered a cubic box with periodic boundary conditions designed to
house a fcc crystal of $N=32$ atoms at the density $\rho_s=0.0313$\AA$^{-3}$.
In both cases we add standard tail corrections to the potential energy to account for the finite 
size of the system by assuming the medium homogeneous (i.e. $g(r)=1$) beyond $L/2$, where $L$ is 
the size of the box.
Obviously, this is not an accurate assumption specially for the solid phase in such a small box, 
but our main purpose here is to show that PIGS method is able to reach the same results 
independently on the considered initial wave function.
Computations of ground state properties of bulk $^4$He with accurate tail corrections
can be found in the current literature.\cite{boro,moro}

\paragraph{Trial wave functions}
\label{subsec:psit}
The trial wave functions commonly used within the PIGS method\cite{pigs} are the  
variational Jastrow wave function (JWF) for the liquid and the Jastrow-Nosanow (J-NWF) for 
the solid.
A JWF represents the simplest possible choice of wave function for strongly interacting 
Bosons\cite{feen} and it contains only two--body correlations.
Using a McMillan pseudopotential\cite{mcmil}, the unnormalized JWF 
reads as
\begin{equation}
 \label{JWF}
 \psi_{\rm JWF}(R) = \prod_{i<j=1}^Ne^{-\frac{1}{2}\left(\frac{b}{r_{ij}}\right)^m}.
\end{equation}
The physical meaning of this JWF is that, due to the sharp repulsive part of the interaction
potential $V$ in the Hamiltonian $\hat H$, $^4$He atoms prefer to avoid each other.
In the J-NWF the JWF is multiplied by a term like the one in Eq. \eqref{GWF} below, that
localizes the particles in a crystalline order.
In this work, however, in order to explore the convergence properties of the PIGS method, we
have considered two wave functions of ``opposite'' quality: the best available one, that is the 
shadow wave function, and the poorest imaginable one, i.e. the constant wave function.
As we shall see, JWF will be considered only when computing the one--body density matrix in the liquid phase.

The constant wave function is the ground state wave function of the ideal Bose gas, 
\begin{equation}
 \label{CWF}
 \psi_{\rm CWF}(R)=1.
\end{equation}
It carries no correlation at all.
We choose this wave function because, allowing an unrestricted sampling of the full 
configurational space, it results in no importance sampling.
Then the whole imaginary time projection procedure is driven only by the short imaginary time 
Green's function $G(R,R',\delta\tau)$, without any input, and then any bias, from the initial state.
Thus at the starting point the system is made up by free particles; if after a long 
enough imaginary time projection, PIGS turns out to be able to reach a strong correlated 
quantum liquid and quantum crystal by itself we can safely believe that no variational bias 
affects PIGS results.

On the other hand, we choose as $\psi_T$ a SWF optimized with a variational computation in order 
to have as reference results the ones coming from the projection of an initial wave function
that is more accurate as possible, i.e. from a wave function whose overlap with the exact 
ground state is known to be large.
In the SWF, additional correlations besides the standard two body terms are introduced via 
auxiliary variables which are integrated out\cite{swf}.
This is done so efficiently that the crystalline phase emerges as a spontaneously broken 
symmetry process, induced by the inter--particles correlations as the density is increased, 
without the need of any a priori knowledge of the equilibrium positions and without losing the 
translationally invariant form of the wave function.
Thus SWF is able to describe both the liquid and the solid phase with the same functional form
and it is explicitly Bose symmetric.
The standard SWF functional form reads
\begin{equation}
 \label{SWF}
 \psi_{\rm SWF}(R) = \phi_r(R)\int dS\,K(R,S)\phi_s(S)
\end{equation}
where $S=(\vec s_1,\vec s_2,\dots,\vec s_N)$ is the set of auxiliary shadow variables, 
$\phi_r(R)$ is the standard Jastrow two body correlation term \eqref{JWF}, $K(R,S)$ is a kernel 
coupling each shadow to the corresponding real variable, and $\psi_s(S)$ is another Jastrow term 
describing the inter--shadow correlations. Due to its analytical expression, the introduction of 
the SWF defined by Eq.~\eqref{SWF} in a PIGS simulation consists in adding a timeslice at each 
extremity of every polymer. These newly added timeslices have special correlations; namely there 
are {\it{real-shadow}} intrapolymer correlations defined by $K(R,S)$ and {\it{shadow-shadow}} interpolymer 
correlations defined by $\phi_s(S)$. As consequence, the PIGS method has to be extended with
 Metropolis moves that accordingly involve the introduced shadow timeslices.

As usual\cite{mcfar}, we take $K(R,S)$ Gaussian and, as pseudopotential in $\phi_s(S)$, we use the  
He--He potential $V$ rescaled in both amplitude and distances.
The variational parameters we use were chosen in order to minimize the expectation value of the
Hamiltonian $\hat H$ and are reported in Ref.~\onlinecite{mcfar}.
Nowadays the SWF represents the best available variational wave function for $^4$He 
systems.\cite{moro} 
Recently, it has been estimated\cite{vita} that, when describing a two dimensional solid, the 
overlap of the SWF with the true ground state is of about $(0.998)^N$, which ensures a fast convergence 
rate when projected within the PIGS method.
The properties of the SWF are so peculiar that the PIGS method that has a SWF as $\psi_T$ 
deserves an its own name and is dubbed SPIGS: Shadow Path Integral Ground State 
method.\cite{spig1,spig2}

In order to test how robust PIGS is, we consider also a wave function that describes the wrong phase: for the liquid phase 
we consider a Gaussian wave function, where each particle is harmonically localized around  
fixed positions $\{\vec r_{0i}\}$
\begin{equation}
 \label{GWF}
 \psi_{\rm GWF}(R)=\prod_{i=1}^Ne^{-C|\vec r_i-\vec r_{0i}|^2},
\end{equation}
i.e. $\psi_T$ it the wave function of an Einstein harmonic solid.
The parameter $C=8$~\AA$^{-2}$ is arbitrary and it is was chosen to ensure a strong localization 
of the particles around the positions $\{\vec r_{0i}\}$ that were taken over a regular cubic lattice 
within the simulation box.
This wave function is evidently not translationally invariant and not Bose symmetric.
Furthermore it does not contain any correlation between the particles, and all the information 
that it carries is that of a crystalline system, i.e. GWF is an extremely poor wave function for 
the liquid phase.
This ``bad'' initial wave function will provide a stringent test on the convergence properties of
the PIGS methods.

As far as the one--body density matrix computation in the liquid phase is concerned, the values 
of the parameters $b$ and $m$ in the JWF have been chosen equal to the ones of the corresponding
Jastrow term in the SWF.

\paragraph{Small time Green's function}
\label{subsec:G}

One of the fundamental elements of path integral projection Monte Carlo methods is the imaginary
time Green's function $G(R,R',\tau)$, whose accuracy turns out to be crucial to the convergence to the exact 
results.
The functional form of $G$ for a generic $\tau$ is unfortunately not known with exception of few 
particular cases, such as, for example, the free particle and the harmonic oscillator, but 
accurate approximations of $G$ are obtainable in the small $\tau$ regime\cite{m:pimc,boni,sakko}.
In this work, we have chosen the Pair--Suzuki approximation\cite{pilat} for the imaginary time 
propagator, which is a pair--approximation of the fourth--order Suzuki--Chin density 
matrix.\cite{boni}

The Suzuki--Chin approximation is based on the following factorization of the density matrix:
\begin{equation}
 e^{-2\delta\tau\hat H}\simeq e^{-\frac{\delta\tau}{3}\hat V_e}
                       e^{-\delta\tau\hat T}
                       e^{-\frac{4\delta\tau}{3}\hat V_c}
                       e^{-\delta\tau\hat T}
                       e^{-\frac{\delta\tau}{3}\hat V_e}
\end{equation}
where $\hat T$ is the kinetic operator and $\hat V_e$ and $\hat V_c$ are given by
\begin{equation}
 \hat V_e=\hat V +\frac{\alpha\delta\tau^2\lambda}{3}\sum_{i=1}^N(\bf{F}_i)^2
\end{equation}
and
\begin{equation}
 \hat V_c=\hat V +\frac{(1-\alpha)\delta\tau^2\lambda}{6}\sum_{i=1}^N(\bf{F}_i)^2
\end{equation}
respectively, with $\hat V$ the potential operator, $\alpha$ an arbitrary constant in the 
range $[0,1]$, $\lambda=\hbar^2/2m$ and ${\bf F}_i={\bf \nabla}_i V$.
The resulting imaginary time propagator is accurate to order $\delta\tau^4$, and has been
successfully applied to liquid $^4$He in two and three dimensions.\cite{boni}
This approximation offers also the advantage that adjusting the parameter $\alpha$ it is possible
to optimize the convergence, and a standard choice for a quantum system is $\alpha=0$.\cite{boni}
A strategy to obtain a simpler, but equally accurate, approximation consists in applying a 
pair product assumption.\cite{pilat}
For sufficiently short time steps, in fact, the many--body propagator (in imaginary time) is well
approximated by the product of two--body propagators.\cite{m:pimc}
In this approximation, the small time propagator reads
\begin{equation}
 \label{pairS}
 \begin{split}
 G(R_m,R_{m+1};\delta\tau) =& \left(4\pi\lambda\delta\tau\right)^{-3N/2}\times \\
 &\prod_{i=1}^N \exp\left(-\frac{(\vec r_{i,m}-\vec r_{i,m+1})^2}{4\lambda\delta\tau}\right)\times \\
 &\exp\left(-u(r_{ij,m},r_{ij,m+1})\right)
 \end{split}
\end{equation}
where $u$ is given as
\begin{equation}
 \label{uPS}
 u(r_m,r_{m+1})=\left\{\begin{array}{ll}
 \frac{\delta\tau}{3}\left[v_e(r_{m})+2v_c(r_{m+1})\right] & m\quad {\rm odd}\\
 \\
 \frac{\delta\tau}{3}\left[2v_c(r_{m})+v_e(r_{m+1})\right] & m\quad {\rm even.}
 \end{array}\right.
\end{equation}
The potentials $v_e(r)$ and $v_c(r)$ are defined as
\begin{equation}
 \begin{split}
 & v_e(r) = V(r) + \alpha\frac{2}{3}\delta\tau^2\lambda\left(\frac{\partial V}{\partial r}\right)^2 \\
 & v_c(r) = V(r) + (1-\alpha)\frac{1}{3}\delta\tau^2\lambda\left(\frac{\partial V}{\partial r}\right)^2
 \end{split}
\end{equation}
where $V(r)$ is the potential experienced by two $^4$He atoms at a distance $r$.
The advantage is that there is no need to calculate $\bf{F}_i$.
As for the full Suzuki--Chin approximation,\cite{boni} also for the Pair--Suzuki the operators
corresponding to physical observables must be inserted only on odd time slices in the imaginary
time path.

\begin{figure}
 \includegraphics*[width=11cm]{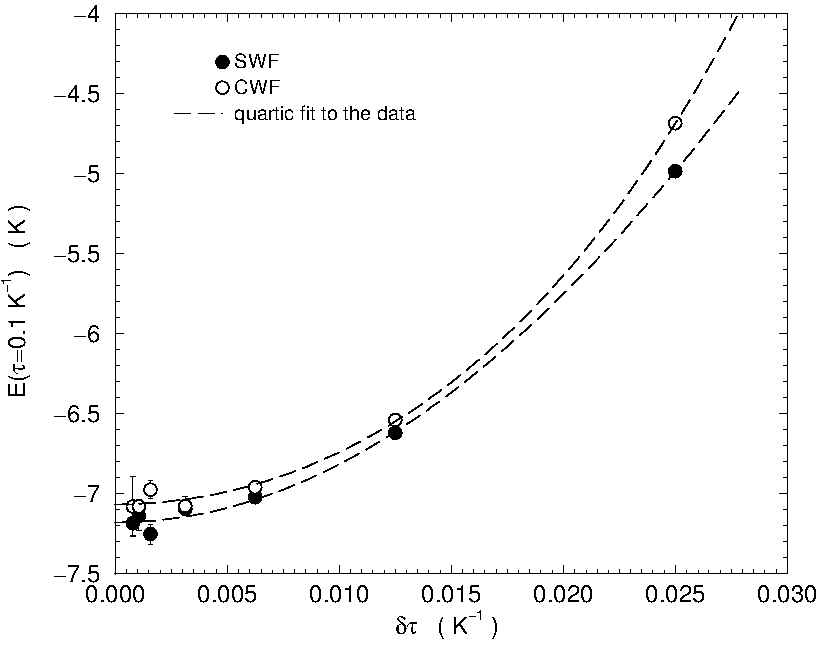}
 \caption{\label{f:dtau} Energy per $^4$He atom $E(\tau)$ vs. imaginary time step $\delta\tau$.
          The total projection time is $\tau=0.1$~K$^{-1}$.
          The calculations were carried out by projecting a SWF and a CWF for a system of 64
          particles at the equilibrium density $\rho=0.0218$~\AA$^{-3}$.
          Dashed lines are quartic fits to the data.
          Error bars, when not shown, are smaller than the used symbols.}
\end{figure}
\begin{figure}
 \includegraphics*[width=11cm]{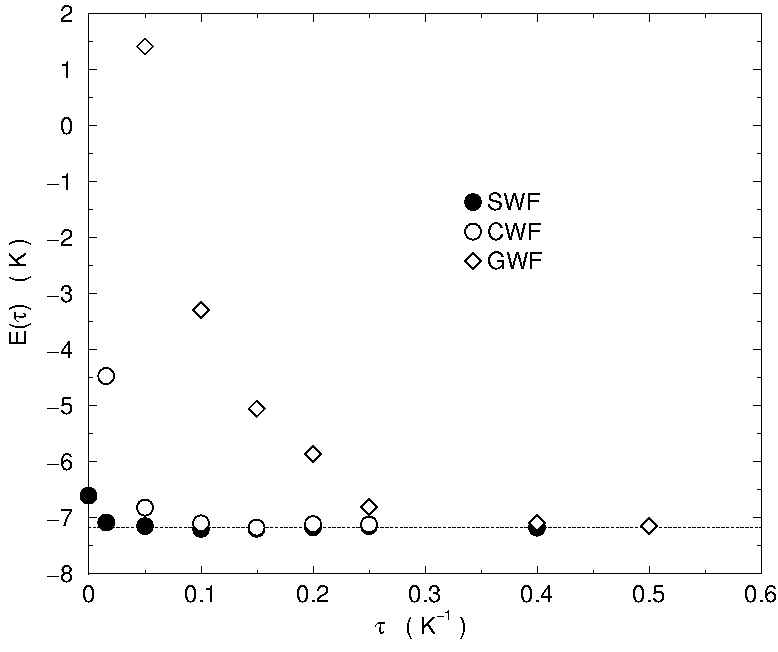}
 \caption{\label{f:enel} Energy per particle $E$ as a function of the total projection time 
          $\tau$ obtained from PIGS simulations for liquid $^4$He at the equilibrium density 
          $\rho=0.0218$~\AA$^{-3}$ by projecting a SWF (filled circles) and a CWF (open circles)
          and a GWF (open diamonds).
          $\tau=0$ result (filled circle) corresponds to the SWF variational estimate of $E$,
          the $\tau=0$ for the GWF is $E=122.08 \pm 0.06$ K and for CWF $E$ is essentially infinite.
          Error bars are smaller than the used symbols.
          Dotted line indicates the convergence value $E=-7.17\pm 0.02$ K.}
\end{figure}

In order to fix the optimal small imaginary time step value, we have performed PIGS simulations
with different initial wave functions.
By considering decreasing $\delta\tau$ values with a fixed total projection time, $\tau$, we 
have taken the energy per particle $E(\tau)$ as observable of reference.
As an example, our results for SWF and CWF in the liquid phase are plotted in Fig.~\ref{f:dtau}.
We choose as optimal value $\delta\tau=1/640$~K$^{-1}$; in fact, further reductions do not 
change the energy in a detectable way, i.e. within the statistical uncertainty.
In Fig.~\ref{f:dtau} SWF and CWF do not converge to the same value simply because the considered
total projection time $\tau$ in this test is not enough to ensure convergence of $E(\tau)$ to the 
ground state energy for CWF (see Fig.~\ref{f:enel}).
Similarly, in the solid phase we take $\delta\tau=1/960$~K$^{-1}$.

Once set the optimal $\delta\tau$ value, we have computed the diagonal properties of the system 
for increasing total projection time $\tau$ until we reached convergence to a value that 
corresponds to the exact ground state result both for the liquid and for the solid phase.
In the liquid phase we have computed also the one--body density matrix.

\paragraph{PIGS results without importance sampling}

For the liquid phase we have projected a SWF and a CWF.
The energy per particle as a function of the total projection time $\tau$ for both the wave 
functions is plotted in Fig.~\ref{f:enel}.
We find that the energy converges, independently from the considered initial wave function, to
the same value $E=-7.17\pm 0.02$~K.
This value, in spite of the small size of the considered system, is close to the 
experimental\cite{roach} result $E=-7.14$~K.
SWF converges very quickly, in fact $\tau=0.05$~K$^{-1}$ is already enough to ensure convergence.
CWF instead, requires a three times larger imaginary time, i.e. $\tau=0.15$~K$^{-1}$.
Nevertheless, the quick convergence of also CWF is a really remarkable result.
In fact, this means that PIGS efficiently includes the exact interparticle correlations through
the imaginary time projections, without any need of importance sampling.
Then, the choice of a good wave function, within the PIGS method, becomes a matter of 
convenience rather than of principle, since better initial wave functions only allow for a 
smaller total projection time $\tau$, and thus less CPU consuming simulations.

\begin{figure}
 \includegraphics*[width=11cm]{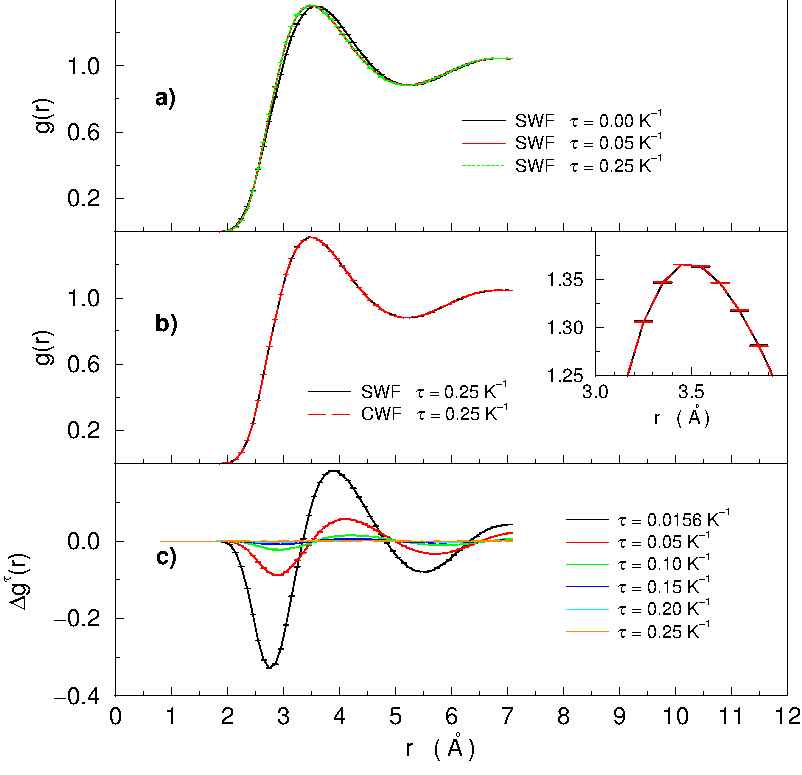}
 \caption{\label{f:grl} Radial distribution function $g(r)$ for bulk liquid $^4$He computed in 
          a cubic box with $N=64$ at the density $\rho=0.0218$~\AA$^{-3}$ with the PIGS method.
          a) $g(r)$ obtained by projecting a SWF for $\tau=0.00$, $0.05$ and $0.25$~K$^{-1}$.
          The $\tau=0.00$ result corresponds to the variational SWF estimate of $g(r)$.
          b) $g(r)$ obtained by projecting a SWF for $\tau=0.25$~K$^{-1}$ and a CWF for
          $\tau = 0.25$~K$^{-1}$.
          In the inset a zoom of the first maximum region.
          c) $\Delta g^\tau(r)=g_{\rm SWF}^\tau(r)-g_{\rm CWF}^\tau(r)$ at different $\tau$ 
          values, where $g_{\rm SWF}^\tau(r)$ is the $g(r)$ computed by projecting a SWF for an 
          imaginary time equal to $\tau$, and $g_{\rm CWF}^\tau(r)$ is the same but by 
          projecting a CWF.
          Note the smaller scale on the vertical axis}
\end{figure}
\begin{figure}
 \includegraphics*[width=11cm]{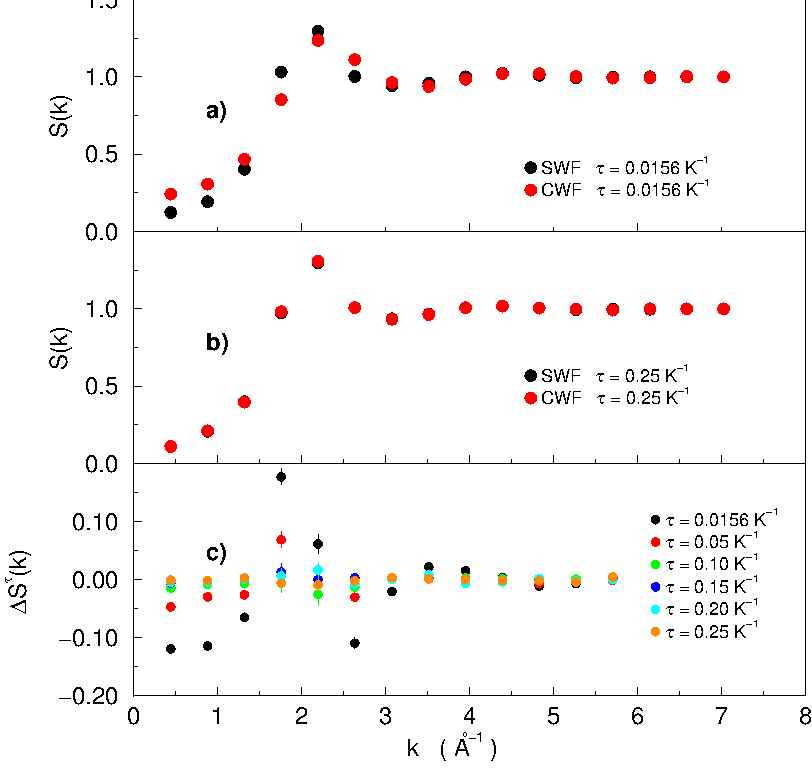}
 \caption{\label{f:skl} Static structure factor $S(k)$ for bulk liquid $^4$He computed in a
          cubic box with $N=64$ at the density $\rho=0.0218$~\AA$^{-3}$ with the PIGS method.
          a) $S(k)$ obtained by projecting a SWF and a CWF for $\tau=0.05$~K$^{-1}$. 
          b) $S(k)$ obtained by projecting a SWF and a CWF for $\tau=0.40$~K$^{-1}$.
          c) $\Delta S^\tau(k)=S_{\rm SWF}^\tau(k)-S_{\rm CWF}^\tau(k)$ at different $\tau$ values,
          where $S_{\rm SWF}^\tau(k)$ is the $S(k)$ computed by projecting a SWF for an
          imaginary time equal to $\tau$, and $S_{\rm CWF}^\tau(k)$ is the same but by
          projecting a CWF.
          Note the smaller scale on the vertical axis.}
\end{figure}

This convergence is confirmed also by the radial distribution function $g(r)$ and the static
structure factor $S(k)$.
For such quantities, the convergence rate is found to be similar to the energy one.
In Fig.~\ref{f:grl} we report the radial distribution function $g(r)$ obtained by projecting 
both a SWF and a CWF at different imaginary time values.
For $\tau>0.05$~K$^{-1}$, SWF results at different $\tau$ are indistinguishable within 
the statistical uncertainty (see Fig.~\ref{f:grl}a).
In fact, with SWF the exact result is reached within very few projection steps and then it is 
no more affected by further projections.
As already pointed out, also CWF displays a fast convergence, as shown in Fig.~\ref{f:grl}c, 
where $\Delta g^\tau(r)=g_{\rm SWF}^\tau(r)-g_{\rm CWF}^\tau(r)$ is shown.
For increasing $\tau$, $\Delta g^\tau$ evolves toward a flat function, meaning that the systems 
described starting from the two different wave functions, i.e the strongly correlated quantum 
liquid of SWF and the ideal gas of CWF, are evolving into the same quantum liquid, which is the
best reachable representation of the exact ground state of the simulated system.
The same conclusion is inferred from the evolution of the static structure factor $S(k)$, 
which is plotted in Fig.~\ref{f:skl}.

\paragraph{PIGS results from a ``bad'' initial function}

In order to put a more stringent check on the PIGS method ability to converge to
the exact ground state without any variational bias, we have considered also a ``bad'' initial
wave function by projecting a GWF.
Thus at the starting point of the imaginary time path there is now a strongly localized 
Einstein crystal.
We note that, differently from the other considered cases, the GWF is {\it not} Bose symmetric; 
as consequence of this choice, the projection relation \eqref{pigs1} is not Bose symmetric and 
thus requires symmetrization. The symmetrization has been introduced with the sampling 
of the permutations between polymers; this is a standard technique used in Path Integral simulations 
and will be described in Sec.~\ref{metromoves}. 

With this initial function, we find even in this case that the energy converges to the same value as before (see 
Fig.~\ref{f:enel}).
Thus PIGS is able not only to drop from the initial wave function the wrong information of
localization, but also to generate at the same time the correct correlations among the particles.
GWF needs $\tau=0.5$~K$^{-1}$ to converge, which is ten times larger than the SWF value.

\begin{figure}
 \includegraphics*[width=11cm]{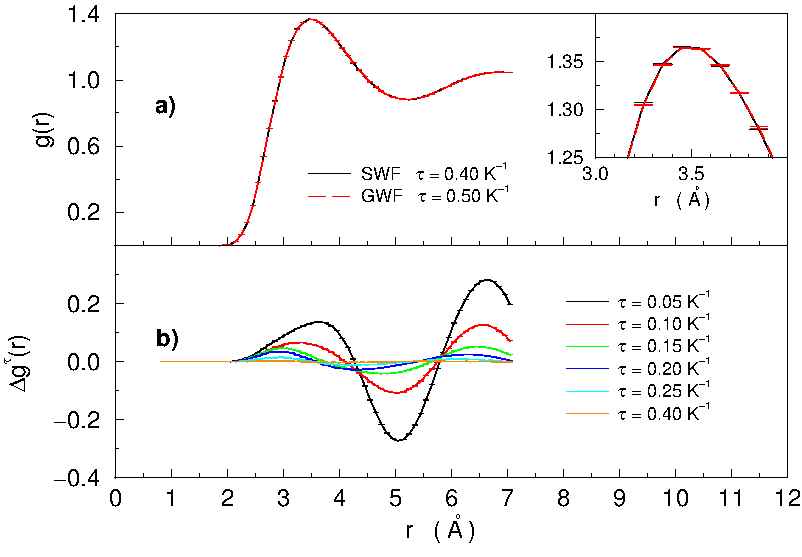}
 \caption{\label{f:grl2} Radial distribution function $g(r)$ for bulk liquid $^4$He computed in 
          a cubic box with $N=64$ at the density $\rho=0.0218$~\AA$^{-3}$ with the PIGS method.
          a) $g(r)$ obtained by projecting a SWF for $\tau=0.40$~K$^{-1}$ and a GWF for
          $\tau = 0.50$~K$^{-1}$.
          In the inset a zoom of the first maximum region.
          b) $\Delta g^\tau(r)=g_{\rm SWF}^\tau(r)-g_{\rm GWF}^\tau(r)$ at different $\tau$ values,
          where $g_{\rm SWF}^\tau(r)$ is the $g(r)$ computed by projecting a SWF for an 
          imaginary time equal to $\tau$, and $g_{\rm GWF}^\tau(r)$ is the same but by 
          projecting a GWF.
          Note the smaller scale on the vertical axis.}
\end{figure}
\begin{figure}
 \includegraphics*[width=11cm]{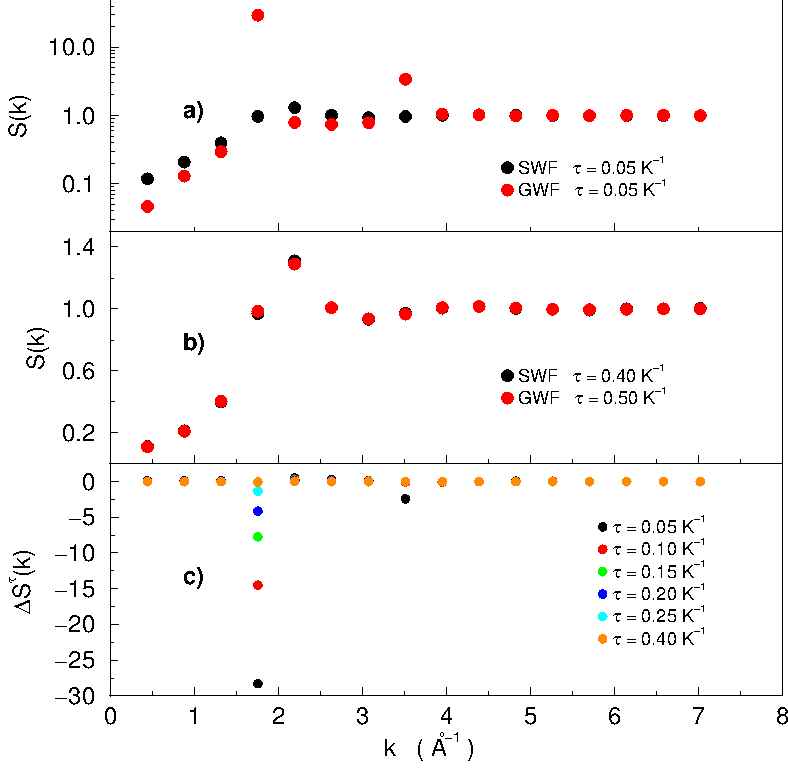}
 \caption{\label{f:skl2} Static structure factor $S(k)$ for bulk liquid $^4$He computed in a
          cubic box with $N=64$ at the density $\rho=0.0218$~\AA$^{-3}$ with the PIGS method.
          a) $S(k)$ obtained by projecting a SWF and a GWF for $\tau=0.05$~K$^{-1}$. 
          It is evident in the GWF result the presence of the Bragg peak.
          Note the logarithmic scale.
          b) $S(k)$ obtained by projecting a SWF for $\tau=0.40$~K$^{-1}$ and a GWF for 
          $\tau=0.50$~K$^{-1}$.
          The Bragg peak is no more present in the GWF result.
          c) $\Delta S^\tau(k)=S_{\rm SWF}^\tau(k)-S_{\rm GWF}^\tau(k)$ at different $\tau$ values,
          where $S_{\rm SWF}^\tau(k)$ is the $S(k)$ computed by projecting a SWF for an
          imaginary time equal to $\tau$, and $S_{\rm GWF}^\tau(k)$ is the same but by
          projecting a GWF. 
          Note the change of the vertical scale.
          Error bars are smaller than the used symbols.}
\end{figure}

Again this convergence is confirmed also by the radial distribution function $g(r)$ and the static
structure factor $S(k)$.
In Fig.~\ref{f:grl2} we report the radial distribution function $g(r)$ obtained by projecting 
a GWF at different imaginary time values compared with the ones coming from the projection of SWF.
It is evident that small imaginary time is not enough to leave out the wrong information 
in the GWF.
For lower $\tau$ values, there are still reminiscences of the starting harmonic solid, which 
are progressively lost as the projection time increases.
This is made clearer in Fig.~\ref{f:grl2}b where we plot the difference $\Delta g^\tau(r)$, 
at fixed imaginary time $\tau$, between the $g(r)$ computed by projecting the SWF and the one 
obtained by projecting the GWF.
A similar behavior is observed in the evolution static structure factor $S(k)$, plotted in 
Fig.~\ref{f:skl2}.
For the GWF, the Bragg peak shown at small $\tau$ values (Fig.~\ref{f:skl2}a), which is typical 
of the solid phase, becomes lower and lower as the projection time is increased 
(Fig.~\ref{f:skl2}b), until convergence is reached (see Fig.~\ref{f:skl2}c).

From the plot of the energy per particle vs. the total imaginary time $\tau$ it is possible to 
estimate the overlap per particle of the initial wave function on the exact ground state.\cite{mora}
By using the results in Fig.~\ref{f:enel} we find that the overlap of SWF is about 99\%, while
the GWF one is about 10\%.
That SWF has an high overlap with the ground state is not a surprise; it was qualitatively 
expected since SWF is presently the best available wave function for $^4$He.\cite{moro}
However a 99\% overlap is really remarkable and provides a further argument on the goodness of
SWF.
On the other hand, a poor overlap of GWF was somehow expected, since the parameter $C$ was 
chosen to strongly localize the atoms of the bulk liquid around fictitious equilibrium 
positions on a regular lattice.

\paragraph{Off-diagonal properties}

\begin{figure}
 \includegraphics*[width=11cm]{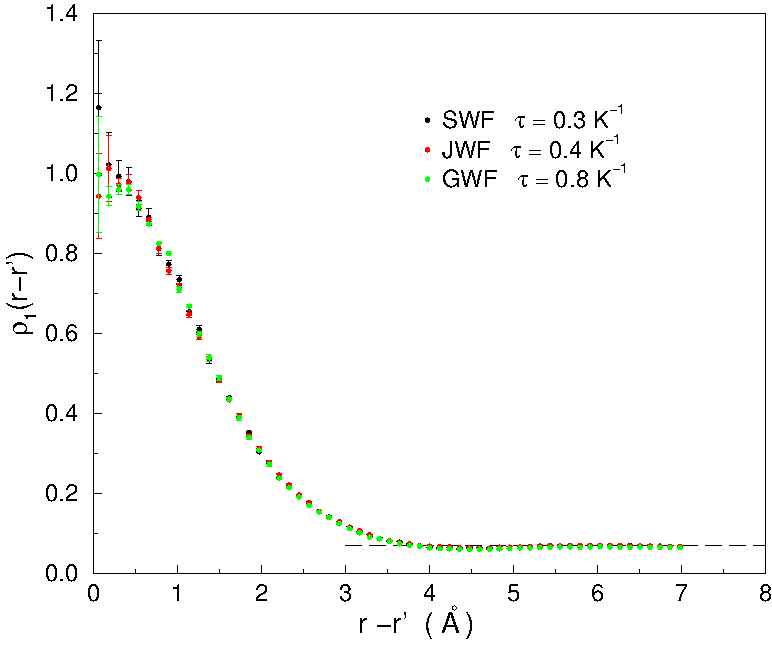}
 \caption{\label{f:odlro} One--body density matrix $\rho_1$ obtained from PIGS simulations for 
          liquid $^4$He at the equilibrium density $\rho=0.0218$~\AA$^{-3}$ by projecting a SWF,
          a JWF and a GWF for an imaginary time $\tau=0.30$, $0.40$ and $0.80$~K$^{-1}$ 
          respectively.
          The dotted line indicates the condensate value $n_0=0.069$ obtained from an independent
          PIGS simulation.\cite{moro2}}
\end{figure}
Besides the diagonal ones, also off--diagonal properties, such as the one--body density matrix,
are accessible within PIGS simulations.
The one-body density matrix $\rho_1(\vec r,\vec r')$ represents the probability amplitude of
destroying a particle in $\vec r$ and creating one in $\vec r'$.
Its Fourier transformation represents the momentum distribution.
In first quantization $\rho_1$ is given by the overlap between the normalized many-body ground
state wave functions $\psi_0(R)$ and $\psi_0(R')$, where the configuration
$R'=(\vec r',\vec r_2,\dots,\vec r_N)$ differs from  $R=(\vec r,\vec r_2,\dots,\vec r_N)$ only
by the position of one of the $N$ atoms in the system.
If $\psi_0(R)$ is translationally invariant, $\rho_1$ only depends on the difference
$|\vec r-\vec r'|$, thus
\begin{equation}
 \label{eq:obdm}
  \rho_1(\vec r-\vec r')=N\int d\vec r_2\dots d\vec r_N\,\psi_0^*(R)\psi_0(R').
\end{equation}
The Bose-Einstein condensate fraction $n_0$ is equal to the large distance limit of
$\rho_1(\vec r-\vec r')$.
In fact, if $\rho_1$ has a nonzero plateau at large distance, the so called
off-diagonal long-range order (ODLRO), its FT contains a Dirac delta function,
which indicates a macroscopic occupation of a single momentum state, i.e. Bose--Einstein
condensation.

The exact $\rho_1$ can be obtained in PIGS simulation by substituting $\psi_0$ in
\eqref{eq:obdm} with $\psi_\tau$ with $\tau$ large enough.
This corresponds to the simulation of a system of $N-1$ linear polymers plus a polymer which is
cut into two halves, called half--polymers, one departing from $\vec r$ and the other from
$\vec r'$.
Thus $\rho_1$ is obtained by collecting the relative distances among the cut ends of the two 
half--polymers during the Monte Carlo sampling.
The present computation of $\rho_1$ has been obtained by implementing a zero temperature version
of the worm algorithm.\cite{worm}
We have worked with a fixed number of particles and not in the grand canonical ensemble, similarly
to what has been done at finite temperature in Ref.~\onlinecite{pilat}.
In practice this corresponds to a usual PIGS calculation of $\rho_1$ where ``open'' and ``close''
moves have been implemented\cite{worm} in order to visit diagonal and off-diagonal sectors within
the same simulation. 
The advantage of doing this does not come from the efficiency of the worm algorithm to explore 
off-diagonal configurations, because similar efficiency is obtained with PIGS when ``swap'' moves 
are implemented.\cite{vita} 
The benefit in using a worm-like algorithm here instead comes from the automatic normalization of 
$\rho_1$ which is a peculiarity of this method.\cite{worm}
In Fig.~\ref{f:odlro} we report $\rho_1$ obtained in PIGS simulations of bulk liquid $^4$He
at $\rho=0.0218$~\AA$^{-3}$ by projecting either a SWF, a JWF and a GWF.
All the simulations give the same result, shown in Fig.~\ref{f:odlro} which turns out to be
compatible with the recent estimate obtained with PIGS given in Ref.~\onlinecite{moro2} of 
$n_0=0.069 \pm 0.005$.

\paragraph{Results on the solid system}
\label{sub:sol}

\begin{figure}
 \includegraphics*[width=11cm]{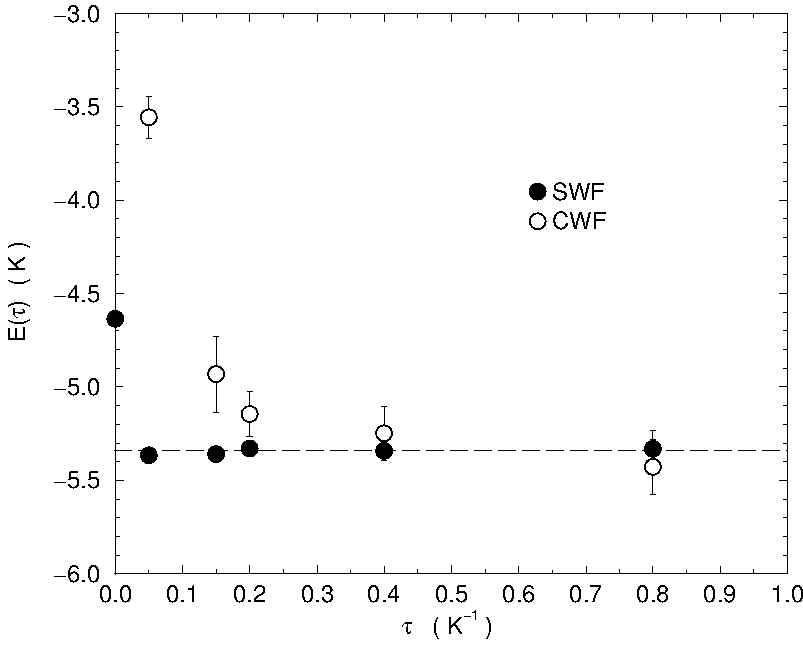}
 \caption{\label{f:enes} Energy per particle $E$ as a function of the total projection time
          $\tau$ obtained from PIGS simulations of an fcc $^4$He crystal at the density
          $\rho=0.0313$~\AA$^{-3}$ by projecting a SWF (filled circles) and a CWF (open circles).
          Dashed line indicates the convergence value $E=-5.34\pm 0.02$~K.}
\end{figure}

\begin{figure}
 \includegraphics*[width=11cm]{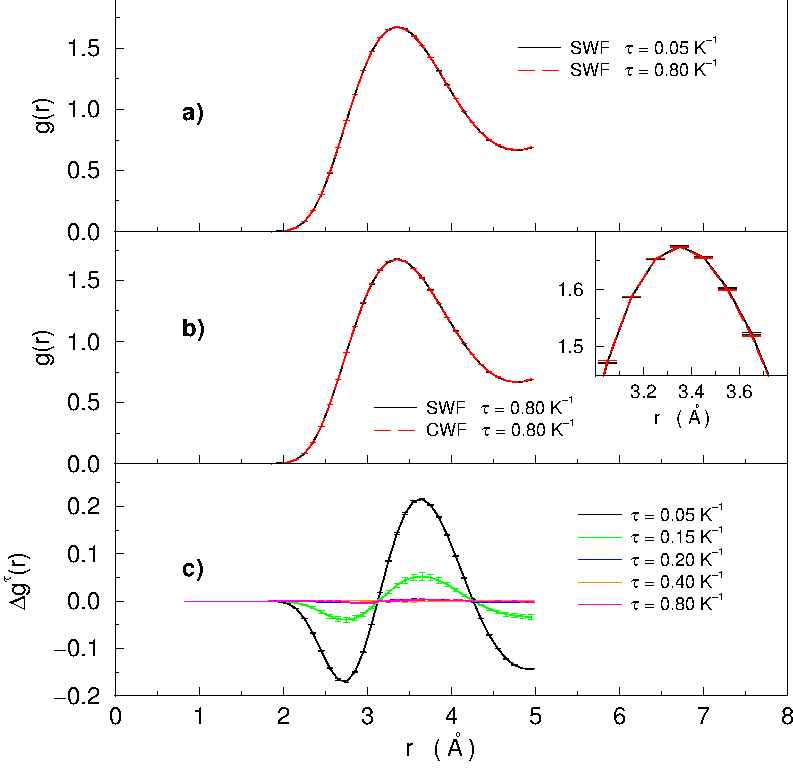}
 \caption{\label{f:grs} Radial distribution function $g(r)$ for bulk solid $^4$He computed in a
          cubic box with $N=32$ at the density $\rho=0.0313$~\AA$^{-3}$ with the PIGS method.
          a) $g(r)$ obtained by projecting a SWF for $\tau=0.05$ and $0.80$~K$^{-1}$.
          b) $g(r)$ obtained by projecting a SWF and a CWF for $\tau=0.80$~K$^{-1}$.
          In the inset a zoom of the first maximum region.
          c) $\Delta g^\tau(r)=g_{\rm SWF}^\tau(r)-g_{\rm CWF}^\tau(r)$ at different $\tau$ values,
          where $g_{\rm SWF}^\tau(r)$ is the $g(r)$ computed by projecting a SWF for an imaginary
          time equal to $\tau$, and $g_{\rm CWF}^\tau(r)$ is the same but by projecting a CWF.}
\end{figure}

\begin{figure}
 \includegraphics*[width=11cm]{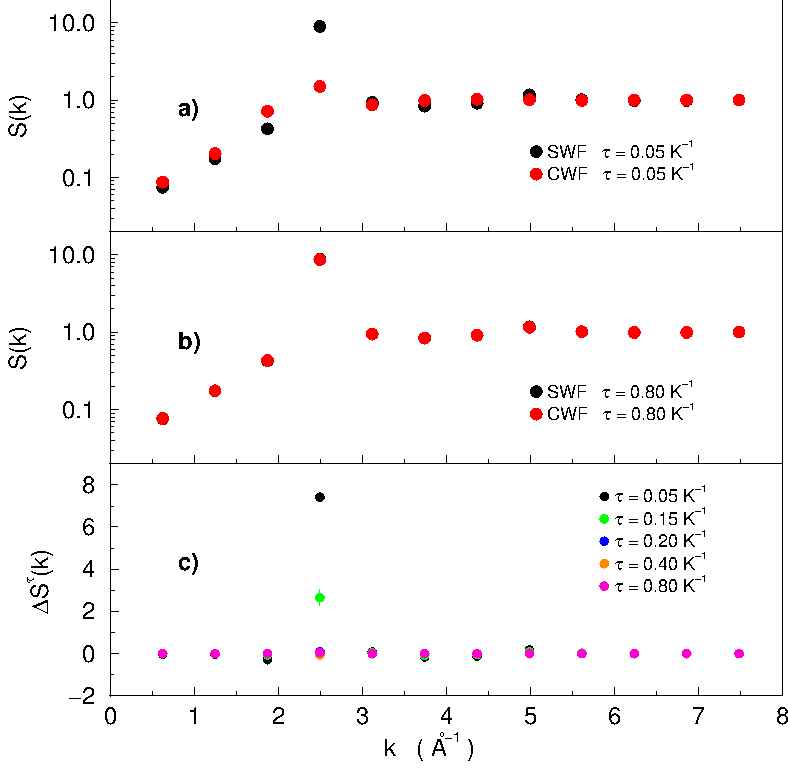}
 \caption{\label{f:sks} Static structure factor $S(k)$ for bulk solid $^4$He computed in a
          cubic box with $N=32$ at the density $\rho=0.0313$~\AA$^{-3}$ with the PIGS method.
          a) $S(k)$ obtained by projecting a SWF and a CWF for $\tau=0.05$~K$^{-1}$.
          b) $S(k)$ obtained by projecting a SWF and a CWF for $\tau=0.80$~K$^{-1}$. The black
             dots are under the red ones.
          c) $\Delta S^\tau(k)=S_{\rm SWF}^\tau(k)-S_{\rm CWF}^\tau(k)$ at different $\tau$ values,
          where $S_{\rm SWF}^\tau(k)$ is the $S(k)$ computed by projecting a SWF for an imaginary
          time equal to $\tau$, and $S_{\rm CWF}^\tau(k)$ is the same but by projecting a CWF.
          Error bars are smaller than the used symbols.
          Notice the logarithmic scale in panels a) and b).}
\end{figure}

We have performed the computation at density $\rho=0.0313$~\AA$^{-3}$, where $^4$He is in the 
solid phase, by projecting a SWF and a CWF.
Our results for the energy per particle are plotted in Fig.~\ref{f:enes} as a function of $\tau$.
In both cases we find convergence to the value $E=-5.34\pm 0.02$~K. 
Even in this phase the convergence of SWF is faster, being $\tau=0.05$~K$^{-1}$ enough to reach 
convergence.
In the case of CWF convergence is reached only for a much larger imaginary time 
$\tau=0.80$~K$^{-1}$.

Also in this case convergence is obtained for the radial distribution function and for the 
static structure factor, reported in Fig.~\ref{f:grs} and Fig.~\ref{f:sks} respectively.
From Fig.~\ref{f:grs}a it is evident that SWF has reached the true ground state with few projection 
steps, since the results for $g(r)$ at $\tau=0.05$~K$^{-1}$ and $\tau=0.80$~K$^{-1}$ are 
indistinguishable.
The evolution toward the correct ground state of the projected CWF is instead detectable.
The presence of the crystalline structure is mainly evident in the static structure factor, 
where a Bragg peak grows with increasing $\tau$ (see Fig.~\ref{f:sks}a,b).
The emerging of the correct solid structure by projecting a really poor wave function such as the 
CWF is made evident by the trend toward a flat function of the differences $\Delta g^\tau(r)$ and
$\Delta S^\tau(k)$ plotted in Fig.~\ref{f:grs}c and Fig.~\ref{f:sks}c respectively.

\paragraph{Conclusions of the test phase} 
 \label{sec:conc}

In this section we have studied with the Path Integral Ground State method
diagonal and off-diagonal properties of a strongly interacting
quantum Bose system like the bulk liquid and solid phases of $^4$He.
We have obtained convergence to the ground state values of quantities like the total energy,
the radial distribution function, the static structure factor and the one-body density matrix
projecting radically different wave functions: equivalent expectation values in the liquid phase 
have been obtained using as initial wave function a shadow wave function, a Gaussian wave function 
with strongly localized particles of an Einstein solid without interparticle correlations and
also a constant wave function where all configurations of the particles are equally probable.
Similarly in the solid phase equivalent expectation values have been obtained by considering a 
shadow wave function, which describes a solid, and a constant wave function which describes
an ideal Bose gas.
The present analysis demonstrates the absence of any variational bias in PIGS; a method that
can be thus considered as unbiased as the finite temperature PIMC.
This remarkable property comes from the accurate imaginary time propagators, exactly the same
used with PIMC, that do not depend on the initial trial state.
It remains true that the use of a good variational initial wave function greatly improves the
rate of convergence to the exact results.

We have addressed here only the case of a realistic interaction potential among Helium atoms. 
However one can reasonably expect that this conclusion holds even for very different kinds of 
interaction, once an accurate approximation for the imaginary time propagator is known (for 
example hard-spheres\cite{hard} or hydrogen plasma\cite{pier}).
As far as pathological potentials like the attractive Coulomb one are concerned, PIGS would suffer
the same limitations of PIMC if inaccurate approximations of the propagator were used.\cite{coul}

\section{Path Integral at finite temperature}\label{sec:pimc}
  Up to now we have focused on the problem of evaluating $T=0$ K expectation values; the formalism of the previous 
  section, however, can be used with very small modifications also for quantum thermal averages, 
  the resulting methodology is named Path Integral Monte Carlo (PIMC). This methodology was developed well before 
  PIGS in the work in Ref.~\onlinecite{pimc_history}.   
The physical properties of the system are obtained from the thermal density matrix
\begin{eqnarray}\label{densop}
\hat{\rho} = \frac{e^{-\beta\hat{H}}}{\mathbf{Z}}
\end{eqnarray}
where $\beta = \frac{1}{k_{b}T}$, $k_b$ is the Boltzmann constant and the 
normalization $\mathcal{Z}=\mathbf{Tr}\left(\hat{\rho}\right)$ is the partition function of the system.
The expectation value of an observable $\hat{O}$ is
\begin{eqnarray} \label{thavg}
<\hat{O}> = \frac{\mathbf{Tr}\left(\hat{O}\hat{\rho}\right)}{\mathcal{Z}}
\quad .
\end{eqnarray}
It is evident in Eq. \eqref{densop} that the unnormalized density matrix operator is formally identical to the 
quantum imaginary time operator appearing in Eq. \eqref{qevoop} if one chooses $\tau = \beta$. 
The density matrix $\hat{\rho}$ in coordinate representation becomes
\begin{eqnarray} \label{densrho}
 G\left(R,R',\beta\right) = \left\langle R\left|e^{\left({-\beta\hat{H}}\right)}\right|R'\right\rangle
 \quad .
 \end{eqnarray}
 Fixed this set of basis, $|R\rangle\langle R| = \int dR$, the trace operator acts on the density matrix as follows
 \begin{eqnarray} \label{expv}
 \mathbf{Tr}\left(\hat{\rho}\right) = \int dR\: G\left(R,R,\beta\right) = \mathcal{Z} \\
 \langle\hat{O}\rangle = \frac{1}{\mathcal{Z}}\int dR\: O\left(R\right)G\left(R,R,\beta\right)
 \end{eqnarray}
 where $\hat{O}$ is diagonal in the coordinate representation.

Using the Path Integral notation and following the same procedure employed for PIGS, Eq. \eqref{thavg} becomes
\begin{eqnarray}\label{thavg2}
 \left\langle\hat{O}\right\rangle = \frac{1}{\mathcal{Z}}\int \prod_{i=1}^{M}dR_{i}\: O\left(R_{k}\right)\prod_{j=1}^{M-1}G\left(R_{j},R_{j+1},\delta\tau\right)G\left(R_{M},R_{1},\delta\tau\right)
 \end{eqnarray}
where here $\delta\tau = \beta /M$; the cyclic property of the trace operation allows $O(R)$ to be evaluated at 
any position in the path integral, 
 in other words $1\le k\le M$. 
 
 Eq.~\eqref{probdist}, however, does not posses the Bose symmetry; in order to introduce
    either the Bose or the Fermi statistics one has to symmetrize the density matrix \eqref{thavg2} over the permutations of the particle labels.
    \begin{eqnarray}
    G_{B}\left(R,R',\beta\right) = \frac{1}{N!}\sum_{P}G\left(R',\hat{P}R,\beta\right) \label{permopB} \\
    G_{F}\left(R,R',\beta\right) = \frac{1}{N!}\sum_{P}\left(-\right)^{n_P}G\left(R',\hat{P}R,\beta\right) \label{permopF}
    \end{eqnarray}
    the first equation holds for the Bose statistics and the second for the Fermi statistics. 
    The permutation operator $\hat{P}$ acts on the many--body coordinate $R=\left\lbrace
    \vec{r}_{i}\right\rbrace_{i=1}^{N}$
     by applying a cycle of $n_P$ exchanges between particle indices, 
     $\hat{P}R=\left\lbrace \vec{r}_{P\left(i\right)} \right\rbrace_{i=1}^{N}$. The sum in the two equations is meant 
     as a sum over all the $N!$ possible permutations. 
      
      The Fermi symmetrization introduces negative density matrices that no longer can be directly interpreted as 
      probability densities; there are different techniques\cite{c1r4} that can re-express 
      the density matrix as a definite positive object with a weight on the sampled configurations that introduces the 
      sign given by the Fermi symmetry, however these techniques yield a signal to noise 
      ratio\cite{c1r1} that dramatically decreases as $e^{-\gamma N}$; as consequence of this {\it sign problem}, 
      these methodologies are restricted to small particle numbers.
   
 Due to its affinity with Eq. \eqref{pigsunsimm}, $\mathcal{Z}$ 
 is also the partition function of a system of classical {\it closed} polymers. 

  \begin{figure}[h]
 \begin{center}
 \includegraphics*[width=11cm]{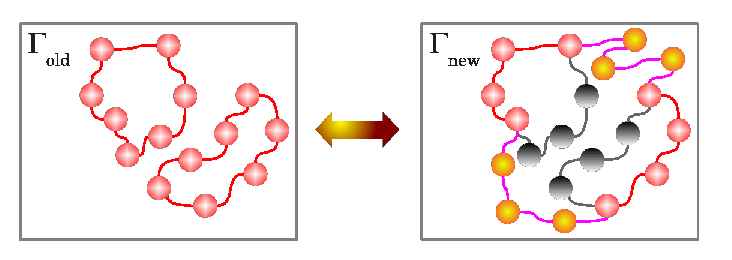}
 \caption{\label{permk} 
Representation of the polymers in PIMC and the effect of a permutation; a permutation between two polymers 
results in a new compound polymer with twice the length in imaginary time. This is a configuration that can not be sampled 
with moves that involve only a single polymer. Grey beads and lines represent the removed segment of the polymer.
This picture can also be viewed from right to left: in this case, grey beads and lines represent the new segment of the 
polymer; this permutation splits a ring polymer of length $2\beta$ into two ring polymers of length $\beta$. 
 }
\end{center}
\end{figure}

 The differences between PIGS and PIMC in ``polymer language" are minimal; first,
 in PIMC does not appear any trial wave function and the quantum imaginary--time is proportional to $\beta= 1/(k_BT)$; 
 a more subtle difference that has deep consequences, however, is the different topology of the polymers.
Due to the trace operation, in fact, the polymers in PIMC are represented by Eq.~\eqref{expv}; these polymers 
close on themselves  (for this reason they are also called {\it ring polymers}, see Fig.~\ref{permk}) and the role of 
permutations becomes more important 
than in the PIGS case. This is so because Eq.~\eqref{expv} is not yet Bose--symmetrized and  
  the effect of the symmetrization ~\eqref{permopB} is that the 
  polymers representing the particles of the system no longer close on themselves but, instead, 
  are allowed to close on another ring polymer; this is explicitly shown in Fig.~\ref{permk}. 
  The relevance of the symmetrization in PIMC is also noted by the 
  fact that permutations explore topologically different configurations of the system that 
  could not be obtained with single polymer sampling. 
  This situation is different from the PIGS case; in PIGS, in fact,    
permutations between polymers will simply yield another configuration of 
open polymers that is topologically identical and can be obtained with single polymer moves; this implies that 
the sampling of permutations is not necessary if the Bose symmetry is already introduced by the 
chosen trial wave function.  Another difference is that open polymers have less constraints on their structure and, 
compared to ring polymers, can be moved more efficiently by the Metropolis sampling; this is especially true when the 
probability distribution to sample has many local maxima: in this case, with open polymers is generally easier to satisfy 
the ergodicity of the sampling.

%% file: ch-polarization/chapter-polarization.tex
\def\onlinecite{\cite}

\chapter{Polarization energy of two--dimensional $^3$He}\label{ch:polarization}

In this chapter, the subject under study is a system of two--dimensional $^3$He  (2$d$ $^3$He) 
in a wide range of densities in the liquid region and up to freezing. 
This subject is of interest because, as it has been shown in Ref.~\onlinecite{3hep:whitlock}, 
a pure 2$d$ $^3$He system is a good approximation of a quasi--two--dimensional $^3$He sample. Such system can be experimentally 
realized over a wide range of liquid densities by adsorbing $^3$He on a variety of 
preplated graphite substrates\cite{3hep:lusher,3hep:morhard,3hep:casey}.
Regarding the effective mass $m^*$ and the spin susceptibility $\chi/\chi_0$,
the system behaves in good approximation like a perfect Fermi liquid\cite{3hep:whitlock}: the 
enhancement of $\chi/\chi_0$ increases with the density and $m^*$ is consistent with a divergence near 
freezing density. This behavior at freezing has been interpreted\cite{3hep:casey} as a signature of a Mott transition 
leading to an insulating crystal. Theoretical studies\cite{3hep:krotscheck_mstar}, however, suggest that the 
singularity of $m^{\star}$ and freezing could not have the same origin and the freezing density is influenced by 
the preplated substrate. In this context, it is relevant the study of the effect of correlations without 
any effect induced by the external potential of the substrate. 

Bulk 2$d$ $^3$He is interesting also from the theoretical point of view because, being a strongly interacting system at high densities,
it provides a severe test case for microscopic calculations\cite{3hep:krotscheck}. Some of the most powerful tools to study 
strongly interacting systems are QMC methods. The so--called fixed--node (FN) approximation\cite{3hep:reynolds}, used in most 
Fermionic QMC calculations, however, has been argued to give a significant bias in the polarization energy of three--dimensional 
liquid $^3$He\cite{3hep:holzmann} at high density. We have thus performed QMC simulations beyond the FN 
level, following a formally exact method\cite{3hep:carleo} that is referred here as {\it Fermionic Correlations} (FC). 
With both the FN and FC methods we have calculated the ground--state energy per particle $e=\frac{E}{N}$ 
of the 2d $^3$He liquid at zero temperature as a function of the number density $\rho$ and the spin polarization $\zeta$.

The FC method is slightly different from the well known transient estimate (TE) technique\cite{3hep:kalos}, 
the basic idea is to perform simulations relying on the basic Hamiltonian 
in an enlarged, unphysical space of states of any symmetry, including those
with Fermi and Bose statistics. The ground state 
energy of the physical fermionic $^3$He is considered as an excitation energy 
of the absolute bosonic ground state, which is sampled exactly with QMC.
In this approach one `trades' (in a sense which we will explain below) 
the sign problem faced by TE\cite{3hep:kalos} for 
the analytic continuation needed to extract excitation energies 
from suitable imaginary--time correlation functions. A mixed approach, 
devised to ease
detection of the asymptotic convergence of TE by a Bayesian analysis of 
imaginary--time correlation functions, was proposed By Caffarel 
and Ceperley\cite{3hep:Bagf}.

We compared our results also with a previous FN QMC calculation\cite{3hep:boronat} that 
however is limited to low densities and only considers the paramagnetic fluid phase.
We find indeed that the FN level of the theory and the exact calculation
predict a qualitatively different behavior. This is rather expected because 
the accuracy of the FN approximation in the
high density regime is questionable\cite{3hep:holzmann}.
In fact, within FN the system becomes
ferromagnetic well before crystallization takes place upon increasing
the density, whereas the unbiased calculation shows that the
spin polarization of the fluid is preempted by freezing, as observed
experimentally. From the estimated curve $e(\zeta)$ we obtain a
spin susceptibility enhancement in quantitative agreement with
the available measurements.

\subsection{QMC simulation}
We simulate $N$ particles with the mass $m_3$ of $^3$He atoms, interacting with
the HFDHE2 pair potential\cite{3hep:Aziz79} in periodic boundary conditions, which is the most 
accurate two--body potential for Helium systems\cite{3hep:Over}.
The simulation box, of area $\Omega$, is a square of side $L$ for the liquid phase;
for the solid it is a rectangle which accommodates a triangular lattice.
The Hamiltonian is

\begin{equation}
\label{hamiltonian}
\hat{H} = - \frac{\hbar^2}{2m_{3}}\sum_{i=1}^{N} \nabla_{i}^2
+ \sum_{i<j=1}^{N}v\left(\vec{r}_i - \vec{r}_j\right)
\end{equation} 

The simulations of the bosonic $^3$He were made with the SPIGS technique, using the Pair Product 
approximation for the propagator at short imaginary times. For this system, we have verified that, using a Shadow Wave Function, 
a projection time of $\tau= 0.2$ K$^{-1}$ is enough to yield an accurate description of the ground state. As for the 
time--step, we used $\delta\tau=1/160$ K$^{-1}$ and verified that it is small enough to be an accurate approximation.




\subsubsection{Fixed--Node approach}
The fixed--node approximation\cite{3hep:reynolds} is one of the most commonly used approaches in the 
QMC simulation of Fermi systems. FN stochastically solves 
the imaginary--time Schr\"odinger equation  subject to the boundary conditions 
implied by the nodal structure of a given trial function $\Psi_T$. This approach 
gives a rigorous upper bound to the ground state energy, which often turns out
to be extremely accurate.

For a given spin polarization, i.e. considering $N_{\uparrow}$ spin--up and 
$N_{\downarrow} = N - N_{\uparrow}$ spin--down $^3$He atoms, $\Psi_T$
is chosen of the form
\begin{equation}
\Psi_T(\mathcal{R})= \mathcal{D}
(\mathcal{R}_{\uparrow})
\mathcal{D}
(\mathcal{R}_{\downarrow})
\Psi_J(\mathcal{R})\chi_{\zeta}
\end{equation}
where $\mathcal{R} \equiv (\vec{r}_1,...,\vec{r}_N)$,
$\mathcal{R}_{\uparrow} \equiv (\vec{r}_1,...,\vec{r}_{N_{\uparrow}})$,
$\mathcal{R}_{\downarrow} \equiv (\vec{r}_{N_{\uparrow}+1},...,\vec{r}_{N})$,
and the whole dependence on the spin degrees of freedom is contained
in $\chi_{\zeta}$,  a spin eigenfunction for the given polarization
\begin{equation}
\zeta = \frac{N_{\uparrow} - N_{\downarrow}}{N}\quad,
\end{equation}

The Jastrow factor,
\begin{equation}
\Psi_J(\mathcal{R})=\prod_{i<j}\exp\left(-\frac{1}{2}u\left(|\vec{r}_i-\vec{r}_j|\right)\right),
\end{equation}
describes pair correlations arising from the interaction potential;
we use a simple pseudopotential of the McMillan form $u(r)=(b/r)^5$. 
Finally, the simplest form of the antisymmetric factors $\mathcal{D}\left(\mathcal{R}_{\uparrow,\downarrow}\right)$
is in the form of Slater Determinants of plane waves:

\begin{equation}
\label{planewaves}
\mathcal{D}\left(\mathcal{R}_{\uparrow,\downarrow}\right)
= \det \left( \left\{\exp(i\vec{k}_i \cdot \vec{r}_j)\right\}_{i,j}\right)
\end{equation}
More accuracy in the FN results is achieved by introducing also backflow 
correlations\cite{3hep:Backflow} via quasi--particles vector positions:

\begin{eqnarray}
\label{backflow}
& \mathcal{D}\left(\mathcal{R}_{\uparrow,\downarrow}\right) = \det \left(
\left\{\exp(i\vec{k}_i \cdot \vec{x}_j)\right\}_{i,j}\right) \\ \nonumber
& \vec{x}_j \buildrel{def} \over {=} \vec{r}_j + \sum_{i\neq j = 1}^{N} \eta(|\vec{r}_j - \vec{r}_i|)
\left(\vec{r}_j - \vec{r}_i\right).
\end{eqnarray}
For the backflow correlation function $\eta(r)$ we adopt the simple form:

\begin{equation}\label{backeta}
\eta(r)=A e^{-B(r-C)^2} \quad .
\end{equation}
We will refer to the two choices respectively as plane waves fixed--node (PW--FN)
and backflow fixed--node (BF--FN).
For each density, the variational parameters $b$, $A$, $B$ and $C$ are optimized
using correlated sampling\cite{3hep:rewate} at $\zeta=0$, and left unchanged at different
polarizations. The backflow parameters, for each density, are shown in Table~\ref{tabbf}

\begin{table}[h]
\begin{center}
\caption{\label{tabbf} Backflow parameters used for each density}
\begin{tabular}{| c | c | c | c |}
\hline
  Density ( \AA$^{-2}$ & A & B ( \AA$^{-1}$ ) & C ( \AA ) \\
\hline
0.020 & 1.90393 & 0.117865 & -1.89877 \\
0.050 & 1.124523 & 0.112559 & -0.94888 \\
0.060 & 1.017654 & 0.147372 & -0.51614 \\
\hline
\end{tabular}
\end{center}
\end{table}

Part of the bias related to the finite size of the simulated system arises 
from shell effects in the filling of single--particle orbitals\cite{3hep:Lin}. 
This bias can be substantially reduced 
adopting twisted boundary conditions\cite{3hep:Lin}, i.e. choosing $\vec{k}$ 
appearing in \eqref{planewaves} and \eqref{backflow} inside the set:

\begin{equation}
\label{wavevectors}
\vec{k}_{\vec{n}} = \frac{2\pi\vec{n} + \vec{\theta}}{L}
\end{equation}
where $\vec{n}$ is an integer vector, $L$ is the side of the simulation box 
$\Omega$ and $\vec{\theta}$ is a {\it twist parameter} $\theta_i \in [0,\pi]$
which, at the end of the calculations, is averaged over. 

In the solid phase, quantum exchanges are strongly suppressed and the
energy difference between a Fermionic and a Bosonic crystal is
negligibly small for the purpose of locating the liquid--solid
transition. We will therefore replace the energy of $^3$He with that of
a fictitious bosonic Helium of mass $m_3$, which can be calculated exactly\cite{3hep:Reptation,3hep:pigs,3hep:spigs}.
The small error incurred by such replacement is bound by
the difference between the fermionic Fixed--Node (FN) energy and the unbiased bosonic energy.
This difference, calculated\cite{3hep:nosanow} as a check at the melting density where it is 
expected to be largest, is indeed in the sub--milliKelvin range. 

 We stress that we actually made a particular choice of trial wave functions; the obtained results 
depend on such a choice: when we will speak about `fixed--node level' of the theory or about `fixed--node approximation'
 we will always implicitly refer to the above mentioned trial wave functions. Naturally it could be possible to improve 
 the fixed--node results using more sophisticated wave functions; instead, we have chosen to follow another way with 
 the FC method; this method is in principle exact and does not depend on a particular choice of the wave function.

\subsubsection{Fermionic correlations approach}

As mentioned before, for the fluid phases the FN approximation may not be accurate enough,
particularly at high density where correlations are stronger and the energy
balance between different polarization states is more delicate.

In order to go beyond the FN level and obtain accurate data, we use 
the FC technique\cite{3hep:carleo} which is in principle 
exact, even if limited to small system sizes.

The idea, with similarities with the transient estimate formalism\cite{3hep:kalos,3hep:Bagf}, is that 
of viewing \eqref{hamiltonian} as an
operator acting inside the Hilbert space $\mathcal{H}(N) \equiv \left(L^2(\Omega)\right)^{\otimes N}$,
that has no constrains on spin and statistics: one can use Quantum Monte Carlo 
to sample the lowest energy eigenfunction $\psi_0(\mathcal{R})$ of $\hat{H}$
among the states of both Bose and Fermi symmetry.

It is known \cite{3hep:Ettore2} that $\psi_0$ must share the {\it {Bose symmetry}} of the Hamiltonian, so that:

\begin{equation}
\label{ebose}
E_0^B \equiv \frac{\langle \psi_0 | \hat{H} \psi_0 \rangle_{\mathcal{H}(N)}}{\langle \psi_0 | \psi_0 \rangle_{\mathcal{H}(N)}}
\end{equation}
is the Ground State energy of a fictitious system of $N$ Bosons of mass 
$m_{3}$ interacting via the potential $v(r)$.

The connection between the fermionic energies is retrieved in the following way: 
let us fix a spin polarization, it is surely a good quantum number since the basic Hamiltonian is spin--independent. 
As discussed in Ref.~\onlinecite{3hep:carleo}, if we are able to define an operator 
$\hat{\mathcal{A}}_{F}$ such that, inside $\mathcal{H}(N)$,

\begin{equation}
\psi_F\left(\mathcal{R}\right) = \left(\hat{\mathcal{A}}_{F}\psi_0\right)\left(\mathcal{R}\right) 
\end{equation}
has {\it non--zero overlap} with the configurational part of any {\it exact
fermionic} Ground State of $\hat{H}$ for the given $\zeta$, we can
define the {\it imaginary--time correlation function}:

\begin{equation}
\label{cfun}
\mathcal{C}_{F}(\tau)
 \equiv \frac{\langle \psi_0 | \left(e^{\tau \hat{H}}\hat{\mathcal{A}}_{F}^{\dagger}e^{-\tau \hat{H}}\right)
\hat{\mathcal{A}}_{F} \psi_0 \rangle_{\mathcal{H}(N)}}{\langle \psi_0 | \psi_0 \rangle_{\mathcal{H}(N)}}, \quad \tau \geq 0
\end{equation}
which can be straightforwardly evaluated in a bosonic QMC simulation\cite{3hep:Reptation,3hep:spigs,3hep:Nava}.
This is readly made because the evaluation of Eq.~\eqref{cfun} at a certain discrete imaginary time $\tau_l=l\delta\tau$ 
can be done with the evaluation of the Slater determinant $\mathcal{A}_F$ on two time sectors located at different 
imaginary times $\tau_0=m\delta\tau$ and $\tau_1=(m+l)\delta\tau$.
 The actual calculation has been done using the Path Integral Ground State with Shadow Wave Functions (SPIGS) 
 that has been described in chapter \ref{ch:methods}. 
The lowest energy contribution in $\mathcal{C}_{F}(\tau)$ provides the {\it exact gap} 
between the fermionic and the bosonic ground states of the two--dimensional Fermi 
liquid; this can be readily seen by formally expressing \eqref{cfun} on the basis 
$\{\psi_n\}_{n \geq 0}$ of eigenvectors of $\hat{H}$ corresponding 
to the eigenvalues $\{E_n\}_{n \geq 0}$:

\begin{equation}
\label{cfun2}
\mathcal{C}_{F}(\tau)
= \sum_{n=0}^{+\infty}
e^{-\tau \left(E_n - E_0^B\right)}
\frac{|\langle \hat{\mathcal{A}}_{F}\psi_0 |
\psi_n \rangle_{\mathcal{H}\left(N\right)}|^2}{\langle \psi_0 | \psi_0 \rangle_{\mathcal{H}\left(N\right)}}
\end{equation}
A quite natural choice\cite{3hep:carleo} is to define $\hat{\mathcal{A}}_{F}$ borrowing
suggestions from the form of the trial wave function for 
the FN calculation, i.e.:

\begin{equation}
\label{operator}
\left(\hat{\mathcal{A}}_{F}\psi_0\right)\left(\mathcal{R}\right) \buildrel{def} \over {=}
\mathcal{D}
(\mathcal{R}_{\uparrow})
\mathcal{D}
(\mathcal{R}_{\downarrow})
\psi_0(\mathcal{R})
\end{equation}
where we can choose either the definition \eqref{planewaves} of $\mathcal{D}$  or 
the definition \eqref{backflow}. We will refer to such choices simply
as the plane waves fermionic correlations (PW--FC) and the backflow fermionic correlations (BF--FC).
Naturally the final results for the Bose--Fermi gap should 
coincide within statistical uncertainties, and the actual comparison
can be a good test for the robustness of the approach. 

 We observe that the sign problem is not really avoided as it manifests itself in two ways: 
on one hand poor choices of the wave functions appearing in the correlation functions imply the necessity to consider 
very large $\tau$ regions of the correlation function; on the other hand, since the gap energy is an extensive quantity, 
the exponential decay of the correlation functions increases in the limit $N\to\infty$, making 
impractical the extraction of meaningful information.

\subsection{The Bosonic System}
\begin{figure}[t]
\begin{center}
\includegraphics*[scale=0.5]{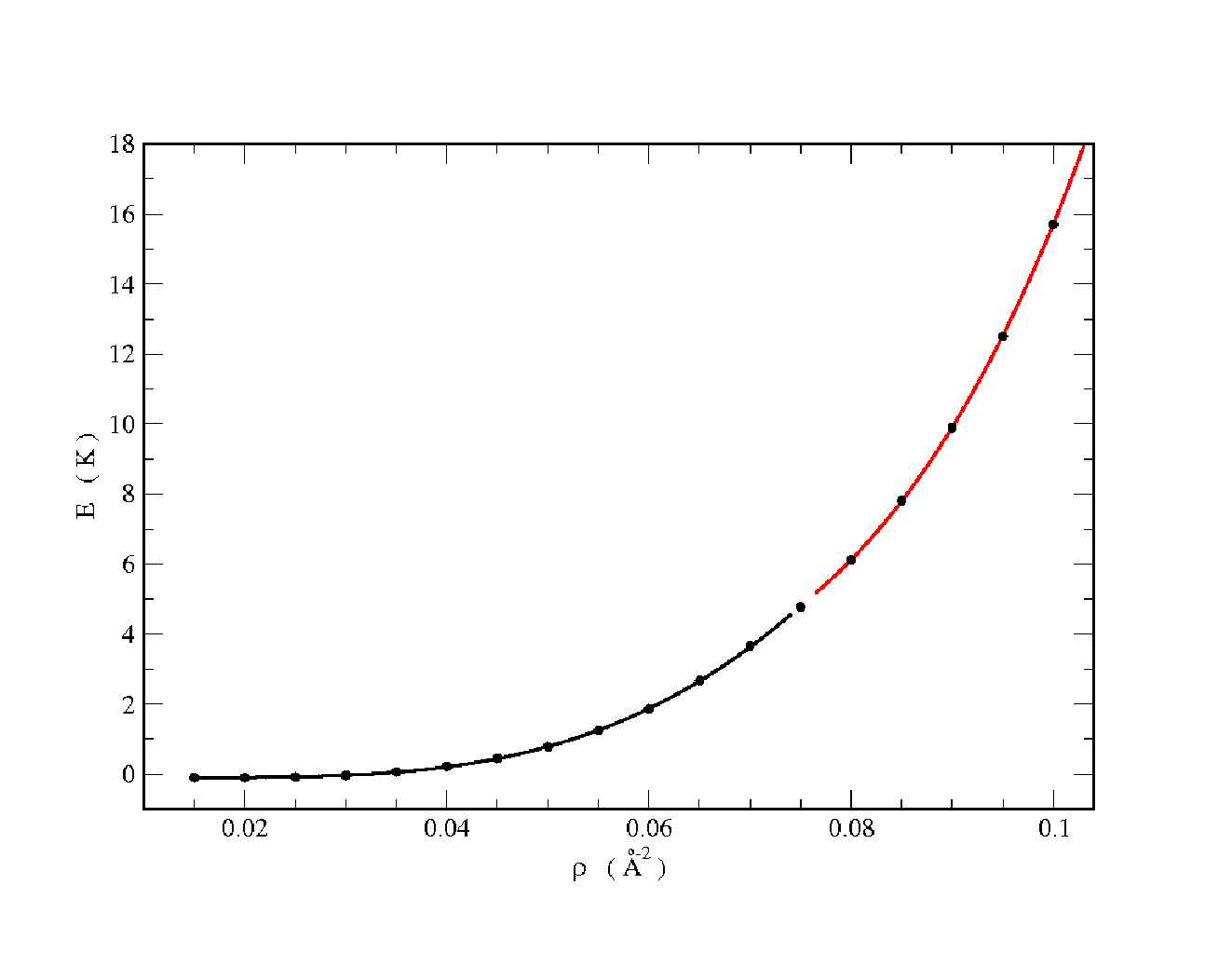}
\caption{State equation of the bosonic system used to calculate the correlation functions. 
The lines represent a fit of the data in both the solid (red line) and the liquid (black line) phase. 
The error bars are hidden inside the symbols. These results are in good agreement with Ref.~\onlinecite{3hep:boronat}}
\label{stateq}
\end{center}
\end{figure}

Figure~\ref{stateq} shows the state equation of both the solid and the liquid phases of the system; in Table~\ref{tabeosbose} the 
values of the energies at each density are shown.

\begin{table}[h]
\begin{center}
\caption{\label{tabeosbose} Potential ($E_{pot}$), Kinetic ($E_{kin}$) and Total ($E_{tot}$) Energy per particle of ``bosonic'' 2$d$ $^3$He at each studied density. 
The system has $N=26$ atoms in a square box of late $L=(N/\rho)^{1/2}$. Tail corrections to the potential energy to account for the finite 
size of the system have been applied only to $E_{tot}/N$; instead, for a more direct comparison, $E_{pot}/N$ is exactly the output of the simulations.
}

\begin{tabular}{| c | c | c | c |}
\hline
  Density ( \AA$^{-2}$ & $E_{pot}/N$  ( K ) & $E_{kin}/N$  ( K ) & $E_{tot}/N$  ( K ) \\
\hline
0.015 & -1.176 (5) & 1.107 (7) & -0.10 (1) \\
0.020 & -1.640 (6) & 1.58 (1) & -0.10 (2) \\
0.025 & -2.150 (6) & 2.12 (1) & -0.09 (2) \\
0.030 & 2.75 (2) & -2.709 (9) & -0.04 (2) \\
0.035 & 3.47 (2) & -3.318 (7) & 0.06 (4) \\
0.040 & 4.29 (2) & -3.97 (1) & 0.22 (4) \\
0.045 & 5.26 (2) & -4.71 (1) & 0.45 (4) \\
0.050 & 6.35 (2) & -5.47 (1) & 0.78 (4) \\
0.055 & 7.62 (3) & -6.28 (1) & 1.25 (4) \\
0.060 & 9.16 (3) & -7.12 (1) & 1.86 (4) \\
0.065 & 10.81 (3) & -7.98 (2) & 2.67 (5) \\
0.070 & 12.75 (4) & -8.91 (2) & 3.66 (7) \\
0.075 & 15.00 (3) & -9.94 (3) & 4.77 (6) \\
0.080 & 17.27 (4) & -10.84 (2) & 6.12 (7) \\
0.085 & 19.70 (4) & -11.65 (2) & 7.81 (7) \\
0.090 & 22.32 (5) & -12.28 (3) & 9.90 (9) \\
0.095 & 25.19 (6) & -12.72 (3) & 12.5 (1) \\
0.100 & 28.31 (7) & -12.88 (3) & 15.7 (1) \\
\hline
\end{tabular}
\end{center}
\end{table}

\begin{figure}[h]
\begin{center}
\includegraphics*[scale=0.5]{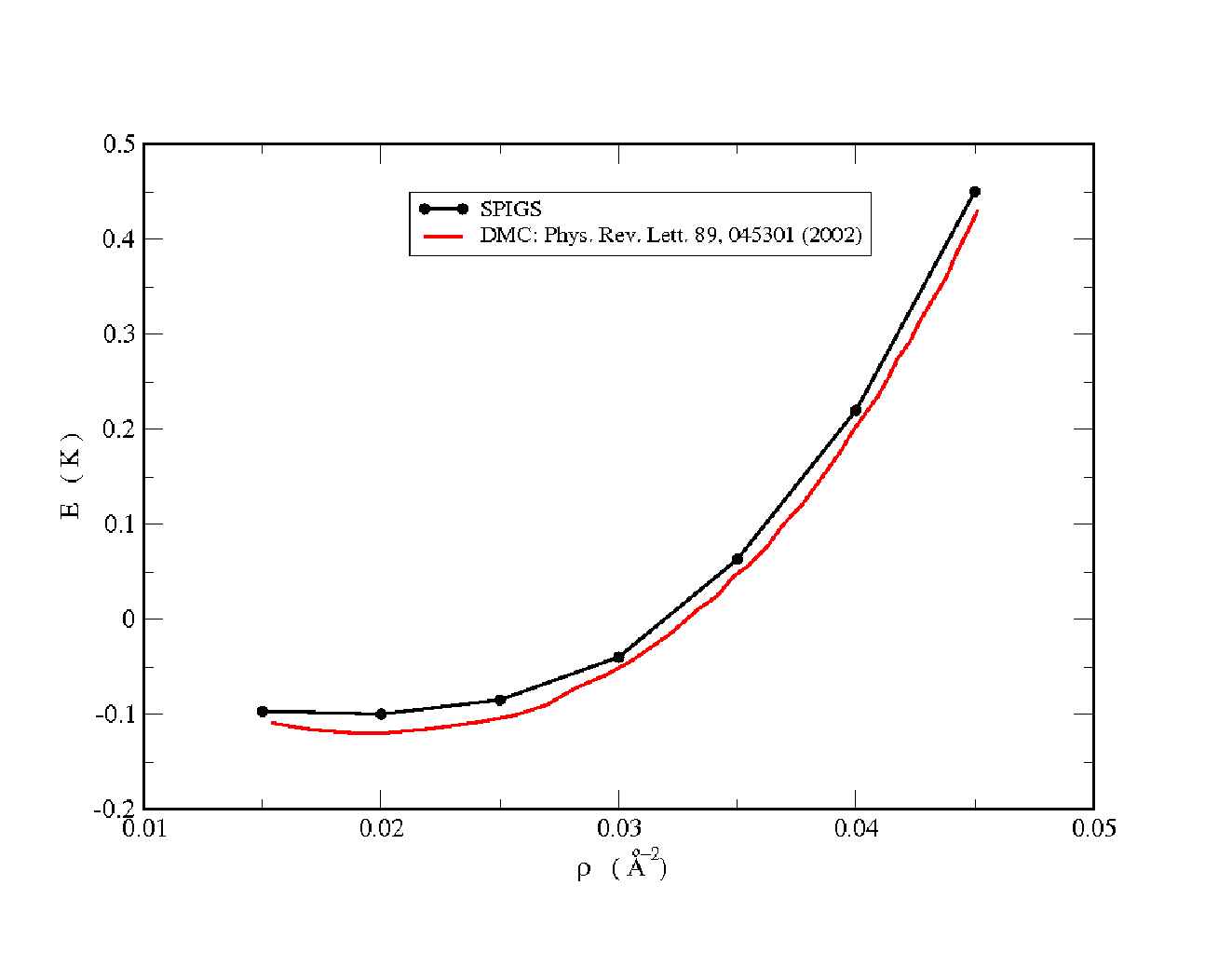}
\caption{Comparison between the equation of state of mass--3 bosons obtained with the Diffusion Monte Carlo 
and that obtained with the SPIGS. 
The systematic difference of roughly 0.03K is due to the different interaction potential that has been employed. }
\label{confronto}
\end{center}
\end{figure}

The agreement with data in Ref.~\onlinecite{3hep:boronat} is good (Fig.~\ref{confronto}), although the comparison 
must take into account the slight difference coming from the interaction potential used in our 
simulation (Aziz `79, Ref.~\onlinecite{3hep:Aziz79}), and that used in Ref.~\onlinecite{3hep:boronat} (Aziz `87, Ref.~\onlinecite{3hep:Aziz87}). 
This difference, $E_{87}-E_{79}$, depends on the density $\rho$ and can be estimated with the following computation if one neglects the 
dependence of the radial distribution function, $g\left(r\right)$, on the interaction potential $v\left(r\right)$:
\begin{eqnarray}
\label{gappotential}
E_{87}-E_{79} \simeq \pi\rho\int_{0}^{\infty}dr\:\left[v_{87}\left(r\right)-v_{79}\left(r\right)\right]g\left(r\right)
\end{eqnarray}

With the Maxwell construction obtained from the polynomial fit of the equation of state in the liquid and the solid 
phase (see Table~\ref{tabmaxwell} for the fitting parameters), the freezing point is estimated at a density of 0.069\AA$^{-2}$, 
while the melting point is approximately at a density of 0.073\AA$^{-2}$.

\begin{table}[h]
\begin{center}
\caption{\label{tabmaxwell} Fitting parameters for the solid and the liquid phases. The polynomial that has been fitted to the data 
is of the form $E/N = a\rho^5 + b\rho^4 + c\rho^3 +d\rho^2 + e\rho$.
}
\begin{tabular}{| c | c | c | c | c | c |}
\hline
  Phase  & $a$  ( K\AA$^{10}$ ) & $b$  ( K\AA$^{8}$ ) & $c$  ( K\AA$^{6}$ ) & $d$ ( K\AA$^{4}$ ) & $d$ ( K\AA$^{2}$ ) \\
\hline
liquid & 1745632.16048 & -66748.2524 & 9553.61311 & -61.70852 & -7.51343\\
solid & 3063129.8758 & -532201.7865 & 58695.29501 & -2598.78493 & 57.3547\\
\hline
\end{tabular}
\end{center}
\end{table}

High accuracy in these computations is very important for two reasons: first, even though the bosonic system is 
unphysical in itself, its energy is required for the evaluation of the energy of the real $^{3}He$ system, and finally, 
high quality of data is essential for a successful inversion of the Laplace Transform.

\subsection{The Twist Averages}
One of the main differences between the TE and the FC methods is the way in which twisted boundary conditions (TBC) are used.
In FC, TBC are not applied whenever a particle crosses a boundary of the simulation box but are taken into account during the 
evaluation of the correlation function; more precisely, the twist angle is introduced during the preparation of 
the Slater matrices, namely:
\begin{eqnarray}\label{sladet}
S_{N_\uparrow}=
\begin{pmatrix} 
e^{i\left(\vec{k}_1+\frac{\vec{\theta_1}}{L}\right)\cdot\vec{\tilde{r}}_{1}} & ... & e^{i\left(\vec{k}_1+\frac{\vec{\theta_1}}{L}\right)\cdot\vec{\tilde{r}}_{N_{\uparrow}}} \\ 
... & ... & ... \\
e^{i\left(\vec{k}_{N_\uparrow}+\frac{\vec{\theta_1}}{L}\right)\cdot\vec{\tilde{r}}_{1}} & ... & e^{i\left(\vec{k}_{N_\uparrow}+\frac{\vec{\theta_1}}{L}\right)\cdot\vec{\tilde{r}}_{N_{\uparrow}}} 
\end{pmatrix}
\end{eqnarray}
where $\theta_1$ is a given twist angle and $N_{\uparrow}$ is the number of spin up particles; naturally the same applies for the spin down particles. 
The coordinates $\vec{\tilde{r}}_{i}$ can be, like in the FN case, with or without backflow correlations, in the latter case we place 
$\vec{\tilde{r}}_{i}=\vec{r}_{i}^{\:abs}$, where the superscript means that the coordinates of the $i$--th particle are obtained without invoking the periodic boundary conditions (pbc); 
this is done in the following way: each particle has two coordinate systems, the first are coordinates inside the simulation box, which we refer here with the superscript ``pbc'',
 these are obtained by invoking the pbc whenever the particle crosses a boundary of the simulation box, the second coordinate system is made with absolute coordinates that are not constrained 
 in the simulation cell: if a particle moves from a position $\vec{x}_{1}^{\:pbc}$ to a position $\vec{x}_{2}^{\:pbc}$, its absolute coordinate changes accordingly $\vec{x}_{2}^{\:abs}$ = 
 $\vec{x}_{1}^{\:abs} + (\vec{x}_{2}^{\:pbc}-\vec{x}_{1}^{\:pbc})_{pbc}$, where the last subscript means that the displacement $(\vec{x}_{2}^{\:pbc}-\vec{x}_{1}^{\:pbc})$ is calculated 
 with periodic boundary conditions. The other choice for $\vec{\tilde{r}}_{i}$ is with backflow correlations,
 \begin{eqnarray}
 \vec{\tilde{r}}_{i} = \vec{r}_{i}^{\:abs} + \sum_{j=1; j \ne i}^{N}\eta\left(\left|\vec{r}_{j}-\vec{r}_{i}\right|_{pbc}\right)\left(\vec{r}_{j}-\vec{r}_{i}\right)_{pbc}
 \end{eqnarray}
 where $\eta(r)$ is defined in Eq.~\eqref{backeta}.
 
 Another delicate point in the construction of the Slater matrix \eqref{sladet} is the choice of the values $\lbrace \vec{k}_{n}+\theta_1/L\rbrace_{n=1}^{N_\uparrow}$; for $\theta_1=0$ 
 the choice would reduce to the wave vectors $\lbrace\vec{k}_{n}\rbrace$ inside the Fermi surface corresponding to $N_{\uparrow}$ particles, however it may not be the case for certain 
 choices of the twist angle $\theta_1$; the procedure to follow is that, for a given twist angle $\theta_1$, the wave vectors are those that give the first $N_{\uparrow}$ lowest energies $E_n$,
 \begin{eqnarray}
 E_n = \lambda \left|\vec{k}_{n}+\frac{\theta_1}{L}\right|^{2}
 \end{eqnarray}
where $\lambda = \hbar^2/(2m_{^3He})$. 

\begin{figure}[h]\label{figbz}
\begin{center}
\includegraphics*[scale=0.5]{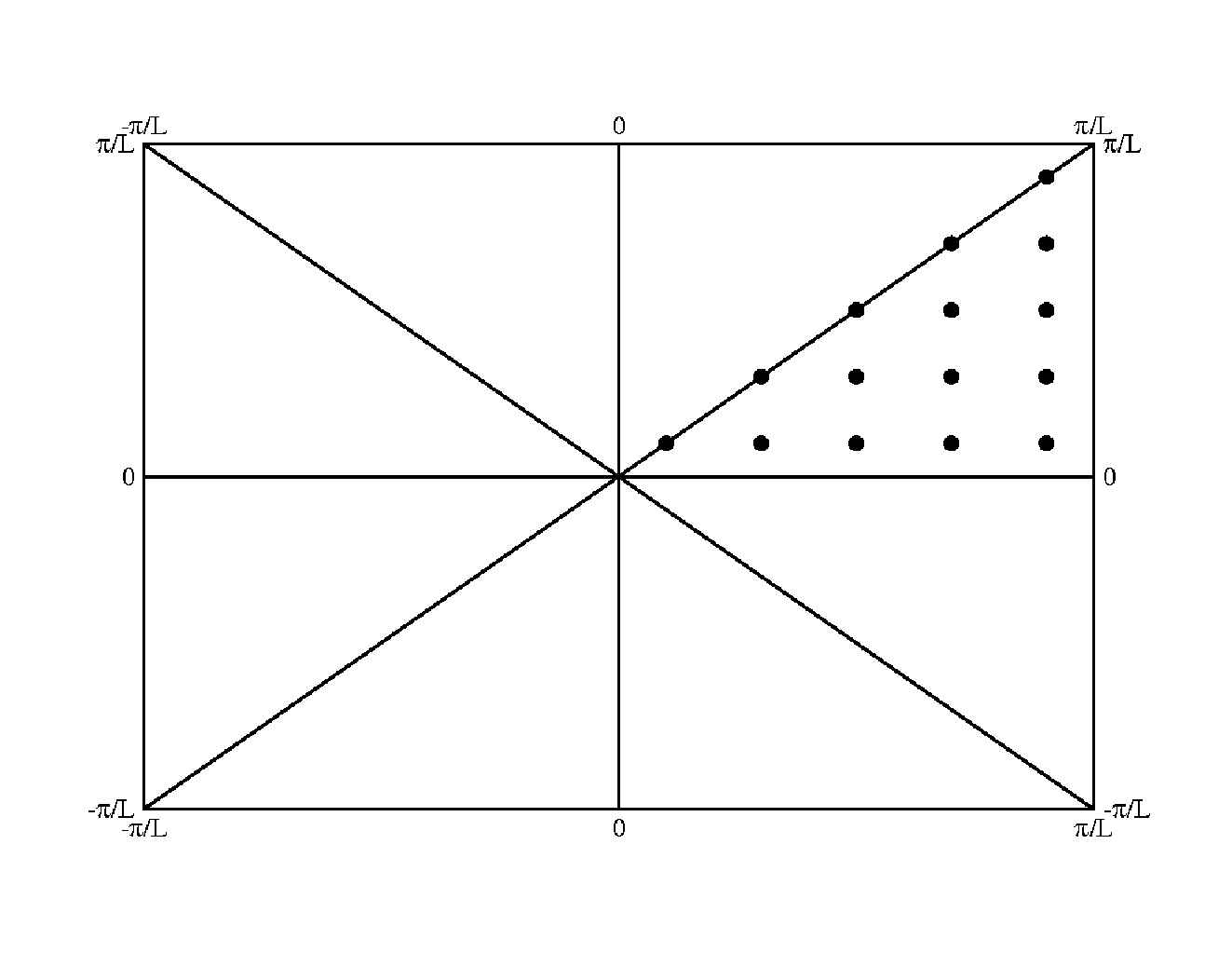} 
\caption{Schematic representation of the chosen twist angles in the first Brillouin zone of the simulation box. The simulation box is a square of late $L$.}
\end{center}
\end{figure}

For the evaluation of a single energy gap, 15 different correlation functions have been used for every Monte Carlo block. 
Each correlation function corresponds to a twist angle. This choice leads to a uniform distribution of of the twist angles in an area 
of the first Brillouin zone of the simulation box that contains no symmetries. This area is shown in Fig.~\ref{figbz}. 

Following the prescription in Eq.~\eqref{operator}, once the Slater matrices have been prepared, the Slater determinant has to be computed. This operation will be invoked 
many times during a Monte Carlo step and an efficient algorithm is advised. Our choice has been the LU decomposition that writes a matrix $A$ as a product of 
an upper triangular matrix $U$ and a lower triangular matrix $L$. The determinant of $A$ is then the product of the diagonal elements of $U$ and $L$. The LU decomposition 
is extensively described in Ref.~\onlinecite{3hep:numrec} for real matrices, the case of complex matrices is easily generalizable: from the algorithm in Ref.~\onlinecite{3hep:numrec} it is 
enough to redefine the matrices $L$ and $U$ to complex matrices $L=\Re L + \Im L$ and $U = \Re U + \Im U$; all the algebraic operation have then to be ambiented in the complex field 
and the resulting determinant will be a complex number. A complex number for the Slater determinant will yield a complex imaginary time correlation function; 
however, due to Eq.~\eqref{cfun2}, this correlation function will have, on average, an imaginary part compatible with zero.

The procedure to obtain the polarization curves may be schematized in the following steps:
\begin{itemize}
\item Consider all of the correlation functions corresponding to a given twist angle that have been computed 
for each block of the simulation. Calculate the error using the central limit theorem.
\item From the previously calculated error, infer the error of the correlation function relative to a single block.
\item Apply the inversion method and localize the position of the first peak.
In this work we used the GIFT method but we obtained compatible results also with a fit of the long imaginary--time part of the imaginary--time correlation function.
\item Repeat the previous steps for every twist angle and hence perform a weighted average according with the 
symmetries of the first Brillouin zone. This yields a single-block estimate of the energy gap.
\item The final energy-gap value for a given polarization is the block average of the single-block estimates that have 
thus been obtained.
\end{itemize}

\begin{figure}[t]
\begin{center}
\includegraphics*[width=12cm]{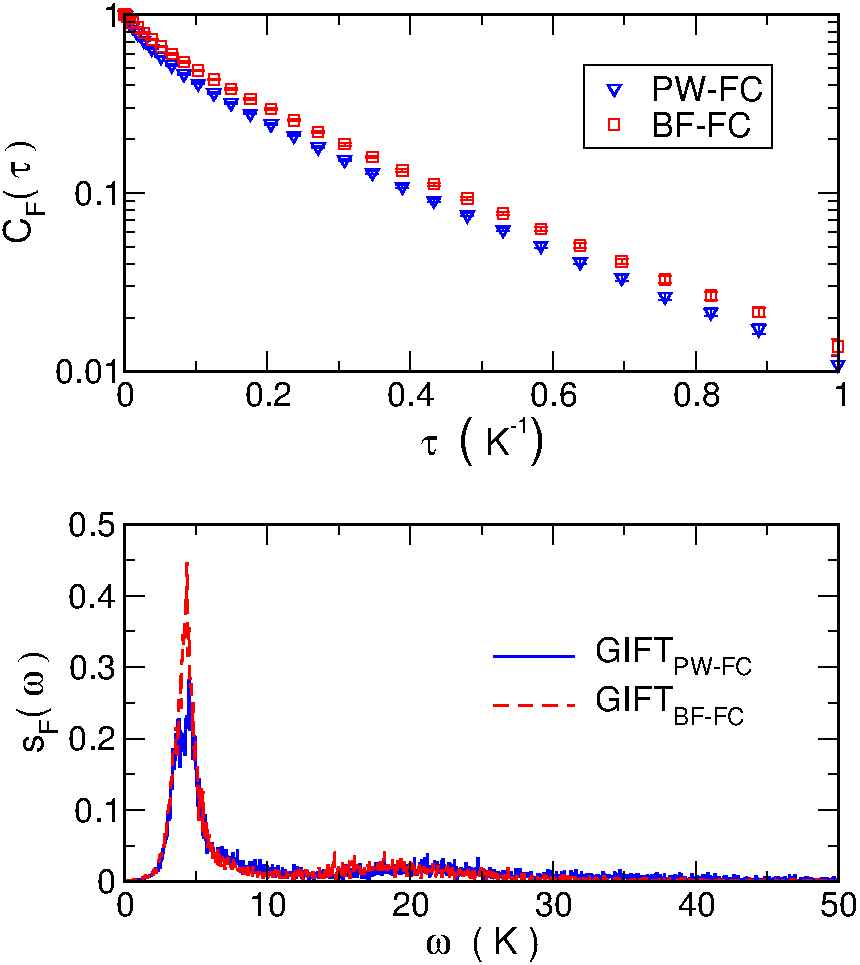}
\caption{(Color online) Upper panel: Imaginary time correlation functions, $\mathcal{C}_{F}(\tau)$,
corresponding to the two different choices of the operator in \eqref{operator}.
Lower panel: reconstructed spectral functions $s_F(\omega)$ obtained with the GIFT method. }
\label{pwbf}
\end{center}
\end{figure}

The described procedure has been applied for every polarization. 
This implies thousands of Laplace transforms to be inverted with relative peaks to be localized; 
it is a task that is unlikely to be done manually and an automated procedure was implemented. 
However, particular attention must be paid in the automatic localization of the peaks in order to avoid false values 
and thus systematic errors. Possibly, different localization algorithms must be applied and a check by eye of a 
random selection of the inversion results is highly advised. We would like to emphasize again that the most delicate 
part of this method lies in the numerical inversion of the Laplace transform of the correlation functions but also in the data analisys 
of the results.

\subsection{Analytic Continuation}
Once we have achieved a QMC evaluation of $\mathcal{C}_{F}(\tau)$, the information 
about the Bose--Fermi gap $\Delta_{BF}= E_0 - E_0^B$ is contained in the
resulting correlation functions. The results for $\mathcal{C}_{F}(\tau)$ appear as 
simple smooth decreasing functions, whose values can be evaluated only in correspondence
with a finite number of imaginary--time values, say $\{\tau_0,\tau_1,\tau_2,...,\tau_l\}$;
moreover the data are perturbed by unavoidable statistical uncertainties.
The Bose--Fermi gap $\Delta_{BF}$ is thus hidden inside the sets of limited and noisy data. 
How can we extract it?

In the upper panel of Fig.~\ref{pwbf} we show two imaginary time correlation functions
$\mathcal{C}_{F}(\tau)$, respectively a PW--FC and a BF--FC,
corresponding to the same spin polarization and twist parameter.
The long--$\tau$ tails of the two curves tend towards a
linear behavior (in logarithmic scale) with the same slope, and this
is a general feature shared by all the functions we have evaluated.
This indicates that, because of the finite--size of the system (and selection rules on the total momentum), the fermionic spectrum has a sufficiently defined gap, i.e. the
lowest--energy term $\exp(-\Delta_{BF}\tau)$ in the correlation function \eqref{cfun2}
appears to be quite well resolved with respect to contributions from higher
fermionic energies.
The difference between the two curves (in particular the rigid shift between their
asymptotic tails) arises from the spectral weight
of the Ground State contribution, which is higher when backflow correlations
are taken into account, as expected.

In this favorable situation, the Bose--Fermi gap can be reliably extracted by 
simply fitting an exponential to the long--time tail of the correlation function.

This key result is strongly supported by a more sophisticated approach,
which evaluates $\Delta_{BF}$ by 
performing the full Laplace transform inversion of $\mathcal{C}_{F}(\tau)$,
i.e. solving 
\begin{equation}
\label{problem}
\mathcal{C}_{F}(\tau) = \int_{0}^{+\infty}d\omega e^{-\tau \omega} s_{F}(\omega) \quad ,
\end{equation}
for the unknown {\it spectral function} $s_{F}(\omega)$.
Recently a new method, the genetic inversion via falsification of theories (GIFT) 
method\cite{3hep:Ettore}, has been developed to face general inverse problems and in 
particular it has allowed to reconstruct the excitation spectrum of superfluid 
$^4$He starting from QMC evaluations of the intermediate scattering function 
in imaginary--time\cite{3hep:Ettore}; the results were in close agreement with 
experimental data\cite{3hep:Ettore}. Moreover the method has allowed to extract 
also multiphonon energies with a good accuracy level.
When applied to the two curves depicted in the upper panel of Fig.~\ref{pwbf},
we find the two spectral functions in the lower panel of Fig.~\ref{pwbf}; it is
apparent that the lowest--$\omega$ peak is indeed well resolved from higher--energy
contributions. Crucially, its position does not depend on the actual choice of the operator $\hat{\mathcal{A}}_F$, and it is in excellent agreement with the smallest
decay constant found by the simple exponential fit. 
The spectral weight instead is different, consistently with the differences
between PW--FC and BF--FC.


\subsection{Results} 

\begin{figure}[h]
\begin{center}
\includegraphics*[width=12cm]{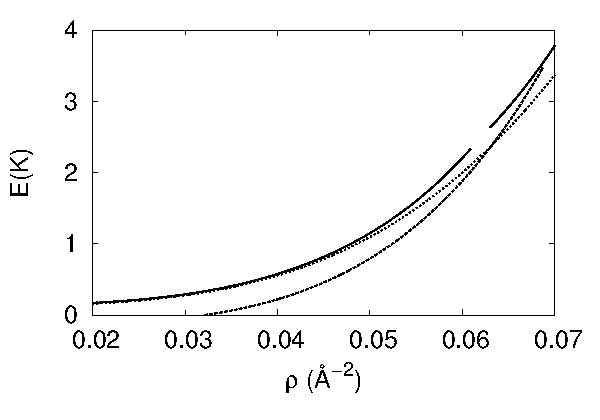}\label{polstateq}
\caption{Equation of state of $^3$He in two dimensions. Solid line (broken across the coexistence region): liquid and solid
         $^3$He; dashed line: mass--3 boson fluid; dotted line: liquid $^3$He, after Ref.~\onlinecite{3hep:boronat}. The latter is
         only reliable at low densities.}
\end{center}
\end{figure}

We fit a fifth order polynomial to the density dependence of the energies of the triangular crystal and of the 
paramagnetic and the ferromagnetic fluids, listed in Table~\ref{tab1}.
\begin{table}[h] \label{tab1}
\begin{center}
 \caption{The equations of state of $^3$He for the paramagnetic fluid and the solid
(solid lines in Figure~\ref{polstateq})
are of the form $\alpha_1 \rho+\alpha_2 \rho^2+\alpha_3 \rho^3+\alpha_4 \rho^4+\alpha_5 \rho^5$.
This Table lists the values of the parameters $\alpha_i$, with lengths in \AA.}

  \begin{tabular}{ccc}
  & liquid & solid \\
  \hline
$\alpha_1$ & 21.23782 & 57.35474 \\
$\alpha_2$ & -1344.413 & -2598.784 \\
$\alpha_3$ & 45093.37 & 58695.29 \\
$\alpha_4$ & -569306.0 & -532201.7 \\
$\alpha_5$ & 4383507  & 3063129 \\
  \end{tabular}
\end{center}
\end{table}
The resulting equation of state of two--dimensional $^3$He is shown in Figure~\ref{polstateq}.
With the fermionic correlations method, we find a transition between the paramagnetic fluid 
and the triangular crystal around $\rho=0.061$~\AA$^{-2}$, with a narrow coexistence of about $0.002$~\AA$^{-2}$,
while the ferromagnetic fluid is never stable (see Table~\ref{tab2}).
The obtained solidification density can not be directly compared with experimental data
since we are studying a model of an ideal $2d$--system. It could be interesting in future calculations to consider 
an adsorbing external potential representing the interaction of the $^3$He atoms with a substrate.

\begin{table}[t]
\begin{center}
 \caption{\label{tab2} Ground state energy of $^3$He in K, calculated by the FC method for the fluid phases and assumed
to equal the bosonic energy for the solid phase.
         }
  \begin{tabular}{cccc}
 density & liquid $\zeta=0$ & liquid $\zeta=1$ & solid \\
  \hline
0.020 & 0.1707(18) & 0.3218(25) &             \\
0.045 & 0.8168(86) & 0.9075(86) &             \\
0.050 & 1.1500(81) & 1.2123(88) &             \\
0.055 & 1.5972(93) & 1.6574(91) &             \\
0.060 & 2.2069(68) & 2.2493(54) & 2.2506(54)  \\
0.065 & 3.0065(73) & 3.0359(45) & 2.9195(26)  \\
0.070 & 4.0644(33) & 4.0915(34) & 3.7878(35)  \\
0.075 &            &            & 4.8728(44)  \\
0.080 &            &            & 6.2445(35)  \\
0.085 &            &            & 7.9589(39)  \\
0.090 &            &            & 10.0661(46) \\
0.095 &            &            & 12.6739(39) \\
0.100 &            &            & 15.8536(45) \\
  \end{tabular}
\end{center}
\end{table}
The energy of the bosonic mass--3 liquid is also shown. 
This fictitious system, simulated to extract the PW--FC and BF--FC energies, crystallizes at $\rho=0.069$~\AA$^{-2}$.
The freezing density of $^3$He is considerably higher than the highest density simulated in Ref.~\onlinecite{3hep:boronat}. 
Correspondingly, the equation of state given in Ref.~\onlinecite{3hep:boronat} is only reliable at relatively low density.
In particular, while it is only slightly below our results for $\rho \lesssim 0.045$~\AA$^{-2}$ as a consequence of the difference 
of interparticle potential adopted\cite{3hep:Aziz87}, it becomes even lower than the bosonic equation of state near the 
melting density, by an amount far larger than what could be due to the different employed potential.

\begin{figure}[h]
\begin{center}
\includegraphics*[width=12cm]{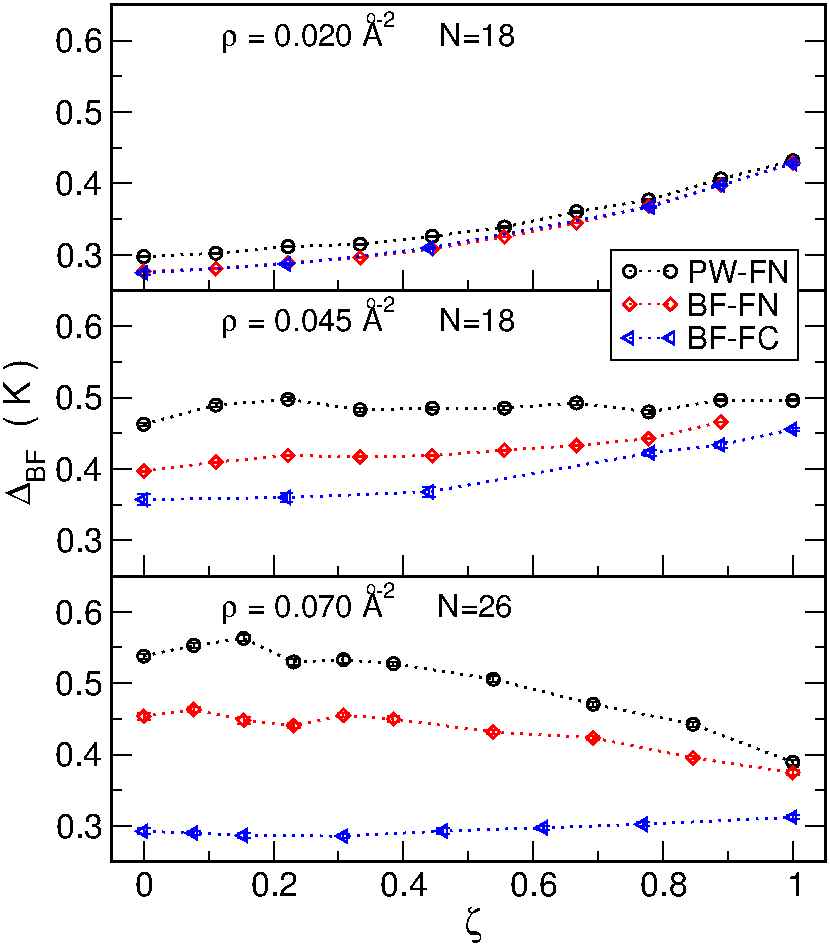}
\caption{(Color online) Upper panel: Bose--Fermi gap, $\Delta_{BF}$,
         as a function of the spin polarization, $\zeta$, at density $\rho=$0.020~\AA$^{-2}$
         evaluated via PW--FN, BF--FN, and BF--FC with $N=18$ particles.
         Middle panel: Bose--Fermi gap, $\Delta_{BF}$,
         as a function of the spin polarization, $\zeta$, at density $\rho=$0.045~\AA$^{-2}$
         evaluated via PW--FN, BF--FN, and BF--FC with $N=18$ particles.
         Lower panel: Bose--Fermi gap, $\Delta_{BF}$,
         as a function of the spin polarization, $\zeta$, at density $\rho=$0.070~\AA$^{-2}$
         evaluated via PW--FN, BF--FN, and BF--FC with $N=26$ particles.
         The statistical uncertainties are below the symbols size.}
\label{pcurves2}
\end{center}
\end{figure}

\begin{figure}[h]
\begin{center}
\includegraphics*[width=12cm]{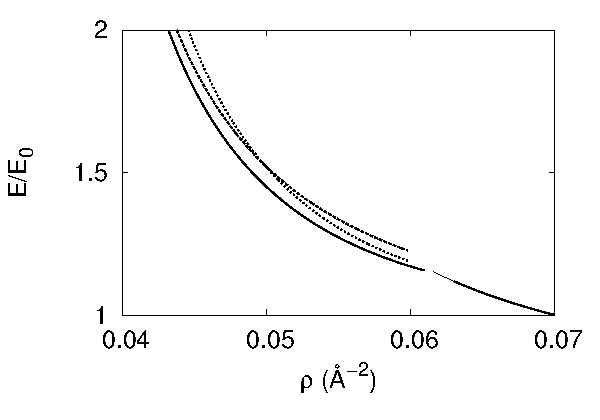}
\caption{Unbiased FC versus Fixed--Node equation of state. Thick solid line (broken across the coexistence region): paramagnetic
         liquid and solid $^3$He (FC); dashed line: paramagnetic liquid (FN); dotted line: ferromagnetic liquid (FN);
         the dashed and dotted lines terminate at the FN freezing density; thin solid line: energy of the solid, down to the
         FN melting density. For each density, the energy is relative to the energy $E_0$ of the mass--3 boson fluid.}
\label{polstateq2}
\end{center}
\end{figure}

\begin{figure}[h]
\begin{center}
\includegraphics*[width=12cm]{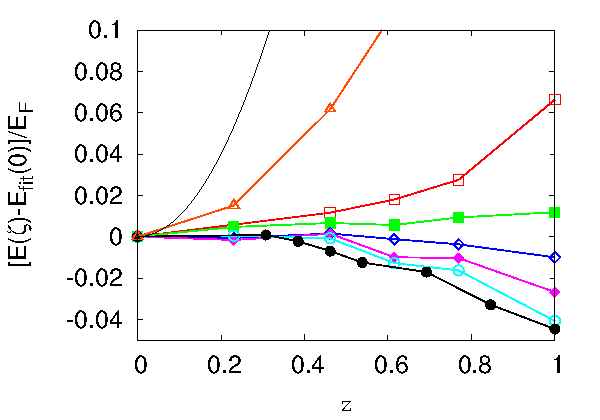}
\caption{(Color online) Fixed--node results for the polarization energy $E(\zeta)-E_{{\rm fit}}(0)$ relative to the Fermi energy
         $E_F$ at $\rho=0.020$ (open triangles), 0.045 (open squares), 0.050 (filled squares), 0.055 (open diamonds),
         0.060 (filled diamonds), 0.065 (open circles), 0.070 (filled circles)~\AA$^{-2}$, i.e. from top to bottom.
         The function $E_{{\rm fit}}(\zeta)$ is a quadratic polynomial in $\zeta^2$ fitted to the simulation data;
         the solid line is the density--independent result for non--interacting particles.}
\label{epol_fn}
\end{center}
\end{figure}

\begin{figure}[h]
\begin{center}
\includegraphics*[width=12cm]{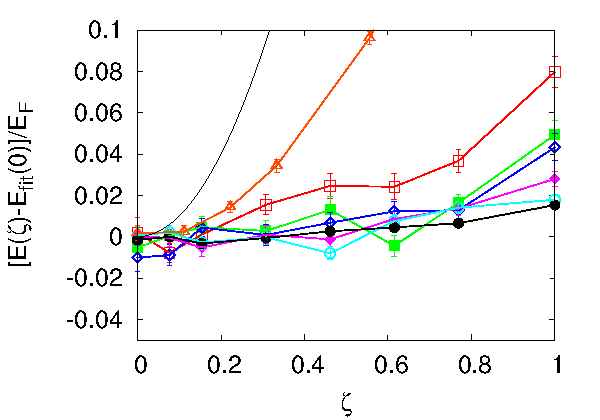}
\caption{(Color online) Exact results for the polarization energy $E(\zeta)-E_{{\rm fit}}(0)$ relative to the Fermi energy $E_F$ at
         $\rho=0.020$ (open triangles), 0.045 (open squares), 0.050 (filled squares), 0.055 (open diamonds),
         0.060 (filled diamonds), 0.065 (open circles), 0.070 (filled circles)~\AA$^{-2}$, in order of decreasing dispersion.
         The function $E_{{\rm fit}}(\zeta)$ is a quadratic polynomial in $\zeta^2$ fitted to the simulation data;
         the solid line is the density--independent result for non--interacting particles.}
\label{epol}
\end{center}
\end{figure}

The treatment of the spin polarization state requires a special care\cite{3hep:ceperley_3d,3hep:holzmann,3hep:drummond,3hep:carleo}.
In contrast to Ref.~\onlinecite{3hep:boronat}, we find that the BF--FN energy can be significantly higher than the unbiased
Fermionic correlations (FC) energy. Starting from negligible values at low density,
the BF-FN error quickly increases approaching the strongly correlated
regime. As expected\cite{3hep:holzmann}, it is larger for the paramagnetic than for the ferromagnetic fluid. This happens, 
because the available variational wave function for ferromagnetic states are more accurate than those for paramagnetic states.  
These findings are exemplified in Figure~\ref{pcurves2}.
The inadequacy of the BF--FN is striking in the phase diagram: Figure~\ref{polstateq2} shows that BF--FN incorrectly
predicts a transition to a ferromagnetic fluid well before crystallization takes place. 
Such a transition is also evident from Figure~\ref{epol_fn}, which shows the BF--FN results for the polarization
energy $e(\zeta)$ at various densities. The unbiased results, shown in Figure~\ref{epol}, display instead a paramagnetic
behavior even in a metastable fluid phase well beyond the freezing density.

\begin{figure}[h]
\begin{center}
\includegraphics*[width=12cm]{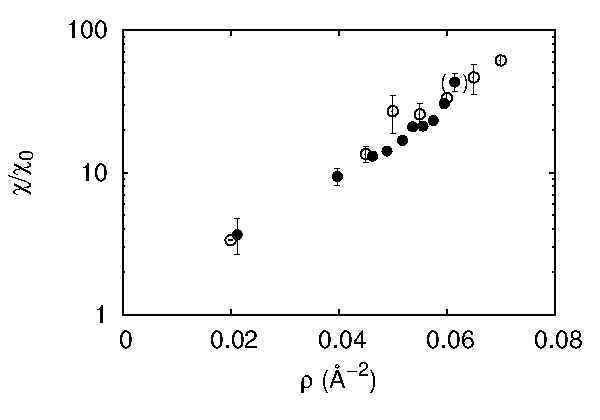}
\caption{Enhancement of the spin susceptibility as a function of the density:
         (filled circles) as measured in the second layer of $^3He$ on graphite\cite{3hep:morhard};
         (open circles) as calculated assuming a quadratic dispersion over the whole polarization range in Fig.~\ref{epol}.
         The corresponding Fixed--node result from Fig.~\ref{epol_fn} would diverge at $\rho\simeq 0.050$~\AA$^{-2}$.}
\label{chi}
\end{center}
\end{figure}

From the FC polarization energy $e(\zeta)$ we can estimate the spin susceptibility enhancement $\chi/\chi_0$. Assuming 
a quadratic dependence over the whole polarization range, which is generally consistent with the data
of Figure~\ref{epol}, we find an excellent agreement with the
measured susceptibility. Figure~\ref{chi} shows the comparison between the calculated $\chi/\chi_0$  
and the experimental data. We display only the results obtained in the second layer 
of $^3$He on graphite\cite{3hep:morhard} since they extend to the highest density in the 
fluid phase, but experiments carried on with differently preplated substrates
lead to equivalent results in their respective density ranges. The agreement among
the results obtained using different substrates induces us to expect that our ideal
model actually captures the physical mechanisms underlying the behavior of $\chi/\chi_0$. 

 The results for $\chi/\chi_0$ evaluated with BF--FN calculations diverge at a density around 
$0.050$~\AA$^{-2}$, consistently with the BF-FN prediction of a phase transition taking place around the above mentioned density.
The need for an exact QMC approach is thus witnessed by the failure of the  BF--FN approximation to predict the lack of a 
polarization transition experimentally observed in the fluid phase, let alone an accurate value for
the spin susceptibility.  We emphasize that in principle it is possible to improve the fixed node results 
working on the choice of the trial wave function. Our purpose in this work was to follow a methodology which is unbiased, 
that is independent on the choice of the wave function; such methodology gives access only to the energy and its derivatives.
 Here we have shown the improvements with respect to the results 
obtained with a particular fixed--node approximation that has already been used\cite{3hep:boronat} for $^3$He.

In conclusion, we have calculated the equation of state and the polarization energy of 
$^3$He in two dimensions by means of an unbiased QMC method.
The system crystallizes into a triangular lattice from the paramagnetic fluid at 
a density of $0.061$~\AA$^{-2}$, with a narrow coexistence region of about $0.002$~\AA$^{-2}$; the ferromagnetic
fluid is never stable. From the polarization energy we obtain a spin susceptibility
enhancement in excellent agreement with the experimental values. 

We remark that, although the Fermionic correlation technique is in principle unbiased, to obtain the estimation of 
the Bose--Fermi gap one has to face an ill--posed inverse 
problem: the inversion of the Laplace transform at the presence of a limited set of noisy data; the quality of the results
 of the inversion procedure cannot be guaranteed {\it a priori}, in this work we have found empirically that the obtained 
 correlation functions could be safely inverted obtaining robust results, which has been checked using different techniques.

Moreover, the estimation of the Bose--Fermi gap via the Fermionic correlation method is 
limited to relatively small systems: the present results are obtained with either 18 or
(in most cases) 26 particles. While the size effect remains the main source of
uncertainty of the present calculation, the agreement of the calculated and measured
spin susceptibility suggests that finite--size errors are relatively small.

%% file: ch-dynamics/chapter-dynamics.tex
\def\onlinecite{\cite}


\chapter{Dynamics of two--dimensional $^3$He\label{ch:dynamics}}

In the previous chapter we have studied static properties of 2$d$ $^3$He; in particular 
we have inspected the dependence of the energy versus spin polarization at different densities, showing 
that 2$d$ $^3$He remains a paramagnetic fluid up to the freezing density. In this chapter 
we focus on the dynamical properties of this system. In fact, the low energy dynamics of $^3$He is 
 of outstanding importance in condensed matter physics to 
understand the thermodynamic behavior of quantum strongly correlated systems\cite{dyn:krotscheck2}.

Recently inelastic neutron scattering experiments
have been performed on a monolayer of liquid $^3$He adsorbed on suitably preplated graphite:
for the first time the collective {\it zero-sound} mode has been detected as a well defined
excitation crossing and possibly reemerging from the particle--hole continuum typical of a Fermi
fluid\cite{dyn:krotscheck1,dyn:nature}.
In this chapter, we undertake an {\it ab--initio} study of the
low-energy collective excitations, in particular the zero-sound mode, of a strictly
two--dimensional (2$d$) $^3$He sample relying on Quantum Monte Carlo (QMC) methods.
This is particularly appealing since it has been shown that the 
strictly 2$d$ model is often a realistic representation of the adsorbed liquid layer,
as far as the liquid phase properties are concerned\cite{dyn:whitlock,dyn:prbnoi}. 
The key quantity to be computed to compare with neutron scattering experiments is
the dynamical structure factor $S(\vec{q},\omega)$, which, apart from kinematical factors, is related to the differential cross section 
and contains informations about low--energy dynamics of the sample; in the case of $^3$He, the dynamic structure factor is a sum of 
a coherent term $S_c(\vec{q},\omega)$ and an incoherent contribution due to the coupling of the nuclear spin with the neutron beam\cite{dyn:glyde}, $S_i(\vec{q},\omega)$ 
\begin{eqnarray}
S(\vec{q},\omega) = S_c(\vec{q},\omega) + (\sigma_i/\sigma_c)S_I(\vec{q},\omega)\\
S_c(\vec{q},\omega)= \frac{1}{2\pi N b} \int_{-\infty}^{+\infty}dt\,e^{i\omega t} \langle e^{i\frac{t}{\hbar} \hat{H}}\,\hat{\rho}_{\vec{q}}\, e^{-i\frac{t}{\hbar} \hat{H}}\,\hat{\rho}_{-\vec{q}} \rangle\label{sqw} \\
S_i(\vec{q},\omega)= \frac{1}{2\pi N } \int_{-\infty}^{+\infty}dt\,e^{i\omega t} \langle e^{i\frac{t}{\hbar} \hat{H}}\,\hat{\rho_z}_{\vec{q}}\, e^{-i\frac{t}{\hbar} \hat{H}}\,\hat{\rho_z}_{-\vec{q}} \rangle
\end{eqnarray}
The brackets indicate a Ground state or thermal average, $\hat{H}$ is the Hamiltonian operator,  
$\hat{\rho}_{\vec{q}} = \sum_{i=1}^{N}\, e^{-i \vec{q} \cdot \vec{r}_i}
$ and $\hat{\rho_z}_{\vec{q}} = \sum_{i=1}^{N_{\uparrow}}\, e^{-i \vec{q} \cdot \vec{r}_i^{\:\uparrow}} - \sum_{i=1}^{N_{\downarrow}}\,e^{-i \vec{q} \cdot \vec{r}_i^{\:\downarrow}}$
are respectively the local particle and spin densities in Fourier space. The parameter $b$ is the coherent scattering length and 
$\sigma_c$ and $\sigma_i$ are the scattering cross sections for the coherent and incoherent scattering.
Similarly to the previous chapter, we are interested here only 
in zero temperature properties.
The excitations of the system manifest themselves in the shape of $S(\vec{q},\omega)$, appearing
either as sharp peaks if they are long-lived or as broad structures if strong damping is present.
In particular, the zero sound mode, which is the main target of this work, is related with $S_c(\vec{q},\omega)$; the ratio $\sigma_i/\sigma_c$ has been shown\cite{dyn:ratiosigma} to be 
0.20(5); moreover, in the experimental data in Ref.~\onlinecite{dyn:nature} 
there is a well defined signal from the zero--sound mode but possible excitations from the incoherent part of $S(\vec{q},\omega)$ (i.e. spin waves) are much harder to discern. 
Given these considerations, the data in Ref.~\onlinecite{dyn:nature} is dominated by $S_c(\vec{q},\omega)$; we focus here on the coherent dynamical structure factor, 
$S_c(\vec{q},\omega)$. QMC methods may indeed give access indirectly to the dynamic structure factor, $S_c(\vec{q},\omega)$, because they allow to
evaluate the intermediate scattering function:

\begin{equation}
\label{fqt}
F(\vec{q},\tau)= \langle e^{\tau \hat{H}}\,\hat{\rho}_{\vec{q}}\, e^{-\tau \hat{H}}\,\hat{\rho}_{-\vec{q}} \rangle
\end{equation}
by simulating a stochastic dynamics in imaginary time driven by the simple Hamiltonian:

\begin{equation}
\label{d:hamiltonian}
\hat{H} = - \frac{\hbar^2}{2m_{3}}\sum_{i=1}^{N} \nabla_{i}^2
+ \sum_{i<j=1}^{N}v\left(\vec{r}_i - \vec{r}_j\right) \quad .
\end{equation}
Here $m_3$ is the mass of $^3$He atoms and the pair interaction $v(r)$ is a realistic effective potential among $^3$He atoms\cite{dyn:Aziz79}.

Since the ground state is not known, a QMC calculation of \eqref{fqt}
requires an additional time $\tilde{\tau}$ to project a trial wave function $\psi_T^F$
onto the exact ground state $\psi_0^F$ (see chapter \ref{ch:methods}):
\begin{equation}
\label{fqt_qmc}
F(\vec{q},\tau) = \frac{\langle \psi_T^F| e^{-\tilde{\tau} \hat{H}} \, {\rho}_{\vec{q}} \, e^{-\tau \hat{H}}\, \hat{\rho}_{-\vec{q}} e^{-\tilde{\tau} \hat{H}} | \psi_T^F \rangle}{\langle \psi_T^F|
e^{-(2\tilde{\tau}+\tau)} |\psi_T^F \rangle}
\end{equation}

The correlation function \eqref{fqt} is the Laplace transform of $S_c(\vec{q},\omega)$. 
Despite the well known difficulties related to the inversion of the Laplace
transform in ill-posed conditions, the evaluation of $S_c(\vec{q},\omega)$ starting
from the QMC estimation of $F(\vec{q},\tau)$ \eqref{fqt_qmc} has been proved to 
be fruitful for several bosonic systems\cite{dyn:Bagf,dyn:gift,dyn:prlsaverio}.

For a Fermi liquid, the difficulty is further enhanced by the famous {\it sign problem}, 
thereby the computational effort grows exponentially with the projection
time (as well as with the number of particles). The total projection time
$2\tilde{\tau}+\tau$ in Eq. \eqref{fqt_qmc} is too large for all practical purposes.

While accurate approximations exist to circumvent this problem in the calculation of static
ground-state properties\cite{dyn:reynolds}, we are aware of no applications of approximate 
schemes such as the restricted path\cite{dyn:rpimc} or constrained path\cite{dyn:cpmc}
methods to the calculation of imaginary-time correlation functions.

We thus resort to the following approximation:
\begin{equation}
\label{approximation}
\psi_0^F=e^{-\tilde{\tau} \hat{H}}\psi_T^F \simeq \mathcal{D} e^{-\tilde{\tau} \hat{H}}\psi_T^B = \mathcal{D} \psi_0^B
\end{equation}
where a superscript $F(B)$ indicates Fermi(Bose) statistics and $\mathcal{D}$ is a Slater determinant.
In the resulting approximate correlation function 
\begin{equation} \label{sqwvar}
F_{A}(\vec{q},\tau) = \frac{\langle \psi_0^{B}|\mathcal{D}^{\star} \, \hat{\rho}_{\vec{q}} \, \, e^{-\tau \hat{H}}\, \, \hat{\rho}_{-\vec{q}} \,\mathcal{D}| \psi_0^{B} \rangle}{\langle \psi_0^{B}|\mathcal{D}^{\star} e^{-\tau \hat{H}}\,\mathcal{D}| \psi_0^{B}\rangle}
\end{equation}
the projection time between the determinants, which determines the severity of the sign problem,
is limited to $\tau$; $F_A$ is an approximation of the intermediate scattering function in imaginary 
time \eqref{fqt}, which would be exact if $\mathcal{D} \psi_0^B$ were the exact Fermi ground state.
Its inverse Laplace transform is an approximation of the dynamical structure factor \eqref{sqw}.
For a given wave vector, the positions of the peaks in $S(\vec{q},\omega)$ provide the energy of the 
excitations, while their shape is related to the life--time of the excited states.
In general, the approximation \eqref{approximation} introduces biases both in the excitation energies 
and in the shape of $S_c(\vec{q},\omega)$. In order to enhance the robustness of this approach, we 
introduce also another correlation function, $F_B$, which is defined on the bosonic ground state:
\begin{equation}
\label{ftex}
F_{B}(\vec{q},\tau) = \frac{\langle \psi_0^{B}| e^{\tau \hat{H}} \, \mathcal{D}^{\star} \, \hat{\rho}_{\vec{q}} \, \, e^{-\tau \hat{H}}\, \, \hat{\rho}_{-\vec{q}} \,\mathcal{D} | \psi_0^{B} \rangle}{\langle \psi_0^{B}| \psi_0^{B} \rangle}.
\end{equation} 
Despite $F_B$ is not directly related to the dynamical 
structure factor of the Fermi liquid, this function has some useful features: on one hand, it contains the 
{\it exact} fermionic spectrum, as can be seen from the spectral resolution:
\begin{equation}
\label{ftex2}
F_{B}(\vec{q},\tau)
= \sum_{n=0}^{+\infty}
e^{-\tau \left(E^{F}_n - E_0^B\right)} b_n, \quad b_n =
\frac{|\langle \hat{\rho}_{-\vec{q}} \,\mathcal{D} \,\,\psi_0^{B} |
\psi_n^F \rangle|^2}{\langle \psi_0^{B} | \psi_0^{B} \rangle}
\end{equation}
On the other hand, it is a bosonic correlation function and thus it can be evaluated with great accuracy 
by means of exact bosonic QMC methods. If, moreover, the approximation \eqref{approximation} is 
accurate enough, the coefficients $b_n$ become, apart from an unessential normalization, 
the spectral weights $f_n$ of the exact intermediate scattering function \eqref{fqt}.
\begin{equation}
f_n = 
\frac{|\langle \hat{\rho}_{-\vec{q}} \,\psi_0^{F} |
\psi^F_n \rangle|^2}{\langle \psi_0^{F} | \psi_0^{F} \rangle}
\end{equation}
We note finally that $F_{B}$ arises as a natural generalization of the Fermionic correlations
method: in fact, $F_{B}$ has the same form of Eq.~\eqref{cfun} that has been used in the previous chapter about the ground 
state of an $^3He$ film: in that context, the Fermionic correlations method provided results for the magnetic properties of the system in impressive agreement
with experimental data.

We argue that a comparison between dynamical properties evaluated with $F_A$ and $F_B$ 
might provide a strong indication of the robustness of our approach.

We studied a system of $N=26$ structureless $1/2$-spin fermions
of mass $m_3$, interacting via the Aziz potential described by ref. \onlinecite{dyn:Aziz79},
a very accurate model for the effective interactions among $^3$He atoms. 
The choice of the particle number was inspired by our previous work in Ref. \onlinecite{dyn:prbnoi}:
such number of atoms was chosen to be a closed--shell number; this choice minimizes
size effects related to the discrete Fermi sea, but still allows to extract physical
information from the imaginary time correlation functions, which rapidly become
steeper as the size of the system is increased.

Differently from the work described in the previous chapter, we have 
not used twisted boundary conditions (TBC). This choice is motivated by the fact that the effect of TBC enters 
in the estimation of both the Fermi--Bose gap and the energy of the excited state with respect to the bosonic ground state; 
being the energy of the excitation a difference between these these two quantities, we assumed that, 
as a first approximation,
 the effects of TBC cancel out. In conclusion, considering that the evaluation of the necessary correlation 
functions are required with high quality data, neglecting TBC is a good compromise between 
accuracy and practical computing times.

We have focused on a density around $0.047$ \AA$^{-2}$, close to the
experimental conditions\cite{dyn:krotscheck1}.
Moreover, we have explored the behavior of the sample at the densities $0.038$ and
$0.060$ \AA$^{-2}$ in order to investigate the possible density-dependence of the excitations
of the system. In particular, the highest density was chosen very close to the freezing point.
The QMC evaluation of $F_B$ requires a simple generalization of the methodology that we have
followed in the previous Chapter: a fictitious system of {\it{bosons}} of mass $m_3$ is 
simulated with the Shadow Path Integral Ground State method.
The imaginary--time propagation was 1.3125 K$^{-1}$ and the density matrix approximation was a Pair Product\cite{dyn:ceperley_pp}
with imaginary--time--step of $1/160$ K$^{-1}$.

The Shadow Path Integral Ground State\cite{dyn:spigs} technique was chosen in
the computation of both the bosonic ground--state energy and the correlation functions.

Performing such a QMC simulation, we have computed $F_B(\vec{q},\tau)$ for each wave--vector $\vec{q}$,
together with the correlation function:

\begin{equation}
F_0(\tau) = \frac{\langle \psi_0^B | \mathcal{D}^{\star} e^{-\tau \hat{H}} \mathcal{D} | \psi_0^B \rangle}
{\langle \psi_0^B | \psi_0^B \rangle}
\end{equation}
which is precisely the correlation function that was used in the previous chapter in order to estimate the energy gap between 
the bosonic fictitious system and the fermionic ground state.
$F_A$ has been then estimated from the exact identity:

\begin{equation} \label{sqwrapport}
F_{A}(\vec{q},\tau) = 
\frac{F_{B}\left(\vec{q},\tau\right)}{F_{0}\left(\tau\right)}
\end{equation}


It is well known that, in order to extract information from imaginary time correlation function,
an inversion of the Laplace transform in ill-posed conditions is necessary.
This can be carried out by means of the Genetic Inversion via Falsification of Theories (GIFT)
\cite{dyn:gift}, which has already provided very accurate results in the study of low energy excitations
of Bose superfluids\cite{dyn:gift} and supersolids\cite{dyn:prlsaverio}.
Naturally the problem is unavoidably ill-posed: the quality of the results of the inversion
procedure cannot be guaranteed {\it{a priori}}; however, a test of reliability of the inversion procedure 
can be obtained by comparing our estimations of the dynamic structure factor with experimental data; Fig.~\ref{fig3} 
shows a remarkable agreement.

\begin{figure}[h]
\includegraphics*[scale=0.6]{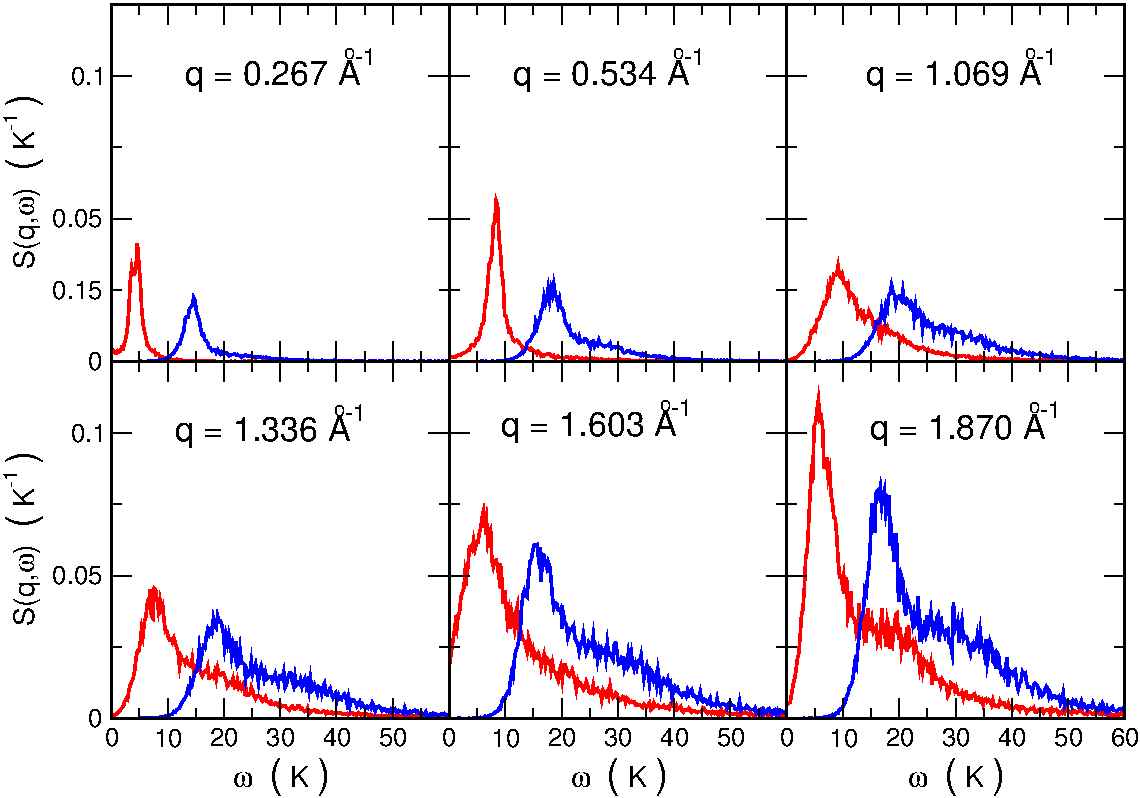}
\caption{Comparison between the spectral functions obtained from $F_A$ and those obtained from $F_B$ for 
some wave--vectors $q$. The two spectral functions have a compatible shape, with a shift in energy of $E_0^F-E_0^B$.
The data for $F_A$, differently from Fig.~\ref{fig3}, has been obtained from the average of six different evaluations of 
$F_A(\tau)$.
}
\label{fig1}
\end{figure}

In Fig.\ref{fig1} we show a comparison between the estimated inverse Laplace transforms of
$F_A$ and $F_B$: apart from a rigid shift in energy, due to the difference
$E_0^F - E_0^B$ which we have estimated in the previous chapter, the
reconstructions coincides within the ``algorithmic resolution'' of the GIFT methodology.
This represents a confirmation for the robustness of our approach.
We remark that $F_B$ is much easier to handle than $F_A$, since it
does not suffer of long-$\tau$ large fluctuations due to the presence of
the $\tau$-dependent denominator in \eqref{sqwrapport}.

The shape of the reconstructed spectral functions depends on the lowest-energy fermionic exact eigenstates 
not orthogonal to the wave function $\hat{\rho}_{-\vec{q}} \,\mathcal{D} \,\,\psi_0^{B}$, a state
containing a density modulation of wave vector $\vec{q}$.

\begin{figure}[t]
\includegraphics*[scale=0.6]{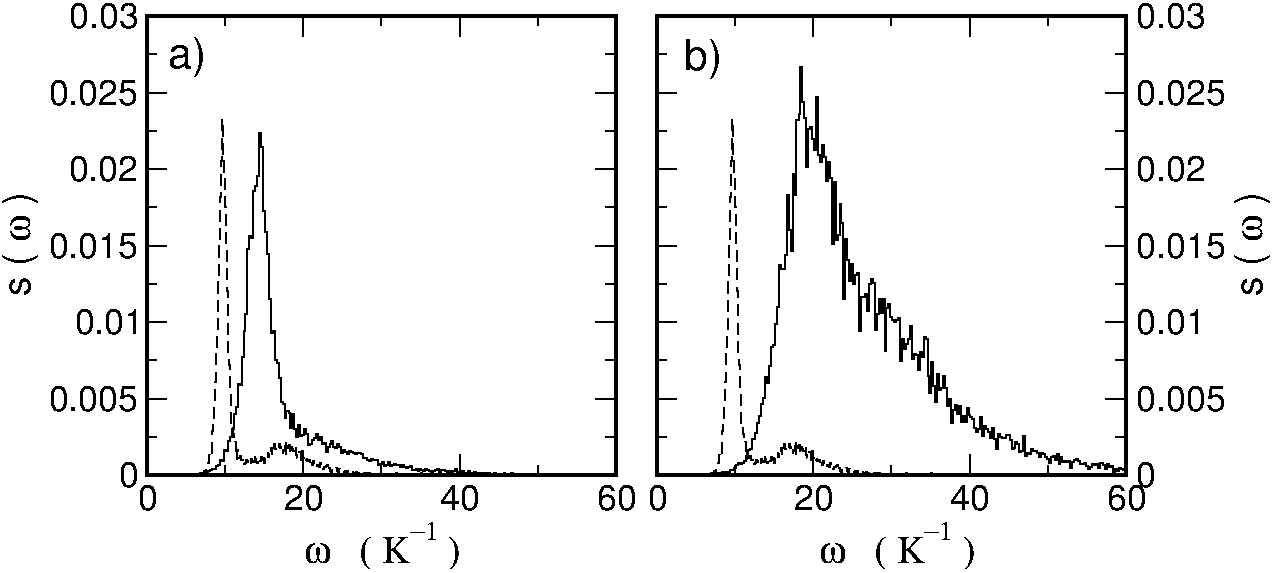}
\caption{Panel a) the gap between the Bosonic and the Fermionic ground state (dashed line) and the inversion
of eq. (\ref{ftex2}) for $q=0.0267$~\AA$^{-1}$ (filled line). Panel b) the same of panel a), but with $q=0.801$~\AA$^{-1}$}
\label{fig2}\end{figure}

Our assumption \eqref{approximation} asserts that such wave function is very similar to $\hat{\rho}_{-\vec{q}} \,\psi_0^{F}$.      
As appears evident in Fig.\ref{fig2}, at low wave vectors the inversions of $F_{B}(\vec{q},\tau)$ 
provide spectral functions with sharp peaks; this provides our microscopic estimation of the {\it zero-sound mode}
dispersion relation, with an ``algorithmic resolution'' similar to that found in 
the previous chapter, where the Fermi Ground State signal was detected, at higher wave vectors the
peaks become much broader. 
We interpret such a broadening as a damping of density fluctuations due to the presence of other excitations,
in particular the particle-hole excitations.

In Fig.\ref{fig3} we show the comparison between our estimation of the dynamic structure
factor of the $^3$He film and the experimental data\cite{dyn:krotscheck1}. The agreement is impressive and gives 
a strong support to the approximation \eqref{approximation}; it is also clear from this comparison that Eq. \eqref{approximation} 
describes accurately also the mechanisms which give rise to 
a broadening of the dynamic structure factor; this is displayed even better 
\begin{figure}[h]
\includegraphics*[scale=0.6]{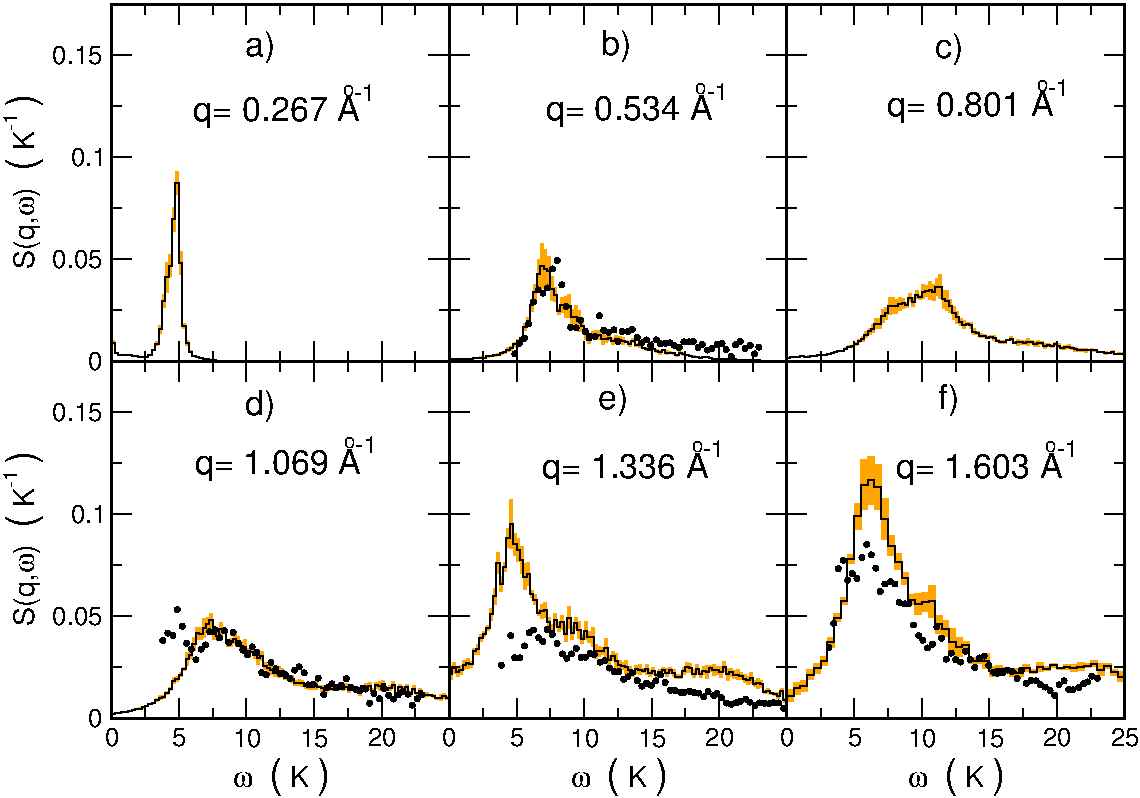}
\caption{From left to right the obtained dynamic structure factor
for increasing wave vectors at $\rho=0.047$ ~\AA$^{-2}$.
The yellow shadow represents statistical uncertainties obtained from six different evaluations of the dynamic structure factor at each wave--vector; filled circles are the 
available experimental data from Ref.~(\onlinecite{dyn:krotscheck1}). The wave--vector shown in picture are those accessible from our simulation, 
the experimental wave vectors are $q = 0.55$~\AA$^{-1}$ (b), $q = 1.15$~\AA$^{-1}$ (d), $q = 1.25$~\AA$^{-1}$ (e) and $q = 1.65$~\AA$^{-1}$ (f).
}
\label{fig3}\end{figure}
in Fig.\ref{fig4}, where we report, in a color plot, the estimated $S_{B}(\vec{q},\omega)$. At low $q$ we
find well defined excitation energies, while, as the wave vector increases,
we observe the sharp mode becoming damped.

\begin{figure*}[t]
\begin{center}
\includegraphics[width=10cm]{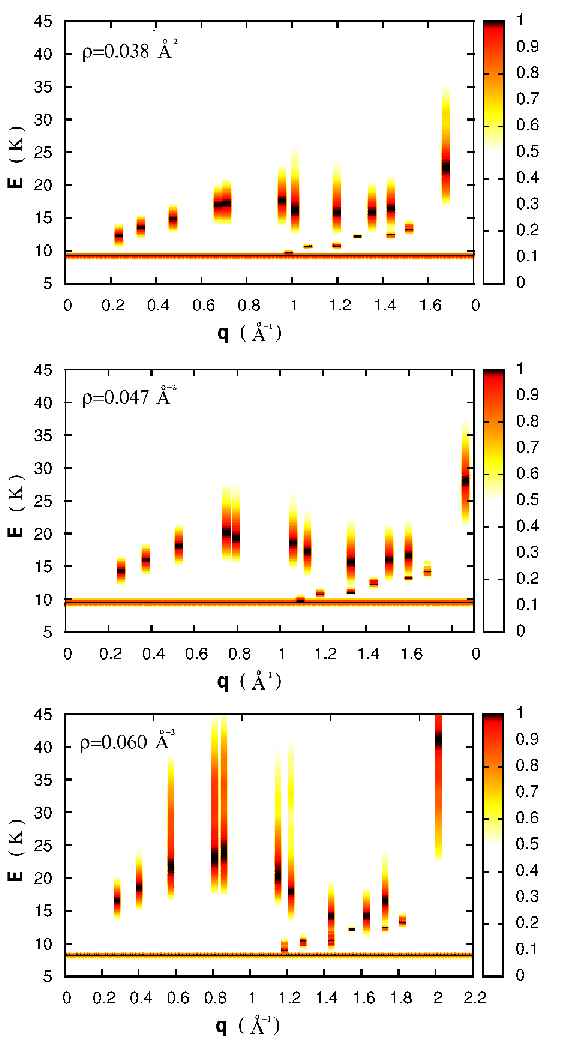}
\caption{The horizontal continuous line represents the energy gap between the Fermionic and the Bosonic ground state,
slightly above - at wave vectors ranging roughly between 1 and 2 \AA$^{-1}$ - the discrete particle--hole
right boundary is shown; the remaining and larger vertical bands are the density--density collective excitations. The bands are centered on the corresponding values and their width
has been enlarged for a better visibility.
 }
\label{fig4}
\end{center}
\end{figure*}
\begin{figure}[h]
\includegraphics*[scale=0.6]{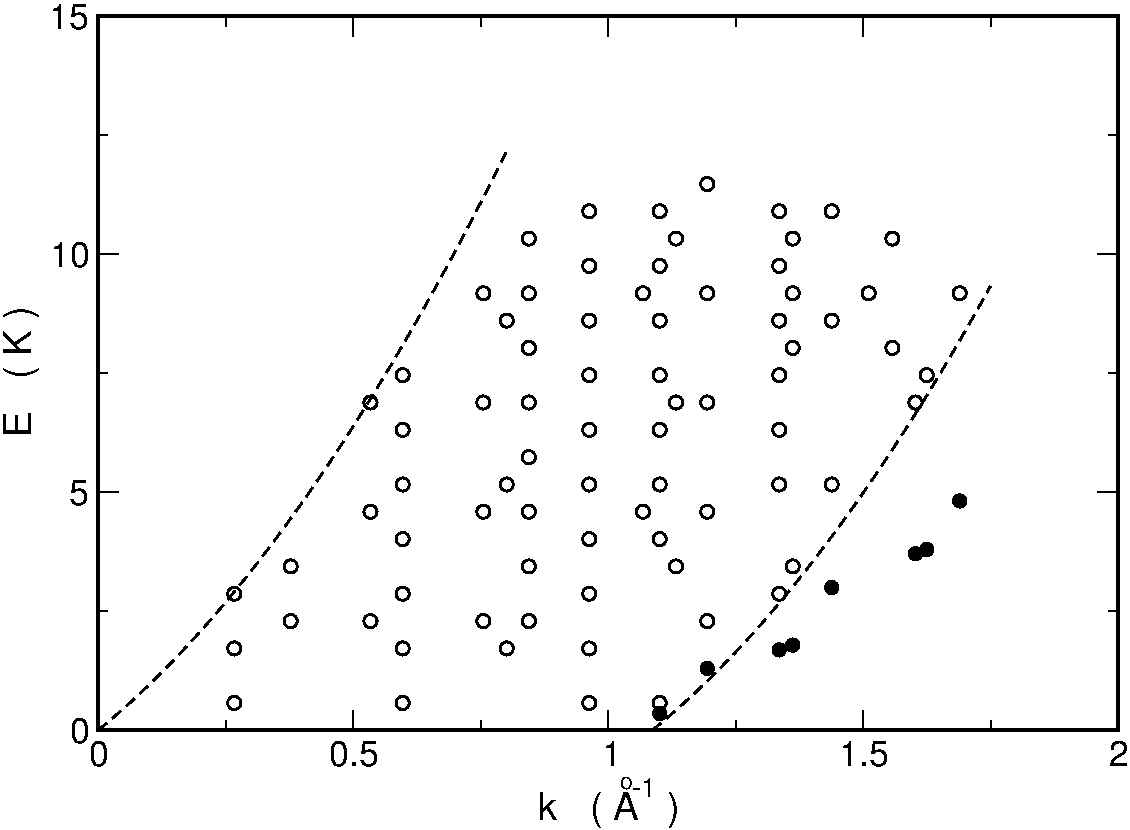}
\caption{(Circles) Particle--hole excitations for $N$=26 non--interacting atoms of $^4$He mass. (Dashed lines) Particle--hole 
band for the ideal gas in the thermodynamic limit. 
 (Filled circles) Particle--hole excitations for $N$=26 atoms of interacting $^4$He. 
}
\label{figcf}\end{figure}

What kind of excitation provide the damping of the zero-sound mode?
Any expert in Fermi liquid theory would immediately answer: the particle-hole
continuum. But what does it mean in an {\it ab-initio} approach to a strongly correlated fermion fluid?
Our idea is to exploit the correlation functions formalism to build up
a ``Fermi--liquid like'' function:

\begin{equation}
\label{cfunph}
F_{ph}(\tau) = \frac{\langle \psi_0^{B}| \mathcal{D}_{ph}^{\star} \, e^{-\tau \hat{H}}\,\mathcal{D}_{ph}| \psi_0^{B} \rangle}{\langle \psi_0^{B}| e^{-\tau \hat{H}}\,| \psi_0^{B} \rangle}
\end{equation}

where $\mathcal{D}_{ph}$ is simply a Slater Determinant like that employed in the previous chapter  
(see Eq. \eqref{sladet}), with $\theta_1 = 0$. Differently from Eq. \eqref{sladet}, in the enumeration 
of the wave--vectors $\lbrace \vec{k}_n \rbrace$, one element has
been taken out of the ideal gas Fermi sea; the bosonic ground state and the backflow
correlations provide a ``dressing'' for the {\it{ph-wave function}} $\mathcal{D}_{ph}\psi_0^{B}$.
In Fig.~\ref{figcf} we show both the particle-hole excitation energies for the ideal Fermi gas,
which do not form a continuum since the system is finite, and the estimated energies
extracted from the inversions of $F_{ph}(\tau)$, which have been evaluated for wave vectors
at the high-$q$ borderline of the structure; we focused on the high-$q$ borderline in order to 
verify whether the roton states reemerges from the particle--hole band or not. 
The particle--hole band of the ideal Fermi gas has been computed following the 
definition of particle--hole energy: from an ideal Fermi gas of $N$ particles of mass $m$ in a square box of late $L$, 
the Fermi Sea is labeled by quantum numbers that define the wave vector of each state, $\vec{k}_{ij}=(2\pi i/L, 2\pi j/L)$;
a particle--hole excitation  characterized by a hole at $(i_0,j_0)$, corresponding to a wave--vector $\vec{k}_{0}$ inside the Fermi sphere, and a particle at $(i_1,j_1)$ 
with wave--vector $\vec{k}_1$ outside the Fermi sphere, has a wave--vector $\vec{q}$ and an energy $E_{\vec{q}}^{ideal}$ defined as: 
\begin{eqnarray}
\vec{q}= \frac{2\pi}{L}\sqrt{(i_1 - i_0)^2 + (j_1 - j_0)^2} \\
E_{\vec{q}}^{ideal} = (E_F + \frac{\hbar^2}{2m}\left|\vec{k}_{2}\right|^2) - (E_F - \frac{\hbar^2}{2m}\left|\vec{k}_{1}\right|^2) = \frac{\hbar^2}{2m}\left(\left|\vec{k}_{2}\right|^2+\left|\vec{k}_{1}\right|^2\right)
\end{eqnarray}
where $E_F$ is the Fermi energy of the system. The particle--hole band of the ideal Fermi gas in Fig.~\ref{figcf} has been obtained following this 
prescription; all the possible particle--hole combinations which gave a wave vector $\vec{q}$ in the displayed range were considered. 
Comparing the ideal particle--hole band with that of the interacting system one can also estimate the effective mass $m^{\star}$; in this context, the effective mass is a parameter 
of the Landau Fermi liquid theory which gives the mass of the quasi--particle; $m^{\star}$ is obtained from the ratio of 
the particle--hole energies of the interacting and the non--interacting systems at the same wave--vector,
\begin{eqnarray}
E_{\vec{q}}^{ideal} = \frac{\hbar^2}{2m}\left|\vec{q}\right|^2 \\
E_{\vec{q}}^{int} = \frac{\hbar^2}{2m^{\star}}\left|\vec{q}\right|^2 \\
\frac{m^{\star}}{m} = \frac{E_{\vec{q}}^{ideal}}{E_{\vec{q}}^{int}} 
\end{eqnarray}
This relation, however, is valid only in the range of applicability of the Landau Fermi liquid theory; in particular, this evaluation of the effective--mass 
holds for small wave--vectors; in this work we give a rough estimate of the effective mass as the average of $E_{\vec{q}}^{ideal}/E_{\vec{q}}^{int}$ 
for each particle--hole wave--vector $\vec{q}$ computed in our simulations. 
In the interacting system, we find in general that the particle-hole energies
become smaller, resulting in an higher effective mass (see Tab.~\ref{tabmstar}). In particular this
has important consequence as far as the re-emergence of the zero-sound mode
from the particle-hole band is concerned: in contrast to what the authors
of Ref.~\onlinecite{dyn:nature} argue, using the non-interacting estimation of the
particle-hole band, we do {\it{not}} observe such re-emergence in the roton region
at any density (see Fig.\ref{fig2}). 

\begin{table}[h]
\begin{center}
\caption{\label{tabmstar} Effective to bare mass ratio estimated from the computed particle--hole excitations in a system of $N=26$ particles at 
the studied densities.}
\begin{tabular}{| c | c |}
\hline
  Density ( \AA$^{-2})$ & $\frac{m^{\star}}{m}$  \\
\hline
0.038 & 1.3(3)   \\
0.045 & 1.8(1)  \\
0.060 & 2.0(1) \\
\hline
\end{tabular}
\end{center}
\end{table}

We point out that this evaluation of the particle hole excitations gives only a first evidence 
that the roton mode is still inside the particle-hole band; a further step on this topic 
consists in a size scaling analysis that has been planned for future works.

\begin{figure}[h]
\includegraphics*[scale=0.6]{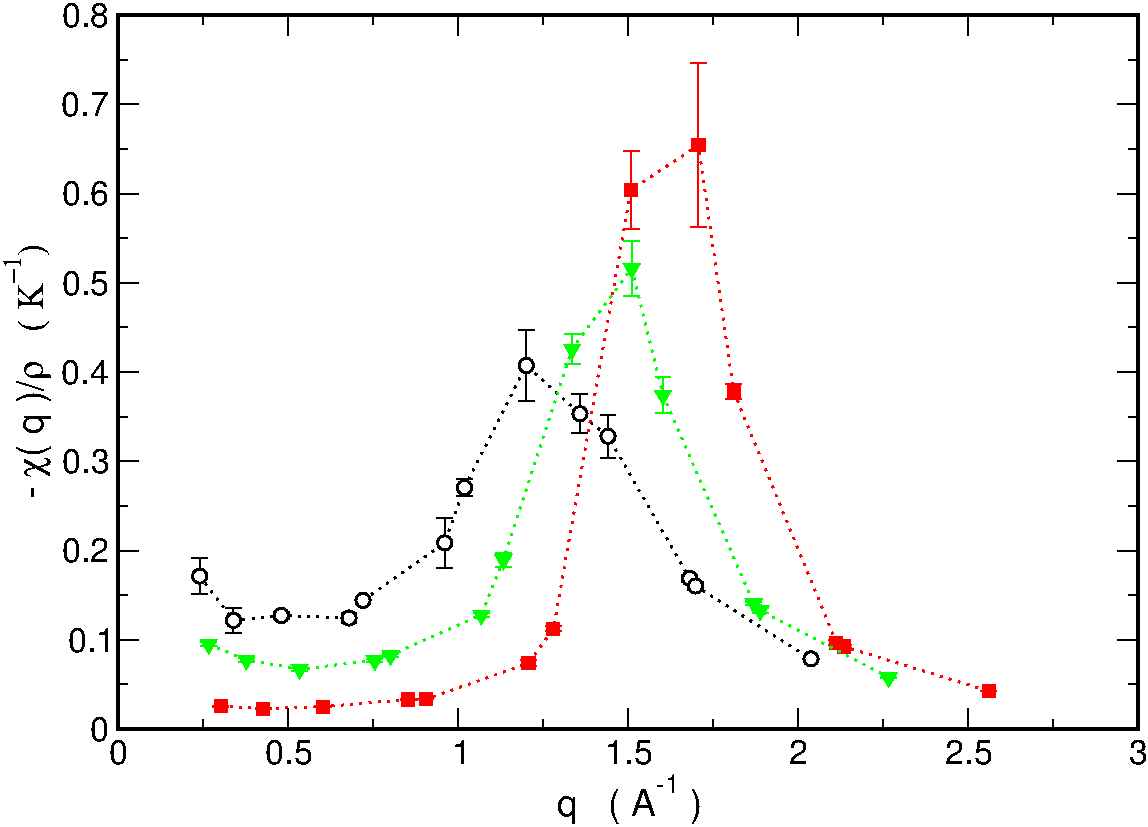}
\caption{The static response function of $^3$He obtained from eq. (\ref{ftex2}). (Circles) $\rho=0.038$ ~\AA$^{-2}$.
(Triangles) $\rho=0.047$ ~\AA$^{-2}$. (Squares) $\rho=0.060$ ~\AA$^{-2}$.
}
\label{ssfcomp}\end{figure}
The static response function can be obtained from the knowledge of the dynamic structure function.
The described method can thus be viewed also as a new and very accurate way to compute
the static response function of a fermion system.
From six independent evaluations of the dynamic structure factor
we computed the static response function defined as
$\chi_{\vec{q}} = -2\rho\int d\omega \:\frac{S\left(\vec{q},\omega\right)}{\omega}$
and the result is displayed in figure \ref{fig3}. 
To our knowledge, this is the first microscopic {\it ab--initio} computation of the static response
function of two--dimensional $^3$He.

As a last result, we note that Eq. \eqref{sqwvar}, for $\tau = 0$ is the definition of 
the static structure factor evaluated on a Fermi state $|\mathcal{D}\Psi_0^B\rangle$. 
This state, as shown in Fig. \ref{fig3}, is a good approximation of the true fermionic ground state; 
this is at least true for the low--energy dynamical properties; we assume that this holds also for the static 
structure factor.
In Fig. \ref{ssfcomp} we show the static structure factor of 2$d$ $^3$He compared with that of the 
fictitious ``bosonic'' $^3$He. The similarity between the two static structure factors is evident; 
this indicates that the structural properties of Helium are dominated by the inter--atomic 
potential rather than the quantum symmetry.

\begin{figure}[h]
\includegraphics*[scale=0.6]{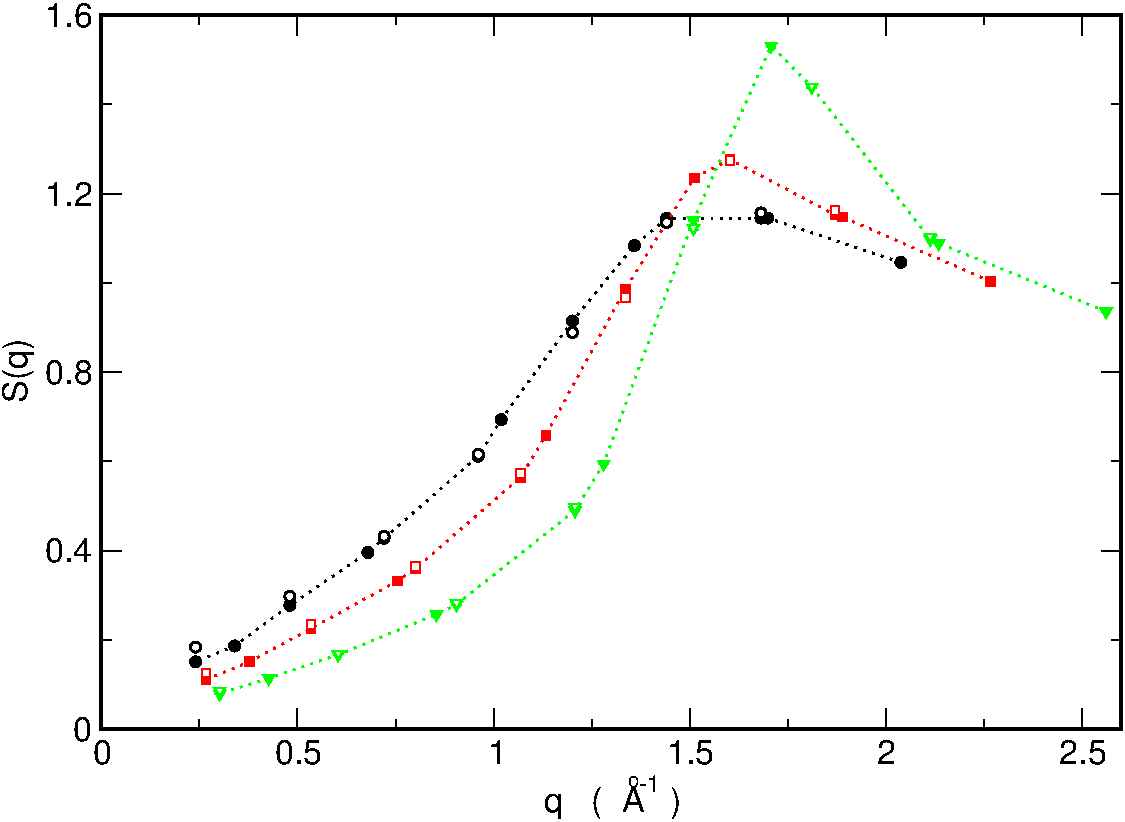}
\caption{ The Static Structure Factor of 2$d$ $^3$He (filled symbols) compared with that of ``bosonic'' 2$d$ $^3$He
 (empty symbols). Data is relative to three densities: $\rho=0.038$ ~\AA$^{-2}$ (Circles);
 $\rho=0.047$ ~\AA$^{-2}$ (Squares) and $\rho=0.060$ ~\AA$^{-2}$ (Triangles). The dashed lines are guides to the eye.
}
\label{fig5}\end{figure}

In conclusion, we have presented the first {\it ab--initio} computation of the zero-sound
excitation energy. We have also proposed an approximate evaluation from first principles of the dynamic
structure factor that is found to be in very good agreement with experimental data\cite{dyn:krotscheck1}.
We employed a well tested methodology involving the Laplace inversion of imaginary--time correlation functions 
\cite{dyn:EttoreSav,dyn:prbnoi} and extended it in order to handle the excited states.
Our results are in agreement with the experimental data and show that our variational 
estimation of the dynamic structure factor is accurate enough to represent the broadening
of the zero--sound mode in the particle--hole band.
At the studied densities we did not observe the re-emergence of the roton mode from the particle--hole band, 
however our data on particle--hole excitations is not yet conclusive: possible finite size effects on 
the particle--hole band have still to be studied with simulations of bigger systems.

%% file: ch-fluorographene/chapter-fluorographene.tex
\def\onlinecite{\cite}

\chapter{Study of $^4$He adsorbed on Graphene--Fluoryde and Graphene--Hydrate}
\label{sec:fluorographene}
At the forefront of current research in condensed matter physics is the study of strongly interacting systems,
with a remarkable variety of phase transitions \cite{prl:ref1}.
The effects of fluctuations are enhanced in low dimensions and in the presence of frustration \cite{prl:ref2}.
These represent some of the motivations for studying adsorption phenomena, where important roles are played
by the gas--gas interaction and the ``tunable'' effect of the substrate.
The surface of graphite has long been a playground for studying two--dimensional (2D) monolayer phases of
classical and quantum gases \cite{prl:ref3}.

Probably the best understood adsorption system is the He monolayer on graphite \cite{lt26:ref1}.
Experiments carried out at the University of Washington ca. 1970 revealed {\it for the 
first time} behavior corresponding to a two--dimensional (2D) gas. More dramatic was 
the appearance of a spectacular peak in the specific heat of $^4$He near $T_c=3$ K. 
This peak, well described by the 3 state Potts model, 
manifested a 2D transition from a high $T$ fluid to a low $T$ commensurate
($\sqrt{3}\times\sqrt{3}$ R30$^o$) phase, providing a benchmark measure of coverage, 
not seen in previous adsorption experiments. This ordered phase 
(at density $\rho_{\sqrt{3}}=0.0636$\AA$^{-2}$) corresponds to atoms localized on second--nearest 
neighbor hexagons. At higher densities near completion of the first 
monolayer ($\rho=0.11$\AA$^{-2}$) an incommensurate 2D triangular solid phase is present; 
the phase diagram at intermediate densities is not yet completely determined. 
A quantitative understanding of the He--graphite interaction was made possible by 
precise scattering measurements of surface bound states and band structures \cite{lt26:ref2,lt26:ref2b}.

The availability of graphene (Gr) and its derivatives like graphane (GH) \cite{lt26:ref3} and 
graphene--fluoride (GF) \cite{lt26:ref4} offers the prospect of novel adsorption phenomena. 

Since Gr is just a single plane of graphite, the symmetry and corrugation are expected to be very similar
in the two cases.
If Gr is rigid, no new phenomena are expected for adsorption on one side of Gr, in comparison with graphite,
\cite{lt26:ref2} and this has been verified by recent quantum simulations of $^4$He \cite{lt26:ref4}.
The situation is different for the derivatives of Gr, graphene-fluoride (GF) \cite{lt26:ref3} and
graphane (GH) \cite{lt26:ref3,lt26:ref3b} that have been recently obtained experimentally.
Because GF and GH have surface symmetries and compositions which are quite different from Gr,
adsorbed gases will have very different properties.

In the next section we show a model adsorption potential for He on GF and GH. We will then show the study of 
 a single $^4$He and $^3$He atom on these substrates, as well as submonolayer films of  $^4$He at coverages 
similar to that ($\rho=0.064$\AA$^{-2}$) of the $\sqrt{3}\times\sqrt{3}$ R30$^o$ state on graphite. In the last section 
the properties at high coverages will be described.

\subsection{Adsorption potential}
\begin{figure}[h]
\begin{center}
\includegraphics*[width=6cm]{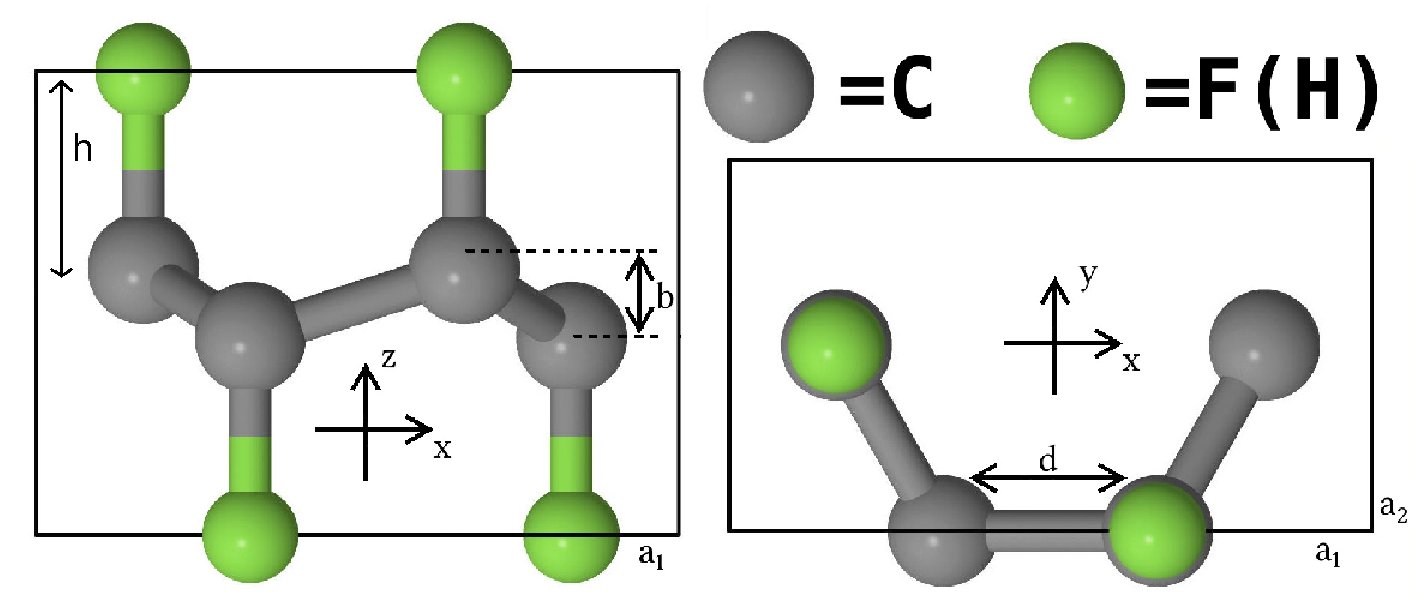}
\caption{Geometry of the substrate with the definitions of the buckling parameter $b$, 
the interplane distance $h$ and $d$, the carbon--carbon distance on the plane}\label{fg:fig1}
\end{center}
\end{figure}

\begin{figure}[h]
\begin{center}
\begin{minipage}{6cm}
\includegraphics*[width=6cm]{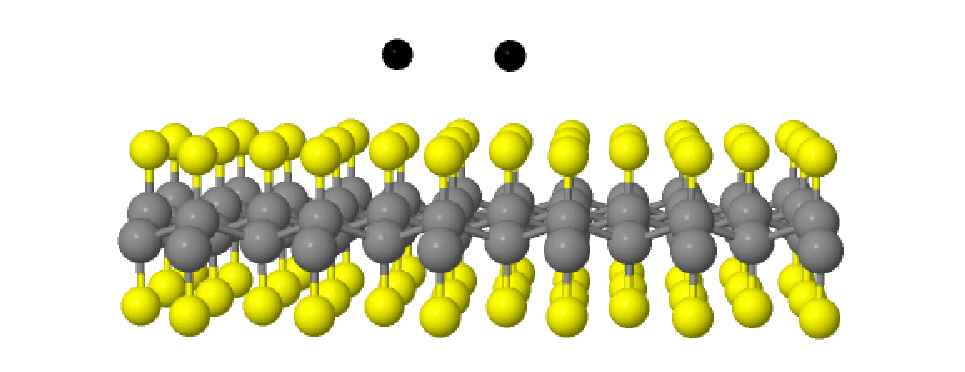}
\end{minipage}\hspace{1cm}
\begin{minipage}{4cm}
\includegraphics*[width=4cm]{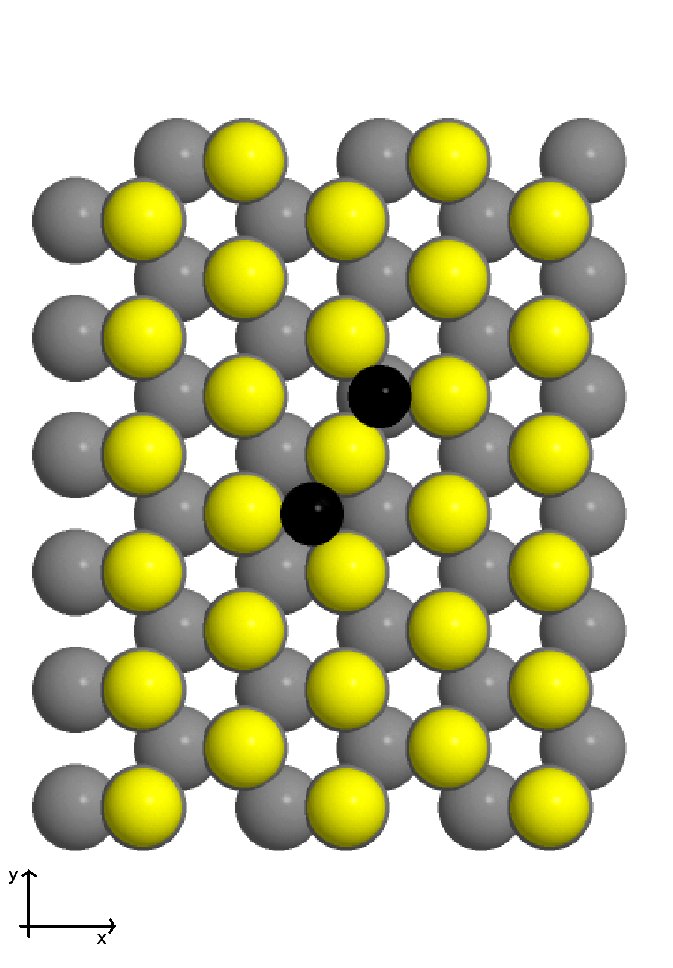}
\end{minipage}\caption{\label{lt26:fig1}
Two schematic views of GF. F (C) atoms are light (dark) gray. Positions of atoms
are to scale but their sizes are arbitrary. The black balls represent two
adsorption sites for He, one of each kind. GH is similar.
}
\end{center}
\end{figure}

Graphane and graphene--fluoride have a similar geometry; half of the H (F) 
atoms are attached on one side of the graphene sheet to the carbon atoms forming 
one of the two sublattices of graphene. 
The other half are attached on the other side to the C atoms forming the other sublattice. 
The H (F) atoms are located on two planes (see left of Fig.~\ref{lt26:fig1}); 
one is an overlayer located at a distance $h$ above the pristine graphene plane 
while the other is an underlayer at a distance $h$ below the graphene plane. 
In addition, as seen in Fig.~\ref{fg:fig1} there is a buckling of the C--plane with the C atoms of one 
sublattice moving upward by a distance $b$ while the other sublattice moves 
downward by the same amount. A He atom approaching GH (GF) from above will 
interact primarily with the H (F) overlayer, but it will interact also with 
the C atoms and the H (F) atoms of the underlayer. 

We have adopted a traditional, semi--empirical model to construct the potential 
energy $V(\vec{r})$ of a single He atom at position $\vec{r}$ near a surface \cite{lt26:ref5,lt26:ref5b,lt26:ref6,lt26:ref7}. 
The potential is written $V(\vec{r}) = V_{\rm rep}(\vec{r})  + V_{\rm att}(\vec{r})$, a sum of a Hartree--Fock 
repulsion derived from effective medium theory, and an attraction, $V_{\rm att}(\vec{r})$, 
which is a sum of damped He atom van der Waals (VDW) interactions and the 
polarization interaction with the surface electric field. The first term 
is $V_{\rm rep}(\vec{r})=\alpha \rho(\vec{r})$.
Here $\alpha=364$ eV--bohr$^3$ is a value derived by several 
workers as the coefficient of proportionality between the repulsive 
interaction and the substrate's electronic charge density $\rho(\vec{r})$ {\it prior} to adsorption. 
The geometry of GH and GF, their electronic charge density and the electrostatic 
potential have been obtained using Density Functional Theory with an all--electron 
triple numerical plus polarization basis set with an orbital cutoff of 3.7 \AA$\:$
as implemented in the DMol3 code \cite{lt26:ref8}. The exchange and correlation potential 
was treated in a Generalized Gradient Approximation parameterized by 
Perdew, Burke, and Ernzerhof \cite{lt26:ref9}. We use a tetragonal unit cell containing 
four C atoms and four H (F) atoms for GH and GF, respectively. 
The cell dimensions for GF are $a_1=2.59$
\AA, $a_2=4.48$
\AA, and $a_3=12$
\AA, while for GH we use $a_1=2.52$
\AA, $a_2=4.36$
\AA, and $a_3=12$
\AA. The Brillouin 
zone was sampled with a Monkhorst--Pack grid of $6\times3\times1$ $\vec{k}$ points in both cases. 
The self--consistent cycles were run until the energy difference was less 
than 10$^{-6}$ eV. The atomic positions were relaxed until the forces on all 
atoms were lower than $0.01$
eV/\AA. As a result, the C--F distance is $1.38$
\AA, the C--C distance $1.57$
\AA, the C--C distance projected on the $x-y$ plane 
is $d=1.495$
\AA $\,$ and the buckling displacement $b=0.484$
\AA; while in GH, 
the C--H distance is $1.11$
\AA , the C--C distance $1.52$
\AA, $d=1.453$
\AA $\,$ and $b=0.45$
\AA. 

The attraction is a sum of contributions; for GH,
\begin{equation}\label{vpotte}
V_{\rm att}(\vec{r}) = V_{\rm H+}(\vec{r}) + V_{\rm gr}(\vec{r}) + V_{\rm H-}(\vec{r})
- \alpha_{\rm He} {\rm E}^{2}(\vec{r})/2 
\end{equation}

\begin{figure}[h]
\begin{center}
\includegraphics*[width=10cm]{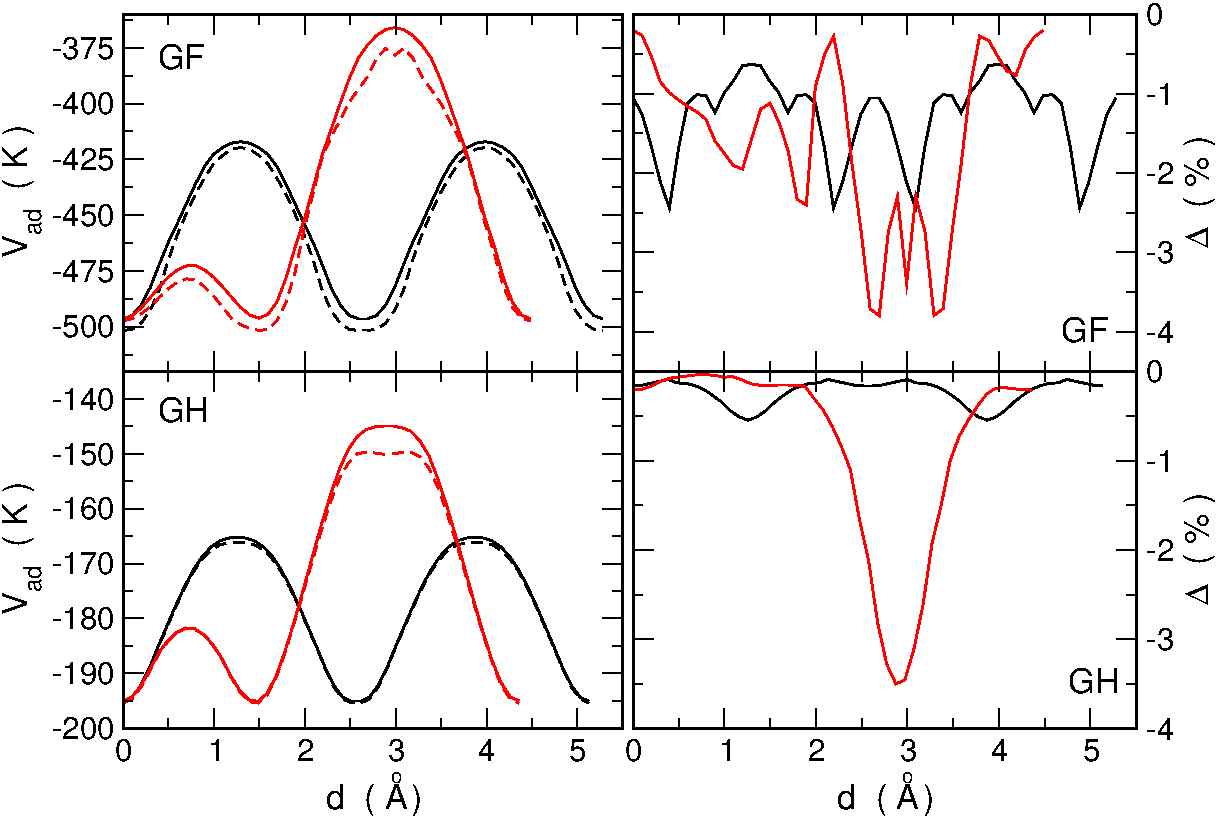}
\caption{
Upper figures: $^4$He on GF. (Left) minimum value with respect to $z$ of $V_{ads}(\vec{r})$ along the direction $(x,0)$ (black) and $(0,y)$ (red). Full lines
 represent the adsorption potential, $V_{a}$, obtained using Eq.~\eqref{vpotte} for the attractive part, dashed lines the adsorption potential, $V_b$, obtained neglecting the induced dipole energy.
 (Right) the relative difference (in percentage) between $V_a$ and $V_b$ with respect to $V_a$, namely $100*(V_b-V_a)/V_a$.
 Lower figures: the same for GH.
}\label{confrnoee}
\end{center}
\end{figure}

The right--most term is the induced dipole energy, where $\alpha_{\rm He}=0.205$ \AA$^3$
is the static polarizability of the He atom and $\vec{E}(\vec{r})$ is the electric field 
due to the substrate. This term gives a minor contribution to the adsorption potential (see Fig.~\ref{confrnoee}) and has 
been neglected in this work.

The three VDW terms for GH originate from the H overlayer, 
the graphene sheet (we are neglecting in this term the small buckling of the 
graphene sheet) and the H underlayer, respectively. These terms are 
described by the attractive part of a Lennard--Jones potential,
\begin{eqnarray}
V_{\rm H-}(\vec{r}) =-\sum_{j}\frac{C_{6H}}{\left|\vec{r}-\vec{r}_j^{\:H-}\right|^6} \label{vdw2}\\
V_{\rm gr}(\vec{r}) =-\sum_{j}\frac{C_{6C}}{\left|\vec{r}-\vec{r}_j^{\:gr}\right|^6} \label{vdw1}\\
V_{\rm H+}(\vec{r}) =-\sum_{j}\frac{C_{6H}}{\left|\vec{r}-\vec{r}_j^{\:H+}\right|^6} \label{vdw3}
\end{eqnarray}
where the sum spans over the carbon or hydrogen positions; $C_{6C}$ and $C_{6H}$ are respectively 
the Helium--Carbon and the Helium-Hydrogen VDW coefficients.
Due to the distance between the helium monolayer and the graphene plane, Eq. \eqref{vdw2} 
can be integrated over the $x$--$y$ plane
\begin{eqnarray} \label{grcoe}
V_{\rm H-}(\vec{r}) = -\theta_{H-} C_{6H}\int d^2\vec{R}\frac{1}{\left((z+\tilde{h})^2+\left|\vec{R}\right|^2\right)^3} = \nonumber \\
-\left(\frac{\theta_{H-}\pi C_{6H}}{2}\right)(z+\tilde{h})^{-4} = -\frac{A_C}{(z+\tilde{h})^4}
\end{eqnarray}
where $\theta_{H-}$ is the Hydrogen density of the sublayer on the $x$--$y$ plane and 
$\tilde{h}=h+b/2$ is the distance between the Hydrogen underlayer and the mean Carbon plane (see Fig.~\ref{fg:fig1}).
The He--H VDW coefficient, $C_{6H}$, is obtained from the VDW interaction of He--H$_2$. 
In Ref.~\onlinecite{vidaliprl}, the potential energy between He and H$_2$ is written as a sum of 
potential terms regarding the interaction of Helium with each single Hydrogen,
\begin{eqnarray}
U_{He-H_2}(\vec{r},\vec{R}_1,\vec{R}_2) = U_{He-H}(|\vec{r}-\vec{R}_1|)+U_{He-H}(|\vec{r}-\vec{R}_2|)
\end{eqnarray}
where $\vec{R}_1$ and $\vec{R}_2$ are the positions of the Hydrogen atoms. 
Considering an attractive VDW term, $-C_{6H}/r^6$,
 for each Hydrogen atom and neglecting the structure of the H$_2$ molecule, we approximate the attractive 
 part of $U_{He-H_2}$ as an isotropic VDW interaction $C_{6H_2}/r^6$. We then have:
 \begin{eqnarray}
 C_{6H} = \frac{C_{6H_2}}{2}
 \end{eqnarray}
The value of $C_{6H_2}$ is obtained from Ref.~\onlinecite{tonnies84}; in that work the anisotropy of 
$U_{He-H_2}$ is also studied. The work shows that the leading term of the long range attractive 
part of the He--H$_2$ potential is 
\begin{eqnarray}\label{anisotropicheh2}
U_{He-H_2}^{att}(\vec{R},\gamma) = -\sum_{n\ge3}\frac{C_{2nH_2}}{|\vec{R}|^{2n}}\left[1+\Gamma_{2n}P_2(\cos \gamma)\right]
\end{eqnarray}
where $\vec{R}$ is the distance from the center of mass of H$_2$ and $\gamma$ the angle between $\vec{R}$ and the 
axis of the molecule. The leading term of Eq.~\eqref{anisotropicheh2} is exactly the VDW term $C_{6H_2}/r^6$;
the anisotropy is described by the Legendre term $P_2$; in particular, Ref.~\onlinecite{tonnies84b} shows that  
the leading anisotropic correction involves $\Gamma_6$ which is of order 0.1. This quantity, even though 
might have some relevance, has been neglected: its effects can be taken into account; however, the 
results in this chapter may not change qualitatively; this has been checked with an arbitrary change of the VDW 
parameters for the interaction potential of both He--GF and He--GH. Moreover, this work is based on a semi--empirical 
adsorption potential and it is not aimed to obtain quantitative data.
\begin{figure}[h]
\begin{center}
\includegraphics*[width=10cm]{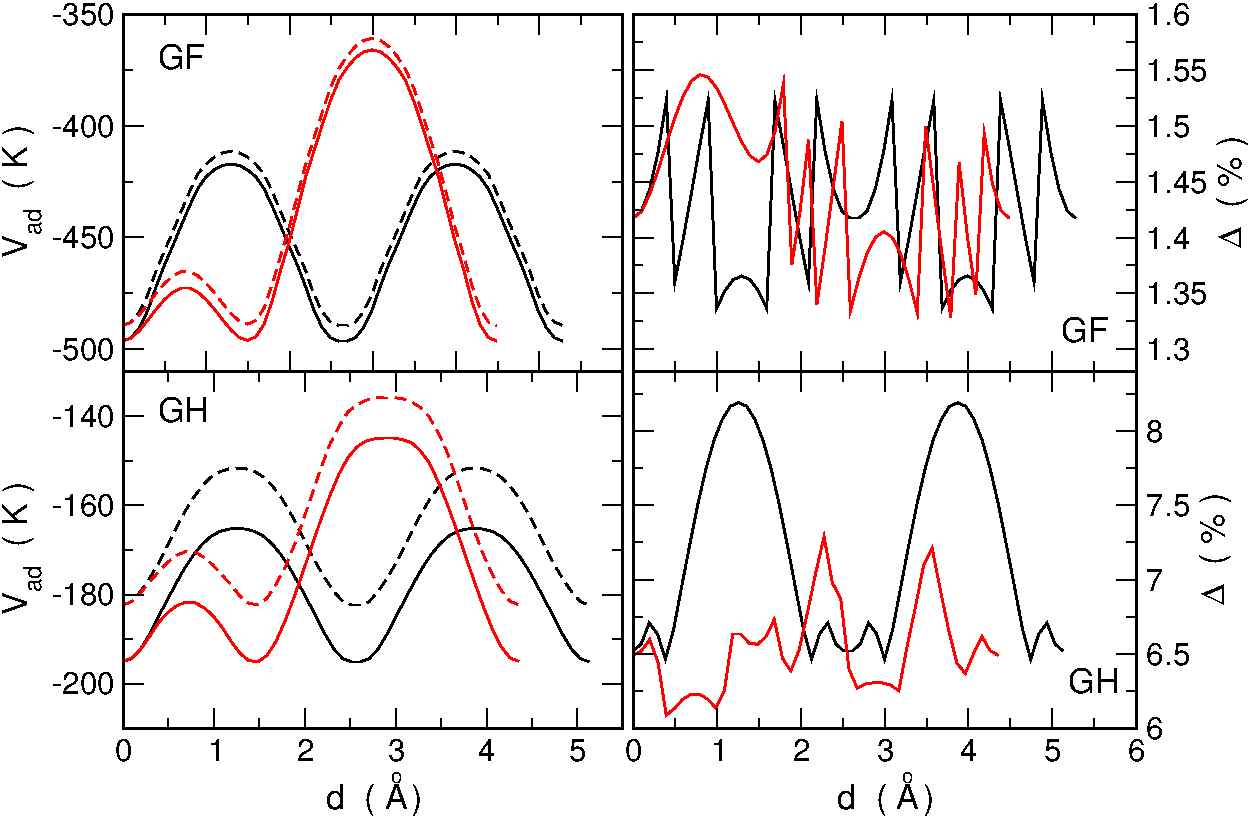}
\caption{
Upper figures: $^4$He on GF. (Left) minimum value with respect to $z$ of $V_{ads}(\vec{r})$ along the direction $(x,0)$ (black) and $(0,y)$ (red). Full lines
 represent the adsorption potential, $V_{a}$, obtained using Eq.~\eqref{vdw1}, dashed lines the adsorption potential, $V_b$, obtained with the approximation~\eqref{fmcoe}.
 (Right) the relative difference (in percentage) between $V_a$ and $V_b$ with respect to $V_a$, namely $100*(V_b-V_a)/V_a$.
 Lower figures: the same for GH.
}\label{figconspal}
\end{center}
\end{figure}

The term $V_{\rm gr}(\vec{r})$ can be treated in an analogous way of the term $V_{\rm H-}(\vec{r})$,
\begin{eqnarray}\label{fmcoe}
V_{\rm gr} (\vec{r}) = -\theta_{gr} C_{6C}\int d^2\vec{R}\frac{1}{\left(z^2+\left|\vec{R}\right|^2\right)^3} = 
-\left(\frac{\theta_{gr}\pi C_{6C}}{2}\right)z^{-4} = -\frac{A_C}{z^4}.
\end{eqnarray}
This approximation, however, neglects the buckling of the Carbon atoms and gives rise to variations of 8\% in the 
case of GH (see Fig.~\ref{figconspal}). Apart from a shift in energy, we don't expect that this approximation 
would change the qualitative behavior of the system; in fact, as can be seen in Fig.~\ref{figconspal}, the larger differences 
are at the maxima of the adsorption potential, where the Helium density is lower. Nevertheless, in this work we preferred to use 
directly Eq.~\eqref{vdw1}. The Helium--Carbon VDW coefficient appearing in this equation, $C_{6C}$, 
can be determined from the known\cite{lt26:ref6} VDW coefficient $C_3 = 180$ meV--\AA$^3$ 
for an Helium atom interacting with a half--space of graphite, $V(z) \simeq -C_3 z^{-3}$. 
This is connected to the VDW interaction $\simeq -C_{6C}r^{-6}$ by an integral over the half--space; 
in a way similar to Eq. \eqref{grcoe} we find the relationship between $C_3$ and $C_{6C}$,
\begin{eqnarray}\label{eqc3}
C_3 = \frac{\pi}{6}nC_{6C}
\end{eqnarray}
where $d=3.4$~\AA~ is the distance between two carbon planes of Graphite; $n=\theta/d$ 
is the three dimensional density of the half plane of Graphite.
Using Eq.~\eqref{vdw1} and Eq.~\eqref{eqc3} one obtains that $A_C = 3C_{3} d =1.84$
eV--\AA$^4$.

The term in Eq.~\eqref{vdw3}, $V_{\rm H+}(\vec{r})$, gives the main attractive contribution; not only this term 
can not be integrated along the $x$--$y$ plane but its proximity with the Helium monolayer requires  
the use of damping; we have adopted the Tang--Toennies damping procedure for this situation \cite{lt26:ref7}; 
the term in Eq.~\eqref{vdw3} thus becomes
\begin{eqnarray}
V_{\rm H+}(\vec{r})= -\sum_j V_{damp}(|\vec{r}-\vec{r_j}|) \nonumber \\
V_{damp}(x) = C_{6H}\frac{1-e^{-\beta x}\sum_{n=0}^{6}\frac{\left(\beta x\right)^n}{n!}}{x^6}.
\end{eqnarray}
 The function $V_{damp}$ has an asymptotic $x^{-6}$ behavior; for small values of $x$ the divergence is cured; this can 
 be seen with a Taylor expansion of $e^{-\beta x}$. Following Ref.~\onlinecite{lt26:ref7}, the parameter $\beta$ is the 
 decay coefficient in the asymptotic charge density $\rho_H(r) \simeq e^{-\beta r}$ due to the H atom; for 
 Hydrogen $\beta= 3.78$~\AA$^{-1}$.

In the case of GF, $V_{\rm att}(\vec{r})$ has an expression similar to Eq.~\eqref{vpotte} with $V_{\rm H+}$ and 
$V_{\rm H-}$ replaced by $V_{\rm F+}$ and by $V_{\rm F-}$. The same procedure used for GH applies with the coefficient 
$C_{\rm 6F}=4.2$
eV--\AA$^6$
as given by Frigo et al \cite{lt26:ref11},
$\beta=3.2$
\AA$^{-1}$
and $A_{\rm F}=1.1$
eV--\AA$^4$.

\begin{table}[h]
\caption{\label{ad:tabpar} Parameters for the adsorption potential of He on GH and GF}

\begin{center}
\begin{tabular}{ | c | c | c |}
\hline
  Parameter & Value & Type \\
\hline
$C_{6F}$ & 4.2 $eV$\AA$^6$& GF \\
$C_{6H}$ & 1.206 $eV$\AA$^6$& GH \\
$C_{6C}$ & 3.447 $eV$\AA$^6$& GF/GH \\
$A_{F}$ & 1.1 $eV$\AA$^4$& GF \\
$A_{H}$ & 0.3455 $eV$\AA$^4$& GH \\
$\beta_{F}$ & 3.2125 \AA$^{-1}$& GF \\
$\beta_{H}$ & 3.77945 \AA$^{-1}$& GH \\
$\gamma$ & 53.9392 $eV$\AA$^{3}$& GF/GH \\

\hline

\end{tabular}
\end{center}
\end{table}

The obtained adsorption potential relies on the electron density of the substrate. This quantity is 
the output of a DFT computation and is in the form of a 3$d$ table formatted as $i\delta x,j\delta y,k\delta z,value$, 
where $\delta x, \delta y, \delta z$ is the spatial discretization of the electron density table. 
The adsorption potential that enters as input in the QMC simulations is thus a 3$d$ table; this 
table is read in the simplest way: if the simulation box is a cube in the region with positive coordinates, 
a position $\vec{r}=(x,y,z)$ corresponds to a bin 
in the table defined by $(a,b,c)$, where $a=x/dx$, $b=y/dy$ and $c=z/dz$.

\begin{figure}[h]\label{figeq} 
\begin{center}
\begin{minipage}{7cm}
\includegraphics*[width=7cm]{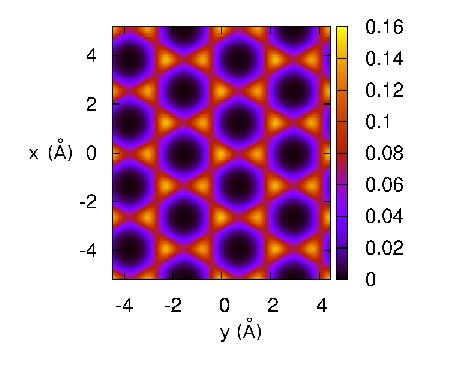} 
\end{minipage}
\begin{minipage}{7cm}
\includegraphics*[width=7cm]{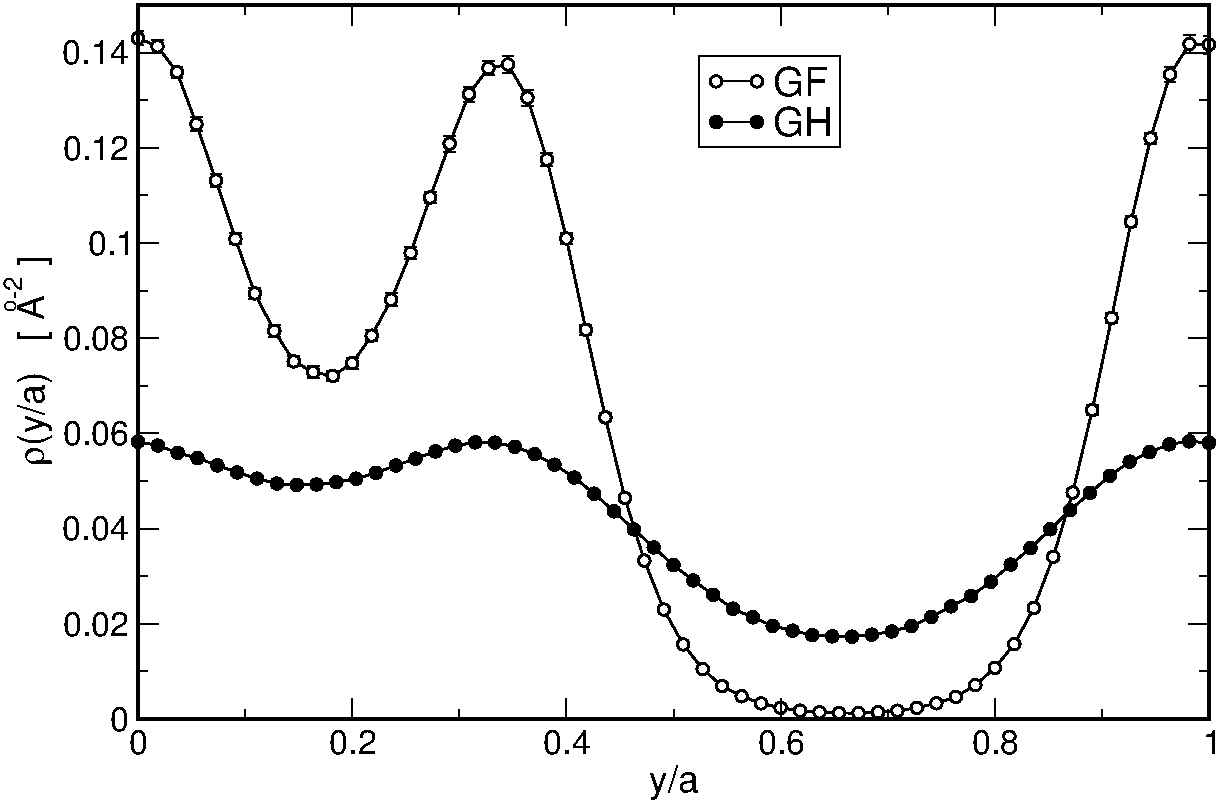} 
\end{minipage}
\caption{Upper panel: local density $\rho (x,y)$ (in units of \AA$^{-2}$) as function of
$x$--$y$ for $^4$He at equilibrium density on GF.
Lower panel: local density $\rho (x = 0, y)$ (in units of \AA$^{-2}$) in the
unit cell (with side $a=4.486$~\AA in the GF case and $a=4.347$~\AA in the GH
case) along the $y$ direction for $^4$He at equilibrium density on GF and GH;
note the logarithmic scale used for $\rho(x=0,y)$. Error bars are below the 
symbol size; lines are guides to the eye.
}
\end{center}
\end{figure}

\begin{figure*}[h] 
\begin{center}
\includegraphics*[width=10cm]{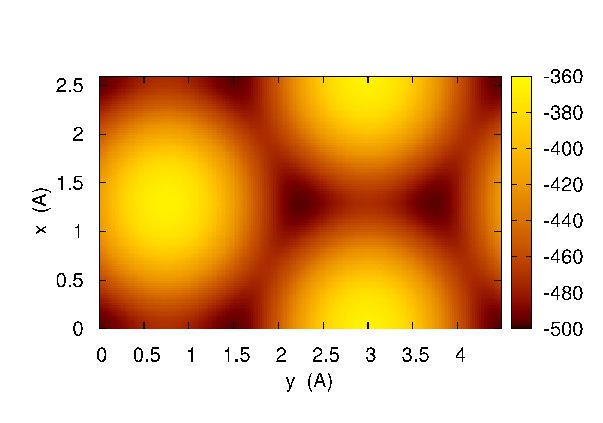}\\
\includegraphics*[width=10cm]{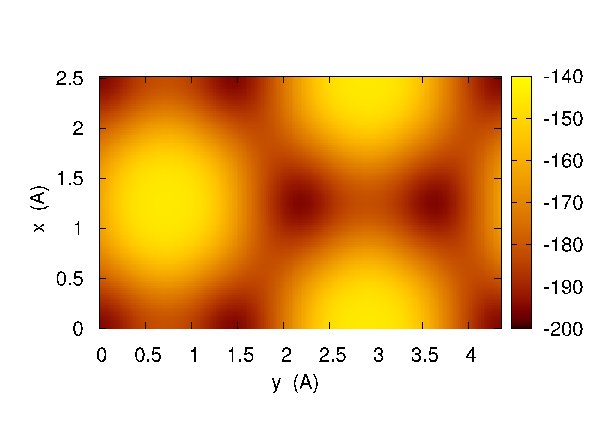}
\caption{\label{ad:adpotl}
Upper panel: plot of the minimum value with respect to $z$ of the adsorption potential He--GF, $V\left({\vec r}\right)$, in K as function of $x$--$y$.
Lower panel: the same for GH.}
\end{center}
\end{figure*}

\begin{figure*}[th] 
\includegraphics*[width=12cm]{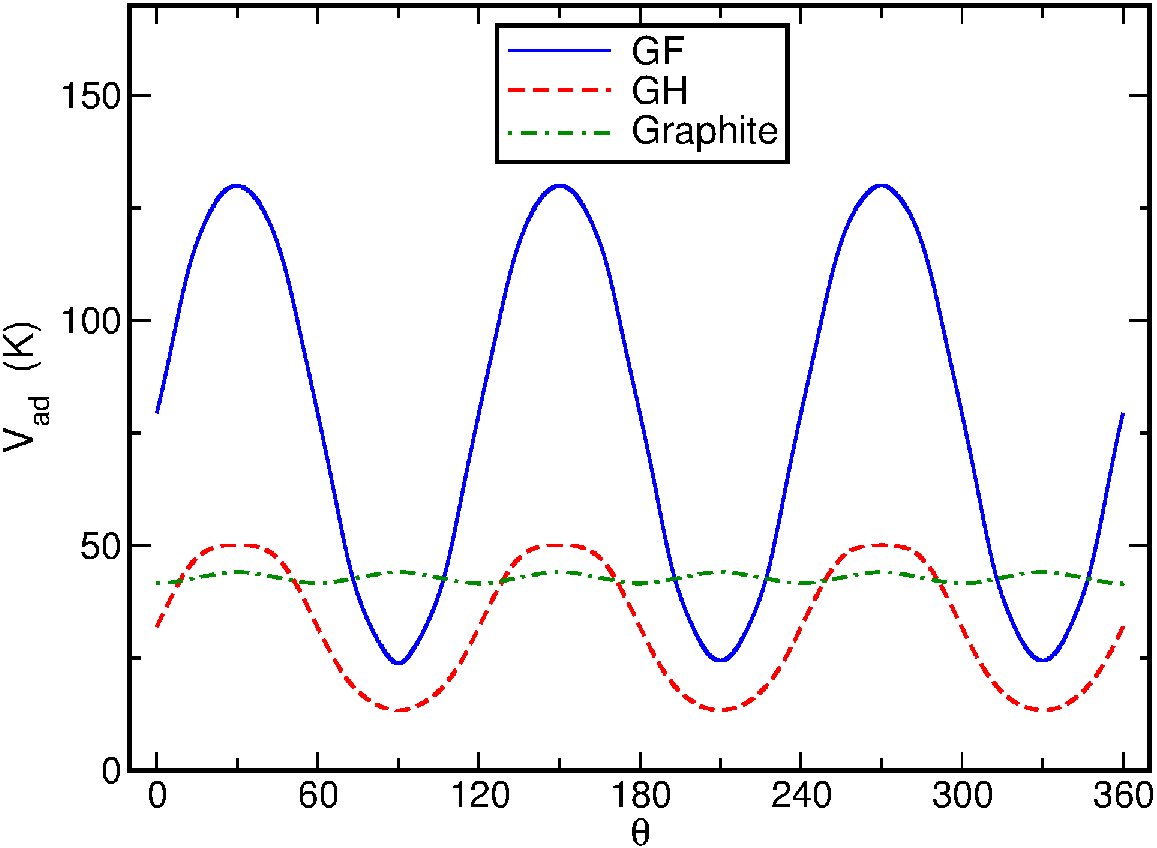}
\caption{\label{ad:corrug}
Energy barrier in GF, GH and graphite as atom moves along a line making an angle $\theta$ with the
horizontal direction and following the height $z(x,y)$ giving the minimum of $V\left({\vec r}\right)$.
Plotted energy is relative to energy at the adsorption site.} 
\end{figure*}

With such model potentials the adsorption sites (see Fig.~\ref{ad:adpotl}) are above the 
centers of each triplets of H (F) atoms of the overlayer, forming a honeycomb 
lattice with the number of sites equal to the number of C atoms, twice as 
many as those on Gr. Half of the sites are above H (F) of the underlayer 
but the difference between the well depths for the two kinds of adsorption 
sites is very small, below 1\%.  For GF the well depth is 498 K and for 
GH it is 195 K. These values do not include the induced dipole energy which 
gives a contribution below 1\%. The inter--site energy barrier is 24 K for 
GF and 13 K for GH. Both values are significantly smaller than the barrier 
height 41K for graphite. In this last case, as shown in Fig.~\ref{ad:corrug}, the energy barrier does not 
depend much on the direction in the $x-y$ plane whereas in the case of GF and 
GH the ratio between maximum and minimum barrier height in the $x-y$ plane is 
of order of 4--5: the energy landscape of the two last substrates is 
characterized by a very large corrugation with narrow channels along which 
low potential barriers are present. The motion of the He atom, especially 
in the case of GF, essentially visits only these channels , as though the atom moves 
in a multiconnected space; this is seen in Fig.~\ref{figeq}. Another significant difference is that the 
distance between two neighboring sites is $1.49$
\AA $\,$ for GF and $1.45$
\AA $\,$ for GH whereas it is $2.46$
\AA $\,$ for graphite and for Gr.
Prior to these studies, graphite was believed to be the most attractive 
surface for He, with a well--depth a factor of 10 greater than that on the 
least attractive surface (Cs). The present results reveal GF 
to replace graphite, since its well is a factor of 3 more attractive.

\subsection{The QMC parameters}
The computations in this chapter are based on the Path Integral Ground State\cite{lt26:ref12} (PIGS) 
and the Path Integral Monte Carlo (PIMC) methods\cite{pimccep}.  
As widely explained in the first chapter, with these methods we can compute quantum averages  
of the system at respectively zero and finite temperature; the PIGS
method uses the quantum evolution in imaginary--time $\tau$ of a trial 
wave function $\Psi_t$. If $\Psi_t$ is not orthogonal to the ground state, and $\tau$
is sufficiently long, the quantum evolution purges from $\Psi_t$ the 
contributions of the excited states, yielding the ground state energy and 
wave function. A valuable feature of the PIGS method is that it is exact, 
in principle; the results are independent of $\Psi_t$ \cite{lt26:ref13} and systematic errors 
may be reduced below the statistical uncertainty. The PIMC method applies the Path Integral formalism 
to the quantum thermal average expressed in coordinates representation; this expression is then evaluated 
with Monte Carlo methods; these methods are extensively explained in Chapter~\ref{ch:methods}.

\begin{figure}[h] 
\includegraphics*[width=12cm]{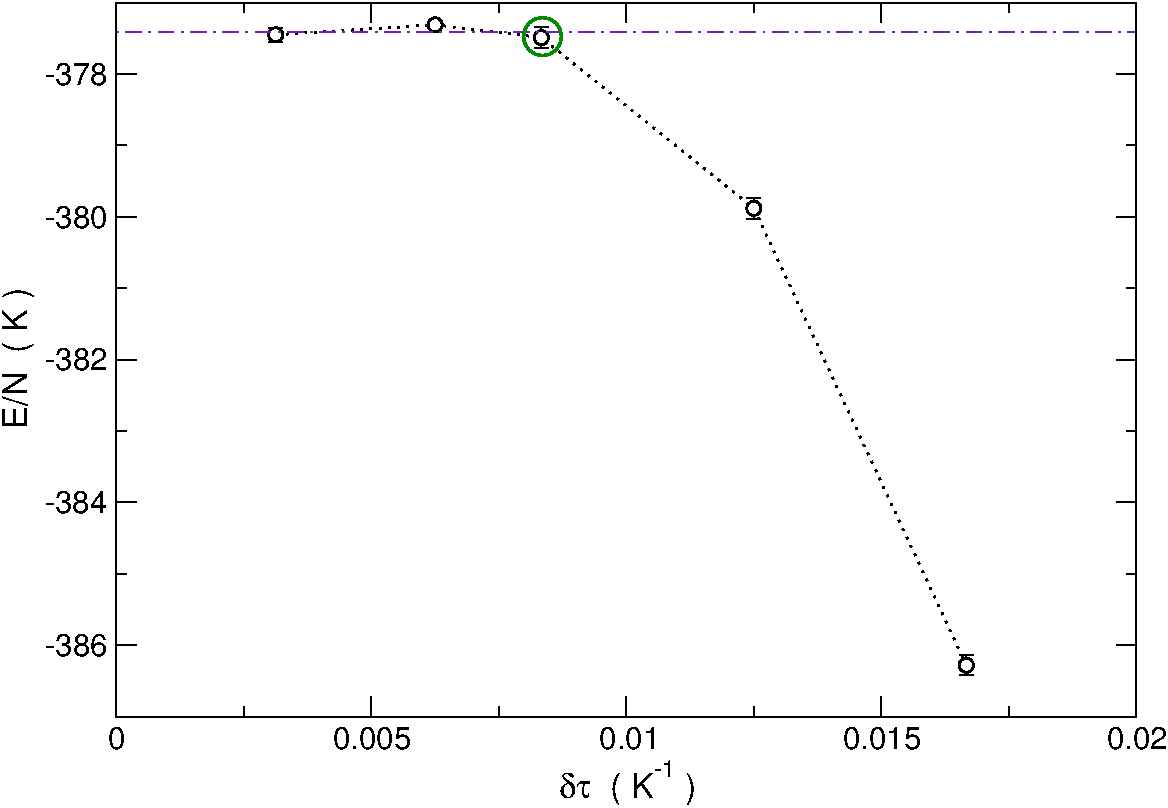}
\caption{\label{ad:figconv}
Convergence of the total energy per particle versus time--step $\delta\tau$ for a system of 
$N=26$ atoms of $^4$He on GF at equilibrium density. The horizontal line is the convergence value taken as the 
average of the energy at the three smallest timesteps. The green circle represents the used 
time--step, $\delta\tau=1/160$~K$^{-1}$.} 
\end{figure}

Both the zero and finite temperature simulations on GF and GH have been performed with the eight order Multi Product 
Expansion of the small imaginary--time propagator; the imaginary--time discretization is $\delta\tau=1/160$~K$^{-1}$, 
which gives a sufficiently accurate approximation of the propagator; an example in the case of GF is given in 
Fig.~\ref{ad:figconv}. 
Due to the computational complexity of PIMC, especially at low temperatures 
(i.e. 0.5 K), the simulations at finite temperature performed in order to obtain 
an estimation of the superfluid density were carried out with $\delta\tau = \frac{1}{60}$ K$^{-1}$;
this choice does not lead to quantitative results, however we note once again that this work is based 
on a semi-empirical adsorption potential; moreover, a quantitative study of the superfluid fraction 
would require extensive size scaling in order to determine the normal--to--superfluid transition; 
the aim of these simulation is thus to give a preliminary estimate 
of the superfluid fraction in order to determine {\it if} there is superfluidity rather than the exact 
density.

The trial wave function $\Psi_t$ that we have used in PIGS is the product of a Jastrow-McMillan wave function 
and a Gaussian along the $z$--direction
\begin{eqnarray}
\Psi_{t} = e^{-\sum_{i<j=1}^{N}\left(\frac{b}{r_{ij}}\right)^{m}}e^{-A\sum_{i=1}^{N}\left(z_0-z_i\right)^2}
\end{eqnarray} 
where $N$ is the particle number and $r_{ij}=\left|\vec{r}_{i}-\vec{r}_{j}\right|$ is the distance between two 
atoms labeled $i$ and $j$.
 he Jastrow parameters are $b=2.84$\AA~ and $m=5$. The Gaussian along the $z$--direction (i.e. the direction perpendicular 
 to the substrate plane) was used only far away from the layer promotion density; its parameters have been obtained with
  a fit of the density along the $z$--direction, for GF $A=5.6$\AA$^{-2}$, $z_{0}=3.72$\AA, for GH $A=3.0$\AA$^{-2}$ and
   $z_{0}=3.85$\AA.  
   
At high densities, where the probability to occupy the second layer is not negligible, a Jastrow wave function 
has been used,
\begin{eqnarray}
\Psi_{t}^{hd} = e^{-\sum_{i<j=1}^{N}\left(\frac{b}{r_{ij}}\right)^{m}.}
\end{eqnarray} 
    
With these trial wave functions, we allowed a $\delta\tau=0.15$ K$^{-1}$ imaginary--time projection before
computing the ground--state expectation values. The total imaginary--time sampled in our calculations 
was $\tau=$0.4 K$^{-1}$. The value 
of $\tau$ has been chosen following a convergence test of the total energy versus the imaginary--time projection.

The Worm algorithm\cite{prokofev} was used at both finite and zero temperature respectively for the sampling of the 
permutations and the computation of the one body density matrix.
 
The computations required on average 10$^5$ Monte Carlo steps, the heaviest computations were those made for the superfluid
fraction at zero temperature and required approximately 10$^7$ Monte Carlo steps.

\subsection{A single Helium atom on the substrates}

\begin{figure}[h]
\begin{center}
\includegraphics*[width=10cm]{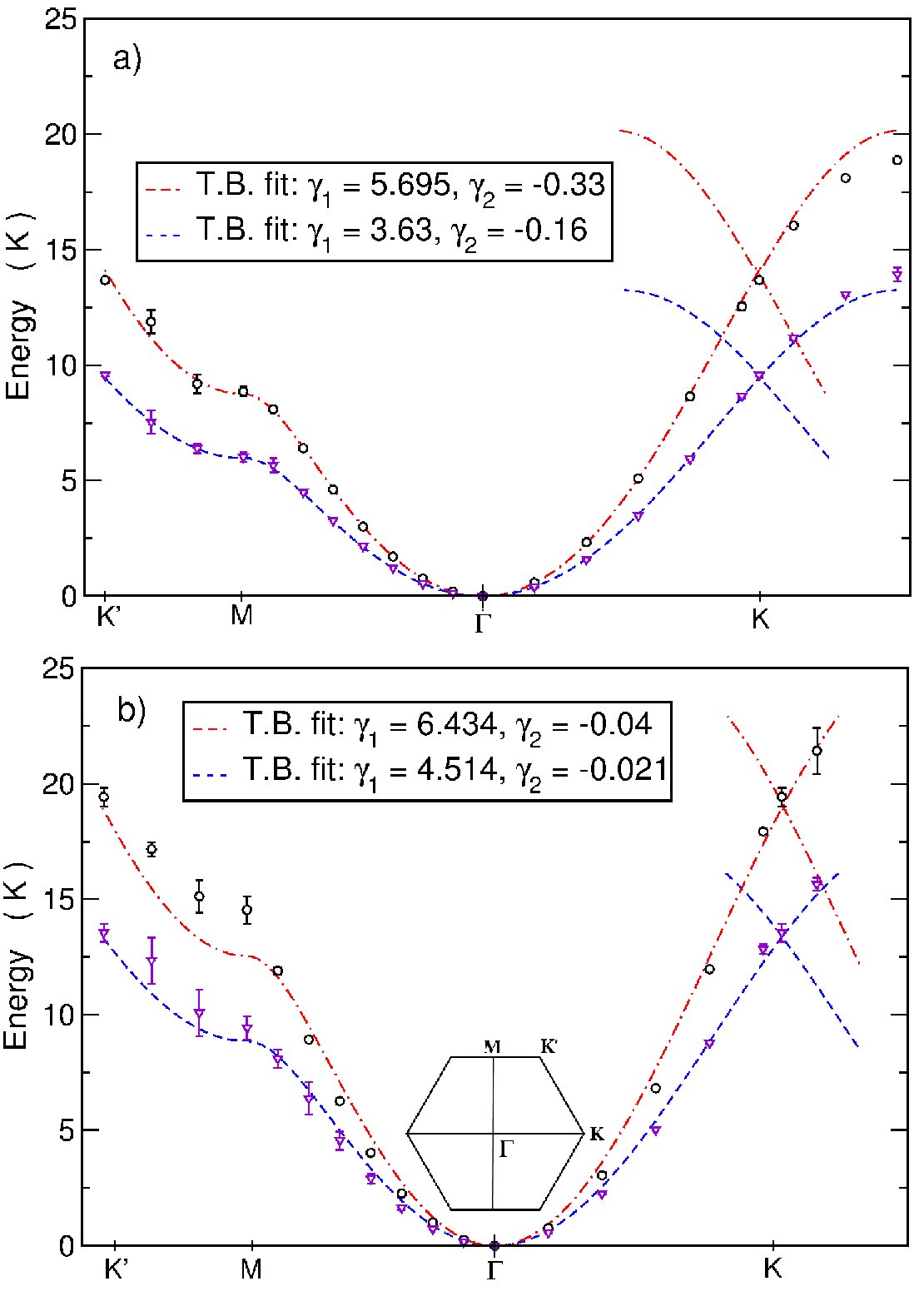}
\caption{\label{lt26:fig2}
Panel a) Energy of the first band along two directions of the first Brillouin zone 
for $^4$He (triangles) and $^3$He (circles) on GF. Data along $\Gamma K$ beyond the 
Dirac point $K$ give the results in the II Brillouin zone. The dashed lines are fits 
made with the tight binding model on a honeycomb lattice with nearest neighbor parameter 
$\gamma_1$ and next nearest neighbor parameter $\gamma_2$ as in legend.
Panel b) Same as for panel a) for $^4$He and $^3$He on GH.
}
\end{center}
\end{figure}

We computed the exact ground 
state energy of one $^4$He atom or one $^3$He atom on GF and GH, see 
Table \ref{lt26:table1}. The binding energy on GH is similar to that on graphite, 
whereas that on GF is about three times that on graphite.  In both cases 
the ground state is delocalized over the full substrate and both kinds 
of adsorption sites are occupied with comparable probability.

\begin{table}[h]
\caption{\label{lt26:table1}
Kinetic, potential and total energies for the ground state of He on GF, 
on GH and on graphite. In the last column the bandwidth $\Delta$ is shown. 
Numbers in parentheses represent statistical uncertainty in the last digit.
}

\begin{center}
\begin{tabular}{*{7}{l}}
\hline
System  &$E_{kin}$ (K)& $E_{pot}$ (K)&$E_{tot}$ (K)&$\Delta$ (K) \\
\hline
$^4$He+GF & 46.78(4) & -422.94(1) & -376.15(2) &  9.6(1) \\
$^3$He+GF & 51.08(1) & -413.41(1) & -362.33(1) & 13.7(1) \\
\hline
$^4$He+GH & 20.51(1) & -153.58(1) & -133.06(1) & 13.6(4) \\
$^3$He+GH & 22.53(2) & -149.50(1) & -126.97(2) & 19.4(4) \\
\hline
$^4$He+Gr & 25.30(4) & -168.49(1) & -143.19(4) &  9.6(2) \\
$^3$He+Gr & 27.05(2) & -162.87(1) & -135.82(2) & 15.7(4) \\
\hline
\end{tabular}
\end{center}
\end{table}

We have also computed the density--density imaginary time correlation function 
in Fourier space; in the case $N=1$, at an imaginary--time $\tau$, this function 
takes the form:
$F(\vec{k},\tau) = \langle \rho_{\vec{k}}(\tau) \rho_{-\vec{k}}(0) \rangle$,
$\rho_{\vec{k}}(\tau) = \exp \left[ i \vec{k} \cdot \vec{r}(\tau)\right]$.
Here $\vec{r}(\tau)$ is the position of the 
atom at imaginary time $\tau$. 
$F(\vec{k},\tau)$ contains  informations  on the excited  
states of the system; these informations can be extracted 
through an inversion of the Laplace transform that gives the 
dynamic structure factor $S(\vec{k},\omega)$:
\begin{eqnarray} \label{invtr}
F(\vec{k},\tau) = \int d\omega \: e^{-\omega \tau} S(\vec{k},\omega). 
\end{eqnarray}
 However, $F(\vec{k},\tau)$ is known only at discrete imaginary--times $\tau_m$ with 
a statistical uncertainty; the inversion of the Laplace transform in such conditions is an 
ill--posed inverse problem; as consequence, the quality of the extracted informations 
can not be guaranteed. The inversion of the Laplace transform has been computed 
with the GIFT method explained in Ref.~\onlinecite{lt26:ref14}. Basically, the GIFT method 
uses a Genetic Algorithm to explore a space of solutions $\lbrace S_n(\vec{k},\omega)\rbrace$; the 
solutions that can reproduce $F(\vec{k},\tau)$ with an user--defined accuracy are averaged together to 
give the solution. 
\begin{figure}[h]
\begin{center}
\includegraphics*[width=10cm]{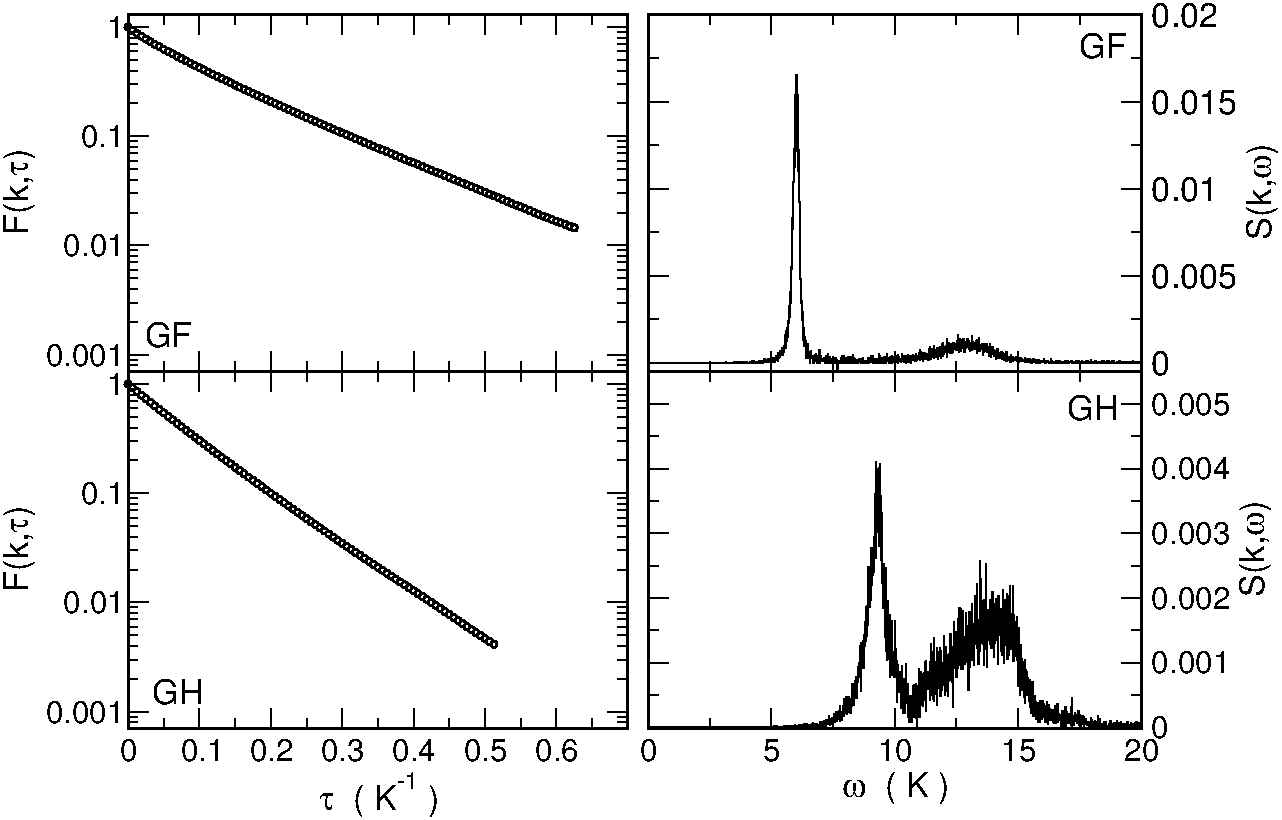}
\caption{\label{ad:figinvband}
Left side: density--density correlation functions $F(\vec{k},\tau)$ for a wave vector $\vec{k}$ corresponding to the 
point $M$ in the first Brillouin zone (see Fig.~\ref{lt26:fig2}). On the right side the respective 
inversions of Laplace transform, $S(\vec{k},\omega)$, are shown.
}
\end{center}
\end{figure}
In Fig.~\ref{ad:figinvband} an example of $F(\vec{k},\tau)$ and its $S(\vec{k},\omega)$ obtained 
with GIFT is given: in these functions $F(\vec{k},\tau)$, the main contribution comes from the lowest energy 
band; moreover, the excitation appears in $S(\vec{k},\omega)$ as a well defined peak; it is thus possible to obtain 
the energy spectrum of the first energy band; we interpret the width of the peaks as the uncertainty associated 
to the excitation energy at that wave--vector. The computed 
energy spectrum along the directions $\Gamma$K and $\Gamma$M for He on GF and on 
GH is shown in Fig.~\ref{lt26:fig2}.
These bands are represented rather accurately by a tight binding model with
nearest and next nearest coupling \cite{lt26:reftb}.

\begin{table*}[h]
\begin{center}
\small
\resizebox{9cm}{!} {
\begin{tabular}{l c c }
\hline
\hline
\multicolumn{3}{c}{\textbf{substrate and $N$=1 properties}} \\
\hline
Property\hspace{1cm}                &\hspace{1cm}GF\hspace{1cm}            &\hspace{1cm}GH\hspace{1cm} \\ 
$U_0$                   &498 K         &195 K        	\\
BA                      &24 K          &13 K          \\
$d_{s}$                 &1.49 \AA      &1.45 \AA     	\\
$E_{0}$ for $^4$He      &-376.15(2) K  &-133.06(1) K  \\
$E_{0}$ for $^3$He      &-362.33(1) K  &-126.97(2) K 	\\
BW of $^4$He            &9.6 K         &13.6K         \\
BW of $^{3}$He          &13.7 K        &19.4 K        \\
$m^*/m$ of $^4$He       &1.40          &1.05         	\\
$m^*/m$ of $^3$He       &1.26          &1.01          \\
\hline
\hline
\end{tabular}}
\caption{Substrate and $N$=1 properties, with: $U_o$) Depth of potential well; BA) Inter--site energy barrier;
$d_s$) Inter--site distance; $E_0$) Ground state energy; BW) Bandwidth; $m^*/m$ effective mass to bare mass ratio.}\label{fgtab1}
\end{center}
\end{table*}

For comparison we have computed with this same 
method the band energy for He on graphite finding substantial agreement 
with the Carlos and Cole result for the lowest band \cite{lt26:ref17}. 
The bandwidths $\Delta$ of He on these three substrates are given in Table \ref{lt26:table1}.

From the first energy band it is possible to obtain an estimate of the 
effective mass of one atom of Helium on GF(GH); this is done with 
a fit of the energy band at small wave--vectors (in Fig.~\ref{lt26:fig2} it is the 
region around the point $\Gamma$); in fact, for small wave--vectors 
the first energy band $E_k \simeq \hbar^2k^2/(2m^{\star})$.
The effective masses $m^{\star}$ of the various systems reflect the varying 
corrugations of the potentials. For $^4$He ($^3$He), the ratios 
of $m^{\star}$ to the bare mass are 1.40 (1.25), 1.10 (1.08) and 1.05 (1.01) 
on GF, graphite and GH, respectively. The smaller mass enhancement 
of $^3$He than $^4$He reflects the smaller ratio of the corrugation 
potential to the translational zero--point energy.

\subsection{Equilibrium density of submonolayer $^4$He on GF}

\begin{figure}[h]
\begin{center}
\includegraphics*[width=10cm]{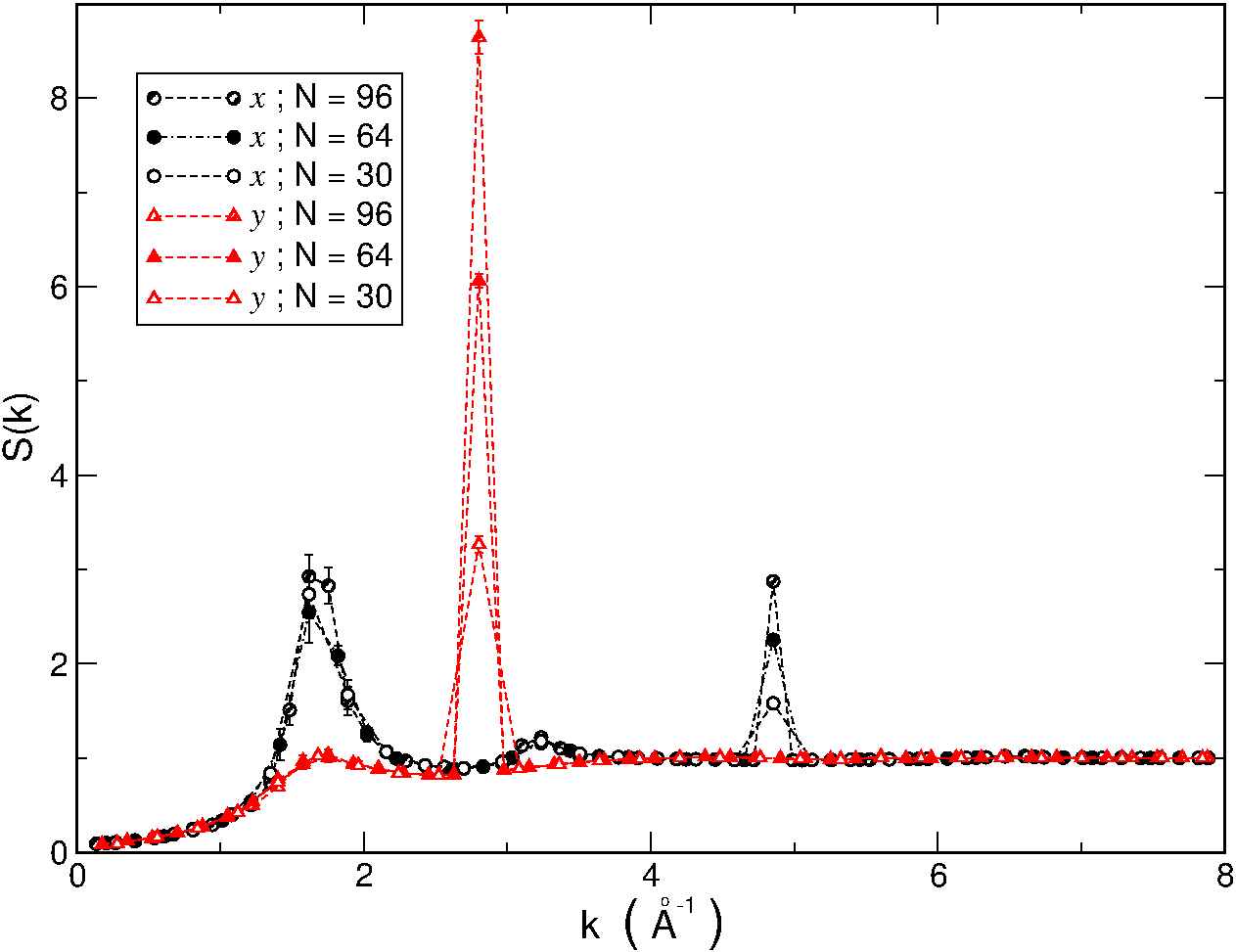}\\
\includegraphics*[width=10cm]{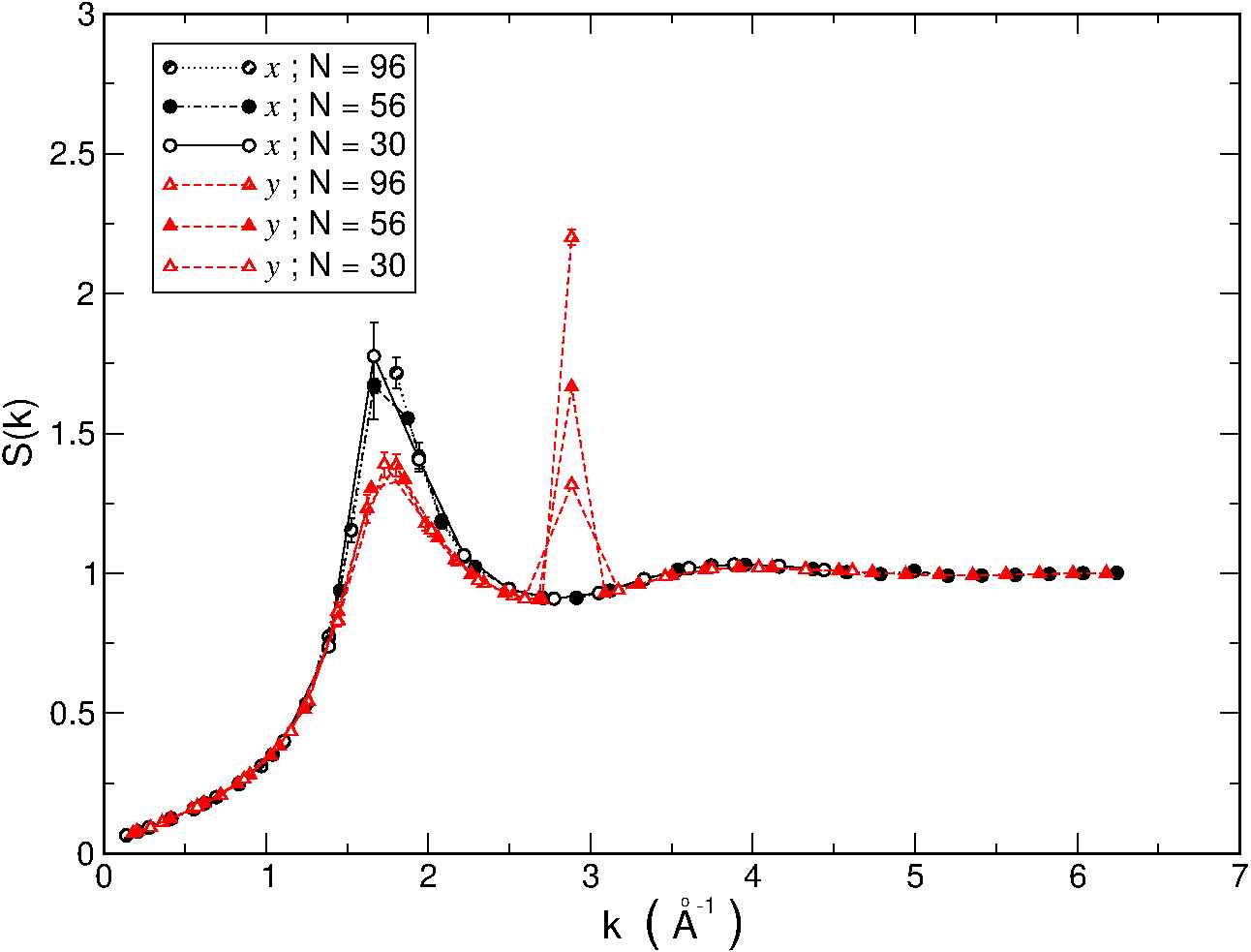}
\caption{Upper panel: static structure factor for GF at density $\rho^{\rm GF}_{1/6}$.
Lower panel: the same for GH, at density $\rho^{\rm GH}_{1/6}$. 
Lines are guides to the eyes.}\label{lt26:sdk1o3}
\end{center}
\end{figure}

We have studied a $^4$He submonolayer on GF. Some of the obtained properties are in Tab.~\ref{fgtab1s}.
As He--He interaction we 
have used an Aziz potential \cite{lt26:ref15}. The ground state has been computed 
for a number of $^4$He atoms from 22 to about 120 spanning the density 
range $\rho$=0.04--0.09 \AA$^{-2}$.
On graphite the ground state is the commensurate
$\sqrt{3}\times\sqrt{3}$ R30$^o$ state with filling factor 1/3 of the 
adsorption sites. A similar state on GF is obtained by populating 
fourth neighbor sites (this corresponds to second neighbors in one 
of the sublattices of the honeycomb at a distance 4.482 \AA) 
with a filling factor of the adsorption sites equal to $1/6$
and it corresponds to a density $\rho^{\rm GF}_{1/6}=0.0574$ \AA$^{-2}$.
Notice that this density is smaller than the 
$\rho_{\sqrt{3}}=0.0636$ \AA$^{-2}$
on graphite due to the dilation of the C plane in GF. 

\begin{figure}[h]
\begin{minipage}{6cm}
\includegraphics*[width=6cm]{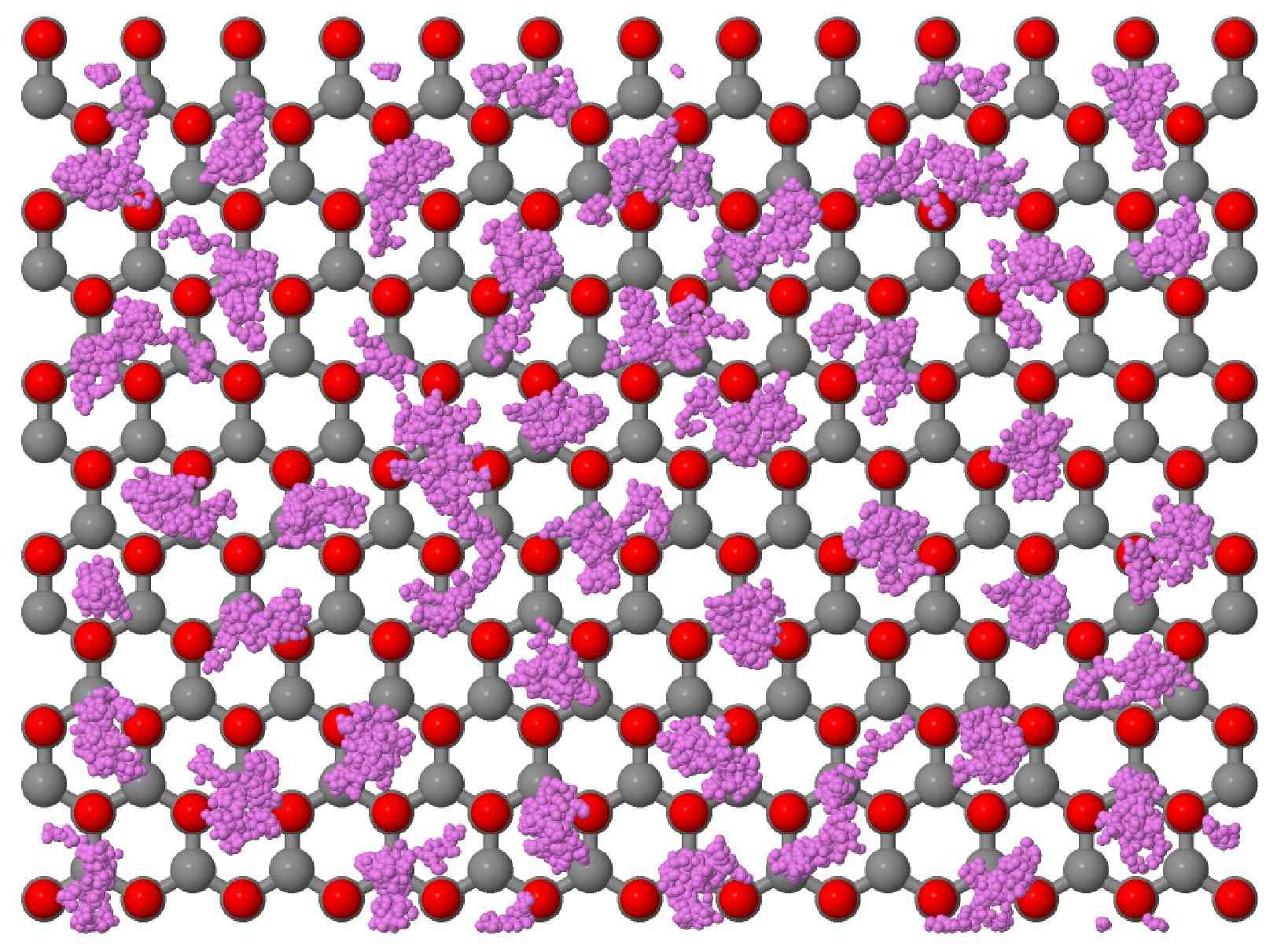}
\end{minipage}
\hspace{1pc}
\begin{minipage}{6cm}
\includegraphics*[width=6cm]{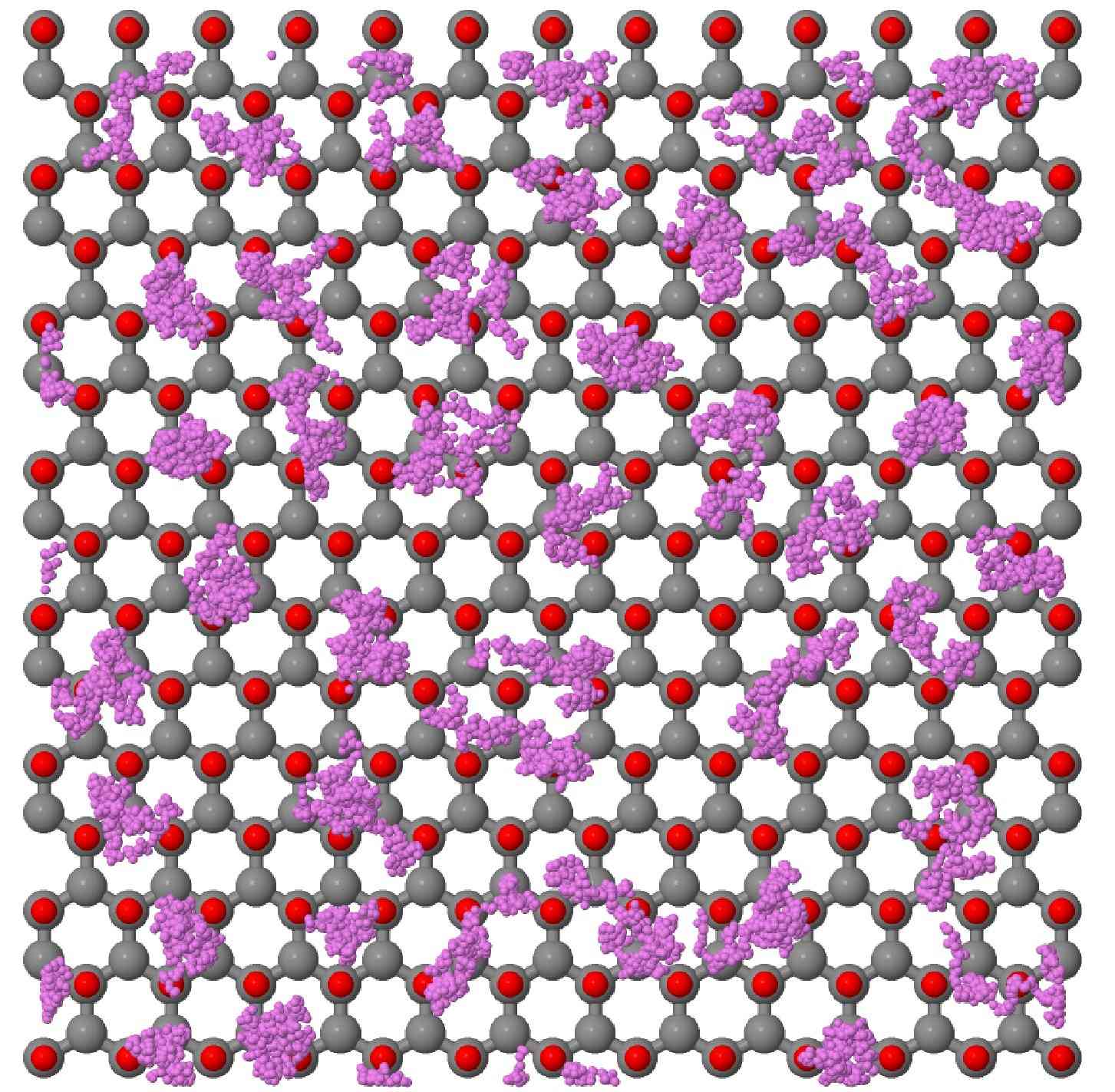}
\end{minipage}
\caption{\label{ad:cfgliq}
Polymer configurations for a liquid system of $N=43$ atoms of $^4$He on GF (left) and $N=41$ 
atoms of $^4$He on GH (right) at equilibrium density.
Each polymer represents the evolution up to 0.4 K$^{-1}$ in imaginary time 
of its correspondant $^4$He atom. The substrate is represented in the background with carbon atoms marked in gray and the F(H) 
overlayer marked in red.
}
\end{figure}

A simple consideration suggests the instability of a similar commensurate state. Using the 
curvature of the He--substrate potential at an adsorption site, the two--dimensional zero point 
energy is estimated to be 55(40) K on GF (GH), much larger than the minimum potential barrier 
23(13) K, so that such a localized state might be unstable. 
We find indeed that this ordered state is unstable: starting the simulation from an 
ordered configuration after a short Monte Carlo evolution the Bragg peaks corresponding to 
the $\sqrt{3}\times\sqrt{3}$ R30$^o$ state disappear and the system 
evolves into a disordered fluid state modulated by the substrate 
potential. $S(k)$ at this density is plotted in Fig.~\ref{lt26:sdk1o3} as function of 
$k_x$ and $k_y$ for two numbers $N$ of particles: the intensity of some of 
the peaks do not depend on $N$ so they are due to short range order, 
others scale roughly as $N$ and arise from the modulation of the density 
due to the adsorption potential.

\begin{table*}[h]
\begin{center}
\small
\resizebox{9cm}{!} {
\begin{tabular}{l c c}
\hline
\hline
\multicolumn{3}{c}{\textbf{Many--body properties}}\\
\hline
Property\hspace{1cm}             &\hspace{1cm}GF\hspace{1cm}  &GH \\ 
$\rho_{eq}$	&0.049 \AA$^{-2}$   &0.042 \AA$^{-2}$	 \\
$x$  		&0.142  	    & 0.115	       \\
$E_{0}$		&-377.71(4) K	    &-134.02(5) K	 \\
$E_b$		&1.55(6) K	    &0.95(6) K         \\
$n_0$		&11 $\pm$ 1 \%      &22.6 $\pm$ 1.3 \%   \\
$\rho_s/\rho$	&0.60(3)	    &0.95(3)	       \\
$T_{c}$		&0.2--0.3 K	    &1.0--1.2 K        \\
$\rho_{sat}$	&0.136 \AA$^{-2}$   &0.108 \AA$^{-2}$	 \\
			&		    &		       \\
\hline
\hline
\end{tabular}}
\caption{Many--body properties, with: $\rho_{eq}$) Equilibrium density; $x$) coverage;
$E_0$) Ground state energy per particle; $E_b$) Binding energy; $n_0$) Condensate fraction; $\rho_s/\rho$) $T$=0 K superfluid fraction;
$T_c$) Transition temperature; $\rho_{sat}$) Completion density }\label{fgtab1s}
\end{center}
\end{table*}

Fig.~\ref{ad:cfgliq} shows a sampled configuration of polymers for a system of $^4$He on GF and GH at 
equilibrium density. The spread in space of a single polymer is related to the zero point motion, whereas the center of mass of 
each polymer gives an idea of the spatial order of that configuration. As expected, the polymers stay on average over the adsorption sites but, apart from 
the modulation of the external potential, there is not spacial order due to the He--He interaction. It is interesting to note that sometimes zero point motion 
allows a polymer to stretch toward a near adsorption site by crossing a saddle point; this is a dynamic in imaginary time that eventually leads a polymer to 
connect with an adjacent one implementing quantum exchanges phenomena in a multi connected geometry. On average, such exchanges are more frequent in the configuration 
on the GH substrate rather than that on the GF, this is expected because of the strongest confinement given by the GF adsorption potential.

\begin{figure}[ht]
\begin{center}
\includegraphics*[width=12cm]{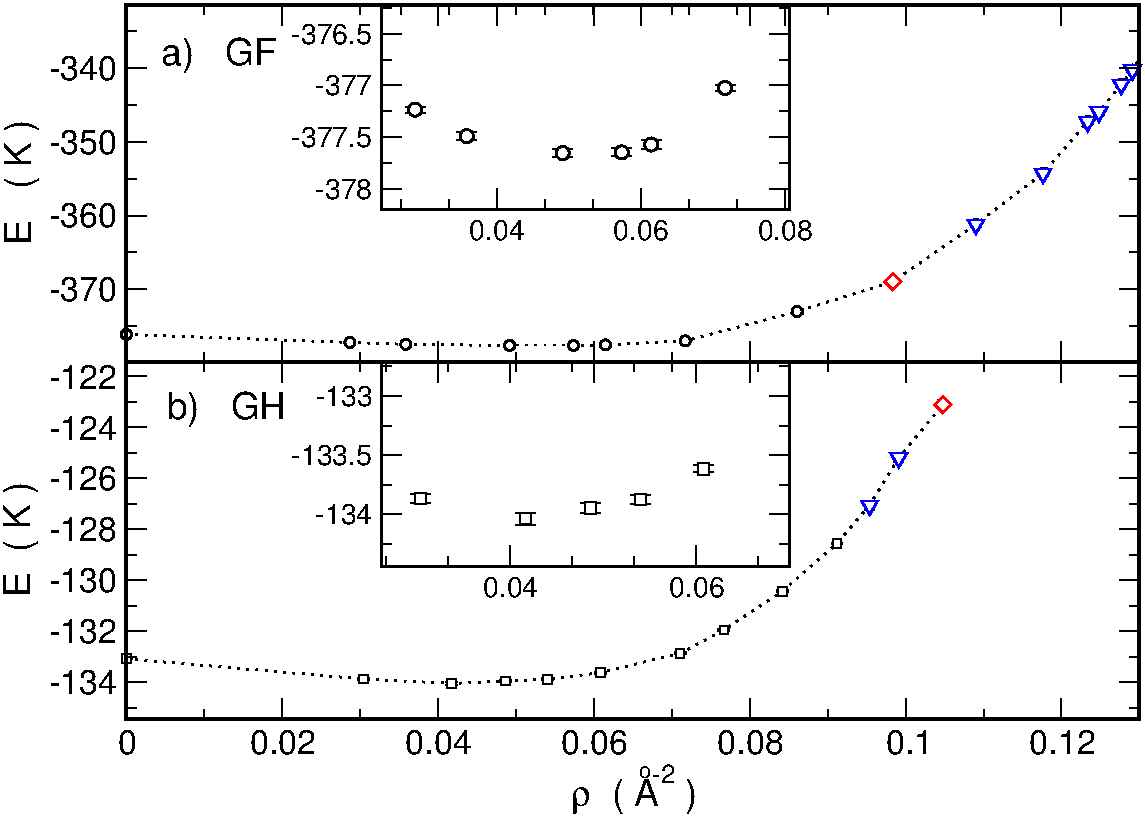}\label{ad:eos4he}
\caption{\label{lt26:eos4he}
Panel (a): Energy per particle of $^4$He on GF at $T=0$K.
Panel (b): Equation of state of $^4$He on GH at $T=0$K. The inset represents a zoom in the region around the 
energy minimum. In both cases, the used particle numbers ranged between $N=60$ and $N=120$. The dashed line represents a guide
to the eye. Circles are liquid densities, Squares represent the commensurate $2/7$ phase and Triangles are incommensurate densities.
}
\end{center}
\end{figure}

In Fig.~\ref{ad:eos4he} the energy per particle of $^4$He on GF and $^4$He on GH at the studied densities are reported. 
In the GF case, the energy per particle has a minimum value $E_0=-377.71 \pm 0.04$ K at the density $\rho_{eq}=0.049$ \AA$^{-2}$. This lies 1.55(6) K below the single particle 
energy, implying that the ground state is a self--bound liquid. In the GH case a similar state is obtained, with $E_0=-134.02 \pm 0.05$ K and a binding energy per atom of 0.95(6).
For comparison, we note that the strictly 2D cohesive energy of 
$^4$He \cite{lt26:ref16} is just 0.84 K and the equilibrium density is $\rho=0.0436$ \AA$^{-2}$. 
In both the cases a liquid phase has been found at least for densities up to filling factors $x= 1/4$ that 
for the GF case correspond to a density $\rho_{1/4}^{GF}=0.0861$~\AA$^{-2}$, and for the GH case 
to a density $\rho_{1/4}^{GF}=0.0912$~\AA$^{-2}$.

\begin{figure}[h]
\begin{center}
\includegraphics*[width=11cm]{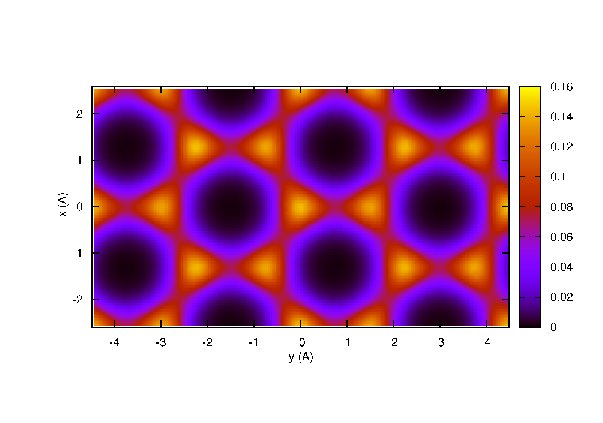}\label{ad:rdfeq}
\includegraphics*[width=11cm]{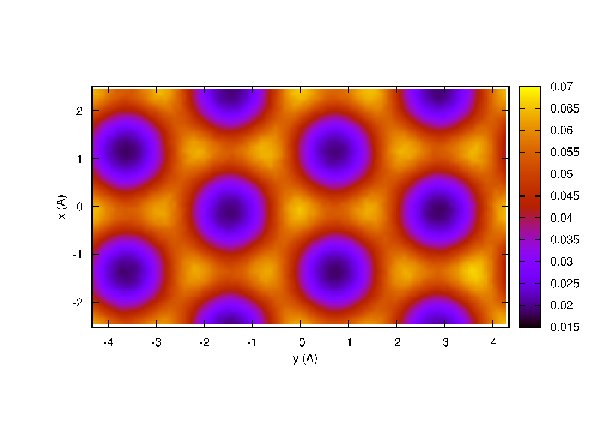}
\caption{Local density (in \AA$^{-2}$ units) on the x--y plane integrated along the 
z direction of $N=33$ atoms of $^4$He on GF (left) and $N=41$ atoms of $^4$He on GH (right) at 
equilibrium density.}
\end{center}
\end{figure}

In Fig.~\ref{ad:rdfeq} the local density on the $x$--$y$ plane is shown for $^4$He on GF (left) and 
GH (right) at equilibrium density. These local densities clearly reflect the geometry of the 
adsorption potential shown in Fig.~\ref{ad:adpotl}. The system is energetically allowed 
to stay in a multi--connected space in which adsorption minima are reachable through channels that
 cross a saddle point of the adsorption potential. Note that although the geometry in the 
 two cases is the same, the potential barrier above the F (H) overlayer is much lower in the
  GH case (see Fig.~\ref{ad:corrug}), this produces an higher degree of anisotropy in the GF case
    and reflects in the local density as a non--zero
  probability to occupy an adsorption maximum in the GH case.

\begin{figure}[h]
\begin{minipage}{7cm}
\includegraphics*[width=7cm]{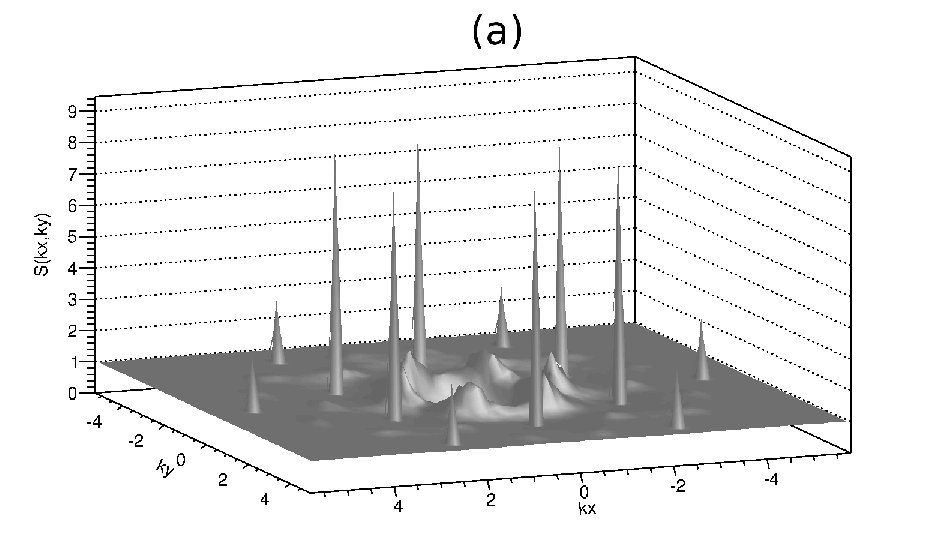}\label{ad:sdkeq}
\end{minipage}\hspace{0.1cm}
\begin{minipage}{7cm}
\includegraphics*[width=7cm]{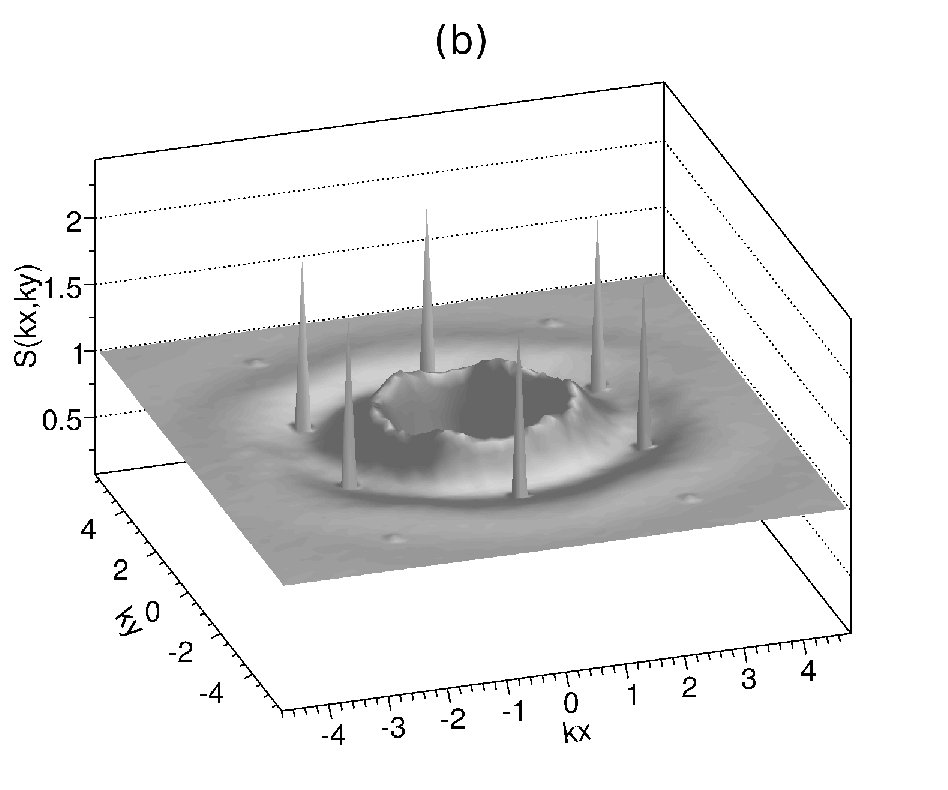}
\end{minipage}\\
\vspace{0.5cm}
\begin{minipage}{7cm}
\includegraphics*[width=7cm]{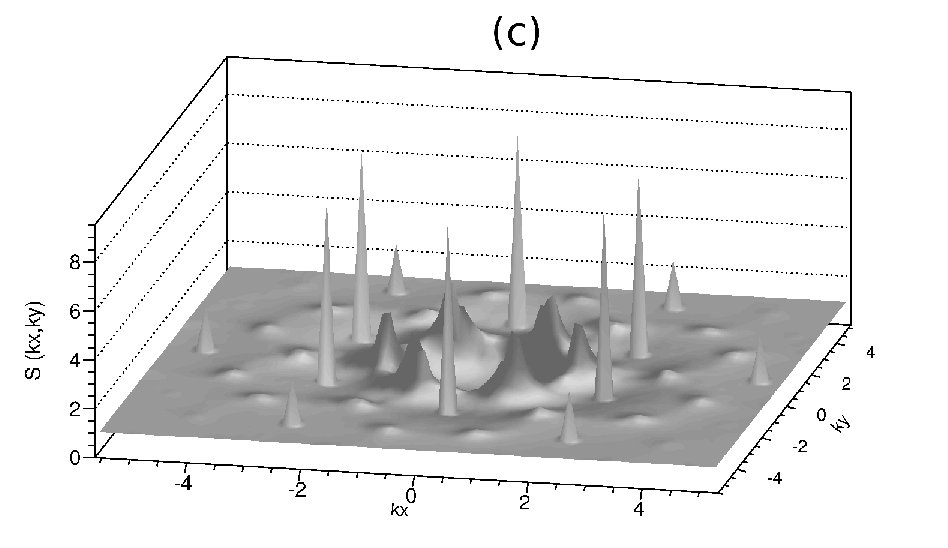}
\end{minipage}\hspace{0.1cm}
\begin{minipage}{7cm}
\includegraphics*[width=7cm]{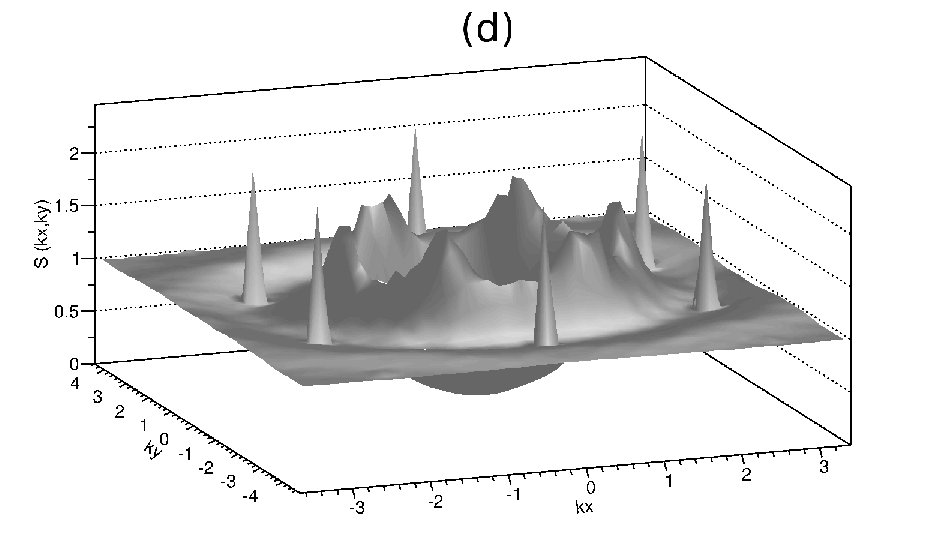}
\end{minipage}\caption{
Static structure factor on the x--y plane of $N=96$ atoms of $^4$He on GF at equilibrium density (upper left), on GH at equilibrium density (upper right),
on GH at filling factor $x=1/6$ (lower left),on GH at $x=1/6$ (lower right).
}
\end{figure}

\begin{figure}[h]
\begin{center}\label{lt26:obdmeq}
\begin{minipage}{10cm}
\includegraphics*[width=10cm]{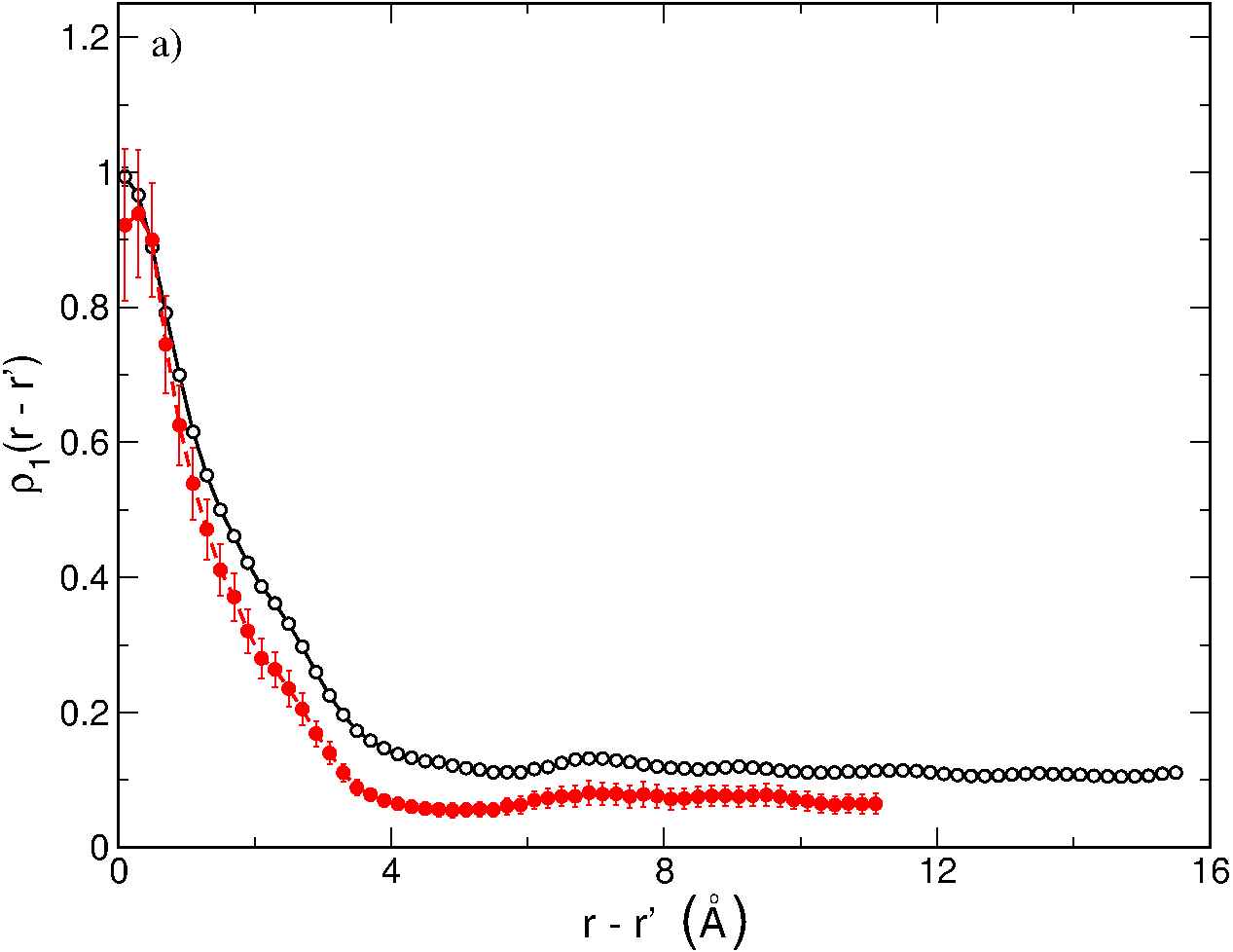}
\end{minipage} \\
\begin{minipage}{10cm}
\includegraphics*[width=10cm]{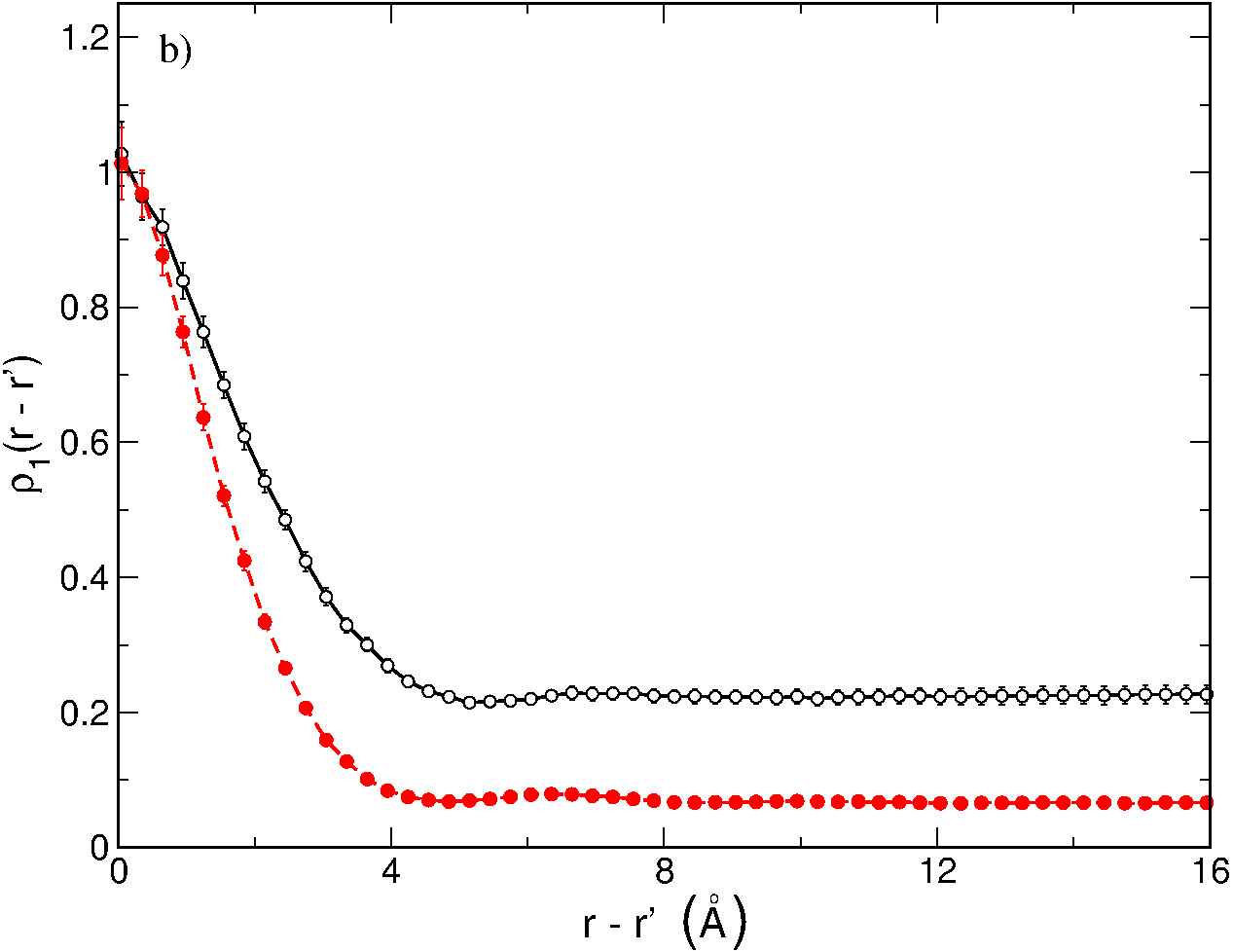}
\end{minipage}
\caption{Panel a)
Off diagonal one body density
matrix for $^4$He on GF at $\rho^{\rm GF}_{eq}=0.049$ \AA$^{-2}$ (open circles) and
$\rho^{\rm GF}_{1/6}=0.0574$ \AA$^{-2}$ (filled circles).
Panel b) the same for GH, with $\rho^{\rm GH}_{eq}=0.042$ \AA$^{-2}$ and
$\rho^{\rm GH}_{1/6}=0.0608$ \AA$^{-2}$. 
Lines are guides to the eyes.
}
\end{center}
\end{figure}

The static structure factor of $^4$He on GF and GH at equilibrium density as well as that at 
filling factor $x=1/6$ are shown in Fig.~(\ref{ad:sdkeq}). The sharp peaks reflect the density modulation due to the 
corrugation of the adsorption potential. The crater like structure at smaller $k$ 
represents short range He--He correlations. It can be noticed that short range correlations in 
the GH case are much less anisotropic than that in the GF case, this reflects the smaller 
corrugation of the adsorption potential of GH.

\begin{figure}[h]
\begin{center}
\includegraphics*[width=10cm]{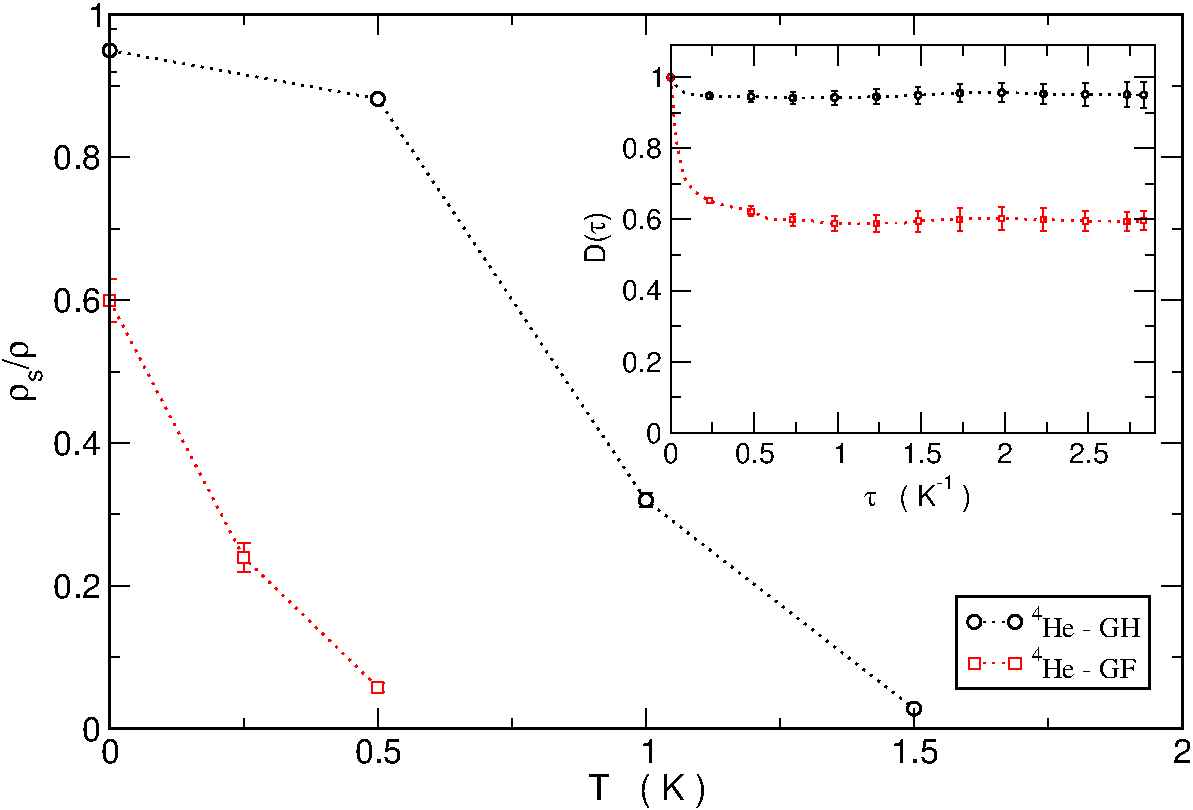}\label{ad:rhosrhoeq}
\caption{
The superfluid fraction as function of temperature for a system of $N=20$ atoms 
of $^4$He on GH and a system of $N=26$ atoms of $^4$He on GF. The transition temperature 
can be roughly estimated in the range 1.0 -- 1.2 K for $^4$He--GH and 0.2--0.3 K for 
$^4$He--GF. In the inset is displayed the diffusion of the center of mass from which 
the zero--temperature estimation of the superfluidity has been extrapolated.	
}
\end{center}
\end{figure}

We have computed the off diagonal one body density matrix $\rho_1(r-r')$.
As can be seen in Fig.~\ref{lt26:obdmeq} $\rho_1$ reaches a plateau at large $r-r'$ and the 
Bose Einstein condensate (BEC) fraction is $10.3 \pm 0.4$ \% at $\rho=0.049$ \AA$^{-2}$
and $7.3 \pm 1.5$ \% at $\rho_{1/6}$; the system is superfluid.  
We reach a similar conclusion in the case of the GH substrate: 
the ground state is a liquid with density $\rho_{eq}=0.045$ \AA$^{-2}$ 
and $E_0=-134.28 \pm 0.02$ K per atom and the BEC fraction is $22.6 \pm 1.3$ \% at the 
equilibrium density and $6.8 \pm 0.5$ \% at $\rho^{\rm GH}_{1/6}=0.0608$ \AA$^{-2}$. Note that this condensate 
fraction is significantly smaller than the value ($\simeq 40$ \%) for $^4$He 
in 2D \cite{lt26:ref16}. The smaller value is a consequence of the spatial order, 
albeit imperfect, induced by the substrate potential and of the 
smaller effective surface available to the atoms due to the strong 
channeling induced by that potential.

In Fig.~(\ref{ad:rhosrhoeq}) the superfluid fraction $\rho_s/\rho$ is shown in function
of the temperature for a system of $N=26$ atoms of $^4$He on GF and $N=20$ atoms of $^4$He 
on GH at their respective equilibrium densities. At finite temperature, the superfluid 
fraction has been estimated with the winding number method. The data at zero temperature 
has been obtained with the evolution of the center of mass of the system for sufficiently 
large imaginary--time\cite{qfs:rnew} $\tau$:
\begin{eqnarray}\label{ad:cmasslim}
\frac{\rho_s}{\rho} = \lim_{\tau\rightarrow +\infty} D(\tau) \nonumber \\
D(\tau)=
\frac{N}{2d\lambda}\frac{\left\langle\left[\vec{R}_{CM}\left(\tau\right)-\vec{R}_{CM}\left(0\right)\right]^{2}\right\rangle}{\tau}
\quad ,
\end{eqnarray}
where $\lambda=\hbar^2/2m$, $N$ is the number of particles, $d$ is the number of dimensions along which the contribution
to the superfluid fraction is considered, and
the squared distance $[\vec{R}_{CM}(\tau)-\vec{R}_{CM}(0)]$ is evaluated without invoking periodic boundary
conditions, i.e. including boundary crossing if $\vec{R}_{CM}(\tau)$ leaves the main simulation box.
Note that in general the estimator for the superfluid fraction
in equation \ref{ad:cmasslim} can be used with a PIGS algorithm only when
the Hamiltonian of the system explicitly breaks the translational symmetry as in the present case.
This is a necessary condition because even if one starts from a trial state $\Psi_T$ in which the
translational symmetry is broken, the imaginary--time evolution of $\Psi_T$ obtained via PIGS
(i.e. $\Psi_\tau=\hat{G}(\tau)\Psi_T$) immediately, for every imaginary time $\tau$ restore the
translational invariance unavoidably disturbing the estimation of $D(\tau)$.
This is due to th fact that the imaginary time evolution depends only on the Hamiltonian via
$\hat{G}(\tau)=\exp(-\tau\hat{H})$.

It is noticeable that $\rho_s/\rho$ for $^4$He on GH joins 
smoothly with the $T=0$K value. This is a strong test on our algorithms since these values 
come from completely different computations, however, in the case of GF, the low transition 
temperature does not allow to reach the $T << T_c$ regime.
From $\rho_s/\rho$ at finite $T$ and taking into account size effect we estimate the
superfluid transition temperature $T_c \simeq 0.2-0.3$ K for GF and $1.0-1.2$ K for GH.
From the zero temperature computation of the superfluid density we predict that a 
submonolayer of $^4$He film on GF (GH) is an {\it anisotropic} superfluid with superfluid 
fraction $\rho_s/\rho=0.95(3)$ for GH and $0.60(3)$  for GF. Remarkably, this quantity 
is less than unity and this is in agreement with the predictions by Leggett for a 
nonuniform superfluid \cite{prl:ref20}.

\begin{figure}[h]
\begin{center}
\includegraphics*[width=12cm]{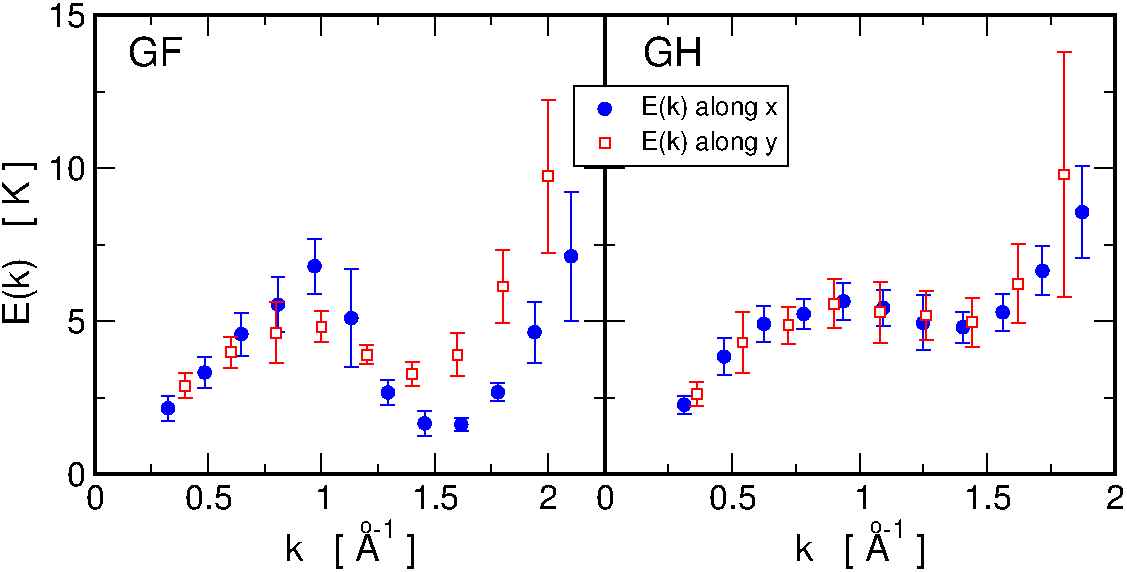}
\caption{Left: Excitation spectrum, $E(k)$, of $^4$He on GF at equilibrium density 
along $x$ and $y$ directions extracted
from the position of the quasi-particle excitation peaks in the dynamical structure factors
obtained via the GIFT algorithm. The error--bars represent the 1/2--height widths.
The Bijl--Feynman spectrum, $E_F(k)$, is also shown.
Right: The same for $^4$He on GH. Lines are guides to the eye.
}
\label{ad:fig8}
\end{center}
\end{figure}

\subsubsection{Dynamics at equilibrium density}

\begin{figure}[h]
\begin{center}
\includegraphics*[width=12cm]{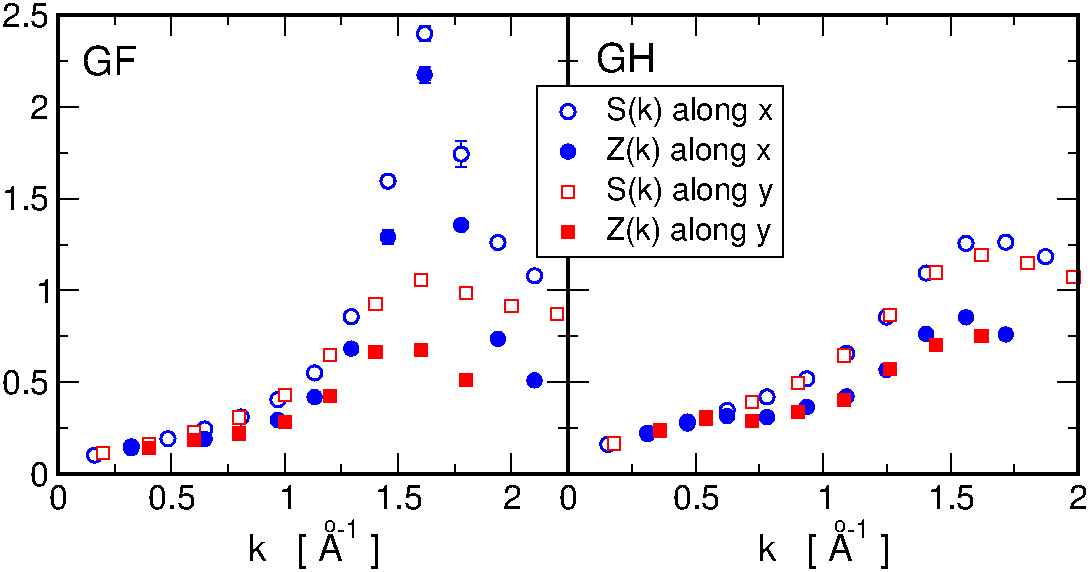}
\caption{Left: Static structure factors $S(\vec{k})$ and strength of the quasi particle
peak $Z(\vec{k})$ as function of $q$ of $^4$He on GF at equilibrium density along $x$ and $y$ directions.
Right: The same for $^4$He on GH.
}
\label{ad:fig9}
\end{center}
\end{figure}

Information about dynamical properties can, in principle, be extracted from imaginary time
correlation functions, without relying on any approximation, focusing on
an ill--posed inverse problem, i.e. the inversion of the Laplace transform which connects
a suitable imaginary time correlations function $f(\tau)$ to the relevant spectral function.
If one consider the dynamical structure factor $S(\vec{k},\omega)$, which is measurable
in an inelastic neutron scattering experiment, the related imaginary--time correlation function is the so called
``intermediate scattering function'' $F(\vec{k},\tau)$:
\begin{equation}
F(\vec{k},\tau)={1\over N} \langle e^{\tau\hat{H}} \hat{\rho}_{\vec{k}} e^{-\tau\hat{H}} \hat{\rho}_{-\vec{k}} \rangle
= \int d\omega \, e^{-\omega\tau} S(\vec{k},\omega) \quad .
\end{equation}
The expression of
$F(\vec{k},\tau_i)=\langle e^{\tau_i\hat{H}}\hat{\rho}_{\vec{k}} e^{-\tau_i\hat{H}}\hat{\rho}_{-\vec{k}}\rangle/N$
can be estimated via ``exact'' Quantum Monte Carlo methods for a discrete set of imaginary time
instants $\tau_i$. However, the extraction of
$S(\vec{k},\omega)$ from the above integral equation, based on the limited and noisy
knowledge of $F(\vec{k},\tau)$, is an ill--posed inverse problem; in fact,
the kernel $e^{-\omega\tau}$ is strongly smoothing
and infinite dynamical structure factors turn out to be
compatible with the information on the correlation function, i.e. with $F(\vec{k},\tau_i)$
for the different $\tau_i$.
Recently, we have developed
a technique to face such problems quite in general:
the Genetic Inversion via Falsification of Theories (GIFT) method\cite{qfs:gift}.
GIFT extracts information
on spectral functions by averaging among models found compatible with observations
(i.e. the correlation function for
a discrete set of imaginary time instants: $f(\tau_i)$)
via a genetic--algorithm--exploration of a given wide space of model spectral functions.
When applied to bulk liquid $^4$He at $T = 0$ K,
GIFT has been found able to extract more information about $S(\vec{k},\omega)$, separating quantitatively the
elementary excitation peak from the multiphonon contributions\cite{qfs:gift,qfs:gift2}.

\begin{figure}[h]
\begin{center}
\includegraphics*[%
  width=0.95\linewidth,
  keepaspectratio]{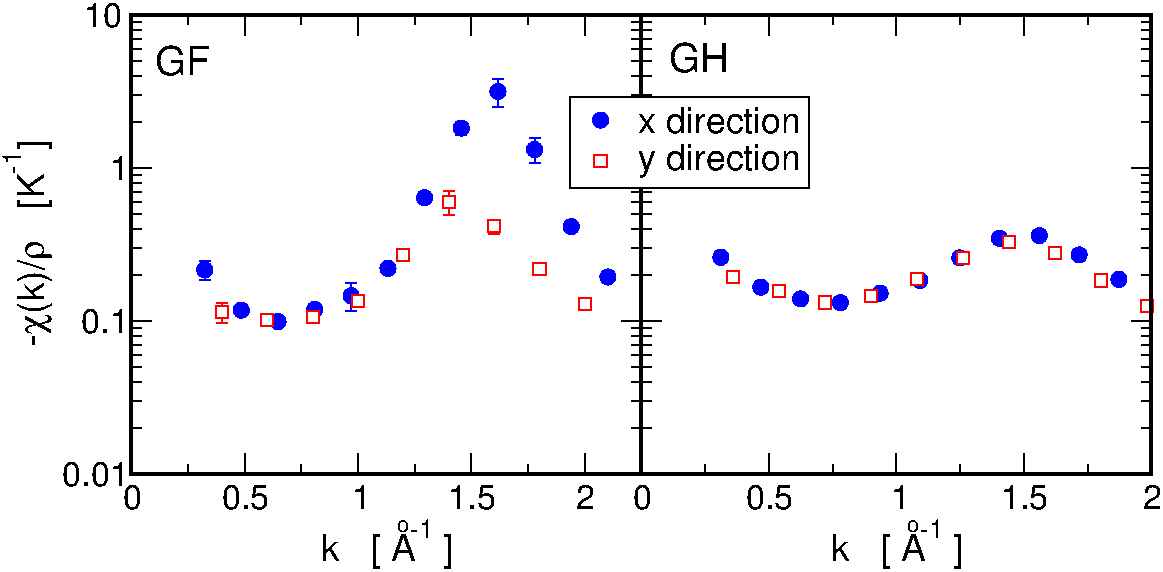}
\caption{Left: Static density response function, $\chi(\vec{k})$,
as function of $q$ of $^4$He on GF at equilibrium density along $x$ and $y$ directions.
Right: The same for $^4$He on GH.
}
\label{ad:fig10}
\end{center}
\end{figure}

Here we have applied the GIFT algorithm to extract information on excited state properties
of the equilibrium superfluid phases of $^4$He on GF and GH.
Intermediate scattering functions have been computed for different wave vectors with the
PIGS method.
Well--defined single excitation peaks and multiphonon contribution are present in the
reconstructed dynamical structure factors via the GIFT algorithm.
In Fig.~\ref{ad:fig8} the position of these peaks as a function of the wave vectors in two different directions
are shown. In the left panel, which correspond to the GF case, the spectrum is found to be highly anisotropic
with roton excitations lower than 2 K along the $x$ direction and of about 3.5 K along $y$.
This is again a consequence of the strong and anisotropic corrugation of the GF substrate respect
to the GH case where we found a much more isotropic spectrum, with a shallow roton minimum near 5 K.

\begin{figure}[t]
\begin{center}
\begin{minipage}{9cm}
\includegraphics*[width=9cm]{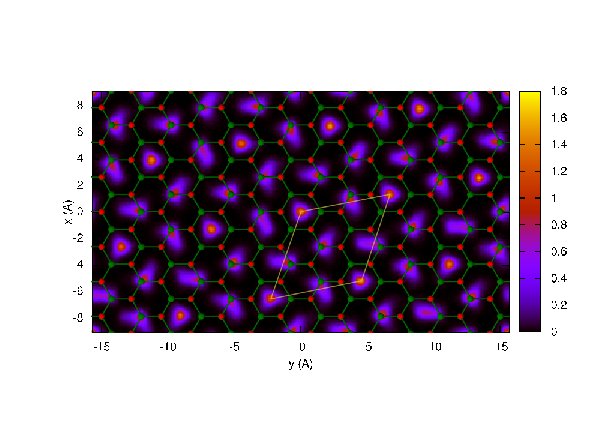}
\end{minipage}\\
\begin{minipage}{9cm}
\includegraphics*[width=9cm]{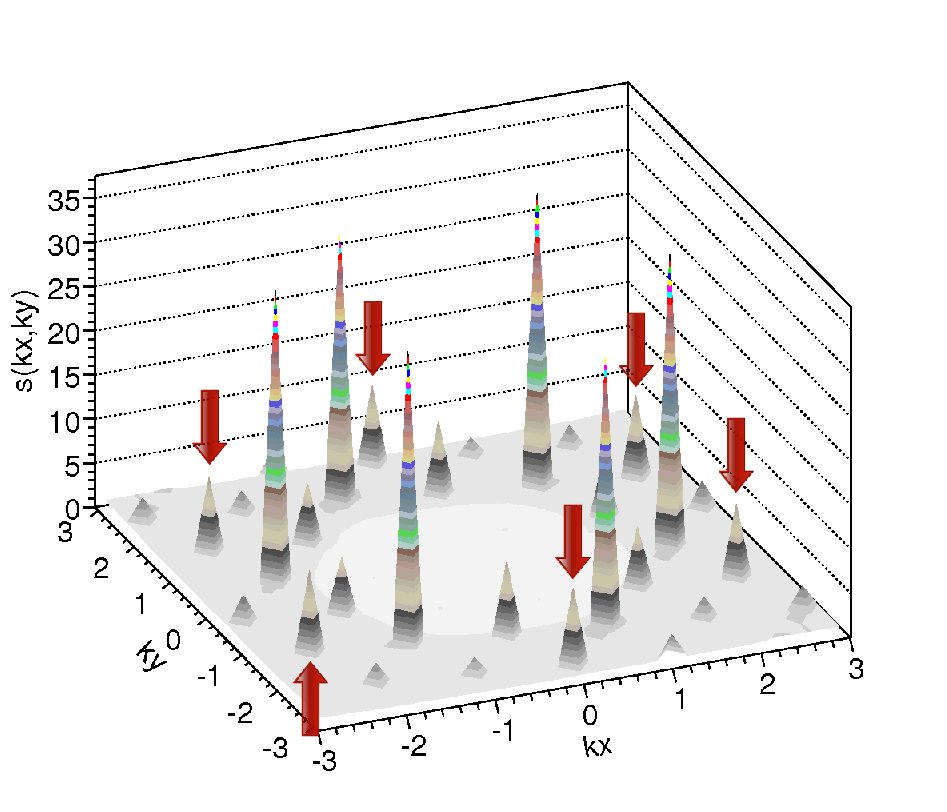}
\end{minipage}
\caption{(Upper panel) Local density (in \AA$^{-2}$ units) on the $x$--$y$ plane of the $2/7$ phase of $^4$He on GF compared with the geometry of the substrate.
Red balls are centered on the position of Fluorine atoms and the green ones on the Carbon atoms. Thin white lines enclose the unit cell of the commensurate 2/7 phase. 
(Lower panel) Static structure factor on the k$_x$--k$_y$ plane of the $2/7$ phase of $N$=112 atoms of $^4$He on GF. k$_x$ and k$_y$ axis are expressed in \AA$^{-1}$.
Red arrows point to the peaks corresponding to the density modulation imposed by the adsorption potential. 
}\label{ad:denssdk47fg}
\end{center}
\end{figure}

\begin{figure}[t]
\begin{center}
\begin{minipage}{9cm}
\includegraphics*[width=9cm]{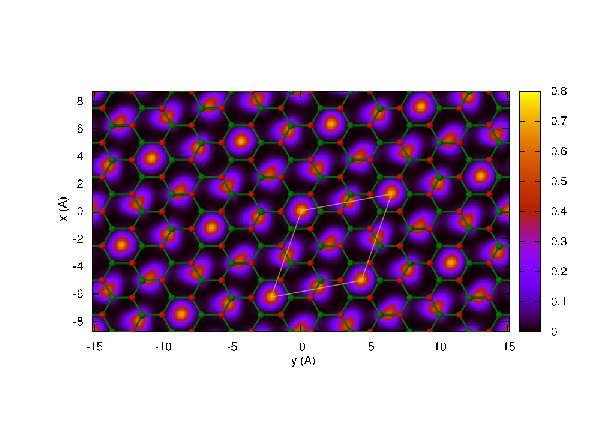}
\end{minipage}\\
\begin{minipage}{9cm}
\includegraphics*[width=9cm]{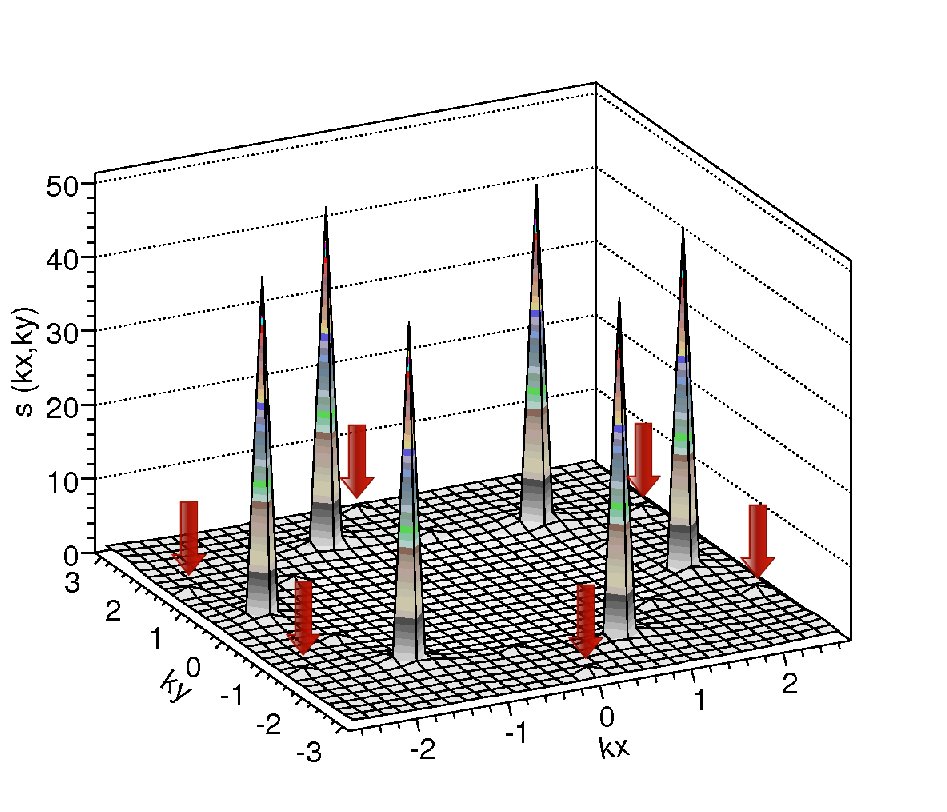}
\end{minipage}
\caption{(Upper panel) Local density (in \AA$^{-2}$ units) on the $x$--$y$ plane of the $2/7$ phase of $^4$He on GH compared with the geometry of the substrate.
Red balls are centered on the position of Hydrogen atoms and the green ones on the Carbon atoms. Thin white lines enclose the unit cell of the commensurate 2/7 phase. 
(Lower panel) Static structure factor on the k$_x$--k$_y$ plane of the $2/7$ phase of $N$=112 atoms of $^4$He on GH. k$_x$ and k$_y$ axis are expressed in \AA$^{-1}$.
Red arrows point to the peaks corresponding to the density modulation imposed by the adsorption potential. 
}\label{ad:denssdk47ch}
\end{center}
\end{figure}

Given the extracted single quasi--particle energies of the excitation spectrum
one can estimate the Landau critical velocity, $v_c=\min(E(k)/\hbar k)$, for both cases and directions;
in the GF case, these turn
out to be $v_c \simeq 13$ m/s along $x$ and $v_c \simeq 31$ m/s along $y$, in the
GH case we obtain $v_c \simeq 45$ m/s along $x$ and $v_c \simeq 51$ m/s along $y$.

By integrating $S(\vec{k},\omega)$ with respect to $\omega$  in the range of the sharp peak
and in the remaining frequency range we can determine the
strength of the single quasi--particle peak, $Z(\vec{k})$, and to the contribution to
the static structure factor, $S(\vec{k})$, coming from multiphonon excitations.
The results for $Z(\vec{k})$ are shown in Fig.~\ref{ad:fig9} for both cases, together with the
static structure factor along the same direction; from the ratio between $Z(\vec{k})$ and $S(\vec{k})$
one can measure the efficiency of the single quasi--particle excitation channel.

The efficiency is specially high along the $x$ direction of the GF case where we found the roton with the
lower energy.

Also, through the relation
\begin{equation}\label{ad:eq1}
\chi(\vec{k}) = -2 \rho \int_0^{\infty} d\omega {S(\vec{k},\omega) \over \omega}
\end{equation}
one can compute the zero temperature static density response function $\chi(\vec{k})$.
In Fig.~\ref{ad:fig10} we present our results for $\chi(\vec{k})$ of $^4$He on GF and GH
computed along different directions respect to the substrates.

\begin{figure}[h]
\begin{center}
\begin{minipage}{6cm}
\includegraphics*[width=6cm]{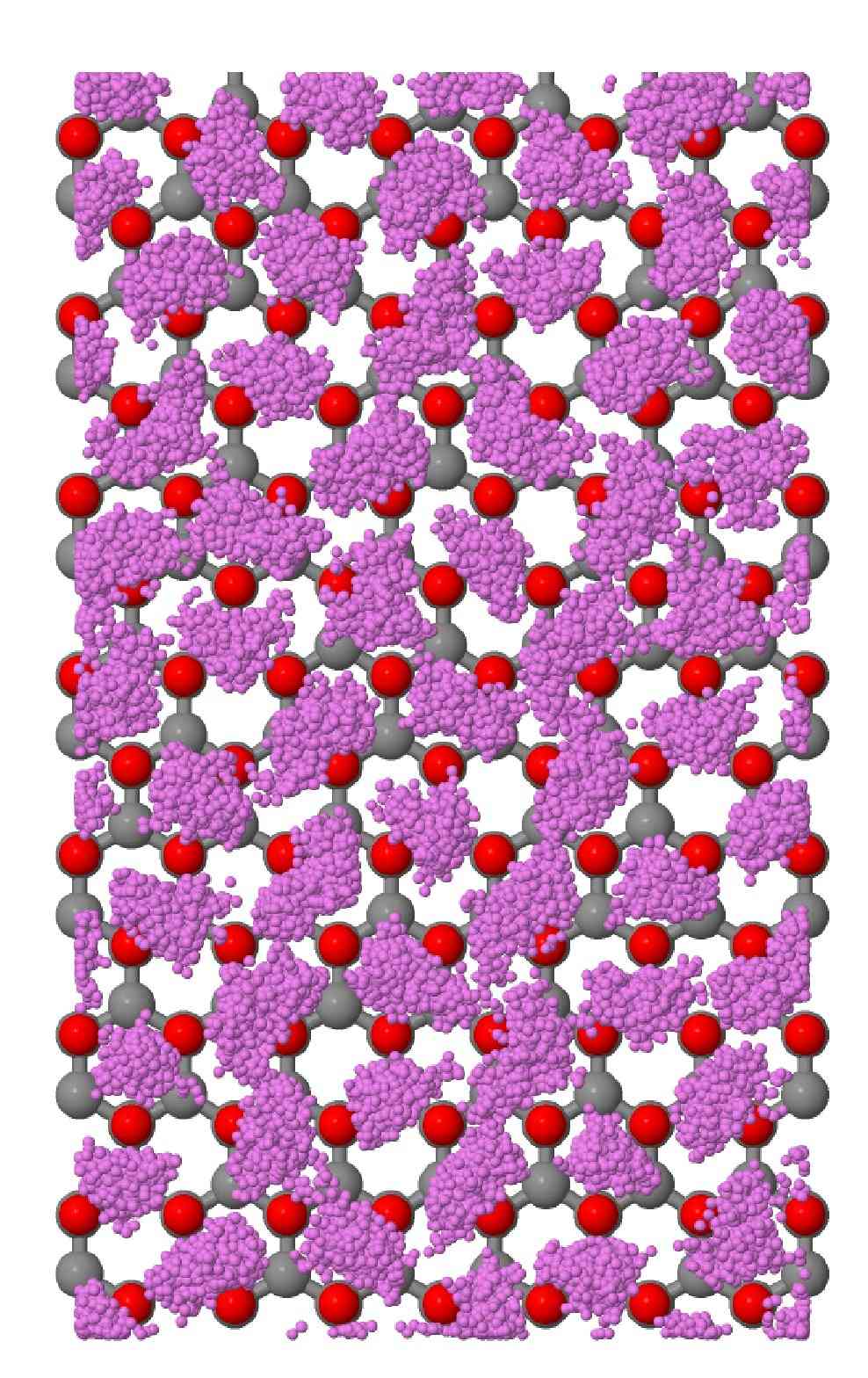}
\end{minipage}
\hspace{1pc}
\begin{minipage}{6cm}
\includegraphics*[width=6cm]{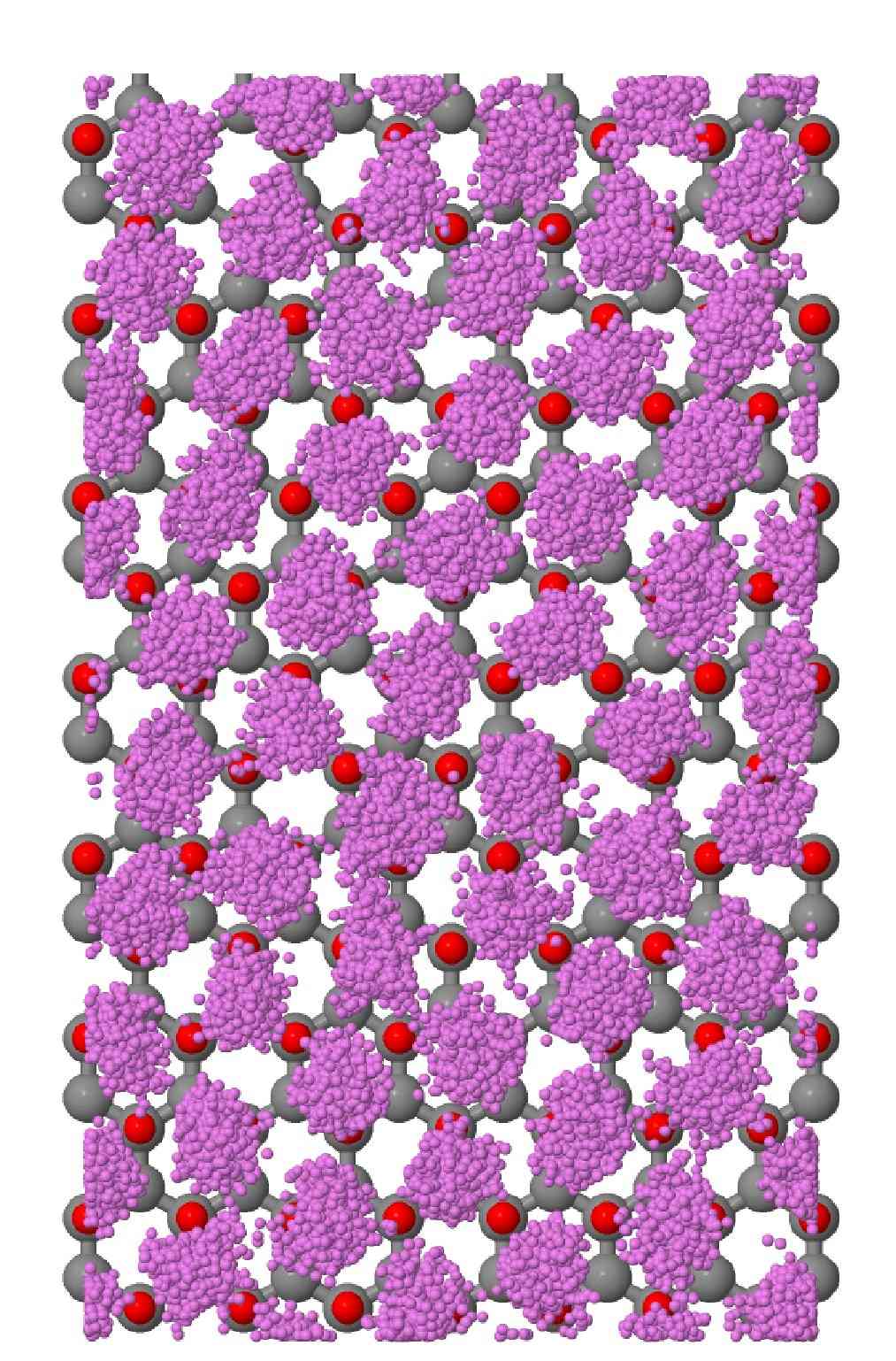}
\end{minipage}
\caption{Polymer configurations for a system of $N=56$ atoms of $^4$He on GF (left) and GH (right) at filling $x=2/7$.
Each polymer represents the evolution up to 0.4 K$^{-1}$ in imaginary time 
of its correspondent $^4$He atom. The substrate is represented in the background with carbon atoms marked in gray and the F(H) 
overlayer marked in red.
}\label{ad:cfg47}
\end{center}
\end{figure}

\begin{figure}[h]
\begin{minipage}{6cm}
\includegraphics*[width=6cm]{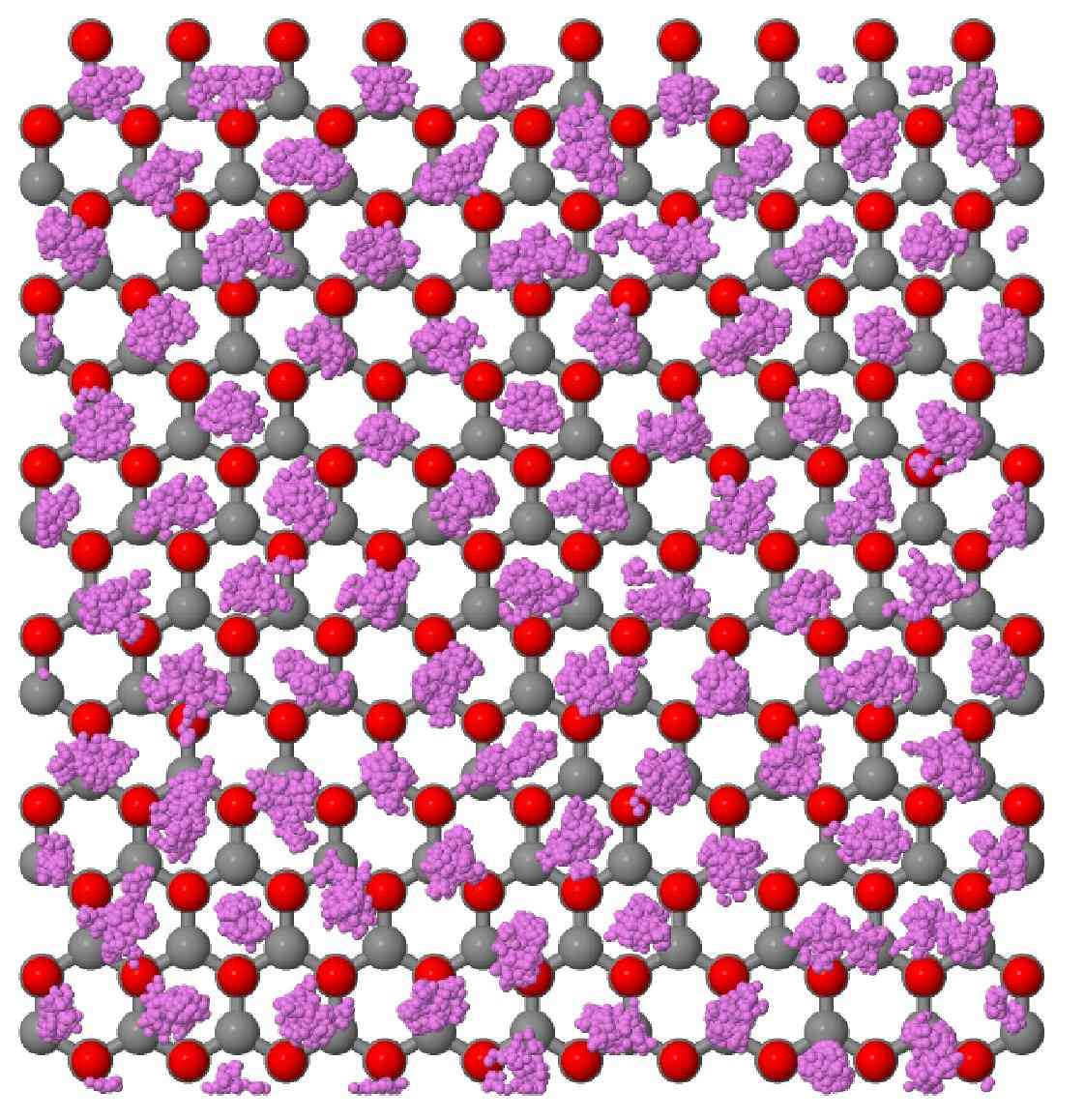}
\end{minipage}
\hspace{1pc}
\begin{minipage}{6cm}
\includegraphics*[width=6cm]{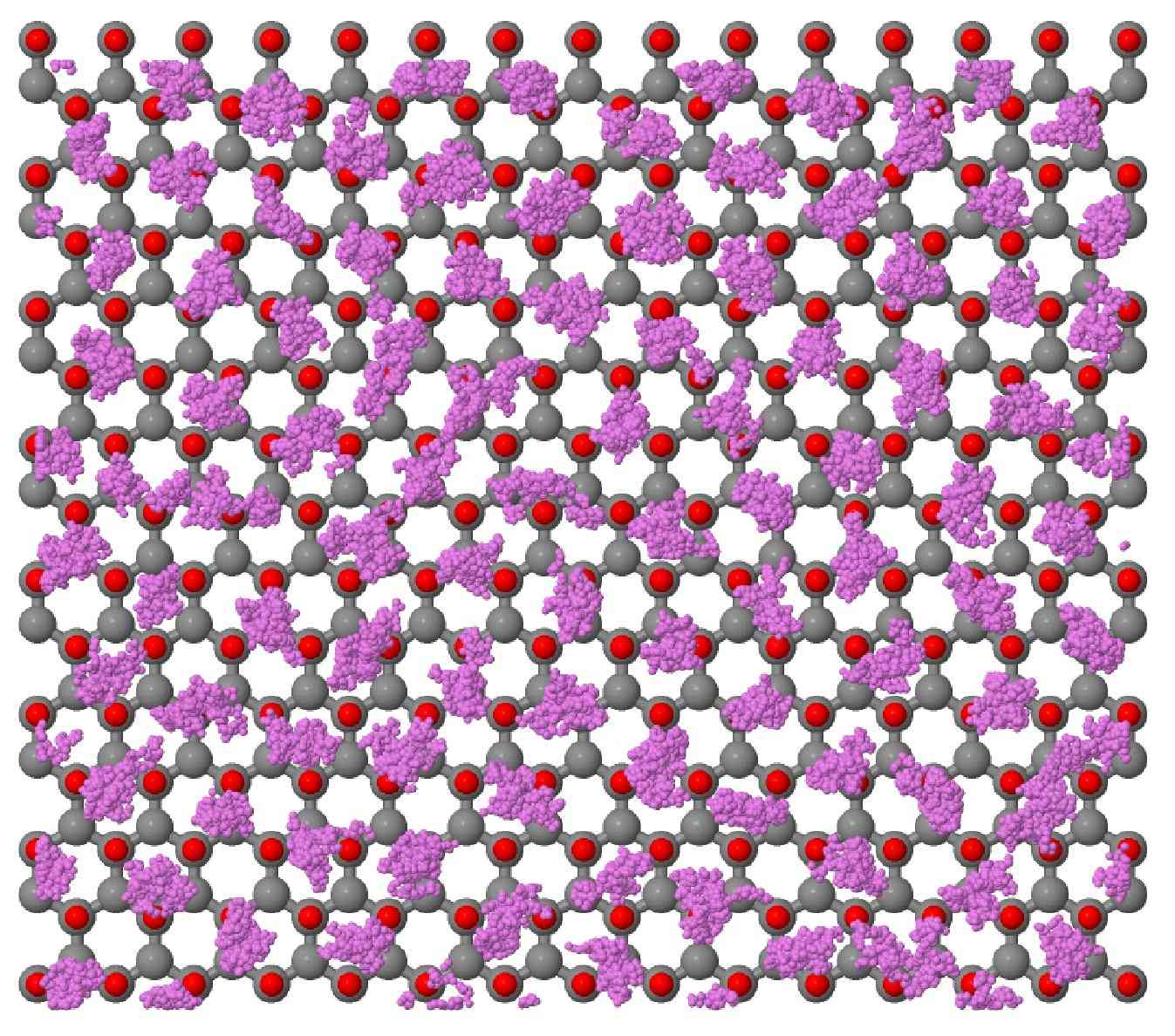}
\end{minipage}
\caption{Polymer configurations in the incommensurate solid density range for a system of $N=82$ atoms of $^4$He on GF at density $\rho=0.118$~\AA$^-2$  (left) 
and a system of $N=98$ atoms of $^4$He on GH  at density $\rho=0.0916$~\AA$^-2$ (right). Each polymer represents the evolution up to 0.4 K$^{-1}$ in imaginary time 
of its correspondent $^4$He atom.
The substrate is represented in the background with carbon atoms marked in gray and the F(H) 
overlayer marked in red.
}\label{ad:cfginc}
\end{figure}

\subsection{Properties at high coverages}

The properties of the first layer at high density have been studied. 

At $x=2/7$ ($\rho=0.0984$~\AA$^{-2}$ on GF,$\rho=0.105$~\AA$^{-2}$ on GH) on both substrates we find that a commensurate triangular solid is stable,
or at least metastable, containing 4 atoms in the unit cell of the triangular lattice
rotated by 19.1$^o$ with respect to the substrate potential.
In the unit cell one of the $^4$He atoms is localized on an adsorption site in the middle of a graphene
hexagonal ring, other two atoms approach adsorption sites of the other kind and finally the
fourth one is centered on a saddle point of the potential. This state has some similarity with the 4/7
commensurate state found for $^3$He in the second layer on graphite\cite{prl:ref5}.
The local density (Fig.~\ref{ad:denssdk47fg} for $^4$He on GF) displays the presence of a superlattice with four 
  atoms in the unit cell of the triangular lattice. The static structure factor $S(k_x,k_y)$ has prominent Bragg peaks forming three stars.
   $S(k_x,k_y)$ for $^4$He on GF is shown in Fig.~\ref{ad:denssdk47fg}. The star of the six highest peaks is the one of a triangular
 lattice with lattice parameter equal to that of a triangular lattice at this density.
 Another star represents the density modulation due to the adsorption potential. The third star formed by six less intense 
 peaks at a smaller wave vector is a combination of vectors of the two previous stars, thus corresponding to interference 
 between the triangular and the honeycomb modulation. The intensity of all these peaks scale with the number of particles 
 (data not shown). Additional peaks are present reflecting the superlattice but these peaks are very weak and hardly visible
  in the figure. Returning to the local density in Fig.~\ref{ad:denssdk47fg} it can be noticed that some of the spots, those located
  at a saddle point, are elongated indicating that the atoms visit also the neighboring adsorption sites. $S(k_x,k_y)$ and the
  local density of $^4$He at coverage 2/7 on GH are shown in Fig.~\ref{ad:denssdk47ch}. The results are similar
  to those on GF, it can be noticed the much smaller intensity of the Bragg peaks due to 
  the adsorption potential in the case of GH as it can be expected due to the weaker corrugation of the adsorption potential
  of GH.

\begin{figure}[h] 
\begin{center}
\includegraphics*[width=11cm]{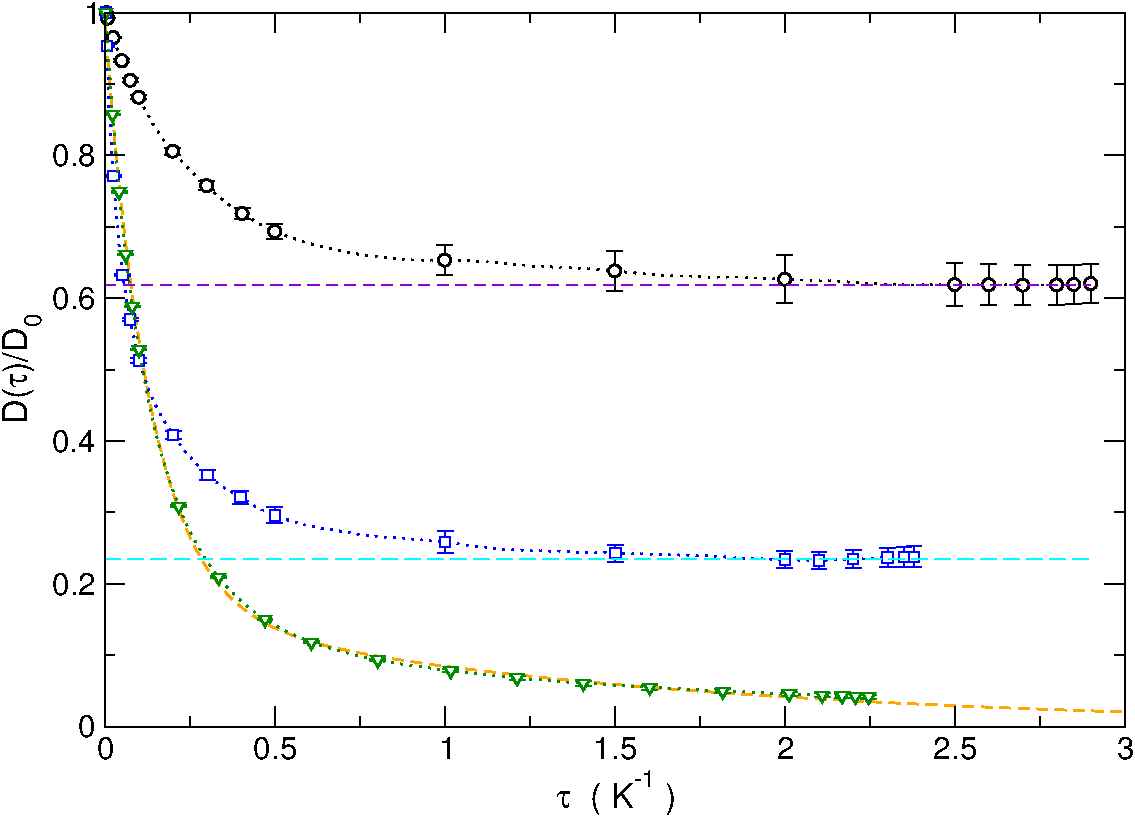}
\caption{Center of mass diffusion for the $2/7$ phase of GF (Circles) and GH (Squares) and the $1/3$ phase of Graphite (Triangles). 
The superfluid fractions are obtained as the long--$\tau$ limit of the plotted functions.
The horizontal dashed lines represent the value of the superfluid fraction, 0.23 for GF and 0.61 for GH. The dashed exponential curve 
is a fit to Graphite data and has a vanishing long $\tau$ behavior.
}\label{ad:supersolid}
\end{center}
\end{figure}

It should be noted that in this 2/7 state not all atoms are localized around a single adsorption site but some
atoms visit two or three neighboring sites, as consequence 
there is spatial order but the atoms are rather mobile and exchange easily so these
solids might be supersolid. Evidence of this is indeed what we find at $T=0$K for both the substrates with the 
superfluidity estimation through the diffusion of the center of mass of the system in imaginary time (see Fig.~\ref{ad:supersolid}). 
At the commensurate $2/7$ phase we estimate a superfluid fraction of 0.23 for GF and of 0.61 for GH.

\begin{figure}[h] 
\begin{center}
\includegraphics*[width=11cm]{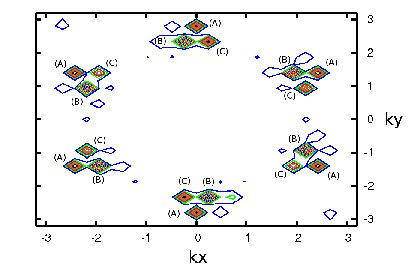}
\caption{Level curves of the static structure factor of the $N$=86 incommensurate solid phase of $^4$He on GF (Figure~\ref{ad:sdk1140}).
The six peaks marked (A) reflect the density modulation due to the corrugation of the adsorption potential. The six peaks
 marked (B) represent the periodicity of a triangular lattice and the peaks marked (C) are interference patterns from
  the density modulations represented by (A) and (B). k$_x$ and k$_y$ axis are expressed in \AA$^{-1}$.
}\label{ad:sdkz}
\end{center}
\end{figure}

\begin{figure}[h]
\begin{center}
\begin{minipage}{6cm}
\includegraphics*[width=6cm]{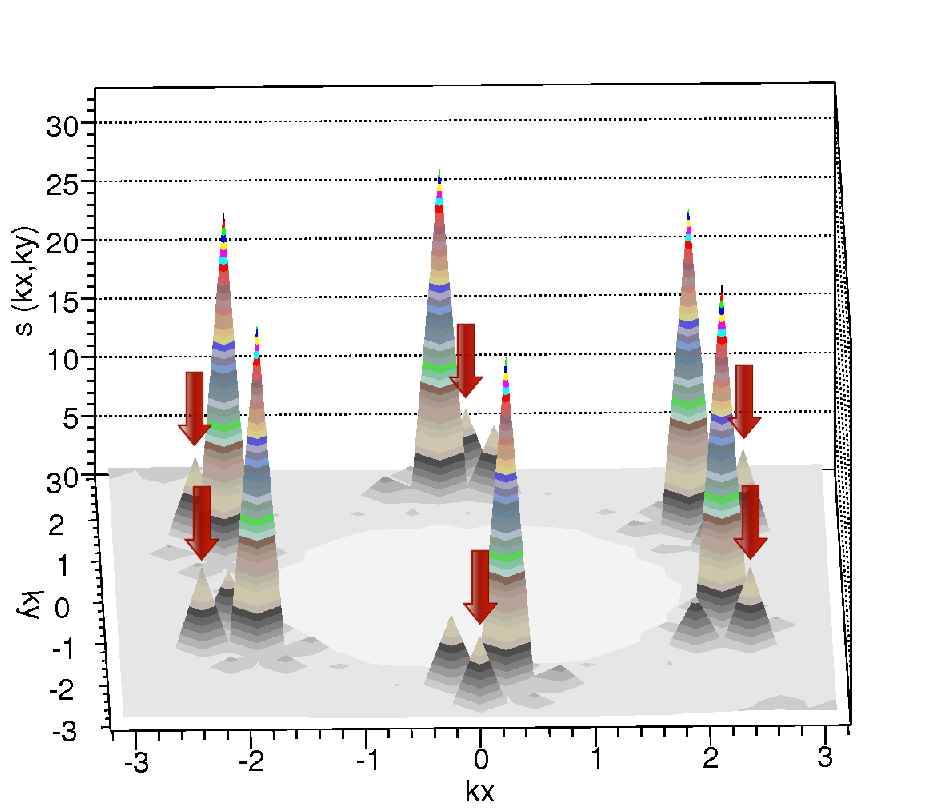}
\end{minipage}
\hspace{1pc}
\begin{minipage}{6cm}
\includegraphics*[width=6cm]{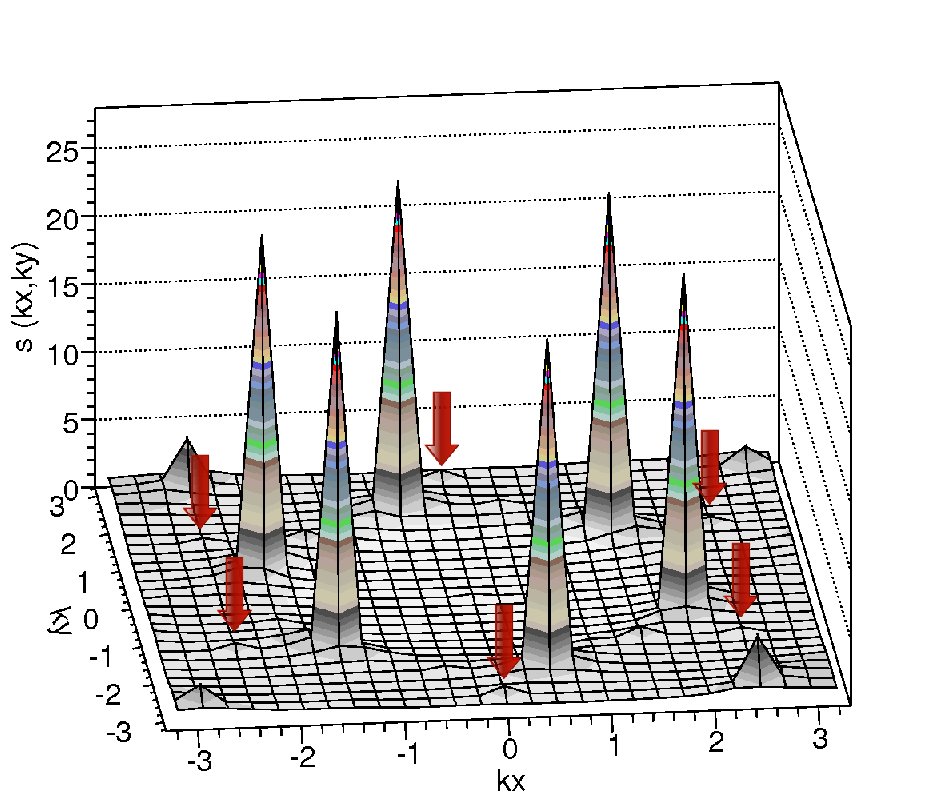}
\end{minipage}
\caption{(Left) Static structure factor on the k$_x$--k$_y$ plane of the incommensurate solid phase of $N$=86 atoms of $^4$He on GF 
at density $\rho=$0.123~\AA$^{-2}$. k$_x$ and k$_y$ axis are expressed in \AA$^{-1}$.
Red arrows point to the peaks corresponding to the density modulation imposed by the adsorption potential. 
(Right) Static structure factor on the k$_x$--k$_y$ plane of the incommensurate solid phase of $N$=66 atoms of  $^4$He on GH at density 
$\rho=$0.102~\AA$^{-2}$. k$_x$ and k$_y$ axis are expressed in \AA$^{-1}$.
Red arrows point to the peaks corresponding to the density modulation imposed by the adsorption potential.
}\label{ad:sdk1140}
\end{center}
\end{figure}

 At coverages around 2/7 we find that $^4$He has an incommensurate triangular order deformed by the substrate potential and defected because
 such order is not compatible with the periodic boundary contitions at the box sides. We discuss first $^4$He on GF.
 We have investigated the density range between $\rho_{2/7}^{GF}=0.0984$~\AA$^{-2}$ and $\rho_{sat}^{GF}=0.136$~\AA$^{-2}$ 
 and as an example $S(k_x,k_y)$ at $\rho=0.123$~\AA$^{-2}$ is shown in Fig.~\ref{ad:sdk1140} and \ref{ad:sdkz}.
  As initial configuration we have used a disordered one as well as an ordered triangular configuration. The runs converge to the same results: $S(k_x,k_y)$
   is dominated by a star of six peaks as expected for a triangular solid. The wave vectors of these peaks are not exactly 
   equal to the value of an ideal triangular solid at this density implying that the triangular order is deformed in order to better fit within the simulation box.
   $S(k_x,k_y)$ has additional Bragg peaks corresponding to the modulation of the substrate potential and to the interference
    between the previous two sets of peaks. Additional smaller peaks are present presumably as consequence of the 
    presence of defects. The modulus of the main Bragg peaks increases in a smooth way as the density is increased as expected 
    for a triangular solid. The observed deviations from the value $k_B = 4\pi\left(\rho/2\sqrt{3}\right)^{1/2}$ of an ideal 
    triangular solid are explained by the deformations of the lattice and by the presence of some defects, mainly dislocations,
     that can be observed from the configuration of the atoms (data not shown).\\
     In the case of GH we have investigated the density range 0.0916~\AA$^{-2}$ -- $\rho_{2/7}^{GH}=0.105$~\AA$^{-2}$.
     Again $S(k_x,k_y)$ is dominated by the Bragg peaks of a triangular lattice (see Fig.~\ref{ad:sdk1140} for S(k) at density $\rho=0.102$~\AA$^{-2}$)
      that is incommensurate with respect to the substrate periodicity.

In Fig.~(\ref{ad:cfg47}) and (\ref{ad:cfginc} we show a sampled configuration of polymers for a system of $^4$He on GF and GH at respectively 
commensurate and incommensurate density. In the commensurate density case, Fig.~\ref{ad:cfg47}, the adsorption potential has a greater influence in the GF case and 
causes the polymers on the saddle point to spread to neighboring adsorption sites; this effect is much less evident in the GH case where the shape of the polymer is more isotropic 
and the occupation of adsorption maxima on top of the H overlayer is more likely. In the incommensurate case, Fig.~\ref{ad:cfginc}, the presence of point dislocations 
are clearly visible in both the cases.

The static structure factors in the solid phase show a characteristic structure of three sets of six peaks 
that is represented in figure (\ref{ad:sdkz}).
The set that has the higher intensity represents the Bragg peaks of a triangular lattice. The six peaks that in 
figures \ref{ad:denssdk47fg}, \ref{ad:denssdk47ch} and \ref{ad:sdk1140} are marked by the red arrows 
represent the density modulation induced by the adsorption potential like the peaks in Fig.~\ref{ad:sdkeq}. 
The third set of peaks is merely an interference pattern of the first two sets.

\begin{figure}[t] 
\begin{center}
\includegraphics*[width=12cm]{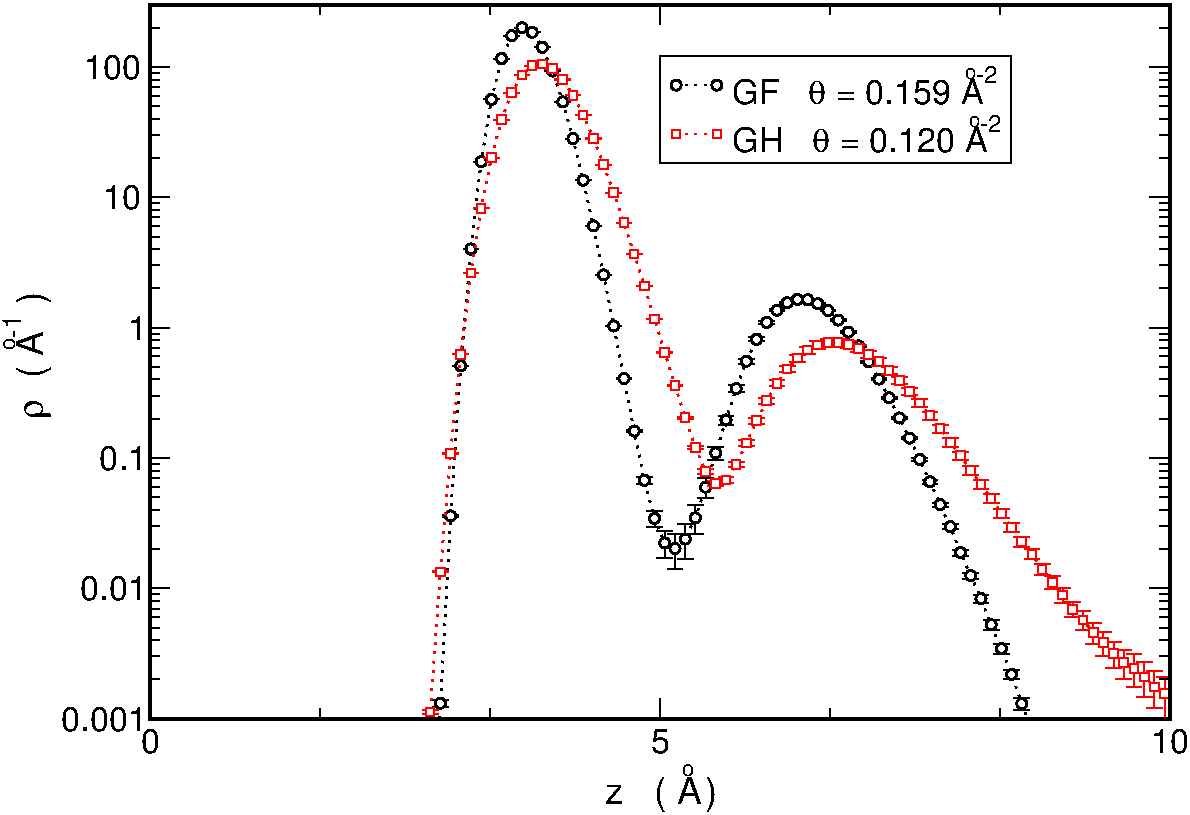}\label{ad:promotion}
\caption{Local density along the $z$--direction of $^4$He on GF (with $N$=111) and $^4$He on GH (with $N$=79) at a density beyond the promotion density.  
The occupation of the first and the second layer are clearly visible as two peaks. The area under the peaks represents the
 number of $^4$He atoms in the corresponding layer. 
}
\end{center}
\end{figure}

Increasing the number $N$, at some point some atoms spill out of the first layer and the density profile in 
the direction normal to the surface develops two well separated peaks. We have thus estimated the first layer's 
completion density, $\rho_{sat}^{GF(GH)}$.
The promotion to the second layer takes place at a density 
$\rho_{sat}^{GF}=0.136$~\AA$^{-2}$ for the GF case and a density 
$\rho_{sat}^{GH}=0.108$~\AA$^{-2}$ for the GH case. Beyond such densities, 
the occupation of the second layer is clearly visible as a secondary peak in 
the local density along the $z$--direction  displayed in Fig.~\ref{ad:promotion}.

\subsection{Equation of state of $^3$He on GF and GH}

\begin{figure}[t] 
\begin{center}
\includegraphics*[width=10cm]{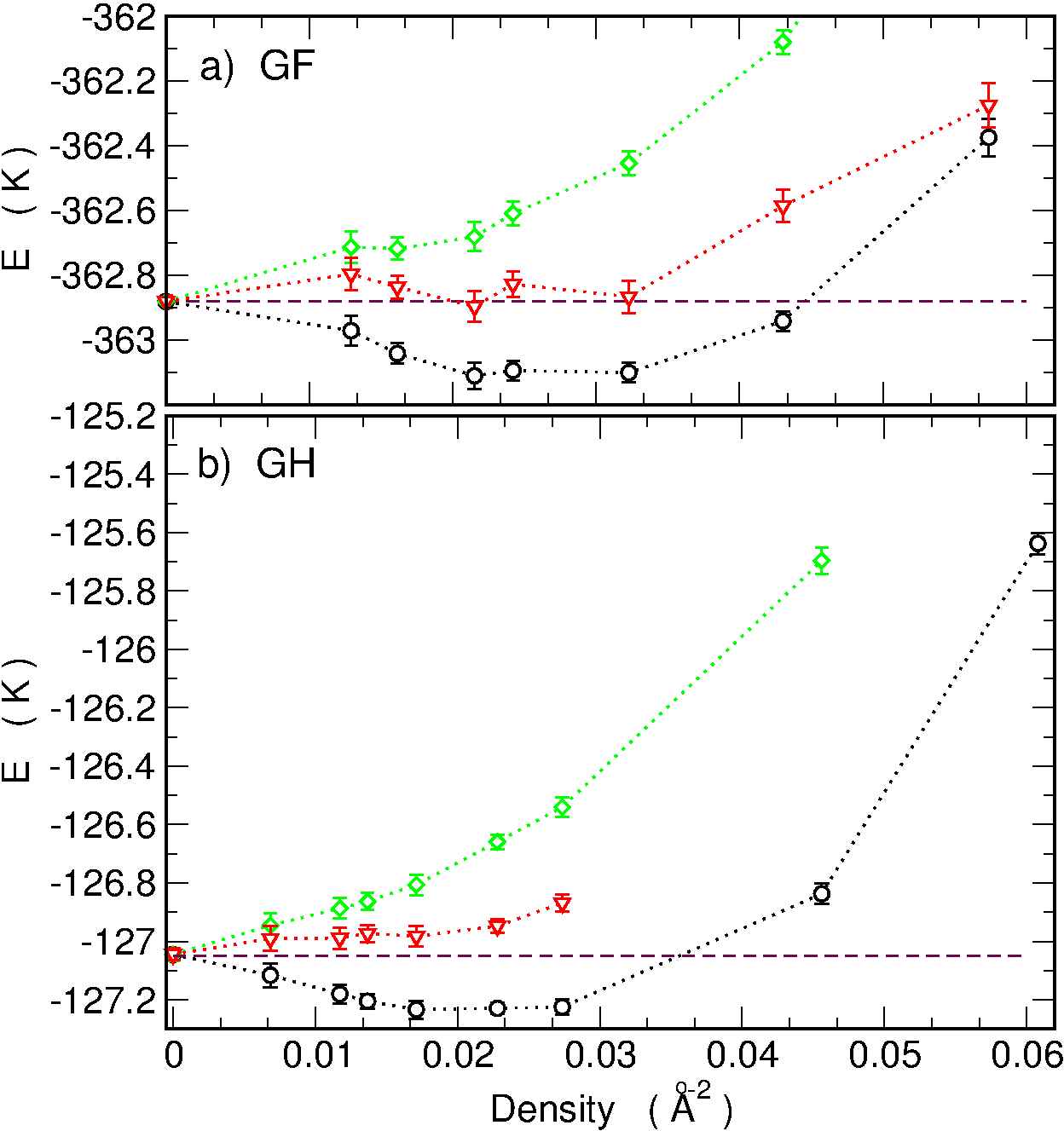}
\caption{ 
Panel a). Ground state energy as function of density of mass 3 bosons on GF (circles),  
Fermionic $^3$He--GF obtained via the Fermionic correlations method (triangles) and  
Fermionic $^3$He--GF obtained by approximating the Bose--Fermi gap with 
the kinetic energy of the free fermion gas (diamonds).  
Panel b). Same as for panel (a) for $^3$He on GH.
}\label{ad:stateqhe3}
\end{center}
\end{figure}

The ground state of $^3$He on graphite is the $\sqrt{3}\times\sqrt{3}$ R30$^o$ state.
We expect that the analogous commensurate state on GF and GH is unstable, as for $^4$He,
because the smaller mass makes $^3$He localization more expensive.

The ground state energy as function of density of mass 3 bosons and of the Fermionic $^3$He on GF and on GH are plotted in Fig.~\ref{ad:stateqhe3} as function of density.
In both cases, the system under study was composed of $N=$18 atoms of $^3$He.\\ 
In the figure we plot also the $^3$He energy based on the crude approximation of taking as Fermi--Bose gap the kinetic energy
 $K_{free}=\hbar^2\pi\rho/2m^*$ of free fermions, where $m^*$ is the effective mass of a $^3$He atom on the substrate
  ($m^*/m=1.26$ for GF,$m^*/m=1.01$ for GH).\\

As shown in Fig.~(\ref{ad:stateqhe3}), the $\sqrt{3}\times\sqrt{3}$ R30$^o$  commensurate state for a mass 3 boson system is indeed unstable toward a fluid state
on both substrates, in fact, the energy at the density corresponding to the
$\sqrt{3}\times\sqrt{3}$ state is well above the energy at lower densities implying that this ordered
commensurate state is indeed unstable and the system is in a fluid state.
As a consequence we predict the existence of two new {\em anisotropic} Fermi fluids, in the sense 
that the local density is non--uniform and anisotropic, with a tunable density
depending on the $^3$He coverage.
The density range depends on whether the $^3$He atoms form a self--bound state.
Such a self--bound state seems unlikely to occur for $^3$He on GH on the basis of our computations.
On the contrary a self--bound state might be present on GF.
For mass 3 Bosons we find a bound state with a binding energy $E_0=-0.22$ K at density 
$\rho_{eq}=0.03$ \AA$^{-2}$. Adding to the boson energy the Fermi--Bose gap the energy yields a
shallow minimum in the density range 0.015--0.025 \AA$^{-2}$.
The energy per particle at this minimum is equal within the statistical error to the energy of a single
adsorbed $^3$He on GF so that the existence of a self--bound state on GF is an unresolved possibility.

\paragraph{Remark:}
An accurate approximation for the energy per particle for $^3$He on GF and GH has been obtained via the Fermionic Correlations technique\cite{prl:fermi,prl:fermico}.
This methodology has been explained in Chapter~\ref{ch:polarization}: given a specific Hamiltonian, the Fermionic Correlations technique extracts the energy gap 
between the symmetric and antisymmetric ground state 
from a suitable Fermionic imaginary--time correlation function computed as an exact average on the Bose ground 
state:
\begin{equation}
\label{fgcfun}
\mathcal{C}_{F}(\tau)
 \equiv \frac{\langle \psi_0^B | \left(e^{\tau \hat{H}}\hat{\mathcal{A}}_{F}^{\dagger}e^{-\tau \hat{H}}\right)
\hat{\mathcal{A}}_{F} \psi_0^B \rangle_{\mathcal{H}(N)}}{\langle \psi_0^B | \psi_0^B \rangle_{\mathcal{H}(N)}}, \quad \tau \geq 0
\end{equation}

where $\hat{\mathcal{A}}_{F}$ is, typically, a Slater determinant.
The lowest energy contribution in $\mathcal{C}_{F}(\tau)$ yields the {\it exact gap} 
between the Fermionic and the Bosonic ground states, provided that one is able to obtain the inverse Laplace transform of 
$\mathcal{C}_{F}(\tau)$; this can be readily seen by formally expressing \eqref{fgcfun} on the basis 
$\{\psi_n^F\}_{n \geq 0}$ of eigenvectors of $\hat{H}$ corresponding 
to the eigenvalues $\{E_n^F\}_{n \geq 0}$:

\begin{equation}
\label{fgcfun2}
\mathcal{C}_{F}(\tau)
= \sum_{n=0}^{+\infty}
e^{-\tau \left(E_n^F - E_0^B\right)}
\frac{|\langle \hat{\mathcal{A}}_{F}\psi_0^B |
\psi_n^F \rangle_{\mathcal{H}\left(N\right)}|^2}{\langle \psi_0^B | \psi_0^B \rangle_{\mathcal{H}\left(N\right)}}
\end{equation}

We have shown that this analytic continuation procedure can be handled efficiently with statistical inversion procedures, like the 
GIFT algorithm introduced in Ref.\cite{qfs:gift}.The Fermi--Bose gap $E_0^F-E_0^B$ is an extensive quantity, so this method can be applied 
 provided that the system is not too large.

\subsection{Discussion}
He adsorption on new substrate materials is valuable because of the 
fundamental importance of helium in many--body physics, with a variety of 
phases seen in both 2D and 3D. Our results indicate that the GF substrate 
provides the strongest binding of any surface (since the previous record 
was held by graphite). Moreover, the novel symmetry, the smaller 
intersite distance and large corrugation imply that quite novel properties 
may be anticipated for this system. This is indeed the case. 
When many $^4$He atoms are adsorbed on GF and on GH a very striking result 
is that the ground state is a low density liquid modulated by the substrate 
potential and the system has BEC, i.e. it is a superfluid. 
This is qualitatively different from graphite for which the lowest energy 
state is the $\sqrt{3}\times\sqrt{3}$ R30$^o$ commensurate one with no BEC \cite{lt26:ref18}. 
We have verified that such an ordered state on GF and GH is unstable 
relative to the liquid phase.  
It should be noticed that some of the parameters in the adsorption potential 
are not known with high precision or they have been adopted from other systems. 
We have verified that even a change of parameters like $\alpha$, $C_{\rm 6F}$ and $\beta$
by 20\% does not modify the qualitative behavior of the adsorbed He atoms 
even if there can be a sizable change in the value, for instance, of the 
adsorption energy.
Measurement of thermodynamic properties and 
He atomic beam scattering experiments from GF and GH will be 
important to test the accuracy of our model potentials.
A remarkable result  is the superfluid behavior of the 2/7 phase that, however, might be a 
property of the system at strictly $T=0$K and is non reachable by experiments; on the other hand 
there might truly be a ``supersolid'' phase transition at a temperature in the $m$K range that is not accessible 
by QMC computations. This, together with all the novel phenomena for He atoms on GF and GH that have been 
predicted in this work, calls for experimental verification.
There is also an important aspect that should be considered in view of experiments; it might be difficult 
to have a 100\% reacted graphene sheet with fluorine. However, the presence on GF of small regions of unreacted 
graphane should not affect the properties of the adsorbed film because the He or H$_2$ atoms are preferentially 
adsorbed on the F covered regions of graphene. This behavior, however, may change when coverages beyond the 
first layer completion on GF and GH are considered; in such cases it could be that the adsorbed atoms begin to 
populate the unreacted regions. The QMC techniques here employed may be used to investigate also these 
interesting cases given that a suitable interaction potential is provided.

From the theoretical point of view many extensions of the present computations can be foreseen,
for instance the characterization of the commensurate 2/7 phases on GF and GH, of the system under rotation
and the study the phase diagram of p--H$_2$ on GF.
As a perspective of future work we plan to provide predictions 
concerning the phase diagrams and thermodynamic properties for both He/GF and He/GH,
hoping to stimulate experimental studies of these systems.

%% file: ch-conclusion/chapter-conclusion.tex
\chapter{Conclusions} \label{ch:conclusions}

The idea underlying this work has been the study of strongly interacting quantum systems along with the development of 
new methodologies in the field of QMC. Strongly interacting quantum systems are indeed a 
fascinating field of research, much is yet unknown and indeed a proof of this assertion might be found in our 
results: we have studied new adsorbed phases of $^4$He and also predicted the presence of a {\it modulated} superfluid 
given by the interplay between interatomic potentials and quantum tunneling. Strongly interacting 
Fermi systems are even more unexplored due to the sign problem. The methodological aspect of this work has been thus 
focused to develop a technique that can study the dynamics of such systems. This technique is an evolution of the 
Fermionic Correlations method; we have shown that, even though this methodology becomes unpractical for 
big numbers of particles, it indeed can compute an {\it ab--initio} low--energy 
excitation spectrum of two--dimensional $^3$He.

In the conclusions of this work, we remind the main results obtained and presented throughout this work; 
we have already drawn conclusions in each chapter, here we will comment mostly the computations that are still 
in progress and even the ``failed attempts''.

\paragraph{2$d$ $^3$He.}Our simulation of two--dimensional $^3$He gave a spin susceptibility as function of density that is in very good
agreement with experimental data; our obtained polarization curves indicates that the ferromagnetic fluid is 
never stable and the system crystallizes into a triangular lattice from the paramagnetic fluid 
at a density of 0.061 \AA$^{-2}$. With an extension of the Fermionic Correlation (FC) technique, we have been able to 
obtain the first {\it ab--initio} evaluation of the zero--sound mode and the dynamic structure factor of 2$d$ $^3$He that 
is in remarkably good agreement with experiments. This excitation spectrum, moreover, turned out to have 
striking similarities with the phonon-maxon-roton spectrum of $^4$He; this indicates that 
the effects of the inter--atomic potential, in particular its strong repulsive part, dominate 
over the effects of the quantum symmetry.
Another interesting question is whether the zero--sound mode, which is known to enter the particle--hole band,  reemerges 
at wave vectors corresponding to the ``roton'' minimum of the spectrum; it is possible, with the FC method, 
to compute the particle--hole excitations and indeed we have presented preliminary results that show 
that the ``roton'' is still in the particle--hole band. However, the re-emergence of the roton is a rather difficult question to answer 
with a simulation of a finite system. This is because there are relevant size effects on the particle--hole; 
these effects are due to the fact that we are far from the thermodynamic limit and the particle--hole is not a continuum; in order to 
obtain more conclusive data, a scaling analysis on bigger systems is in order. Such a study that is very demanding in term of computing 
resources and has been planned for future work. 

We also attempted a study of the spin--waves excitations but in this case we found that the results were 
highly dependent on the direction of the wave vector. This anisotropy is a clear sign of size effects; 
unfortunately, these size effects for spin--waves are stronger than in the zero--sound or even the particle hole case 
and thus require the study of systems with particle numbers for which, like for most QMC methods, FC becomes unpractical.

An interesting perspective is the application of the FC technique to the study of elementary excitations of the 
2$d$ electron gas; possibly this would provide an {\it ab--initio} evaluation of the 
plasmon excitations.

\paragraph{$^4$He on Graphene-Fluoryde and Graphane.} The study of $^4$He adsorption on Graphene-Fluoryde (GF) 
and Graphane (GH) has been a comprehensive and articulated work. At the early stages of the project we showed that 
the commensurate $\sqrt 3 \times \sqrt 3$ R30$^o$ phase is unstable on both substrates.We then determined the 
equilibrium density at $T=0$ K; our results indicated clearly that on both the substrates the equilibrium density 
has a condensate fraction and is thus a {\it modulated superfluid}. We determined the superfluid fraction at 
zero and finite temperature giving also a rough estimate of the fluid--superfluid transition temperature. The study 
of the equilibrium density at $T=0$ K comprised also the excitation spectrum, and we have shown the 
phono--roton spectrum of $^4$He on GF and GH. We focused then on high coverages of the monolayer and found a density 
range, not yet precisely determined, in which $^4$He forms an incommensurate triangular solid; a remarkable result 
is that on both GF and GH a commensurate phase at filling factor $x=2/7$ is stable or at least metastable. This result becomes even more 
interesting because we found a first evidence of superfluidity at zero temperature: at this density, the system may 
possibly be both solid and superfluid, in other words this system could posses the long sought property of supersolidity.

For the immediate future, we plan to study further this commensurate density with also a size scaling aimed to better estimate the finite size 
effects on the properties of the $2/7$ phase, in particular on the superfluid fraction. In our further studies, there are mainly two points to inspect: 
first, is the 2/7 phase thermodynamically stable? This far, we have shown that it is mechanically stable, meaning that, at the 
density corresponding to $x=2/7$, the configuration that gives the lowest energy is the $2/7$ triangular lattice; now 
we are planning to search for signatures of possible phase transitions between the incommensurate solid and the 
$2/7$ phase. Second, we are searching for more evidence of superfluidity in the commensurate phase, we already tried 
the computation of the superfluid density at finite temperature down to 0.5 K but we did not find any superfluid signal, indeed, 
the very low rate of exchanges between atoms suggests that, if any, the transition temperature to the supersolid state would be 
too low to be reachable with PIMC with our current computing resources; another approach is the computation of the one body density matrix at zero temperature, 
in fact, the presence of even a small fraction of condensate would be a strong support for the 
supersolidity of this phase; this is very demanding in term of computing resources and is planned for the next future.

Besides the supersolidity, we have scheduled also a deeper study of the incommensurate density range with a quantitative 
characterization of the defects. This will prepare the background for the study of the second adsorbed layer that will be left for 
 future works.

We conclude with a last remark. For the Helium-substrate interaction we have adopted a semi--empirical 
potential: its repulsive part has been obtained from a DFT calculation of the electron density of the substrate, on the 
other hand, the attractive part has been modeled with a Van der Waals type interaction with parameters taken from literature,
 adopted from the interaction potentials of Helium with similar chemical compounds. This study, as consequence, can be considered 
 a semi--quantitative approach but the very fact that we find qualitatively the same behavior on two substrates that have 
 completely different values of energies and corrugation is a strong proof of plausibility of our results; moreover, the 
 robustness of the results has been explicitly tested at the $\sqrt 3 \times \sqrt 3$ R30$^o$ density with a variation 
 up to 20\% of the parameters of the He--substrate potential. 
 More exactly, this is an ``exact'' study on a semi--empirical Hamiltonian aimed to the research of new properties of adsorbed 
 matter, our hope is that this predictive work will encourage, on one side, the development of more accurate Helium--substrate 
 interaction potential and, on the other side, the experimental exploration of this subject so fascinating and full of surprises.

%% file: ch-details/chapter-details.tex
\def\onlinecite{\cite} 
\chapter{Computational details\label{ch:details}}

In this Chapter some technical details of the Monte Carlo techniques introduced in Sec. \ref{ch:methods} 
will be described. 
Monte Carlo sampling will then be applied to the problem of evaluating physical properties of
quantum systems at both zero and finite temperature. 

The basic idea underlying the used path integral methods is that the computation of an expectation value 
in a quantum system can be viewed as an $N$--dimensional integral; in the case of a bosonic system, the ground state 
wave function can be chosen real and non--negative and this integral can be interpreted as the average of a random 
variable over a probability density\cite{c7r1}.

 \section{Monte Carlo integration: the strategy}
An effective way to compute an $N$--dimensional integral is to employ Monte Carlo (MC). 
Monte Carlo basically means ``the use of random numbers in order to solve a problem''. In our case, 
the problem is the $N$--dimensional integral representing an expectation value for a quantum many--body system. 
In order to show how MC is employed in our context, consider, as an illustrative example, 
a function $f(\vec{x})$ that is a product of an arbitrary function $g(\vec{x})$ and a probability 
 density $p(\vec{x})$
 \begin{eqnarray}
 f(\vec{x}) = g(\vec{x})\cdot p(\vec{x}) \:\:\mbox{with} \\
 p(\vec{x}) \ge 0 \forall \vec{x} \in \Gamma ,\,\,\,\,\,\,\,  \int_{\Gamma}d\vec{x}\:p(\vec{x}) = 1
 \end{eqnarray}\label{eq11}
 The integral can be rewritten as\cite{c7r2},
 \begin{eqnarray}\label{c1rpe}
 \int_{\Gamma}d\vec{x}\:g(\vec{x})p(\vec{x}) = \lim_{N\rightarrow +\infty} \frac{1}{N} \sum_{i=1}^{N}g(\vec{x}_{i}) \simeq \langle g\rangle_{p}
 \end{eqnarray}
 where $\vec{x}_{i}$ are elements sampled from the probability density $p(\vec{x})$. 
 The integral value is thus the average of $g(\vec{x})$ over sets of values $\vec{x}$ that are sampled from the probability 
 density $p(\vec{x})$. The advantage of MC is that, given a way to sample $p(\vec{x})$, the computing time required 
 for the evaluation of the integral does not scale with the dimensionality; this is 
 very important since, as will be clear further on, the integrals that are computed in this context have generally 
 a very high dimensionality. 
 
  \begin{figure}[h]
 \begin{center}
 \includegraphics*[width=11cm]{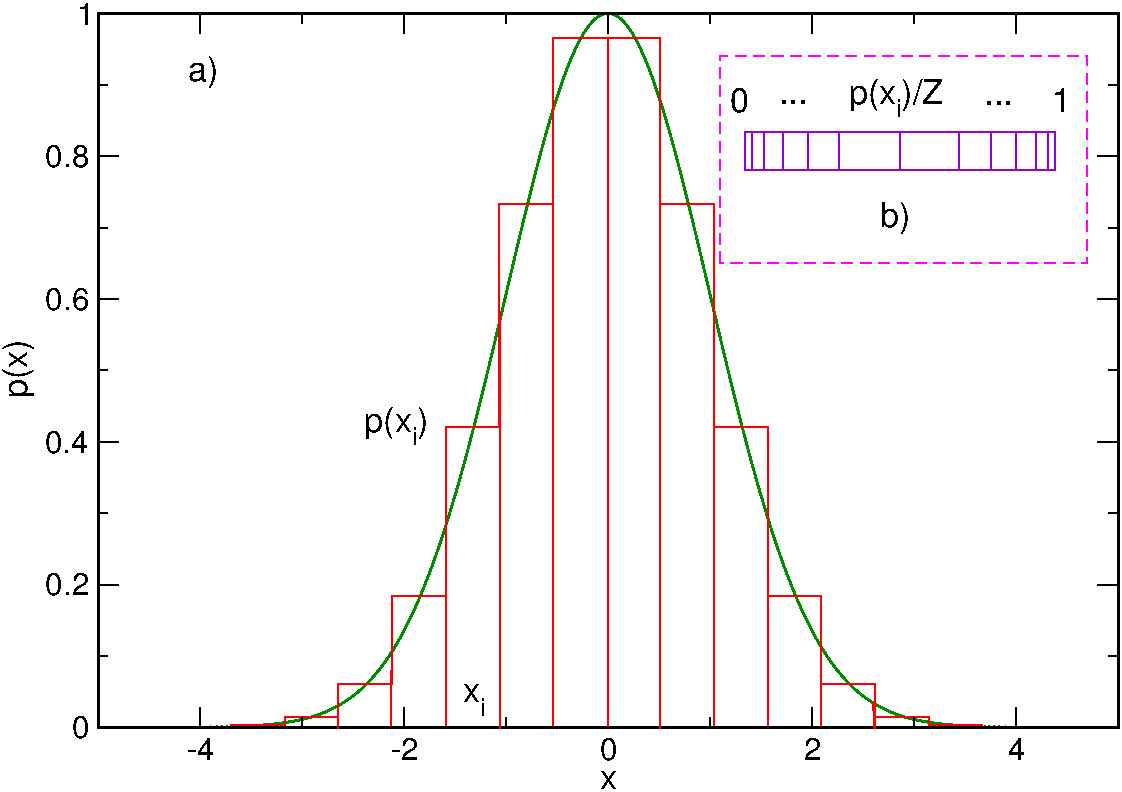}
 \caption{\label{samplingmeans} 
 a) Schematic 1D representation of the sampling of a probability density and its discretization in $N$ bins. b) The partition of the unity used in the faked roulette method: the interval $\left[0;1\right)$ 
 is divided into $N$ bins, the $m$ bin is the interval $I_m=\left[\sum_{j=1}^{m-1}p(x_j)/Z;\sum_{j=1}^{m}p(x_j)/Z\right)$, where the normalization $Z=\sum_{j=1}^{N}p(x_j)$.
 The faked roulette is a method to randomly chose $m$: a random number $r$ uniformly distributed in $[0;1)$ is generated, $m$ corresponds to the interval $I_m$ in which $r$ falls, namely: $I_m$ so that 
 $I_m \cap \lbrace r\rbrace = \lbrace r\rbrace$
 }
\end{center}
\end{figure}

 This however requires that one is able to sample an arbitrary $N$--dimensional probability 
 distribution. Sampling means the generation of a random variable according to an arbitrary probability 
 density $p(\vec{x})$; as sketched in Fig.~\ref{samplingmeans}, sampling can be done by dividing the 
domain of $p(\vec{x})$ in $K^N$ bins with an assigned probability $p'_i=\frac{p(\vec{x}_{i})}{Z}$, with $Z$ a 
normalization constant and $\vec{x}_i$ the central 
coordinate of the $i$--th bin. A simple way to extract a random variable value is through a 
{\it faked roulette} (see Fig.~\ref{samplingmeans}), however, the computational weight of this approach 
increases  exponentially with the number of degrees of freedom and is thus unpractical for the evaluation 
of Eq. \eqref{eq11}. A more sophisticated and efficient way 
to sample an arbitrary probability density is with Markov chains\cite{d:kalos}. As is shown in the next section, 
a Markov chain has at least one invariant probability density and 
 there is a sufficient condition for its uniqueness; the basic idea is thus to build a Markov chain that converges 
 to the required unique invariant probability density.
 In the next section we provide a mathematical demonstration of the properties of the Markov chains used in this context, 
 after that section an algorithm that can be used to build the required Markov chain, namely
  the Metropolis algorithm\cite{d:metropolis}, will be described.
  
  \subsection{Mathematics of Markov chains}
In this section, we follow Ref.~\onlinecite{d:baldimarkov} and show the mathematical basis of the Markov chains. 
Let's consider a given finite set $E=\lbrace 1,...N\rbrace$ and a probability space $(\Omega,\mathcal{F},P)$, where $\Omega$ is a sample space, $\mathcal{F}$ is a 
$\sigma$--algebra on $\Omega$ and $P$ a probability measure. 

{\bf Definition 1.} A Markov chain on a sample space $E$ is a sequence of random variables $\lbrace X_n \rbrace_{n \in \mathbf{N}^{0}}$, $X_n \:\: : \:\: \Omega \rightarrow E$ such that 
there are non negative numbers $\mathcal{P}_{i\rightarrow j}(n)$, $n \in \mathbf{N}^{0}$ and $i, j \in E$ for which,
\begin{eqnarray} 
P(X_{n+1}=j|X_n=i,X_{n-1}=i_{n-1},...,X_0=i_0)= \nonumber \\
P(X_{n+1}=j|X_{n}=i)=\mathcal{P}_{i\rightarrow j} (n) \label{et1}
\end{eqnarray}
whenever the conditional probabilities $P(\cdot|\cdot,...)$ are defined. 
From here on we focus on Markov chain that are independent on time translations, in this case $\mathcal{P}_{i\rightarrow j} (n)$ does not depend on $n$.  

{\bf Transition matrix.} The non-negative numbers $\mathcal{P}_{i \rightarrow j}$ can be represented in an $N \times N$ matrix $\underline{\mathcal{P}}$ that is referred as {\it transition matrix} of 
the Markov chain. Following from the definition of conditional probability, the sum of the element of a row is one, namely $\sum_{j=1}^{N}\mathcal{P}_{i\rightarrow j}=1$ for $i=1,...N$.
The probability distribution of the random variable $X_0$ is the starting probability of the chain and is defined by the numbers $v_k = P(X_0 = k) \:,\:\: k \in E$,
this probability distribution can be identified by a row vector in $\mathbf{R}^N$, $\underline{v}=(v_1,...,v_N)$. 

{\bf Statement:} {\it a Markov chain that is independent on time translations is defined by a starting probability and a transition matrix}.
This can be shown if, from the definition of conditional probability, we first obtain the probability distribution of $X_1$:
\begin{eqnarray}
P(X_1=k) = \sum_{h=1}^{N}P(X_0=h)P(X_1=k|X_0=h)=\sum_{h=1}^{N}v_h\mathcal{P}_{h\rightarrow k}
\end{eqnarray}
that in matrix notation becomes,
\begin{eqnarray}
\underline{v}^{(1)}=\underline{v}\,\underline{\mathcal{P}}
\end{eqnarray}
where we define $\underline{v}^{(1)}=P(X_1=k)$. Iterating to obtain the probability distribution for the next time step, $P(X_2)$,
\begin{eqnarray}
P(X_2=k)=\sum_{l=1}^{N}P(X_1=l)P(X_2=k|X_1=l)= \nonumber \\
=\sum_{l=1}^{N}P(X_1=l)\mathcal{P}_{l\rightarrow k} = \sum_{l=1}^{N}\sum_{h=1}^{N}v_h\mathcal{P}_{h\rightarrow l}\mathcal{P}_{l\rightarrow k}= \nonumber \\
=\sum_{h=1}^{N}v_h\sum_{l=1}^{N}\mathcal{P}_{h\rightarrow l}\mathcal{P}_{l\rightarrow k}
\end{eqnarray}
in matrix form:
\begin{eqnarray}
\underline{v}^{(2)}=\underline{v}\,\underline{\mathcal{P}}^2
\end{eqnarray}
iterating this rule the probability distribution at time step $n$ can be obtained as the $n$--power of the transition matrix applied to the starting probability,
$\underline{v}^{(n)}=\underline{v}\,\underline{\mathcal{P}}^n$.
It will be used later on in this section the {\it $m$ steps transition probability}, $\mathcal{P}_{i\rightarrow j}^{(m)} = P(X_{n+m}=j|X_{n}=i)$. 

{\bf Statement:} {\it $\mathcal{P}_{i\rightarrow j}^{(m)}$ are the matrix elements of $\underline{\mathcal{P}}^m$. }
This, again, is shown by iteration:
\begin{eqnarray}
P(X_{n+m}=j|X_n=i)=\frac{P(X_{n+m}=j,X_n=i)}{P(X_n=i)}= \nonumber \\
= \sum_{h=1}^N \frac{P(X_{n+m}=j,X_{n+m-1}=h,X_n=i)}{P(X_n=i)} = \nonumber \\
= \sum_{h=1}^{N} \frac{P(X_{n+m}=j,X_{n+m-1}=h,X_n=i)}{P(X_{n+m-1}=h,X_n=i)}\,\frac{P(X_{n+m-1}=h,X_n=i)}{P(X_n=i)} = \nonumber \\
= \sum_{h=1}^{N}P(X_{n+m}=j|X_{n+m-1}=h,X_n=i)P(X_{n+m-1}=h|X_n=i)= \nonumber \\
=\sum_{h=1}^{N}\mathcal{P}_{h\rightarrow j}P(X_{n+m-1}=h|X_n=i).
\end{eqnarray}
Repeating this passage with $P(X_{n+m-1}=h|X_n=i)$, and so on until $P(X_{n}=k|X_n=i)$, the statement is demonstrated.

{\bf Invariant probabilities.} Given a probability distribution $\pi$ on the set $E$ identified by a row vector $\underline{\pi}=(\pi_1,...,\pi_N) \: \in \: \mathbf{R}^N$ and a 
Markov chain defined by a transition matrix $\underline{\mathcal{P}}$ and starting probability $\underline{v}$, $\pi$ is {\it invariant} if:
\begin{eqnarray}
\underline{\pi} = \underline{\pi}\,\underline{\mathcal{P}},
\end{eqnarray} 
in the particular case of $\underline{v} = \underline{\pi}$ we say that the Markov chain is {\it stationary}. 

{\bf Theorem 1. (Markov-Kakutani)} {\it There is always at least one invariant probability distribution.} 

\begin{proof} The probabilities on $E$ are mapped onto the set 
\begin{eqnarray}
S=\left\lbrace\underline{x}\,\in\,\mathbf{R}^N\: :\: 0 \le x_i \le 1\, , \, \sum_{i=1}^{N}x_i = 1 \right\rbrace,
\end{eqnarray}
this is a closed and limited set in $\mathbf{R}^N$ and hence, by the Bolzano-Weierstrass theorem, it is a compact set; given a sequence of points in S, it is thus possible to define a 
subsequence that converges to a point in $S$. From a point $\underline{x}$ of $S$, we define the sequence:
\begin{eqnarray}
\underline{x}_n = \frac{1}{n}\sum_{k=0}^{n-1}\underline{x}\,\underline{\mathcal{P}}^k,
\end{eqnarray}
this vector has non negative components because is a product of elements with non negative components; $\underline{x}_n$ also belongs to $S$, this is readily seen with a summation of its components $x_{n,i}$,
\begin{eqnarray}
\sum_{i=1}^{N}x_{n,i}= \frac{1}{n}\sum_{k=0}^{n-1}\sum_{h=1}^{N}\sum_{i=1}^{N}x_h\mathcal{P}_{h\rightarrow i}^{(k)} = \frac{1}{n}\sum_{k=0}^{n-1}\sum_{h=1}^{N}x_h = 1
\end{eqnarray}
where in the last passage the property of the transition probability $\sum_i \mathcal{P}_{h\rightarrow i}^{(m)}=1$ was used. Having proved that $\underline{x}_n \: \in \: S$, there is a subsequence 
$\lbrace\underline{x}_{n_k}\rbrace$ converging to a point $\underline{\pi} \: \in \: S$. Now we write
\begin{eqnarray}
\underline{x}_{n_k}-\underline{x}_{n_k}\,\underline{\mathcal{P}} = \nonumber \\
\frac{1}{n_k}\left(\sum_{h=0}^{n_k-1}\underline{x}\,\underline{\mathcal{P}}^h-\sum_{h=0}^{n_k-1}\underline{x}\,\underline{\mathcal{P}}^{h+1}\right) = \nonumber \\
= \frac{1}{n_k}\left(\underline{x}-\underline{x}\,\underline{\mathcal{P}}^{n_k}\right)
\end{eqnarray}
taking the limit for $k$ to infinity, we observe that while $n_k$ diverges, the quantity $\underline{x}-\underline{x}\,\underline{\mathcal{P}}^{n_k}$ remains finite because 
is the difference between two elements of the limited set $S$. We thus obtain the following relation for $\underline{\pi}$:
\begin{eqnarray}
\underline{\pi} - \underline{\pi}\,\underline{\mathcal{P}} = \lim_{k\rightarrow +\infty}(\underline{x}_{n_k}-\underline{x}_{n_k}\,\underline{\mathcal{P}}) = \lim_{k\rightarrow +\infty}\frac{1}{n_k}(\underline{x}
-\underline{x}\,\underline{\mathcal{P}}^{n_k}) = 0,
\end{eqnarray}
that demonstrates the theorem.\end{proof} 
The invariant probability distributions of a Markov chain can be obtained from the solution of the linear system 
\begin{eqnarray}
\pi_j = \sum_{i=1}^{N}\pi_i\mathcal{P}_{i\rightarrow j},
\end{eqnarray}
however, a {\it sufficient condition} for $\pi$ to be invariant is that it satisfies the {\it detailed balance equation}:
\begin{eqnarray}
\pi_i\mathcal{P}_{i\rightarrow j} = \pi_j\mathcal{P}_{j\rightarrow i}
\end{eqnarray}
for every $i, j \: \in E$. The demonstration of this statement comes directly from the definition of transition matrix
\begin{eqnarray}
\sum_{i=1}^n \pi_i\mathcal{P}_{i\rightarrow j} = \sum_{i=1}^{n}\pi_j \mathcal{P}_{j\rightarrow i} = \pi_j
\end{eqnarray}
The detailed balance condition will be used in the next section when the Metropolis algorithm will be described. This algorithm is used to build a Markov chain that converges to 
an arbitrary invariant probability density. 

{\bf Uniqueness of invariant probability distributions.} Markov chains wouldn't be so much useful in Monte Carlo if there were not conditions of uniqueness of the invariant probability distribution.
The uniqueness property is in fact what guarantees that the Metropolis algorithm converges to the wanted probability distribution. To state and prove this property some definitions are necessary.

{\bf Definition 2.} Given the transition matrix $\underline{\mathcal{P}}=\lbrace\mathcal{P}_{i\rightarrow j}\rbrace_{i,j}$ of a time invariant Markov chain,
\begin{itemize}
\item $\underline{\mathcal{P}}$ is {\it irreducible} if for each $i,j \, \in \, E$ there is a positive integer number $m = m(i,j)$ so that $\mathcal{P}_{i\rightarrow j}^{(m)}>0$.
\item  $\underline{\mathcal{P}}$ is {\it regular} if there is a positive integer number $m$ for which $\mathcal{P}_{i\rightarrow j}^{(m)}>0$ for every $i,j \, \in \, E$.
\end{itemize}
Clearly, a regular transition matrix is also irreducible but the opposite is not generally true, however if an irreducible transition matrix satisfies the following criterion, then 
we will show that it is also regular. 

{\bf Statement.} {\it If a transition matrix is irreducible and there is $h\,\in\,E$ such that $\mathcal{P}_{h\rightarrow h} >0$, then that transition matrix is also regular.} 

\begin{proof}
From the definition of irreducibility, for each $i,j\,\in\,E$ there is $m=m(i,j)>0$ such that $\mathcal{P}^{(m)}_{i\rightarrow j} >0$; defined $s = \max_{i,j\,\in\,E}\:m(i,j)$, then 
$\mathcal{P}_{l\rightarrow k}^{(2s)}>0$ for each $l,k\,\in\,E$. In fact, one can always use iteratively the transition element $\mathcal{P}_{h\rightarrow h} > 0$ to express 
$\mathcal{P}_{l\rightarrow k}^{(2s)}$ as a chain of products: the irreducibility guarantees that given two elements $l,k \,\in\, E$ there are positive integer $n_1=n(l,h)$ and $n_2=n(h,k)$
such that $\mathcal{P}_{l\rightarrow h}^{(n_1)}>0$ and $\mathcal{P}_{h\rightarrow k}^{(n_2)}>0$; the element $\mathcal{P}_{l\rightarrow k}^{(2s)}$ will thus be expressed as:
\begin{eqnarray}
\mathcal{P}_{l\rightarrow k}^{(2s)} \ge \mathcal{P}_{l\rightarrow h}^{(n_1=n(l,h))}\mathcal{P}_{h\rightarrow h}...\mathcal{P}_{h\rightarrow h}\mathcal{P}_{h\rightarrow k}^{(n_2=n(h,k))} > 0,
\end{eqnarray}
and this proves the statement.
\end{proof}
We state now the uniqueness theorem.

{\bf Theorem 2. (Markov)} {\it If a transition matrix is regular, then there is only one invariant probability $\pi$ and, for any starting probability $v$, the following holds
\begin{eqnarray}
\pi_j = \lim_{n\rightarrow +\infty} \left(\underline{v}\,\underline{\mathcal{P}}^n\right)_j
\end{eqnarray}
}
\begin{proof}
From Markov-Kakutani theorem at least one invariant probability $\pi$ exist and, by definition
\begin{eqnarray}
\underline{\pi}=\underline{\pi}\,\underline{\mathcal{P}},\:\:\:\sum_k\pi_k=1,\:\:\:0\le\pi_k\le 1
\end{eqnarray}
Let's consider the one dimension vector subspace of $\mathbf{C}^N$ generated by $\underline{\pi}$:
\begin{eqnarray}
\mathcal{V}_{\pi}=\left\lbrace\underline{u}\,\in\,\mathbf{C}^N|\underline{u}=t\underline{\pi},\:t\,\in\,\mathbf{C}\right\rbrace
\end{eqnarray}
By the hypothesis and this definition follows respectively that $\underline{\pi}$ is an eigenvector of $\underline{\mathcal{P}}$ and $\underline{\pi}\,\in\,\mathcal{V}_{\pi}$. Define also the 
subspace $\mathcal{V}_0$:
\begin{eqnarray}
\mathcal{V}_0 = \left\lbrace\underline{y}\,\in\,\mathbf{C}^N|\sum_{k}y_k=0\right\rbrace.
\end{eqnarray}
This subspace has dimension $M-1$ and $\mathcal{V}_0\, \cap\, \mathcal{V}_\pi = \lbrace\underline{0}\rbrace$ because $\mathcal{V}_\pi$ is made of elements $\underline{u}$ for which $\sum_ku_k=t$.
Following from this conditions, the vector space $\mathbf{C}^N$ decomposes in the direct sum $\mathcal{V}_0 \oplus \mathcal{V}_\pi$. This implies that any element $\underline{v}\,\in\,\mathbf{C}^{N}$ 
can be uniquely written as:
\begin{eqnarray}\label{et2}
\underline{v}=t\underline{\pi}+\underline{y},\:\:\:\underline{y}\,\in\,\mathcal{V}_0
\end{eqnarray}
The eigenvalues equation for $\underline{\mathcal{P}}$ is:
\begin{eqnarray}
\underline{v}\,\underline{\mathcal{P}}=\lambda\underline{v},\:\:\: \lambda\,\in\,\mathbf{C}
\end{eqnarray}
Let's consider also that given an element $\underline{y}\,\in\,\mathcal{V}_0$, then also the element $\underline{y}\,\underline{\mathcal{P}}\,\in\,\mathcal{V}_0$, this is seen in this passage:
\begin{eqnarray}
\sum_k(\underline{y}\,\underline{\mathcal{P}}_k) = \sum_k\sum_i y_i\mathcal{P}_{i\rightarrow k} = \sum_i y_i\sum_k\mathcal{P}_{i\rightarrow k} = \sum_i y_i = 0
\end{eqnarray}
For an eigenvector $\underline{v}\,\in\,\mathbf{C}^N$ of $\underline{\mathcal{P}}$, using Eq. \eqref{et2}, 
\begin{eqnarray}
\underline{v}\,\underline{\mathcal{P}}=t\underline{\pi}\,\underline{\mathcal{P}}+\underline{y}\,\underline{\mathcal{P}}=\lambda t\underline{\pi}+\lambda\underline{y}
\end{eqnarray}
here, $\underline{y}\,\underline{\mathcal{P}}\,\in\,\mathcal{V}_0$, and thus, for the decomposition of the vector space $\mathbf{C}^N$, it must necessarily be 
\begin{eqnarray}
\begin{cases}
t\underline{\pi}\,\underline{\mathcal{P}}=\lambda t\underline{\pi}\\
\underline{y}\,\underline{\mathcal{P}}=\lambda \underline{y}
\end{cases}
\end{eqnarray}
 Let's consider the eigenvalue equation
\begin{eqnarray}
\underline{y}\,\underline{\mathcal{P}}=\lambda \underline{y},
\end{eqnarray}
 with $\underline{y}\,\in\,\mathcal{V}_0$ and $\underline{y}\ne 0$. Explicitating the matrix product and focusing on the $i$--th component,
\begin{eqnarray}
\lambda y_i = \sum_k y_k \mathcal{P}_{k\rightarrow i}
\end{eqnarray}
Taking the absolute value, summing over the components, and considering that $\underline{\mathcal{P}}$ has non--negative elements. 
\begin{eqnarray}\label{et3}
|\lambda|\sum_i |y_i | = \sum_i\left|\sum_k y_k \mathcal{P}_{k\rightarrow i}\right| \le \sum_i \sum_k |y_k |\mathcal{P}_{k\rightarrow i} = \sum_k |y_k |,
\end{eqnarray}
 that implies 
 \begin{eqnarray}
 |\lambda | \le 1
\end{eqnarray}
If the transition matrix $\underline{\mathcal{P}}$ has the elements strictly positive, then Eq. \eqref{et3} is strict inequality; this can be 
understood if we consider a summation of complex numbers $\sum_{n=1}^{N} r_n e^{i\phi_n}$, it is true that 
\begin{eqnarray}
\left|\sum_{i=1}^{N} r_n e^{i\phi_n}\right| = \sum_{i=1}^{N} \left|r_n e^{i\phi_n}\right| 
\end{eqnarray} 
if and only if $\phi_n = a\,\in\,\mathbf{R}\,\forall n\,\in\,[1,N]$ ; however, by hypothesis $\sum_k y_k = 0$ and $\underline{y}\ne 0$ and this necessarily implies that at least one 
component of $\underline{y}$ must have a different phase; moreover, we have assumed that the matrix elements of $\mathcal{P}_{k\rightarrow i}$ are strictly positive 
so in a product they won't change the phases of the vector, meaning that $y_k\mathcal{P}_{k\rightarrow i}$ has the same phase of $y_k$. Hence, under the condition of 
strict positiveness of the matrix elements of $\underline{\mathcal{P}}$, holds that
\begin{eqnarray}
|\lambda| < 1
\end{eqnarray}
This is the crucial point of this demonstration. Using now the regularity condition we demonstrate the following statement.

{\it Statement.} For a regular transition matrix, $|\lambda|<1$. 

By definition of regularity, if $\underline{\mathcal{P}}$ is regular, then there is an integer $m > 0$ such that $\underline{\mathcal{P}}^m$ has strictly positive elements, 
$\underline{\mathcal{P}}^m$ is also a transition matrix, so the passages of the demonstration can be applied also to $\underline{\mathcal{P}}^m$ resulting in 
\begin{eqnarray}
|\lambda^m|<1
\end{eqnarray}
and this immediately implies that $|\lambda|<1$. Consider now an arbitrary initial probability $v$, then
\begin{eqnarray}
\underline{v}\,\underline{\mathcal{P}}^n = (\underline{\pi}+\underline{v}-\underline{\pi})\underline{\mathcal{P}}^n=\underline{\pi}+(\underline{v}-\underline{\pi})\underline{\mathcal{P}}^n
\end{eqnarray}
From Eq. \eqref{et2} it is clear that $\underline{v}-\underline{\pi}\,\in\,\mathcal{V}_0$, but the linear operator $\underline{\mathcal{P}}$ defined on $\mathcal{V}_0$ has 
eigenvalues that in modulus are strictly lesser than 1; from Functional Analisys\cite{c7r3}, given these conditions, it follows that
\begin{eqnarray}
\lim_{n\rightarrow +\infty}\left\lbrace(\underline{v}-\underline{\pi})\underline{\mathcal{P}}^n\right\rbrace=0.
\end{eqnarray}
In conclusion we have,
\begin{eqnarray}
\lim_{n\rightarrow +\infty} \underline{v}\,\underline{\mathcal{P}}^n = \underline{\pi}+\lim_{n\rightarrow +\infty}\left\lbrace(\underline{v}-\underline{\pi})\underline{\mathcal{P}}^n\right\rbrace=\underline{\pi},
\end{eqnarray}
this proves the theorem of uniqueness.
\end{proof}

 \subsection{The Metropolis algorithm}\label{metroalg}
Given a probability distribution $\pi$ defined on a finite set $E=\lbrace 1,...,N\rbrace$ we now show a recipe to obtain a transition matrix $\underline{\mathcal{P}}$ 
for which $\pi$ is the only invariant probability, namely
\begin{eqnarray}
\pi_j = \lim_{n\rightarrow +\infty} (\underline{v}\,\underline{\mathcal{P}}^n)_j \quad .
\end{eqnarray}
The recipe that we are going to show is the Metropolis algorithm\cite{d:metropolis} and it allows to sample any arbitrary probability density that satisfies some conditions. This algorithm is relevant in our context 
because it is used to evaluate $N$--dimensional integrals like the one in Eq. \eqref{c1rpe}. 

{\bf Theorem 3. (Metropolis)}. {\it Given a strictly positive probability distribution $\pi$, $\pi_j > 0 \:\:\: \forall\,j$, that is not the uniform probability density, for each probability distribution 
$v$, there is a Markov chain with initial probability $v$ and {\it regular} transition matrix $\underline{\mathcal{P}}$ that has $\pi$ as invariant distribution probability. The transition matrix 
$\underline{\mathcal{P}}$ is defined as:
\begin{eqnarray}\label{mrtk}
\mathcal{P}_{i\rightarrow j} = 
\begin{cases}
\mathcal{L}_{i\rightarrow j},\:\:\:\:\: i \ne j,\,\,\, \pi_j \ge \pi_i \\
\mathcal{L}_{i\rightarrow j}\frac{\pi_j}{\pi_i},\:\:\:\:\: i \ne j,\,\,\, \pi_j < \pi_i \\
1 - \sum_{j \ne i} \mathcal{P}_{i\rightarrow j},\:\:\:\:\: i = j
\end{cases}
\end{eqnarray}
where $\underline{\mathcal{L}}$ is any symmetric and irreducible transition matrix.
}
\begin{proof}
With this choice of $\underline{\mathcal{P}}$, let's show that $\pi$ satisfies the detailed balance condition; chose two elements $i,j$ of $E$ such that $\pi_j \le \pi_i$, applying equation \eqref{mrtk} 
we obtain,
\begin{eqnarray}
\pi_i\mathcal{P}_{i\rightarrow j} = \pi_i \mathcal{L}_{i\rightarrow j}\frac{\pi_j}{\pi_i}=\mathcal{L}_{i\rightarrow j}\pi_j = \pi_j \mathcal{P}_{i\rightarrow j},
\end{eqnarray}
where we used the symmetry of $\underline{\mathcal{L}}$; this shows that $\pi$ is indeed an invariant probability associated to $\underline{\mathcal{P}}$. To prove the theorem we have to show 
that $\pi$ is also the {\it unique} invariant probability of $\underline{\mathcal{P}}$; to this purpose, we will show then that $\underline{\mathcal{P}}$ is regular. Let's start from the irreducibility.

{\it Statement.} $\underline{\mathcal{P}}$ is irreducible. This follows directly from the irreducibility of $\underline{\mathcal{L}}$. The irreducibility is in fact related to the non--zero element of 
the transition matrix, this implies that if a transition matrix is irreducible, then it is also true for any transition matrix that has at least the same non--zero elements. This is exactly the case 
for $\underline{\mathcal{P}}$ as can be readily seen from its definition in Eq. \eqref{mrtk}. To show that $\underline{\mathcal{P}}$ is also regular, as shown in the previous section, it is enough to verify that 
there is an element $i_0\,\in\,E$ such that $\mathcal{P}_{i_0\rightarrow i_0} > 0$. By hypothesis, $\pi$ is not the uniform probability distribution and thus there is a subset $M \,\subset\, E$ 
in which $\pi$ is maximum, moreover, due to the irreducibility of $\underline{\mathcal{L}}$, there are two elements $i_0\,\in\,M$ and $j_0\,\in\,M^C$ such that $\mathcal{P}_{i_0\rightarrow j_0}>0$;
remembering that, by definition, if $i \ne j$ then $\mathcal{P}_{i\rightarrow j} \le \mathcal{L}_{i \rightarrow j}$, we have,
\begin{eqnarray}
\mathcal{P}_{i_0 \rightarrow i_0} = 1 - \sum_{j \ne i_0}\mathcal{P}_{i_0 \rightarrow j} = 1 - \sum_{j\ne i_0,j_0}\mathcal{P}_{i_0 \rightarrow j} - \mathcal{P}_{i_0 \rightarrow j_0} \ge \nonumber \\
\ge 1 - \sum_{j \ne i_0,j_0}\mathcal{L}_{i_0 \rightarrow j} - \mathcal{L}_{i_0 \rightarrow j_0}\frac{\pi_{j_0}}{\pi_{i_0}} = \nonumber \\
= 1 - \sum_{j\ne i_0}\mathcal{L}_{i_0 \rightarrow j} + \mathcal{L}_{i_0 \rightarrow j_0}\left(1 - \frac{\pi_{j_0}}{\pi_{i_0}}\right) \ge \nonumber \\
\ge \mathcal{L}_{i_0 \rightarrow j_0}\left(1 - \frac{\pi_{j_0}}{\pi_{i_0}}\right) > 0 \quad .
\end{eqnarray}
And this proves the regularity of $\underline{\mathcal{P}}$; from the Markov theorem follows the uniqueness of the invariant probability distribution and this proves the theorem.
\end{proof}

In most practical cases, Eq. \eqref{mrtk} for $i\,\ne\, j$  is written in the form:
\begin{eqnarray}\label{metroacc}
\mathcal{P}_{i\rightarrow j} = \mathcal{L}_{i\rightarrow j} \min\left(1,\frac{\pi_j}{\pi_i}\right) \quad .
\end{eqnarray}
The meaning of this relation is that the entire Markov chain can be built with predetermined moves, $\mathcal{L}_{i\rightarrow j}$ that might be accepted with probability 
$\min\left(1,\frac{\pi_j}{\pi_i}\right)$. These moves, starting from a probability $X_n$ propose a transition to a probability $X_{n+1}$: 
if this transition is accepted, $X_{n+1}$ has been determined, in case of rejection $X_{n+1}=X_{n}$. The condition of irreducibility of $\underline{\mathcal{L}}$ means that the moves must be 
chosen so that their combination is able to explore the whole state space $E$, this property is called {\it ergodicity}. The symmetry of 
$\underline{\mathcal{L}}$ is a detailed balance condition on the Metropolis moves, this however can be dropped in favor of the weaker condition $\mathcal{L}_{i\rightarrow j}>0$ 
whenever $\mathcal{L}_{j\rightarrow i}>0$ if a new definition of $\mathcal{P}_{i\rightarrow j}$ is taken:
\begin{eqnarray}\label{metroaccg}
\mathcal{P}_{i\rightarrow j} = \mathcal{L}_{i\rightarrow j} \min\left(1,\frac{\pi_j \mathcal{L}_{j\rightarrow i}}{\pi_i \mathcal{L}_{i\rightarrow j}}\right)
\quad .
\end{eqnarray}
A good choice of Metropolis moves will enhance the convergence of the Markov chain towards the equilibrium probability distribution. 

 \subsection{Sampling and expectation values}
In the previous section, we have seen a procedure to sample an arbitrary distribution probability 
such that in Eq. \eqref{c1rpe}. In order to evaluate that $N$--dimensional integral, however, there are still two problems 
to take care of. First, the Monte Carlo evaluation of an integral is a statistical method and as such the results have an 
intrinsic statistical error due to the finite number of sampled values, and second, the Metropolis algorithm produces an highly 
correlated sampling of the probability distribution $p(\vec{x})$. To overcome these problems the block average technique
 can be used: if the summation  in Eq. \eqref{eq11} is truncated after $N_s$ terms, a block 
 $I_{N_s} = \frac{1}{N_s}\sum_{i=1}^{N_s}g(\vec{x}_{i})$ is defined. If $N_s$ is long enough, each evaluated $I_{N_s}$ can be 
 considered statistical independent from the others; hence, by the central limit theorem, $I_{N_s}$ is a Gaussian 
 distributed random variable; if the set of values 
  $I_{N}^{1},I_{N}^{2},...,I_{N}^{N_{blocks}}$ are generated 
 with Eq. \eqref{eq11}, then the average value $I_{avg}$ and the standard deviation $\sigma$ can be computed with the usual 
 formulas:
 \begin{eqnarray}
 I_{avg} = \frac{1}{N_{blocks}}\sum_{i=1}^{N_{blocks}}I_{N}^{i} \\
 \sigma^{2} = \frac{1}{N_{blocks}-1} \sum_{i=1}^{N_{blocks}} \left(I_{n}^{i}-I_{avg}\right)^{2}
 \quad ,
 \end{eqnarray}
where the error of the estimation $I_{avg}$ is $\sigma / \sqrt{N_{blocks}}$.
 This method for the evaluation of $N$--dimensional integrals is asymptotic, in fact the correct sampling of a 
 distribution will be given only after that a long enough Markov chain has been built. This means that the probability 
 distribution $p(\vec{x})$ is sampled only after a certain number of equilibration steps. 
   \begin{figure}[h]
 \begin{center}
 \includegraphics*[width=11cm]{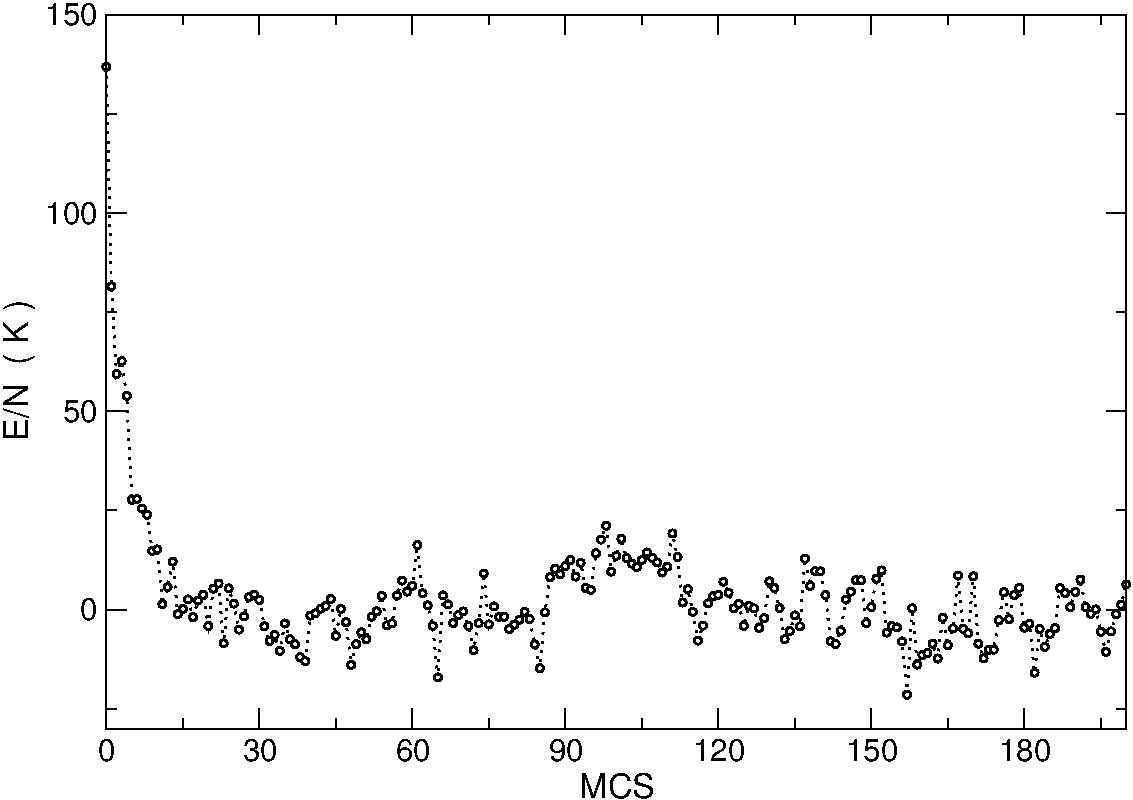}
 \caption{\label{figtransient} 
 The ``istantaneous '' value of an integrand for each Monte Carlo Step. This integral represents the energy of 
 one atom of $^4$He in a 1D model potential defined by $V(x)=\sigma_1 x^4 - \sigma_2 x^2$, where $\sigma_1 = 8$ K\AA$^{-4}$ and 
 $\sigma_2 = 8$ K\AA$^{-2}$. The methodology used to evaluate this integral is the Path Integral Ground State that will be described 
 in Sec.~\ref{secpigs}. 
 }
\end{center}
\end{figure}
 This equilibration number can be evaluated by plotting on a graph the value of the integral averaged 
 within a Monte Carlo block versus the index of the corresponding Monte Carlo block: equilibration is over as soon as 
 transients disappear from the plot, provided that the chosen set of moves can efficiently explore the whole 
 space of events. This might seem to be an easy condition 
 to fulfill but in some cases a 
 long equilibration is required, especially when the distribution density to sample has many local maxima separated by regions 
 of low probability density. In Fig.~\ref{figtransient} we show the value of an integrad evaluated at each MC step: the equilibration 
 transient is clearly visible in the first twenty MC steps; the correlation of the Markov chain manifests here as a pattern in the 
 values of the integrand.

   \subsection{Metropolis sampling}\label{metromoves}
   We now apply the Monte Carlo sampling to the problem of evaluating a quantum expectation value of a local operator $\hat{O}$ introduced in Sec.~\ref{secpigs}; this 
   quantum expectation value, from Eq. \eqref{pigsunsimm}, can be written in compact form as
   \begin{eqnarray} \label{expval1}
   \left\langle\hat{O}\right\rangle = \int d\Gamma \: O(\Gamma)p(\Gamma)
   \end{eqnarray}
   where $\Gamma = \left\lbrace R_1,...,R_M\right\rbrace$. In this section we consider the PIGS case; the adaptation to quantum thermal averages is straightforward and will be considered contextually.   
   The Metropolis algorithm is used to sample the multi--dimensional probability distribution $p(\Gamma)$ in Eq.~\eqref{probdist}, that, 
   explicating the normalization constant $\mathcal{N}$, takes the form
   \begin{eqnarray}
   p(\Gamma)= \frac{\Psi_T\left(R_1\right)\prod_{j=1}^{M-1}G\left(R_{j},R_{j+1},\delta\tau\right)\Psi_T\left(R_M\right)}{\int d\Gamma\:\Psi_T\left(R_0\right)\prod_{j=1}^{M-1}G\left(R_{j},R_{j+1},\tau\right)\Psi_T\left(R_M\right)}
   \quad .
   \end{eqnarray}
     \begin{figure}[h]
 \begin{center}
 \includegraphics*[width=11cm]{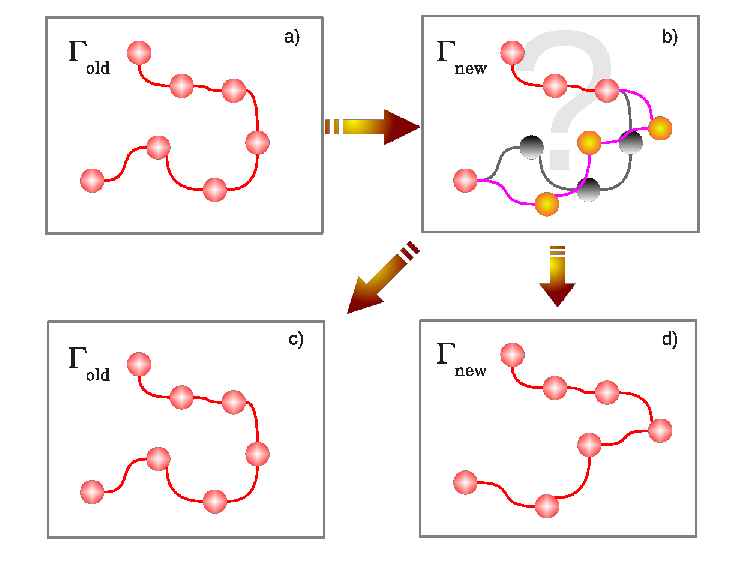}
 \caption{\label{sketchmove} 
 General scheme for a Metropolis move: from a configuration (a), a move is proposed (b). At this point the move can be accepted (c) or rejected (d). If the move is accepted, the new MC step 
 will have a new configuration $\Gamma_{new}$; otherwise the same configuration $\Gamma_{old}$ is sampled again.
 Grey beads and lines represent the removed segment of the polymer.
 }
\end{center}
\end{figure}
 The sampling of $p(\Gamma)$ is made with a sequence of Metropolis ``moves". A move is a two--step process sketched in Fig.~\ref{sketchmove}; this process, from a set of configurations 
 $\Gamma$ proposes a new set $\Gamma_{new}$ and then evaluates whether to accept or reject the 
 new set of configurations. The probability to accept the move is defined by Eq.~\eqref{metroacc}, where the term $\mathcal{L}_{i\rightarrow j}$ represents the probability to try a move that from a 
 configuration $i$ proposes a new configuration $j$. From the {\it a--priori} knowledge of the system under study it is possible to use {\it guided} moves that are more likely able to sample physical 
 configurations rather than highly improbable ones (in this case the probability to accept the move becomes Eq. \eqref{metroaccg}; such a guided approach would enhance the convergence of the 
 sampling, especially if the probability distribution has many local minima. 
 The moves that will be described shortly are unguided, so that $\mathcal{L}_{i\rightarrow j} = \mathcal{L}_{j\rightarrow i}$ and their probability to be accepted simplifies to the following relation
 \begin{eqnarray}
 a\left(\Gamma_{new}\right) = 
\min\left(1,\frac{\Psi_T\left(R_1^{new}\right)\prod_{j=1}^{M-1}G\left(R_{j_{new}},R_{j+1_{new}},\delta\tau\right)\Psi_T\left(R_M^{new}\right)}{\Psi_T\left(R_1\right)\prod_{j=1}^{M-1}G\left(R_{j},R_{j+1},\delta\tau\right)\Psi_T\left(R_M\right)}\right)
 \quad .
 \end{eqnarray}
 In the context of quantum--classical isomorphism, a move can involve one or more different polymers. A move involving a single polymer is represented as a reconfiguration of some or all the beads 
 of the polymer itself; the probability to accept such a move depends on the correlations of the beads at the same imaginary--time discretization (inter--polymer correlations) and on the correlations between adjacent beads that 
  belong to the same polymer (intra--polymer correlation). This latter contribution is also referred as {\it kinetic spring} because comes from the kinetic term of the Hamiltonian (Eq.~\ref{papprox}) and disfavors configurations in 
  which two adjacent beads are placed far away from each other. A move involving many polymers is a generalization of a single polymer move; these moves can also involve permutations between polymers that can be
  employed to take into account the quantum statistics of the system. 
   
  A Monte Carlo simulation consists of a set of Monte Carlo Steps (MCS); in general, after each step the estimators can be evaluated. A MCS consists of a set of Metropolis moves that are tuned so that the effect of all the 
  accepted moves modifies the positions of roughly half the beads that compose the system of polymers. The Metropolis moves that are proposed here are the translation moves, the Brownian bridges and the 
  permutation sampling. Later in this section will be introduced the Worm algorithm in the Canonical ensemble with its Metropolis moves. These moves can be also used for PIMC with a slight 
  adaptation for the translation of a polymer.
  
  In order to simplify the notation, we define the free particle propagator that appears in Eq.~\eqref{papprox} as
  \begin{eqnarray}\label{freep}
  G_0\left(\vec{r},\vec{r}^{\:'},\delta\tau\right) = \left(\frac{1}{4\pi\lambda\delta\tau}\right)^{\frac{d}{2}}e^{-\frac{\left|\vec{r}-\vec{r}^{\:'}\right|^2}{4\lambda\delta\tau}}
  \quad .
  \end{eqnarray}

\subsubsection{Translation of a single bead}
\begin{figure}[h]
 \begin{center}
 \includegraphics*[width=11cm]{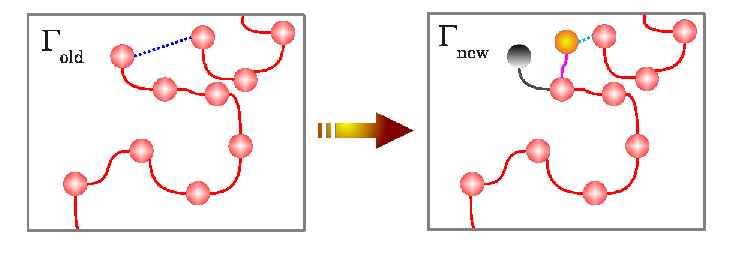}
 \caption{\label{mossasingola} 
 Scheme for the translation of an extremal bead of a polymer. 
 Grey beads and lines represent the old position and kinetic correlation of the bead.
 }
\end{center}
\end{figure}

In a PIGS simulation, this move, represented in Fig.~\ref{mossasingola} is generally applied to the first or the last bead of a polymer, namely $\vec{r}_i^{1}$ or $\vec{r}_i^{M}$. This is the simplest move 
for such beads: there are other, more performant, possibilities that allow to move a certain number of beads including $\vec{r}_i^{1}$ or $\vec{r}_i^{M}$; one of those 
moves is a generalization of the Brownian bridge that takes into account the correlations from the trial wave function. The Brownian bridge will be introduced soon; 
however, in this work we did not implement the mentioned extension. Instead, we used the following move. Let's focus on the first bead of the $i$--th polymer, $\vec{r}_i^{1}$; 
the new configuration will have a translation of a vector $\vec{d}$ applied to $\vec{r}_{i_{old}}^{1}=\vec{r}_i^{1}$, namely $\vec{r}_{i_{new}}^{1}=\vec{r}_i^{1}+\vec{d}$.

The probability to accept this move is
  \begin{eqnarray}
  a\left(\left\lbrace R\right\rbrace_{new}\right) = \min \left(1,P_{tr}\right) \nonumber \\
  P_{sing}=\frac{\Psi_T(R_1^{new})e^{-\frac{\delta\tau}{2}\sum_{k\neq i}v\left(\left|\vec{r}_{i_{new}}^{\:1}-\vec{r}_{k}^{\:1}\right|\right)}}
  {\Psi_T(R_1^{old})e^{-\frac{\delta\tau}{2}\sum_{k\neq i}v\left(\left|\vec{r}_{i_{old}}^{\:1}-\vec{r}_{k}^{\:1}\right|\right)}}
    \quad .
    \end{eqnarray}

For the last bead, $\vec{r}_i^{M}$,
  \begin{eqnarray}
  a\left(\left\lbrace R\right\rbrace_{new}\right) = \min \left(1,P_{tr}\right) \nonumber \\
  P_{sing}=\frac{e^{-\frac{\delta\tau}{2}\sum_{k\neq i}v\left(\left|\vec{r}_{i_{new}}^{\:M}-\vec{r}_{k}^{\:M}\right|\right)}\Psi_T(R_M^{new})}
  {e^{-\frac{\delta\tau}{2}\sum_{k\neq i}v\left(\left|\vec{r}_{i_{old}}^{\:M}-\vec{r}_{k}^{\:M}\right|\right)}\Psi_T(R_M^{old})}
    \quad .
    \end{eqnarray}

  \subsubsection{Translation of a polymer}
\begin{figure}[h]
 \begin{center}
 \includegraphics*[width=11cm]{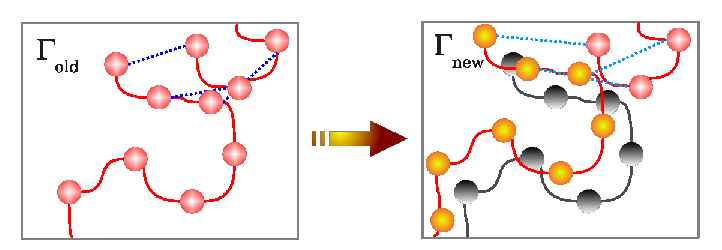}
 \caption{\label{traslazione} 
 Scheme for the translation of an entire polymer. Grey beads and lines represent the old configuration of the polymer.
 }
\end{center}
\end{figure}

  In a translation (Fig.~\ref{traslazione}, a polymer $i$ is rigidly moved by a vector. Kinetic springs remain unchanged and the probability to accept the move depends thus only on the interpolymer 
  correlations. 
  
  \begin{eqnarray}
  a\left(\left\lbrace R\right\rbrace_{new}\right) = \min \left(1,P_{tr}\right) \nonumber \\
  P_{tr}=\frac{\Psi_T(R_1^{new})e^{-\frac{\delta\tau}{2}\sum_{k\neq i}v\left(\left|\vec{r}_{i_{new}}^{\:1}-\vec{r}_{k}^{\:1}\right|\right)}}
  {\Psi_T(R_1^{old})e^{-\frac{\delta\tau}{2}\sum_{k\neq i}v\left(\left|\vec{r}_{i_{old}}^{\:1}-\vec{r}_{k}^{\:1}\right|\right)}} \,\, \times \nonumber \\
 \times\,\, \frac{\prod_{j=2}^{M-1}e^{-\delta\tau\sum_{k\neq i}v\left(\left|\vec{r}_{i_{new}}^{\:j}-\vec{r}_{k}^{\:j}\right|\right)}e^{-\frac{\delta\tau}{2}\sum_{k\neq i}v\left(\left|\vec{r}_{i_{new}}^{\:M}-\vec{r}_{k}^{\:M}\right|\right)}\Psi_T(R_M^{new})}
  {\prod_{j=2}^{M-1}e^{-\delta\tau\sum_{k\neq i}v\left(\left|\vec{r}_{i_{old}}^{\:j}-\vec{r}_{k}^{\:j}\right|\right)}e^{-\frac{\delta\tau}{2}\sum_{k\neq i}v\left(\left|\vec{r}_{i_{old}}^{\:M}-\vec{r}_{k}^{\:M}\right|\right)}\Psi_T(R_M^{old})} \label{trpigs}
 \quad .
  \end{eqnarray}

  In PIMC, the translation move has a probability to be accepted that is slightly different from Eq.~\eqref{trpigs}:
   \begin{eqnarray}
  a\left(\left\lbrace R\right\rbrace_{new}\right) = \min \left(1,P_{tr}\right) \nonumber \\
  P_{tr}=\frac{\prod_{j=1}^{M}e^{-\delta\tau\sum_{k\neq i}v\left(\left|\vec{r}_{i_{new}}^{\:j}-\vec{r}_{k}^{\:j}\right|\right)}}
  {\prod_{j=1}^{M}e^{-\delta\tau\sum_{k\neq i}v\left(\left|\vec{r}_{i_{old}}^{\:j}-\vec{r}_{k}^{\:j}\right|\right)}}
 \quad .
  \end{eqnarray}

  \subsubsection{Brownian bridge}
  
\begin{figure}[h]
 \begin{center}
 \includegraphics*[width=11cm]{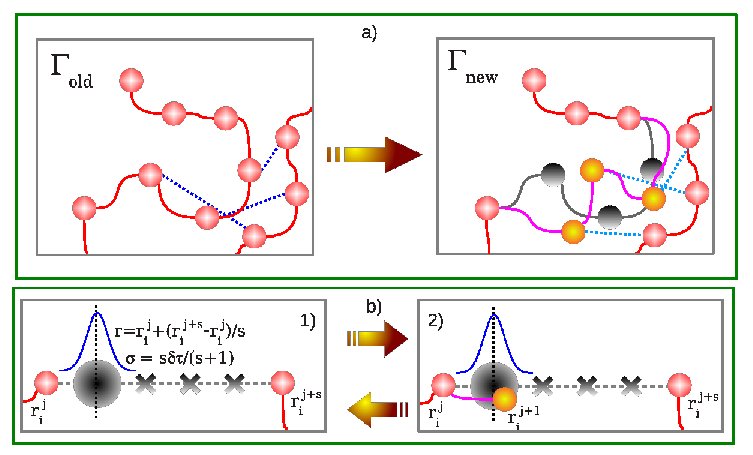}
 \caption{\label{brownianbridge} 
 (Upper panel) The Brownian bridge: a segment of a polymer is reconstructed.
 (Lower panel) Iterative procedure used to sample a free particle propagation between two fixed extremities. 
 Grey beads and lines represent the removed segment of the polymer.
 }
\end{center}
\end{figure}

  The Brownian bridge move is a very efficient way to reconstruct a segment composed of $S$ adjacent beads. A schematic 
  description of this move is shown in Fig.~\ref{brownianbridge}.
  The segment of polymer that the Brownian bridge re--creates represents the sampling of the free particle propagation in imaginary time between two time sectors. This 
  propagation, as will be shown below, can be sampled exactly via the Box--Muller method\cite{c7r6}. 
  This is a useful feature: the kinetic correlations of the reconstructed segment are sampled exactly, therefore the probability of acceptation of the move will depend only on the 
  correlations between different polymers. This is readily seen if one considers the correlations of a segment of the $i$--th polymer $(\vec{r}_i^{\:j},...,\vec{r}_i^{\:j+s})$ that is 
  part of a configuration of polymers $\Gamma$:
  \begin{eqnarray}\label{cseg}
  \pi(\Gamma)=\prod_{m=j}^{j+s} e^{-\frac{1}{4\lambda\delta\tau}\left|\vec{r}_i^{\:m} - \vec{r}_i^{\:m+1}\right|^2}e^{-\delta\tau \sum_{l\ne i}^{N} v(\left|\vec{r}_i^{\:m} - \vec{r}_l^{\:m}\right|)}
  \end{eqnarray}
 the kinetic and potential parts are factorized. Recalling the Metropolis algorithm (Sec.~\ref{metroalg}), the general form of the transition matrix that represents the Markov chain is factorized in a 
 ``move'' $\mathcal{L}_{i\rightarrow j}$  and an acceptance of the move:

\begin{eqnarray}\label{bbma}
\mathcal{P}_{i\rightarrow j} = \mathcal{L}_{i\rightarrow j} \min\left(1,\frac{\pi_j \mathcal{L}_{j\rightarrow i}}{\pi_i \mathcal{L}_{i\rightarrow j}}\right)
\quad .
\end{eqnarray}

In the case of the Brownian bridge, the move $\mathcal{L}_{i\rightarrow j}=\mathcal{L}_{j\rightarrow i}$ is the exact sampling of the free particle propagation in imaginary--time, 
$\pi_j=\pi(\Gamma_{new})$ is the correlation value of the new segment and $\pi_i=\pi(\Gamma_{old})$ is the correlation value of 
the segment before the move: the rest of the system is not changed by this move and the respective correlations, being unchanged, cancel out. 
From Eq.~\eqref{cseg} it is clear that the free particle propagator that defines $\mathcal{L}_{i\rightarrow j}$ is 
a part of $\pi_j$; therefore, the probability to accept a new segment becomes  $a=\min\left(1,P_{bb}\right)$, with: 
  \begin{eqnarray}\label{pbb}
  P_{bb}=\frac{\prod_{m=j}^{j+s}e^{-\delta\tau\sum_{k\neq i}v\left(\left|\vec{r}_{i_{new}}^{\:m}-\vec{r}_{k}^{\:m}\right|\right)}}{\prod_{m=j}^{j+s}
  e^{-\delta\tau\sum_{k\neq i}v\left(\left|\vec{r}_{i_{old}}^{\:m}-\vec{r}_{k}^{\:m}\right|\right)}}
  \end{eqnarray}
  where the reconstruction starts from the $j+1$ bead of the $i$--th polymer and the last reconstructed bead is at position $j+s-1$. 
  
  The following operations, also illustrated in Fig.~\ref{brownianbridge}, are performed during the move:
  \begin{itemize}
  \item Remove the beads between $j$ and $j+s$.
  \item Create a new timeslice at position $j+1$: from the coordinates of the timeslices $j$ and $j+s$, 
  determine the coordinates of the new timeslice $j+1$. These coordinates are determined in the following way:
  we first note that the bead at position $j+1$ is the free propagation from the bead at position $j$, 
   $p_1(\vec{r}_i^{\:j},\vec{r}^{\:\star})=G_0(\vec{r}_i^{\:j},\vec{r}^{\:\star},\delta\tau)$, {\it and} 
   the free propagation from the bead at position $j+s$, 
   $p_2(\vec{r}_i^{\:j+s},\vec{r}^{\:\star})=G_0(\vec{r}_i^{\:j+s},\vec{r}^{\:\star},s\delta\tau)$.
   The probability density from which the position $\vec{r}_i^{\:j+1}$ is sampled is thus the joint probability 
   $p_1p_2$; with straightforward algebraic operations, this joint probability can be reconduced to the Gaussian form 
   of Eq.~\eqref{freep} times a trivial normalization constant $\mathcal{N}_t$ that won't affect the sampling; namely, the new 
   bead is sampled from
   \begin{eqnarray}
   p(\vec{r}^{\:\star}) = p_1(\vec{r}_i^{\:j},\vec{r}^{\:\star})p_2(\vec{r}_i^{\:j+s},\vec{r}^{\:\star}) =
   \mathcal{N}_t G_0(\vec{r}^{\:\star},\vec{r}_{i}^{\:j}+\frac{\vec{r}_i^{\:j+s}-\vec{r}_i^{\:j}}{s},\frac{s}{s+1}\delta\tau)
   \end{eqnarray}

   \item From the newly created bead, $j+1$, and the bead $j+s$, 
   determine the coordinates of the bead $j+2$. This is done by iteration, considering 
   a segment of polymer that starts at $j+1$ and has length $s-1$. The procedure is iterated 
   until the bead at position $j+s-1$ has been determined.
  \end{itemize}

  This move samples the free particle propagation between two given extremities; it may happens that the two extremities are in different polymers that are connected at a timeslice $j_p$ by a permutation cycle $\hat{P}$; 
  in this case the labels $j$, $i$ and $k$ must be permuted, so that $i\rightarrow \hat{P_j}i$ and  $k\rightarrow \hat{P_j}k$, where $\hat{P}_{j}=\hat{I}$ for $j<j_p$ and $\hat{P}_j = \hat{P}$ otherwise.

  The probability of acceptation can be varied modifying the length of the Brownian bridges. 
  As an empirical rule, a good choice of this probability can be between 0.3 and 0.5. 
  This is a reasonable trade-off between long moves and small moves: long moves, on one hand,
   would yield an high rejection rate resulting in poor performance of the simulation;
   small moves, on the other hand, have an high acceptance ratio but might compromise the ergodicity of the simulation;
   this happens because unprobable configurations would rarely be sampled.

  \subsubsection{Permutation sampling}  
  The permutation sampling introduces the Bose symmetry in the system. In the polymer description, 
  it is a move that involves a permutation between a variable number of polymers greater than one. 
  As mentioned in Sec.~\ref{secpatatelesse}, this move is necessary in PIGS when the trial wave function does not possess the Bose symmetry. 
  This move is also necessary at finite temperature: in PIMC, in fact, 
  the thermal average is not expressed through the quantum imaginary--time evolution of a trial wave function; as consequence, the Bose symmetry 
  has to be explicitly introduced through permutation sampling.
  
\begin{figure}[h]
 \begin{center}
 \includegraphics*[width=11cm]{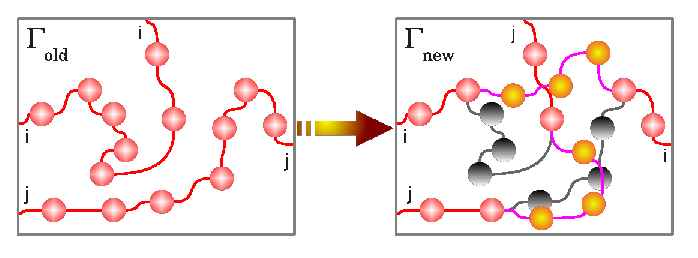}
 \caption{\label{perm0} 
 Permutation of two polymers $i$ and $j$: the resulting configuration has still two polymers 
 of the same length; it is thus topologically similar to the previous one.
 Grey beads and lines represent the removed segment of the polymer.
 }
\end{center}
\end{figure}

  Here we will show the algorithm described in Ref.~\onlinecite{d:permutations_boninsegni}. The polymers involved in the permutation are selected with a kinetic test to be described soon. Once the polymers $i_1$ and $i_2$  have been selected, 
  their beads between two 
  time sectors $j_0$ and $j_0+s$ are removed. At this point, a Brownian bridge is made from the bead at position $j_0$ of the polymer $i_1$, to the bead at position $j_0+s$ of the polymer $i_2$.
  The permutation move follows the scheme in Figure \ref{perm0}: the iteration of this `swap' procedure proceeds between polymers $i_2$ and $i_3$, and so on until a polymer $i_n$ closes on the polymer $i_0$. 

  There are two main steps in permutation sampling: the kinetic test and the reconstruction step.
  \paragraph{Kinetic Test step.}  Given a starting timeslices $j_0$ and a length of reconstructions $s$, this operation selects the polymers that are best suited for permutations and gives in 
  output an ordered sequence of swaps between the polymers. The kinetic test starts from a random polymer $i_1$ that is chosen by generating an integer number between 1 and $N$ from a uniform distribution
  probability. At this point, the following operations determine the next polymer which joins the permutation cycle.
  \begin{itemize}
  \item For the particle $i_1$, build a table as follows
  \begin{eqnarray} \label{per1}
  K_{i_1 \omega}^{1} = G_{0}\left(\vec{r}_{i_1}^{\:j_0},\vec{r}_{\omega}^{\:j_0+s},s\delta\tau\right)\left(1-\delta_{i_1,\omega}\right)
  \end{eqnarray}
  where $\vec{r}_{i_1}^{\:j_0}$ are the coordinates of the polymer $i_1$ at timeslice $j_0$.
  \item From Eq.~\eqref{per1}, the probability to accept the particle $i_1$ in the permutation cycle is
  \begin{eqnarray} \label{per2}
  C^{(1)} = \frac{\sum_{\omega} K^{1}_{i_1 \omega}}{G_{0}\left(\vec{r}_{i_1}^{\:j_0},\vec{r}_{i_1}^{\:j_0+s},s\delta\tau\right) + \sum_{\omega} K^{1}_{i_1 \omega}}
  \quad .
  \end{eqnarray}
  Generate a random number $p$, uniformly distributed between 0 and 1. If $p > C^{(1)}$, the move is rejected; otherwise the process continues to the next step.
  \item From $K_{i_1\omega}^{1}$ a new particle is randomly chosen. The probability to chose a the particle $\nu$ is
  \begin{eqnarray} \label{per3}
  \Pi_{\nu} = \frac{K_{i_1 \nu}^{1}}{\sum_{\omega}K_{i_1 \omega}^{1}}
 \quad .
  \end{eqnarray}
  The new particle is selected with a `faked roulette': the interval $\left[0,1\right)$ is partitioned with bins of width $\Pi_\nu$; a bin $\Pi_\nu$ corresponds to the particle $\nu$; generate an uniformly 
  distributed random number in the interval $\left[0,1\right)$, the bin which contains this number corresponds to the particle that `wins' the faked roulette; this particle is $i_2$.
  \item Make an acceptance test on $i_2$, similarly to that made on $i_1$
  \begin{eqnarray} \label{per4}
  K_{i_2 \omega}^{2} = G_{0}\left(\vec{r}_{i_2}^{\:j_0},\vec{r}_{\omega}^{\:j_0+s},s\delta\tau\right)\left(1-\delta_{i_2,\omega}\right) \nonumber \\
  C^{(2)} = \frac{\sum_{\omega} K^{2}_{i_2 \omega}}{G_{0}\left(\vec{r}_{i_2}^{\:j_0},\vec{r}_{i_2}^{\:j_0+s},s\delta\tau\right) + \sum_{\omega} K^{2}_{i_2 \omega}}
  \quad .
  \end{eqnarray}
  Again, generate a random number $p$, uniformly distributed between 0 and 1. If $p > C^{(2)}$, the move is rejected, else the particle $i_2$ is added to the permutation cycle.
  \end{itemize}
  These operations are repeated until either an acceptation test fails or a particle $i_\alpha=i_1$ is added to the permutation cycle. 
  
  The Dirac's deltas that appears in $K_{i_n \omega}^{n}$ have two purposes:
  \begin{itemize}
  \item Exclude the possibility for a particle to swap with itself
  \item Exclude the possibility for a particle to swap with any other particle already added to the permutation cycle, except for the first particle of the permutation cycle.
  \end{itemize}
  The general definition for $K_{i_n \omega}^{n}$ is thus
  \begin{eqnarray} 
  K_{i_n \omega}^{n} = G_{0}\left(\vec{r}_{i_n}^{\:j_0},\vec{r}_{\omega}^{\:j_0+s},s\delta\tau\right)\left(1-\delta_{i_2,\omega}\right)\left(1-\delta_{i_3,\omega}\right)...\left(1-\delta_{i_n,\omega}\right) 
  \quad .
  \end{eqnarray}
  The output of the kinetic test step is a sequence of particles $\left(i_1,i_2,...,i_\alpha,i_1\right)$; from this output, the reconstruction step begins.
  \paragraph{Reconstruction step.} Starting from the previously obtained permutation cycle $\left(i_1,i_2,...,i_\alpha,i_{\alpha+1}\equiv i_1\right)$, $\alpha$ Brownian bridges are built from the 
  $j_0$ bead of the $i_1$ polymer to the $j_0+s$ bead of the $i_{2}$, an so on until the last Brownian bridge from the  $j_0$ bead of the $i_\alpha$ polymer to the $j_0+s$ bead of the $i_{\alpha+1}$ 
  closes the loop. 
  
  The obtained new configuration has a probability to be accepted that is the product of Eq.~\eqref{pbb} for each reconstructed segment. If the move is accepted, this new configuration is kept, if this 
  acceptance test fails the configuration prior to the permutation move has to be restored.
  The probability to accept exchanges is usually very low and in order to obtain an efficient permutation sampling one has to try thousands of permutation moves in a single MCS; moreover, in most cases, the probability 
  to accept a permutation drops exponentially with the number of polymers involved in the permutation and thus the efficiency of this algorithm for permutation sampling get worse with 
  increasing particle number $N$. 
  Given a permutation cycle, the probability to accept the reconstruction step is roughly the product of the probability to accept each single Brownian bridge of the same length; 
  this suggests that in most cases, a good choice of the length of reconstructions $s$ can be roughly the same as that of a single Brownian bridge; this however may not be true 
  if the polymers are distant each other; in this case the probability to accept a permutation is maximized if it involves a sufficiently large imaginary--time; this holds even though 
  the Brownian bridges in the reconstruction step would have low acceptances.

\subsection{Estimators}
Let's consider again the expectation value of a local operator $\hat{O}$
\begin{eqnarray}\label{thermavg}
\left\langle\hat{O}\right\rangle = \int d\Gamma \: \hat{O}(\Gamma_k)p(\Gamma)
\end{eqnarray}
where $\Gamma = \left\lbrace R_1,...,R_M\right\rbrace$ and $ 1 \le k \le M$ represents the position in the path integral at which the operator is applied. 
This equation holds for both PIGS and PIMC depending on the choice of the multi--dimensional probability distribution 
$p(\Gamma )$, to be more specific, in the PIMC case,
\begin{eqnarray}
p(\Gamma)= p^{PIMC}(\Gamma)= \frac{\prod_{j=1}^{M}G\left(R_{j},R_{j+1},\delta\tau\right)}{\int d\Gamma\:\prod_{j=1}^{M}G\left(R_{j},R_{j+1},\delta\tau\right)}
\end{eqnarray}
where we set $R_{M+1} \equiv R_{1}$. In the PIGS case,
\begin{eqnarray}
p(\Gamma)= p^{PIGS}(\Gamma)= \frac{\Psi_T(R_1)\prod_{j=1}^{M-1}G\left(R_{j},R_{j+1},\delta\tau\right)\Psi_T(R_M)}{\int d\Gamma\:\Psi_T(R_1)\prod_{j=1}^{M-1}G\left(R_{j},R_{j+1},\delta\tau\right)\Psi_T(R_M)}
\quad .
\end{eqnarray}
In the previous Sec  we described a method to sample $p^{PIGS}(\Gamma)$ and $p^{PIMC}(\Gamma)$, here we focus on the application of the operator $\hat{O}$ to the density matrix. 
We have already pointed out in Sec.~\ref{sec:pimc} that in PIMC, due to the cyclic property of the trace operation, one can shift the position of $\hat{O}$ along the path integral without changing the expectation value.
 This is useful because one can use all the configurations $\left\lbrace R_1,...,R_M\right\rbrace$ to compute the expectation values and then average the results. 
Also in PIGS an operator can be evaluated at any imaginary--time $\tau_l$; to be more explicit on the meaning of ``application of an operator at an imaginary time $\tau_l$":
\begin{eqnarray} \label{exv0}
\left\langle \Psi\left(\tau=l\delta\tau\right)|\hat{O}|\Psi\left(\tau=(M-l)\delta\tau\right)\right\rangle \simeq \,\,\,\,\,\,\,\,\,\,\,\,\,\,\,\,\,\,\,\, \\
\frac{\int dR_1...dR_M\:\Psi_T\left(R_1\right)...G(R_{l-1},R_{l},\delta\tau)\hat{O}\left(R_{l}\right)G(R_{l},R_{l+1},\delta\tau)...\Psi_T\left(R_M\right)}{\int
dR_1...dR_M\:\Psi_T\left(R_1\right)G(R_1,R_2,\delta\tau)...G(R_{M-1},R_M,\delta\tau)\Psi_T\left(R_M\right)} \nonumber
\end{eqnarray}
with $2 \le l \le\ M-1$. Here $|\Psi(\tau)\rangle$ represents the evolution of the trial wave function $|\Psi_T\rangle$ at an imaginary--time $\tau$, 
namely $|\Psi(\tau)\rangle=|e^{-\tau\hat{H}}\Psi_T\rangle$. Differently from PIMC, due to Eq. \eqref{conve}, only for $\tau_0 \le \tau_l \le \tau-\tau_0$ it is verified that, to a good approximation, 
$\left|\Psi(\tau=\tau_l=l\delta\tau)\right\rangle \simeq
\left|0\right\rangle$ {\it and} $\left|\Psi(\tau=\tau_{M-l}=(M-l)\delta\tau)\right\rangle \simeq
\left|0\right\rangle$; in this case, Eq.~\eqref{exv0} becomes an expectation value on the ground state of the system.
Outside the interval $\left[\tau_0 ; \tau-\tau_0\right]$ the expectation values are mixed, more specifically, for an imaginary--time index $h$ so that $\tau_h < \tau_0$, 
$\left\langle \Psi(\tau=\tau_h)\left|\hat{O}\right|0\right\rangle$.
 For $\tau_h = 0$ and $\tau_h = \tau$ we obtain respectively the mixed expectation values 
$\left\langle \Psi_{t}\left|\hat{O}\right|0\right\rangle$ and $\left\langle 0\left|\hat{O}\right|\Psi_T\right\rangle$ with the trial wave function $\Psi_T$.
These expectation values are obtained by applying the operator directly on the trial wave function,
\begin{eqnarray} \label{exv1}
\left\langle \Psi_T|\hat{O}|\tilde{\Psi}\right\rangle \simeq \,\,\,\,\,\,\,\,\,\,\,\,\,\,\,\,\,\,\,\, \\
\frac{\int dR_1...dR_M\:\hat{O}\left(R_1\right)\Psi_T\left(R_1\right)G(R_1,R_2,\delta\tau)...G(R_{M-1},R_M,\delta\tau)\Psi_T\left(R_M\right)}{\int
dR_1...dR_M\:\Psi_T\left(R_1\right)G(R_1,R_2,\delta\tau)...G(R_{M-1},R_M,\delta\tau)\Psi_T\left(R_M\right)} \nonumber
\end{eqnarray}
an analogous relation holds for $\Psi_T\left(R_M\right)$. The mixed expectation value is very useful in the evaluation of the total energy. The Hamiltonian $\hat{H}$ ,
commutes with the propagator $e^{-\delta\tau\hat{H}}$ and thus the total energy can be evaluated at any time--step; however, in Eq.~\eqref{exv0}, we are approximating 
$e^{-\delta\tau\hat{H}}$ with a propagator $G(R,R',\delta\tau)$; as consequence, the commutation rule that allows the evaluation of the total energy at any time--step, 
holds only if the small--time approximation of the propagator is accurate enough; this provides a useful check of convergence of PIGS for what concerns 
the choice of $\delta\tau$. Moreover, if an accurate trial wave function 
is available for the system to study, the evaluation of the total energy on the variational wave function is more accurate than that at other imaginary times; this happens because the trial wave function 
introduces correlations that guide the Metropolis sampling in a more efficient way to explore physical configurations.

Now that the effect of the imaginary time evolution on the expectation values has been described, we show the application of 
 the operator $\hat{O}$  on a density matrix $G\left(R_{j},R_{j+1},\delta\tau\right)$ in the most common choices of $\hat{O}$. 
 This, of course, is generally dependent on the particular choice of small imaginary--time approximation for $G$,
 exception made of the case of an operator that is diagonal in the coordinate representation, for example, the static structure factor, the density--density correlation function in imaginary time or 
 the radial distribution function.
 Consider the general case for a density matrix that can be expressed as follows: 
  \begin{eqnarray}\label{densest}
  G\left(R_m,R_{m+1},\delta\tau\right) = G_0\left(R_m,R_{m+1},\delta\tau\right)e^{-U\left(R_m,R_{m+1},\delta\tau\right)}
  \end{eqnarray}
where $G_0$ is the density matrix for free particles. 
Some of these density matrices are illustrated in Appendix.~\ref{sechighorder}. 
Let's also consider, for simplicity, the PIMC case, so that the partition function of the system becomes
  \begin{eqnarray} \label{pfuncest}
  \mathcal{Z} \simeq \int \prod_{m=1}^{M}dR_{m}\:e^{-\frac{\left(R_{m}-R_{m+1}\right)^{2}}{4\lambda\delta\tau}}e^{-\delta\tau U\left(R_{m},R_{m+1},\delta\tau\right)}
 \quad .
  \end{eqnarray}
The following discussion applies also in the PIGS case with at worse slight modifications that will be described contextually.

\subsubsection{Energy}
The Energy per particle, $E/N$, is the expectation value of $\hat{O}=\hat{H}/N$. Apart from the physical importance 
of this quantity, the energy is one of the main expectation values that are 
used to tune the parameters of the simulations.

\paragraph{Hamiltonian estimators}
This estimator is obtained by applying the operator $\hat{H}=\hat{T} + \hat{V}$ to the density matrix \eqref{densest}.  

  \begin{figure}[h]
 \begin{center}
 \includegraphics*[width=11cm]{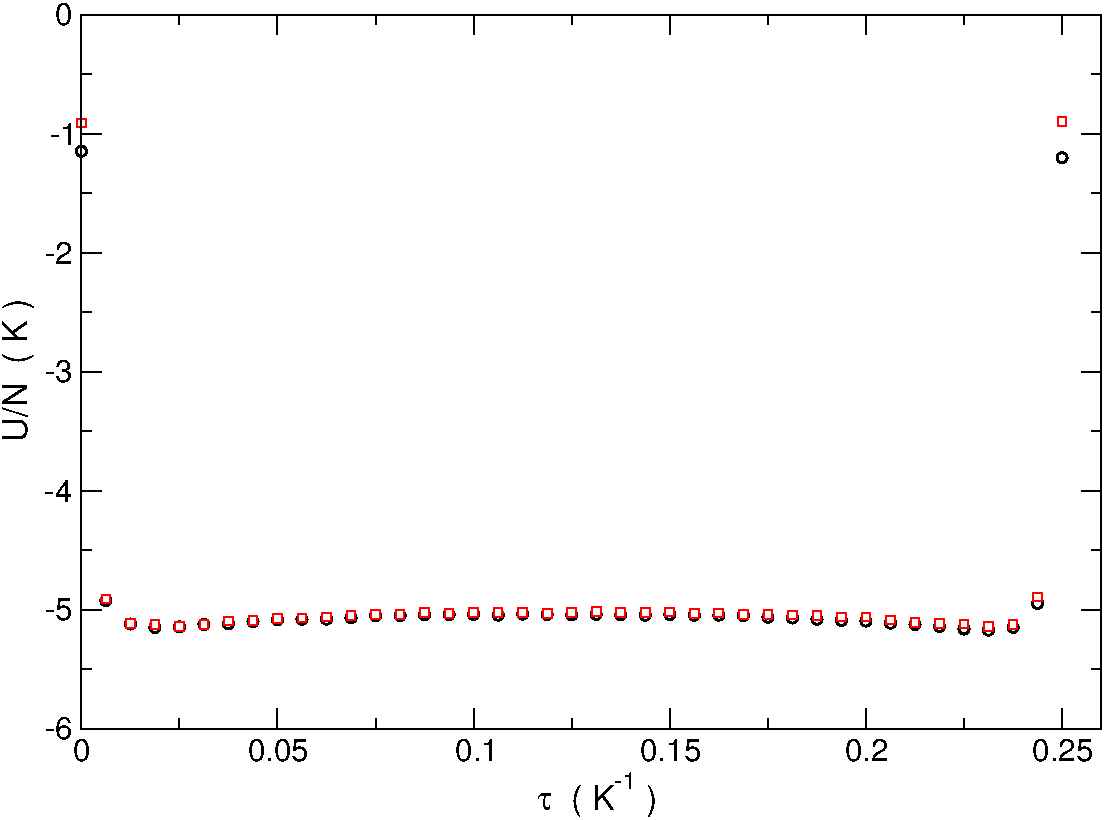}
 \caption{\label{epot} 
PIGS computation of the potential energy per particle versus imaginary--time for a 2D system of $N=16$ atoms of $^4$He at a density $\rho=0.045$~\AA$^{-2}$,
interacting with the Aziz potential described in Ref.~\onlinecite{d:Aziz}.
The trial wave function is $|\Psi_T>=1$ and the total projection time $\tau=0.5$ K$^{-1}$. 
Red squares are obtained with an $8$--th order multi--product expansion (see Appendix \ref{sechighorder}) at a timestep 
$6\delta\tau=1/80$ K$^{-1}$. Black circles are obtained with the primitive approximation at a timestep 
$\delta\tau=1/480$ K$^{-1}$; however, for comparison purposes, only points at $\tau_m = 6m/\delta\tau$ are shown.
}
\end{center}
\end{figure}
The potential term $\hat{V}=\sum_{i<j}v\left(r_{ij}\right)$ is diagonal on coordinates representation, so that
  \begin{eqnarray}
  \hat{V}G\left(R_m,R_{m+1},\delta\tau\right) = V\left(R_m\right)G\left(R_m,R_{m+1},\delta\tau\right) 
 \quad .
  \end{eqnarray}
 In Fig.~\ref{epot} we show the potential energy obtained from a PIGS simulation of two--dimensional 
 $^4$He.
  
The kinetic term is $\hat{T}=-\frac{\hbar^2}{2m}\sum_{i}^{N}\nabla_{i}^{2}$ and
  \begin{eqnarray}
  \nabla_{i}^{2}G\left(R_m,R_{m+1},\delta\tau\right) = G\left(R_m,R_{m+1},\delta\tau\right) \:\:\:\:\:\:\: \\
  \left[\delta\tau^2\left|\vec{\nabla}_{i}U\right|^2 +
  \frac{\left(\vec{r}_{i}^{\:m}-\vec{r}_{i}^{\:m+1}\right)^2}{4\lambda^2\delta\tau^2} + \frac{1}{\lambda}\left(\vec{r}_{i}^{\:m}-\vec{r}_{i}^{\:m+1}\right)\cdot\vec{\nabla_i}U -\delta\tau \nabla_i^{2}U
  -\frac{d}{2\lambda\delta\tau}\right] \nonumber
  \end{eqnarray}
where $U=U\left(R_m,R_{m+1},\delta\tau\right)$ for simplicity, $\lambda=\frac{\hbar^2}{2m}$ and $d$ is the dimensionality of the system.

\paragraph{Thermodynamic estimators}
The total energy per particle can be obtained also from the thermodynamic definition,
\begin{eqnarray}
\frac{E\left(N,V,\beta\right)}{N} = -\frac{1}{N\mathcal{Z}}\frac{\partial \mathcal{Z}\left(N,V,\beta\right)}{\partial\beta}
\end{eqnarray}
where $\mathcal{Z}$ is the partition function defined in Eq.~\eqref{pfuncest}.
The thermodynamic estimator for the energy per particle is thus
\begin{eqnarray}\label{thermtot}
\frac{E}{N}=\left\langle\frac{d}{2\delta\tau} - \frac{1}{4\lambda\delta\tau^2MN}\sum_{m=1}^{M}\left(R_m-R_{m+1}\right)^2+\frac{1}{MN}\frac{\partial U\left(R_m,R_{m+1},\delta\tau\right)}{\partial \delta\tau}\right\rangle
\quad .
\end{eqnarray}
In the same way, the kinetic energy per particle, $K/N$, can be obtained from the thermodynamic relation
\begin{eqnarray}
\frac{K}{N} = \frac{m}{\beta\mathcal{Z}}\frac{\partial \mathcal{Z}\left(N,V,\beta\right)}{\partial m}
\end{eqnarray}
and the estimator becomes
\begin{eqnarray}\label{thermkin}
\frac{K}{N}=\left\langle\frac{d}{2\delta\tau} - \frac{1}{4\lambda\delta\tau^2MN}\sum_{m=1}^{M}\left(R_m-R_{m+1}\right)^2+\frac{m}{\delta\tau MN}\frac{\partial U\left(R_m,R_{m+1},\delta\tau\right)}{\partial m}\right\rangle
\end{eqnarray}
where $\langle ... \rangle$ is the average on the configurations $\left\lbrace R_{m}\right\rbrace_{m=1}^{M}$ that are sampled by the metropolis algorithm. 
  \begin{figure}[h]
 \begin{center}
 \includegraphics*[width=11cm]{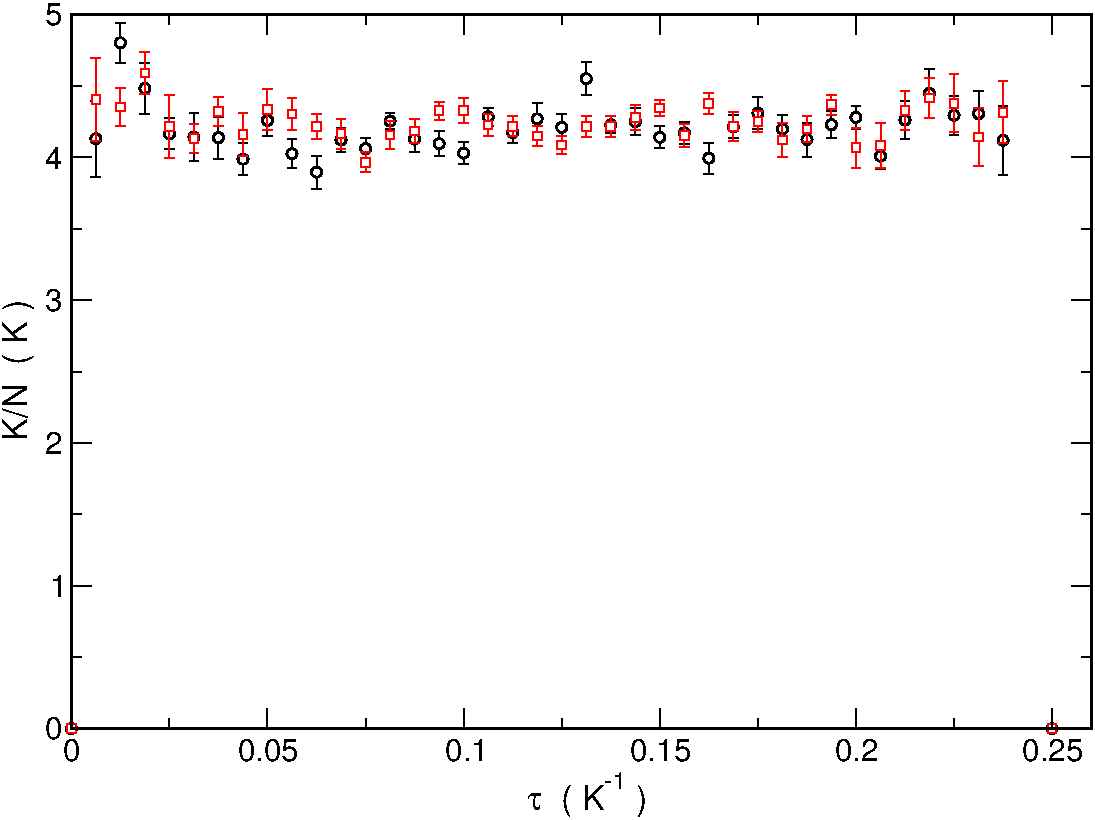}
 \caption{\label{ekin} 
PIGS computation of the kinetic energy per particle versus imaginary--time for a 2D system of $N=16$ atoms of $^4$He at a density $\rho=0.045$~\AA$^{-2}$,
interacting with the Aziz potential described in Ref.~\onlinecite{d:Aziz}.
The trial wave function is $|\Psi_T>=1$ and the total projection time $\tau=0.5$ K$^{-1}$. 
Red squares are obtained with an $8$--th order multi--product expansion (see Appendix \ref{sechighorder}) at a timestep 
$6\delta\tau=1/80$ K$^{-1}$. Black circles are obtained with the primitive approximation at a timestep 
$\delta\tau=1/480$ K$^{-1}$; however, for comparison purposes, only points at $\tau_m = 6m/\delta\tau$ are shown.
}
\end{center}
\end{figure}
An example of application of this estimator is shown in Fig.~\ref{ekin}.
The Hamiltonian and the thermodynamic estimators provide two different ways to obtain the energy, however they suffer from statistical fluctuations that increase with smaller values of $\delta\tau$, 
this is particularly true for the Hamiltonian estimator due to the presence of the laplacian operator, but happens in smaller degree also in the thermodynamic estimator because the first two terms 
in Eq.~\eqref{thermtot} and \eqref{thermkin} are quantities that 
increase when $\delta\tau$ decreases and cancel each others. This requires longer simulations when $\delta\tau$ is small and usually poses a computational limit for the evaluation of the total energy. 
There are at least two possibilities to overcome this problem: one can either use higher order estimators which achieve convergence at higher timestep or introduce a more advanced estimator. A choice
 for the latter possibility is the virial estimator\cite{c7r7}; the derivation of this estimator is shown in appendix \ref{ch:estimators} together
 with an explicit derivation of the thermodynamic estimators for the Pair Suzuki approximation. In this appendix we show also that although the thermodynamic estimators are obtained from a 
 thermodynamic relation, because of the similar formalism of PIGS and PIMC, it is possible to use these estimators also in PIGS.

\subsubsection{Radial distribution function}
The pair correlation function $g\left(\vec{r}_{1},\vec{r}_{2}\right)$  is the probability to have a particle at $\vec{r}_{1}$ and a particle at $\vec{r}_{2}$. Within the path integral formalism,
\begin{eqnarray}\label{gdz}
g\left(\vec{r}_{1},\vec{r}_{2}\right) = \frac{V^2}{\mathcal{Z}}\int d\vec{r}_{3}...d\vec{r}_{N}\: G\left(R,R,\beta\right)
\quad .
\end{eqnarray}  
In a uniform system the pair distribution function depends only on the distance $r=\left|\vec{r}_1-\vec{r}_2\right|$, with a change of integration variables in Eq.~\eqref{gdz} and using the definition of
 thermal average \eqref{thermavg}, the estimator becomes
 \begin{eqnarray}\label{gdzest}
 g\left(r\right) = \frac{V}{N^2M}\left\langle\sum_{m=1}^M\sum_{i\neq j}^N\delta\left(\left|\vec{r}\right|-\left|\vec{r}_{i}^{\:m}-\vec{r}_{j}^{\:m}\right|\right)\right\rangle
 \end{eqnarray}
where we have taken into account the symmetry under particle exchange and the estimator has been averaged over the timeslices $m$ in order to employ larger statistics.
In PIMC the sum over $m$ covers all the timeslices; in PIGS, this sum must be intended only over the central timeslices, where Eq.~\eqref{exv0} gives an accurate description of the ground state.
  \begin{figure}[h]
 \begin{center}
 \includegraphics*[width=11cm]{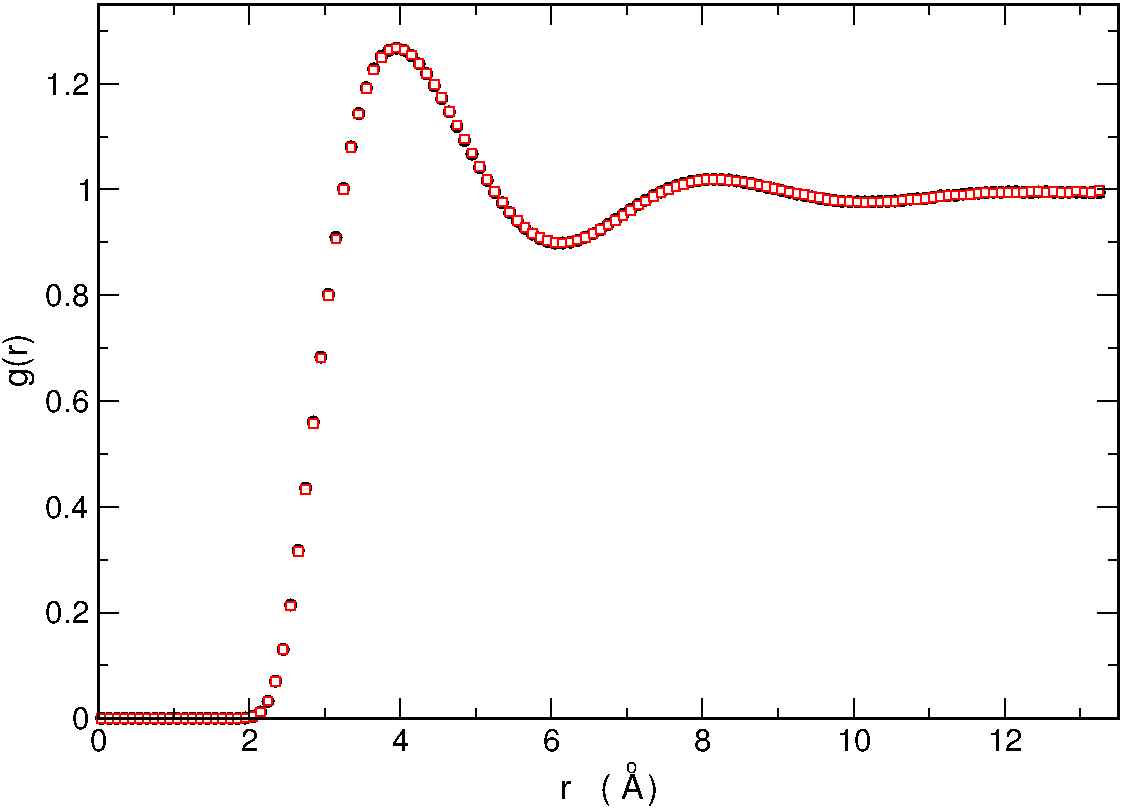}
 \caption{\label{gdr} 
PIGS computation of the radial distribution function of a 2D system of $N=16$ atoms of $^4$He at a density $\rho=0.045$~\AA$^{-2}$,
interacting with the Aziz potential described in Ref.~\onlinecite{d:Aziz}.
The trial wave function is $|\Psi_T>=1$, the total projection time $\tau=0.5$ K$^{-1}$ and the averages 
were taken in the imaginary--time interval 0.2 K$^{-1}$ -- 0.3 K$^{-1}$.
Red squares are obtained with an $8$--th order multi--product expansion (see Appendix \ref{sechighorder}) at a timestep 
$6\delta\tau=1/80$ K$^{-1}$. Black circles are obtained with the primitive approximation at a timestep 
$\delta\tau=1/480$ K$^{-1}$.
The simulation box is a square of late $L$; the radial distribution function has been computed 
also in the range $(L/2;L\sqrt{2})$.
}
\end{center}
\end{figure}
To evaluate this estimator in a computer simulation one defines a partition $P_n$ of the interval $\left[0;L_l/2\right]$ where $L_l$ is the smallest size of the simulation box,
and every element of the partition has a length $\Delta r$; with $P_n=[n\Delta r;(n+1)\Delta r]$ then, construct an histogram of the frequencies of the relative distance $r_{ij}$ between two particles of the
system at the same imaginary--time index. This histogram has to be normalized with the number of particles that a free particles system of the same density would have at a bin $n$, namely for $d=3$,
$V_n = \frac{N}{V}\frac{4}{3}\pi\left[\left(\left(n+1\right)\Delta r\right)^3-\left(n\Delta r\right)^3\right]$.
An example of QMC evaluation of the radial distribution function for a two--dimensional system of $^4$He is provided 
in Fig.~\ref{gdr}.

\subsubsection{Static structure factor}
The static structure factor is useful to study the spatial order of a system in the reciprocal lattice; it is in fact connected to the pair distribution function by a Fourier transform. This estimator is defined
 as a quantum average of the density operator $\hat{\rho}_{\vec{k}}$,
 \begin{eqnarray}
 S\left(\vec{k}\right) = \frac{1}{N}\left\langle\hat{\rho}_{\vec{k}}\hat{\rho}_{-\vec{k}}\right\rangle = \frac{1}{N\mathcal{Z}}\int dR\:
 \rho\left(R,R,\beta\right)\left(\sum_{i=1}^{N}e^{-i\vec{k}\cdot\vec{r}_{i}}\right)\left(\sum_{i=1}^{N}e^{i\vec{k}\cdot\vec{r}_{i}}\right)
 \quad .
 \end{eqnarray}
 Using the Euler identities, the static structure factor can be expressed in a form that is computable by different Monte Carlo methods
 \begin{eqnarray}
 S\left(\vec{k}\right) = \frac{1}{NM}\left\langle \sum_{i \neq j}^{N}\sum_{m=1}^{M} \left[\cos\left(\vec{k}\cdot\vec{r_i}^{m}\right)\cos\left(\vec{k}\cdot\vec{r_j}^{m}\right)+
 \sin\left(\vec{k}\cdot\vec{r_i}^{m}\right)\sin\left(\vec{k}\cdot\vec{r_j}^{m}\right)\right]   \right\rangle \nonumber  
 \left. \right.
\quad .
 \end{eqnarray}
   \begin{figure}[h]
 \begin{center}
 \includegraphics*[width=11cm]{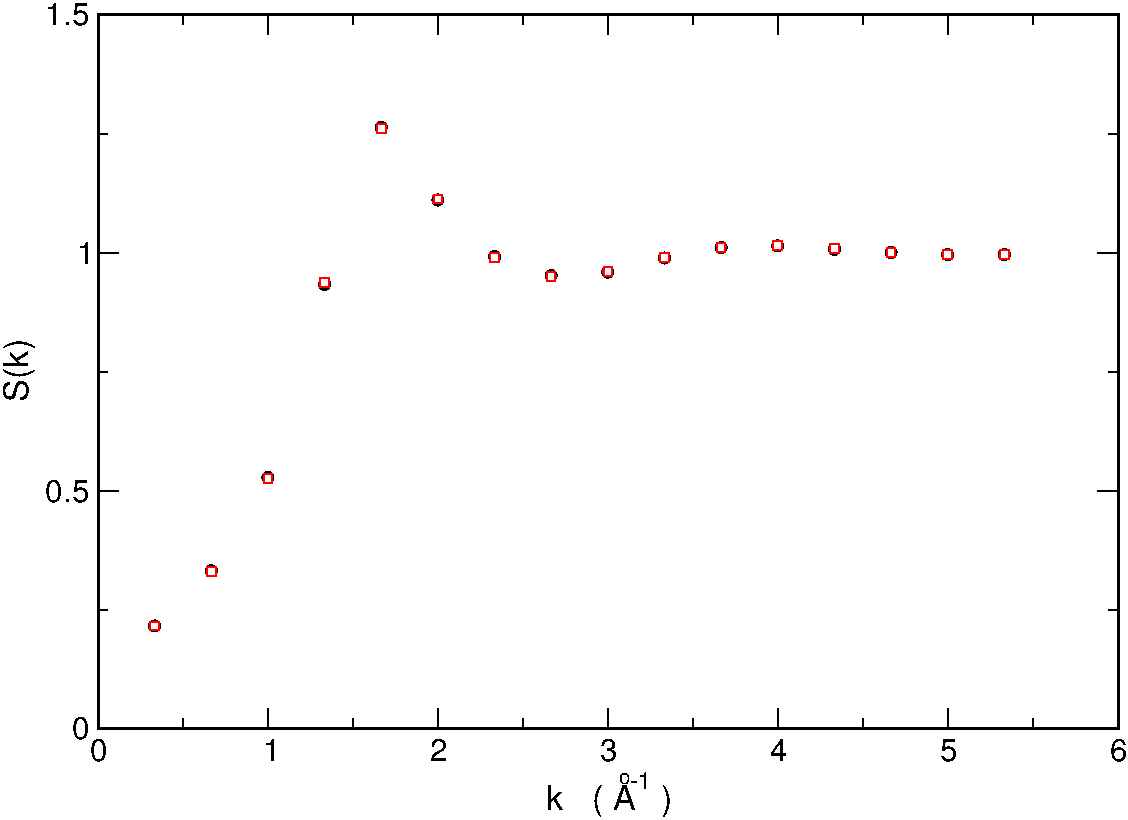}
 \caption{\label{sdk} 
PIGS computation of the static structure factor along the axis direction of the simulation box. The system is 
two--dimensional and consists of $N=16$ atoms of $^4$He at a density $\rho=0.045$~\AA$^{-2}$,
interacting with the Aziz potential described in Ref.~\onlinecite{d:Aziz}.
The trial wave function is $|\Psi_T>=1$, the total projection time $\tau=0.5$ K$^{-1}$ and the averages 
were taken in the imaginary--time interval 0.2 K$^{-1}$ -- 0.3 K$^{-1}$.
Red squares are obtained with an $8$--th order multi--product expansion (see Appendix \ref{sechighorder}) at a timestep 
$6\delta\tau=1/80$ K$^{-1}$. Black circles are obtained with the primitive approximation at a timestep 
$\delta\tau=1/480$ K$^{-1}$.
}
\end{center}
\end{figure}
 where we have averaged over the $M$ equivalent timeslices as before. As in the previous case, in PIMC the sum over $m$ covers all the timeslices; 
 in PIGS, this sum must be intended only over the central timeslices, where Eq.~\eqref{exv0} gives an accurate description of the ground state.
 
  It must be remarked that in a QMC simulation the simulation box has finite dimensions that in the case of
 $d$ dimensions are $\left(L_0,...,L_d\right)$. This implies that the wave vectors that are accessible by the simulation are of the form 
 $\vec{k}=\left(\frac{2\pi}{L_0}n_0,...,\frac{2\pi}{L_d}n_d\right)$, with $n_0,...,n_d \in \mathbb{I}$.
 An example of static structure factor is shown in Fig.~\ref{sdk}

\subsubsection{Imaginary--time correlation functions}
The static structure factor is also called `density--density correlation function'. In general, a correlation function between two operators $\hat{A}$ and $\hat{B}$ is a quantum average
\begin{eqnarray}
c_{AB} = \left\langle \hat{A}\hat{B} \right\rangle
\quad .
\end{eqnarray}
In particular, the operators $\hat{A}$ and $\hat{B}$ can be evaluated at different imaginary--times (different time--sectors) and this is an imaginary--time correlation function. A significative example 
of imaginary--time correlation function is the density--density one:
\begin{eqnarray} \label{itcf1}
F\left(\vec{k},\tau\right)=\frac{1}{N}\left\langle\hat{\rho}_{\vec{k}}\left(0\right)\hat{\rho}_{-\vec{k}}\left(\tau\right)\right\rangle
\end{eqnarray}
that for $\tau=0$ reduces to the static structure factor. This function is related to the dynamical structure factor of the system by a Laplace transform and thus contains informations about the 
excitations of the system. However, these informations are accessible from Eq.~\eqref{itcf1} only by solving a numerical inverse Laplace transform; 
this operation is carried out on a function $F(\vec{k},\tau)$ which is known only for some imaginary--time $\tau_m$ and with a statistical uncertainty. In these conditions, the inversion of the Laplace 
transform is an ill--posed problem. Methodologies to face those problems have been implemented by many research groups\cite{c7r8}. 

\subsubsection{One body density matrix}
An important quantity to sample is the one--body density matrix (OBDM) because it is strictly connected to the Bose Einstein condensation (BEC). 

BEC can be defined as a macroscopic occupancy of a given quantum state. The simple BEC occurring in $^4$He may be described by a momentum distribution of the form
\begin{eqnarray}\label{becmom}
n\left(\vec{p}\right) = N_0\delta\left(\vec{p}\right) + \tilde{n}\left(\vec{p}\right)
\quad .
\end{eqnarray}
The OBDM of the ground state $|0\rangle$ of the system is
\begin{eqnarray}\label{becobdmdef}
\rho_1\left(\vec{r},\vec{r}^{\:'}\right) = \left\langle0\left|\hat{\Psi}^{\dagger}\left(\vec{r}\right)\hat{\Psi}\left(\vec{r}^{\:'}\right)\right|0\right\rangle
\quad .
\end{eqnarray}
The Fourier transform of this equation gives the momentum distribution $n_{\vec{p}}$ of the system at its ground state:
\begin{eqnarray}
n_{\vec{p}} = \left\langle 0\left|\hat{a}_{\vec{p}}^{\dagger}\hat{a}_{\vec{p}}\right| 0\right\rangle \label{nopdef}\\
\hat{a}_{\vec{p}} = \frac{1}{\left(2\pi\hbar\right)^{\frac{3}{2}}}\int d\vec{r}\:e^{\frac{i}{\hbar}\vec{p}\cdot\vec{r}}\hat{\Psi}\left(\vec{r}\right) \label{opcdef}
\quad .
\end{eqnarray}
Placing Eq.~\eqref{opcdef} in Eq.~\eqref{nopdef} yields
\begin{eqnarray} \label{obdmder}
n_{\vec{p}} = \frac{1}{\left(2\pi\hbar\right)^{\frac{3}{2}}}\int d\vec{r}d\vec{r}^{\:'}\:e^{-\frac{i}{\hbar}\vec{p}\cdot\left(\vec{r}-\vec{r}^{\:'}\right)}
\left\langle0\left|\hat{\Psi}^{\dagger}\left(\vec{r}\right)\hat{\Psi}\left(\vec{r}^{\:'}\right)\right|0\right\rangle = \nonumber \\
\frac{1}{\left(2\pi\hbar\right)^{\frac{3}{2}}}\int d\vec{r}d\vec{r}^{\:'}\:e^{-\frac{i}{\hbar}\vec{p}\cdot\left(\vec{r}-\vec{r}^{\:'}\right)}\rho_1\left(\vec{r},\vec{r}^{\:'}\right) = \nonumber \\
\frac{1}{\left(2\pi\hbar\right)^{\frac{3}{2}}}\int d\vec{t}d\vec{s}\:e^{-\frac{i}{\hbar}\vec{p}\cdot\vec{s}}\rho_1\left(\vec{t}+\frac{\vec{s}}{2},\vec{t}-\frac{\vec{s}}{2}\right)
\end{eqnarray}
where $\vec{s}=\vec{r}-\vec{r}^{\:'}$ and $\vec{t}=\left(\vec{r}+\vec{r}^{\:'}\right)/2$. If the system is uniform and isotropic, the OBDM depends only on $s=\left|\vec{s}\right|$, and in the 
thermodynamic limit
\begin{eqnarray} \label{obdmft1}
n_{\vec{p}} = \frac{V}{\left(2\pi\hbar\right)^3}\int d\vec{s}\: e^{-\frac{i}{\hbar}\vec{p}\cdot\vec{s}}\rho_1\left(s\right)
\quad .
\end{eqnarray}
If the momentum distribution in Eq.~\eqref{becmom} is anti--transformed to coordinates, an OBDM with unvanishing tail at high $s$ is obtained. An OBDM that displays such asymptotic behavior, that is 
$\lim_{s\rightarrow\infty}\rho_1\left(s\right)=n_0 > 0$, has an off--diagonal long range order; if that function is normalized so that $\rho_1\left(0\right)=1$, $n_0$ corresponds to the fraction of BEC.

The OBDM in the path integral notation is
\begin{eqnarray}\label{obdmpigs}
\rho_1^{PIGS}\left(\vec{r},\vec{r}^{\:'}\right)=\frac{V}{\mathcal{N}}\int dR_1dR_Md\vec{r}_2^{\:M/2}...d\vec{r}_M^{\:M/2}\:
\Psi_T(R_1)G(R_1,R_{M/2},\tau/2)\: \times \nonumber \\
\times\:G(R_{M/2}^{\:'},R_M,\tau/2)\Psi_T(R_M)
\end{eqnarray}
where $R_m=\left(\vec{r},\vec{r}_2^{\:m},...,\vec{r}_N^{\:m}\right)$, $R_m^{\:'}=\left(\vec{r}^{\:'},\vec{r}_{2}^{\:m},...,\vec{r}_N^{\:m}\right)$ and 
$dR_m = \prod_{i=1}^{N}d\vec{r}_{i}^{\:m}$. In the PIMC case, an analogous procedure gives
\begin{eqnarray} \label{obdmpimc}
\rho_1^{PIMC}\left(\vec{r},\vec{r}^{\:'}\right) = \frac{V\int d\vec{r}_{2}..d\vec{r}_{N}\:G\left(R,R',\beta\right)}{\mathcal{Z}}
\quad .
\end{eqnarray} 
In the classical isomorphism, the $i$--th 
polymer has been split: if the original polymer had $M$ beads, the bead at a position $j$ is removed and two new beads are inserted. 
In PIGS $j$ should correspond to an imaginary--time projection large enough to have an accurate description of the ground state; in 
PIMC $j$ can be anywhere in the path integral.
These beads, $\vec{r}$ and $\vec{r}^{\:'}$, are not linked each other but are respectively
 the last and the first timeslice of the new open polymer that would appear. In the PIMC case, this will be a single open polymer 
 with extremities $\vec{r}$ and $\vec{r}^{\:'}$; in the PIGS case two half polymers will appear, with extremities 
 $\vec{r}_i^{\:1},\vec{r}$ and $\vec{r}^{\:'},\vec{r}_i^{\:M}$.
  
\begin{figure}[h]
 \begin{center}
 \includegraphics*[width=11cm]{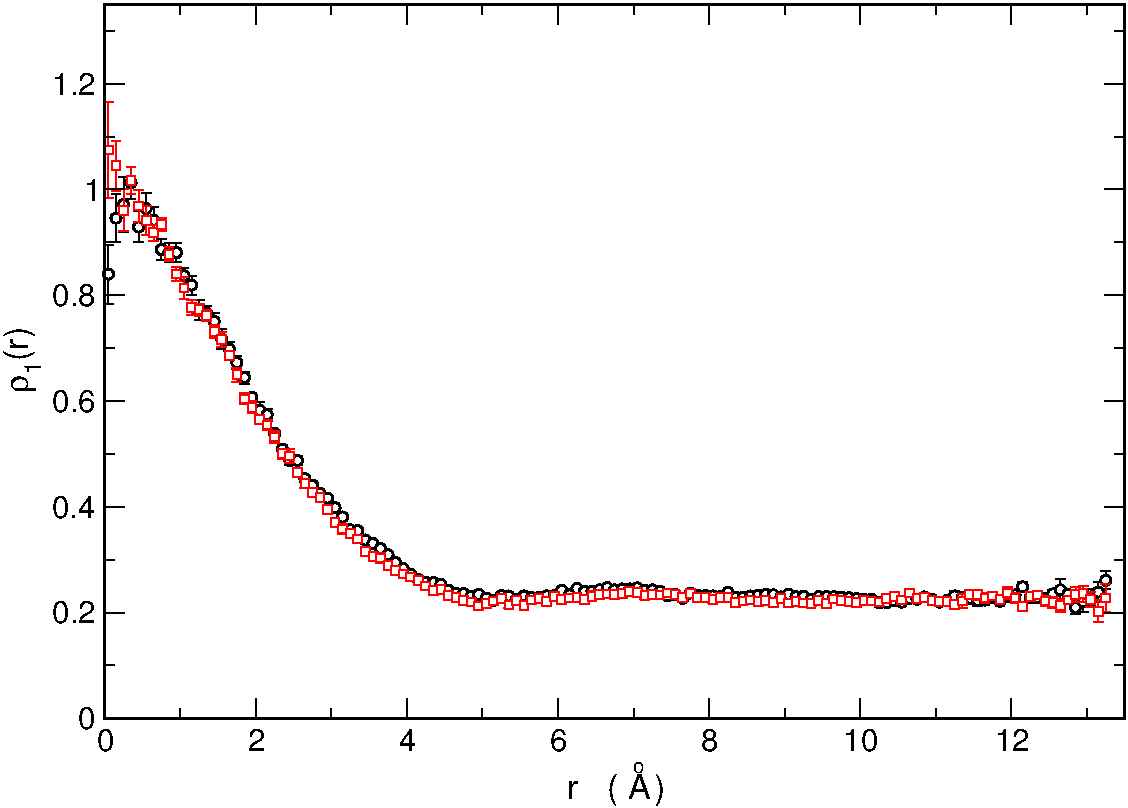}
 \caption{\label{obdm} 
PIGS computation of the one body density matrix of a 2D system of 
 $N=16$ atoms of $^4$He at a density $\rho=0.045$~\AA$^{-2}$,
interacting with the Aziz potential described in Ref.~\onlinecite{d:Aziz}.
The trial wave function is $|\Psi_T>=1$ and the total projection time $\tau=0.5$ K$^{-1}$. 
Red squares are obtained with an $8$--th order multi--product expansion (see Appendix \ref{sechighorder}) at a timestep 
$6\delta\tau=1/80$ K$^{-1}$. Black circles are obtained with the primitive approximation at a timestep 
$\delta\tau=1/480$ K$^{-1}$.
The simulation box is a square of late $L$; the radial distribution function has been computed 
also in the range $(L/2;L\sqrt{2})$. The normalization constant has been computed with the Worm algorithm 
(see Sec.~\ref{wormsec}).
}
\end{center}
\end{figure}

If the system is homogeneous, $\rho_1$ depends only on the distance $r=|\vec{r}-\vec{r}^{\:'}|$; 
the OBDM is then sampled making an histogram of the relative distance that the two newly created beads have at every MCS.
 This histogram should be normalized so that $\rho_1(r=0)=1$; this is done after the QMC evaluation of $\rho_1$ with 
 a Gaussian fit of the small $r$ part of $\rho_1(r)$. In some cases, this might not be the optimal choice for the 
 normalization: the Worm algorithm, described in Sec.~\ref{wormsec}, can provide an {\it a priori} normalization 
 of the OBDM with a QMC evaluation of $\mathcal{N}(\mathcal{Z})$. As example we show in Fig.~\ref{obdm} the one body density matrix 
 for a two--dimensional system of $^4$He.
 We point out that in the PIMC case, the sampling of permutations is crucial to obtain off diagonal long range order: 
 permuting ring polymers with the open polymer will result in a bigger open polymer that allows its extremities to
  get far away each other, eventually contributing to a non vanishing tail in the OBDM. 
  In PIGS, permutations will yield simply other open polymers and in most situations are not essential; however, there are 
  cases in which permutations are essential in order to obtain an ergodic sampling of the configurations; 
  we have indeed shown an example in Sec.~\ref{secpatatelesse} when we studied, with PIGS, the condensate fraction of $^4$He with 
  a particular choice of $\Psi_T$: a Gaussian wave function centered on the equilibrium positions of solid HCP $^4$He; 
  this is a very particular case in which the trial wave function introduces correlations that constrain the 
  terminal beads of the polymer to arbitrary positions. In this context, single polymer moves will have the same 
  pathology of the PIMC case: the extremities of the two half polymers are pinned, as consequence a move that would 
  increase the distance between $\vec{r}$ and $\vec{r}^{\:'}$ will be soon rejected by the stretching of the 
  kinematic correlations; a permutation with another polymer, on the other hand, will allow the sampling of 
  long--range order exactly as in the PIMC case. 
  
  In some situations, the OBDM decays exponentially; this is typical in most solid systems. In such contexts, a common way used to sample the OBDM at large distances 
  $r$ is with the introduction of a repulsive factor $f(r)$ in Eq.~\eqref{obdmpigs} (or Eq.~\ref{obdmpimc} for PIMC)
  \begin{eqnarray}\label{frep1}
  f(r)=\frac{1}{1+Ae^{-Br^2}+Ce^{-Dr}} \:\:.
  \end{eqnarray}
  The parameters $B$ and $D$ are tuned to fit the exponential decay of the OBDM whereas $A$ and $C$ determine the strength of the repulsive factor. The 
  repulsive factor is aimed to modify the probability densities \eqref{obdmpigs} and \eqref{obdmpimc} to a roughly uniform probability density $\tilde{\rho}(r)$ 
  that is easier to sample. Once the histogram of $\tilde{\rho}(r)$ has been obtained, the OBDM $\rho(r)$ is recovered with a reweighting of $\tilde{\rho}(r)$:
  each histogram bin $(\tilde{\rho}(r_i),r_i)$ is divided by the quantity $f(r_i)$.

\subsubsection{Superfluidity}
  It has been derived in Ref.~\onlinecite{d:pimc} that the superfluid density of a system can be expressed through the {\it winding number} $\vec{W}$,
   \begin{eqnarray}
  \frac{\rho_s}{\rho} = \frac{\langle\vec{W}^{\:2}\rangle}{2\lambda\beta N}
  \end{eqnarray}
  and the winding number $\vec{W}$ is defined by
  \begin{eqnarray}\label{wnumber}
  \vec{W} = \frac{1}{L}\sum_{i=1}^{N} \int_{0}^{\beta} dt\: \left[\frac{d\vec{r}_{i}(t)}{dt}\right]
  \quad .
  \end{eqnarray}
  In the path integral notation, the discretized expression for Eq.~\eqref{wnumber} becomes
  \begin{eqnarray} \label{wnumberdisc}
  \vec{W} = \frac{1}{L}\sum_{m=1}^{M}\sum_{i=1}^{N} \left(\vec{r}_{i}^{\:m}-\vec{r}_{i}^{\:m+1}\right)
  \end{eqnarray}
 where $L$ is the late of the simulation box, $\vec{r}_{i}^{\:M+1} = \vec{r}_{\hat{P_1}i}^{1}$ and $\hat{P_1}$ the permutation operator introduced for the sampling of the Bose symmetry. 
 The winding number represents the number of polymers that wind around the periodic boundaries of the simulation box.
  Averaging over the $d$ spatial dimensions, the winding number estimator in periodic boundaries conditions is
  \begin{eqnarray}\label{wnumberfinally}
  \frac{\rho_s}{\rho} = \frac{\left\langle\vec{W}^{\:2}\right\rangle}{2d\lambda\beta N}
  \quad .
  \end{eqnarray}

\paragraph{Superfluid density at $T=0$ K}
At zero temperature the superfluid fraction can be obtained with the center of mass diffusion {\it in imaginary--time}\cite{d:gubernatis}. Eq.~\eqref{wnumberfinally} in fact can be 
viewed as the ratio between the diffusion constant $D_c$ of the center of mass of the system and  the diffusion constant of 
the non--interacting gas, $D_0=\hbar^2/2m$. The diffusion of the center of mass is obtained from the long $\tau$ limit of this relation
   \begin{eqnarray}
   D_c=\lim_{\tau_x\rightarrow\infty}\frac{N}{4}\frac{\left\langle\left[\vec{R}_{CM}\left(\tau_x\right)-\vec{R}_{CM}\left(0\right)\right]^{2}\right\rangle}{\tau_x}
   \end{eqnarray}
where the center of mass at a discrete imaginary time $\tau = m d\tau$ is  $\vec{R}_{CM} (\tau) = \sum_{i=1}^{N}\vec{r}_{i}^{\:m}/N$.

 The superfluid density becomes
   \begin{eqnarray}\label{cmasslim}
   \frac{\rho_s}{\rho} = \frac{D_c}{D_0}=\lim_{\tau\rightarrow\infty}\frac{N}{4\lambda}\frac{\left\langle\left[\vec{R}_{CM}\left(\tau\right)-\vec{R}_{CM}\left(0\right)\right]^{2}\right\rangle}{\tau}
   \quad .
   \end{eqnarray}

In a PIGS calculation one has to make some considerations: first, this estimator can be applied only in an interval of imaginary--time $\left[\tau_0,\tau-\tau_0\right]$ so that, in order to 
achieve long--$\tau$ convergence in Eq.~\eqref{cmasslim} one has to employ sufficiently long imaginary--time projections; secondarily, the PIGS method does not explicitly fix the center of 
mass of the system and thus Eq.~\eqref{cmasslim} cannot be used when the center of mass of the system is allowed to drift; this happens, for instance, in bulk homogeneous systems, where the 
system can translate freely in the simulation box: due to the property of PIGS to give an unbiased sampling, any correlations from the trial wave function that would eventually fix the center 
of mass is removed by the quantum imaginary--time evolution. If the center of mass of the system is allowed to drift, Eq.~\eqref{cmasslim} will then contain an unphysical contribution that 
is usually difficult to consider.

\subsection{The Worm algorithm}\label{wormsec}
Here we present the Worm algorithm in the Canonical ensemble. This method offers an enhanced permutation sampling and also a way to compute, {\it within the same simulation}, 
both diagonal and off-diagonal properties of the system under study. Differently from the worm in the Grand Canonical ensemble, this method can also be applied in PIGS without any further adaptation: 
the Worm algorithm in the Canonical ensemble has not moves that create, destroy or change the length in imaginary--time of the polymers; these moves, in fact, 
would not be of easy interpretation in the context of quantum evolution in imaginary--time; Canonical Worm is based on moves that in PIGS can be applied at the time--slice at position 
$M/2$, and in PIMC can be applied anywhere in the path integral. 
   \begin{figure}[h]
 \begin{center}
 \includegraphics*[width=11cm]{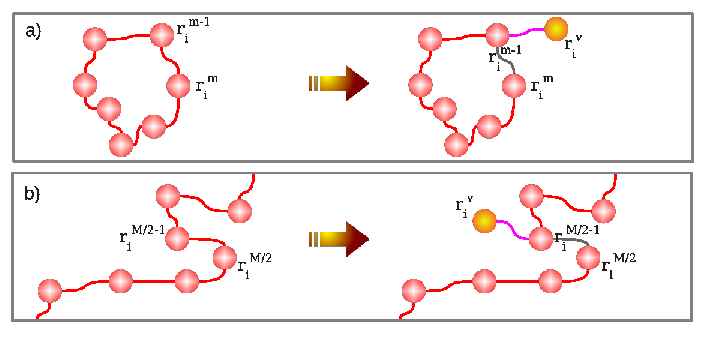}
 \caption{\label{wormdef} 
The worm in PIMC (a) and in PIGS (b). The gray lines represent the removed kinetic correlations. 
The worm in PIGS originates two half polymers that are not connected each other; in PIMC, instead, it originates 
an open polymer.
}
\end{center}
\end{figure} 
 The space of configurations that is sampled by Metropolis is enlarged by including also configurations with one open polymer, see Fig.~\ref{wormdef}; this polymer is called ``worm''. A configuration 
 with a worm is called ``off--diagonal'' (in worm notation: $G$ sector) whereas a configuration without worm is a ``diagonal'' ($Z$ sector) configuration. Starting from a polymer $i$, defined by the set of beads $\left\lbrace
 \vec{r}_{i}^{\:j}\right\rbrace_{j=1}^{M}$, where eventually $\vec{r}_{i}^{\:1} = \vec{r}_{\hat{P}i}^{\:M}$, a worm at position $m$ is constructed with the following operations:
 \begin{itemize}
 \item Remove the kinetic correlation between $\vec{r}_{i}^{\:m-1}$ and $\vec{r}_{i}^{\:m}$
 \item Add a bead $\vec{r}_{i}^{\:\nu}$ that is linked by a kinetic correlation only with its previous bead, $\vec{r}_{i}^{\:m-1}$
 \item The beads $\vec{r}_{i}^{\:\nu}$ and $\vec{r}_{i}^{\:m}$ are the two worm extremities on which off--diagonal properties such as the one body density matrix can be computed.
 \end{itemize}
 In PIMC the timeslice $m$ can be anywhere between $1$ and $M$ because any time sector is an equivalent representation of the system; in PIGS $m$ should be in the time sector range in which the trial wave
 function can be considered an accurate representation of the ground state, in our implementation of the worm algorithm in PIGS, $m$ is fixed at the central timeslice.
 
The implemented worm algorithm consists of two Metropolis moves plus an extension of the Brownian bridge and an extension of the translation move which deal with the presence of the worm extremities.
There are two input parameters: $C$ and $s$. The parameter $C$ sets the ratio $g/z$ between the number of $G$ and $Z$ configurations that are sampled during the simulation, $g+z$ are the total MCS after the 
equilibration. There is not an universal relation between $C$ and $g/z$ but large values of $C$ result in simulations with more off--diagonal sampling. Usually $C$ is of the order of unity, but the best choice
 may vary drastically with the system under study. The parameter $s$ is a tuning for the worm moves and specifies how many time--slices are to be involved by the swap and the open/close moves. 
 
\begin{figure}[h]
 \begin{center}
 \includegraphics*[width=11cm]{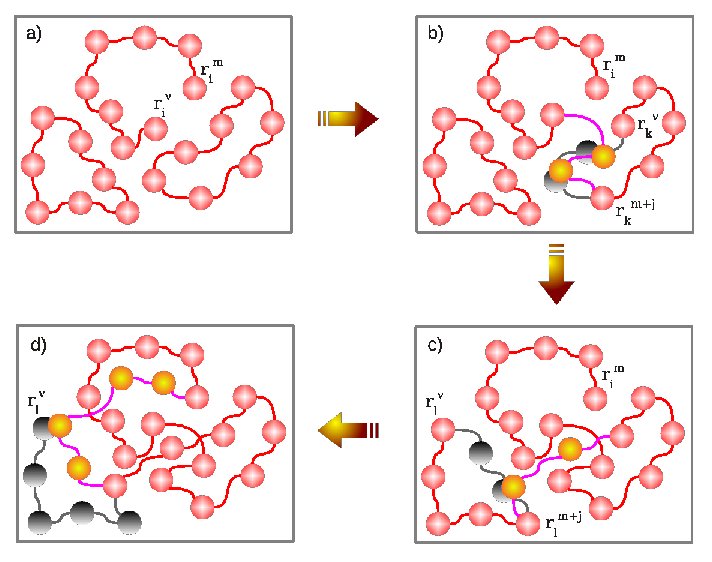}
 \caption{\label{wodyn} 
The sampling of permutations in the Worm algorithm: 
from an off--diagonal configuration (a), the first swap generates an open polymer of length $2\beta$ (b). 
Another successful swap results in a configuration with a polymer of length $3\beta$ (c); at this point, 
a successful close move yield a diagonal configuration with a ring polymer of length $3\beta$ (d).
Grey beads and lines represent the removed segment of the polymer.
}
\end{center}
\end{figure}

\paragraph{Open/Close}
These moves allow to switch from the $Z$ sector to the $G$ sector and vice--versa. The Open move creates a worm from a diagonal configuration while the Close move closes a worm and gives a diagonal
configuration as a result. In order to maintain the detailed balance of the sampling, these moves are coupled meaning that in a MCS there is always one attempt to Open/Close and whether to Open or to Close
 is decided with a random number: the Open and the Close moves should always be equally probable, so, for instance, if a random number uniformly distributed between 0 and 1 is lesser than 0.5 in that MCS an Open
 move will be tried, otherwise a Close will be attempted, no matter whether the system is in $G$ sector or in $Z$ sector.
   \begin{figure}[h]
 \begin{center}
 \includegraphics*[width=11cm]{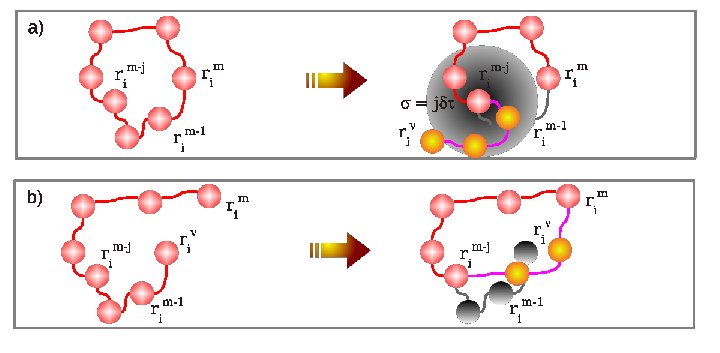}
 \caption{\label{opclose} 
Panel a) The ``open'' move in the worm algorithm. The gray area on the left side represents the Gaussian probability 
distribution from which the coordinates of $r_i^{\nu}$ are sampled.
Panel b) The ``close'' move in the worm algorithm. Grey beads and lines represent the removed segment of the polymer. 
}
\end{center}
\end{figure} 
The Open move is as follows
\begin{itemize}
\item If there is already an open polymer, reject the move. Otherwise select a random integer number $i$ between 1 and $N$, a random integer number $j$ between 1 and $s$ and a random integer number $m$ between 1 and $M$.
\item Create a worm in the $i$--th polymer at the $m$--th bead, the added bead  $\vec{r}_{i}^{\:\nu}$ is sampled from the probability distribution of a free particle propagator with time--step $j \delta\tau$
\begin{eqnarray}
G_0\left(\vec{r}_{i}^{\:m-j},\vec{r}^{\:*},j\delta\tau\right) = \frac{1}{\left(4\pi\lambda j\delta\tau\right)^{-d/2}}e^{\left[-\frac{\left|\vec{r}^{\:*}-\vec{r}_{i}^{\:m-j}\right|^{2}}{4\lambda j\delta\tau}\right]}
\end{eqnarray}
\item From the bead $\vec{r}_{i}^{\:m-j}$ remove the beads $\vec{r}_{i}^{\:m-j+1}$,$\vec{r}_{i}^{\:m-j+2}$,...,$\vec{r}_{i}^{\:m-1}$ and build a discrete free--particle path to the newly created 
bead $\vec{r}_{i}^{\:\nu}$.
\end{itemize}
The probability to accept this move is
\begin{eqnarray}
p_{o} = \min\left\lbrace1,P_o\right\rbrace \hspace{6cm}\nonumber \\
P_o = \frac{NC\sum_{l=m-j}^{m}\left[U\left(R_l^{new},R_{l+1}^{new}\right)-U\left(R_{l}^{old}-R_{l+1}^{old}\right)\right]}{VG_0\left(\vec{r}_{i}^{\:m-j},\vec{r}_{\hat{P}i}^{\:m+1},j\delta\tau\right)}
\quad .
\end{eqnarray}
The Close move is as follows
\begin{itemize}
\item If the configuration is diagonal, reject the move. Otherwise there is a worm, say in the polymer $i$ at the bead $m$. Select a random integer number $j$ between 1 and $s$ 
\item Remove the worm extremity $\vec{r}_{i}^{\:\nu}$ that is linked by a kinetic term only to its previous bead $\vec{r}_{i}^{\:m-1}$
\item Replace the beads between $\vec{r}_{i}^{\:m-j}$ and $\vec{r}_{i}^{\:m}$, with a free particle path 
\end{itemize}
The probability to accept the move is
\begin{eqnarray}
p_{c} = \min\left\lbrace1,P_c\right\rbrace \hspace{6cm}\nonumber \\
P_c = \frac{V}{NC}\sum_{l=m-j}^{m}\left[U\left(R_l^{new},R_{l+1}^{new}\right)-U\left(R_{l}^{old}-R_{l+1}^{old}\right)\right] \cdot \nonumber \\ 
G_0\left(\vec{r}_{i}^{\:m-j},\vec{r}_{\hat{P}i}^{\:m+1},j\delta\tau\right)
\quad .
\end{eqnarray}
These moves can be optimized in term of computing efficiency with an automatic rejection whenever the distance between $\vec{r}_{i}^{\:m-j}$ and $\vec{r}_{\hat{P}i}^{\:m}$ is such that the quantity
\begin{eqnarray}
\frac{\left|\vec{r}_{i}^{\:m-j}-\vec{r}_{\hat{P}i}^{\:m}\right|^{2}}{4\lambda j\delta\tau}
\end{eqnarray}
is larger than some arbitrary quantity of order unity. Our choice was to set it equal to 4. This avoid very small acceptation rates when  the worm extremities are far away. In order to  maintain the detailed
balance, this kinetic test must be applied both to the Open and the Close moves.
\paragraph{Swap}
This move is attempted only in the off--diagonal sector and implements the sampling of permutations. Consider a configuration in $G$ sector with a worm in the $i$--th polymer at bead $m$.
\begin{itemize}
\item Select a random integer $j$ between 1 and $s$
\item Select a bead $\vec{r}_{i_{k}}^{\:m+j+1}$ with probability
\begin{eqnarray}
P_{i_k} = G_0\left(\vec{r}_{i}^{\:\nu},\vec{r}_{i_{k}}^{\:m+j+1},j\delta\tau\right)/\Sigma_T \\
\Sigma_T = \sum_{n=1}^{N}G_0\left(\vec{r}_{i}^{\:\nu},\vec{r}_{n}^{\:m+j+1},j\delta\tau\right)
\quad .
\end{eqnarray}
\item Evaluate the quantity
\begin{eqnarray}
\Sigma_K = \sum_{n=1}^{N}G_0\left(\vec{r}_{i}^{\:m+1},\vec{r}_{n}^{\:m+j+1},j\delta\tau\right)
\quad .
\end{eqnarray}
\item Consider the bead $\vec{r}_{i_{k}}^{\:m+j+1}$ and insert a new bead $\vec{r}_{\hat{P}i_{k}}^{\:\nu}=\vec{r}_{\hat{P}i_{k}}^{\: m}$ that is connected by a kinetic term only to its previous bead.
\item Replace $j$ beads of the polymer $i_k$, namely $\vec{r}_{i_{k}}^{\:m+1}$, $\vec{r}_{i_{k}}^{\:m+2}$, ..., $\vec{r}_{i_{k}}^{\:m+j}$  with a Brownian bridge starting
from $\vec{r}_{i}^{\:\nu}$ and ending at $\vec{r}_{i_{k}}^{\:m+j+1}$. With this operation, the bead $\vec{r}_{i}^{\:\nu}$ is no longer a worm end: the Brownian bridge 
swaps the polymer $i$ with the polymer $i_k$ and the new worm extremities become $\vec{r}_{i_{k}}^{\:\nu}$ and $\vec{r}_{i}^{\:m}$.
\end{itemize}
The probability to accept a swap move is
\begin{eqnarray}
P_{sw} = \min\left\lbrace 1, \frac{\Sigma_T}{\Sigma_K}e^{\sum_{l=m}^{m+j}\left[U\left(R_l^{new},R_{l+1}^{new}\right)-U\left(R_l^{old},R_{l+1}^{old}\right)\right]}\right\rbrace
\quad .
\end{eqnarray}
A remark is necessary here: the interpolymer correlations in the time sector $(m,m+1)$, namely $U(R_m,R_{m+1})$ must take into account the worm extremities 
$\vec{r}_{\hat{P}i}^{\:\nu}$ and $\vec{r}_{i}^{\:m}$. The worm extremities contribute to the interpolymer correlations like the other beads but with a weight factor 
of $0.5$. Using a symmetrized form for the density matrix, such as Eq.~\eqref{propPA} automatically gives the correct weights:
\begin{eqnarray}
U(R_m,R_{m+1})_{PA} =
e^{-\frac{\delta\tau}{2}\sum_{h<k}^{N}v(r_{hk}^{\:m})}e^{\frac{1}{4\lambda\delta\tau}\sum_{h}^{N}\left(\vec{r}_h^{\:m}-\vec{r}_{h}^{\:m+1}\right)^2}e^{-\frac{\delta\tau}{2}\sum_{h<k}v(r_{hk}^{\:m+1})}
\nonumber \\
U(R_{m-1},R_{m})_{PA} =
e^{-\frac{\delta\tau}{2}\sum_{h<k}^{N}v(r_{hk}^{\:m-1})}e^{\frac{1}{4\lambda\delta\tau}\sum_{h}^{N}\left(\vec{r}_h^{\:m-1}-\vec{\tilde{r}}_{h}^{\:m}\right)^2}e^{-\frac{\delta\tau}{2}\sum_{h<k}v(\tilde{r}_{hk}^{\:m})}
\nonumber \\ \label{umworm}
\end{eqnarray}
where 
\begin{eqnarray}
\vec{\tilde{r}}_{h}^{\:m} = \begin{cases} \vec{r}_{h}^{\:m}, & \mbox{if } h \neq \hat{P}i \\ \vec{r}_{h}^{\:\nu}, & \mbox{if } h = \hat{P}i \end{cases}
\end{eqnarray}
and
\begin{eqnarray}
\tilde{r}_{hk}^{\:m} = \begin{cases} \left|\vec{r}_{h}^{\:m}-\vec{r}_{k}^{\:m}\right|, & \mbox{if } h \neq \hat{P}i \\ \left|\vec{r}_{h}^{\:\nu} - \vec{r}_{k}^{\:m}\right|, & \mbox{if } h = \hat{P}i \end{cases}
\quad .
\end{eqnarray}
The swap moves are very efficient in the permutation sampling for two reasons: first, a permutation that involves several polymers is obtained with a certain number of swaps that are more likely to be 
accepted; second, the Worm itself, being an open polymer, has a better probability to avoid overlaps during swap moves.
The mechanism used by the Worm algorithm to sample the permutations space is depicted in Fig.~\ref{wodyn}; 
the basic idea is that two topologically different diagonal configurations are 
connected by at least three successful Worm moves: the open move generates an off-diagonal configuration; 
the swap move (or a series of swap moves) samples the permutations space and, finally, the close move 
returns the system to a diagonal configuration. 

\paragraph{Brownian bridge extension}
The Brownian bridge in a worm computation remains unchanged; however, if the attempted reconstruction involves the worm ends, i.e. starts from a bead $j_1$ that is before the imaginary--time position $m$ 
of the worm and ends to a bead $j_2 > m$, the reconstruction of the segment is split and the position of the worm extremities is updated:
\begin{itemize}
\item The first worm extremity, $\vec{r}_{i}^{\:\nu}$ is updated from the distribution probability
\begin{eqnarray}
G_0\left(\vec{r}_{i}^{\:j_1},\vec{r}^{\:*},(m-j_1)\delta\tau\right) = \frac{1}{\left(4\pi\lambda (m-j_1)\delta\tau\right)^{-d/2}}e^{\left[-\frac{\left(\vec{r}^{\:*}-\vec{r}_{i}^{\:j_1}\right)^{2}}{4\lambda (m-j_1)\delta\tau}\right]}
\quad .
\end{eqnarray}
\item The beads $\vec{r}_{i}^{\:j_1+1}$,$\vec{r}_{i}^{\:j_1+2}$,...,$\vec{r}_{i}^{\:m-1}$ are updated with a Brownian bridge that starts from $\vec{r}_{i}^{\:j_1}$ and ends to the freshly updated bead 
$\vec{r}_{i}^{\:\nu}$
\item The second worm extremity, $\vec{r}_{i}^{\:m}$ is updated from the distribution probability
\begin{eqnarray}
G_0\left(\vec{r}^{\:*},\vec{r}_{i}^{\:j_2},(j_2-m)\delta\tau\right) = \frac{1}{\left(4\pi\lambda (j_2-m)\delta\tau\right)^{-d/2}}e^{\left[-\frac{\left(\vec{r}_{i}^{\:j_2}-\vec{r}^{\:*}\right)^{2}}{4\lambda (j_2-m)\delta\tau}\right]}
\quad .
\end{eqnarray}
\item The beads $\vec{r}_{i}^{\:m+1}$,$\vec{r}_{i}^{\:m+2}$,...,$\vec{r}_{i}^{\:j_2-1}$ are updated with a Brownian bridge that starts from the second worm extremity $\vec{r}_{i}^{\:m}$ and ends at 
$\vec{r}_{i}^{\:\nu}$
\end{itemize}
The new position of every bead is sampled here with a free particle propagator, as consequence the probability to accept this move is the same of that of a Brownian bridge between $j_1$ and $j_2$ 
that is expressed in Eq.~\eqref{pbb} with $U(R_{m-1},m)$ and $U(R_m,R_{m+1})$ defined as in Eq.~\eqref{umworm}.

\paragraph{Translation move extension}
The translation move for polymers with a worm has been extended as follows. 
In the PIGS case, a polymer $i$ with Worm extremities $\vec{r}_{i_{old}}^{\:\nu}$ and $\vec{r}_{i_{old}}^{\:m}$ is defined by a set of coordinates 
$S=(\vec{r}_{i_{old}}^{\:1},...,\vec{r}_{i_{old}}^{\:\nu},\vec{r}_{i_{old}}^{\:m},...,\vec{r}_{i_{old}}^{\:M})$.
Here, for a polymer without worm, we define $\vec{r}^{\:\nu}=\vec{r}^{\:m}$.
The translation of this polymer is parameterized by two displacement vectors $\vec{d}_1,\vec{d}_2$; from $S$, the move will generate a new polymer 
$S^{new} = (\vec{r}_{i_{new}}^{\:1}=\vec{r}_{i_{old}}^{\:1}+\vec{d}_1,...,\vec{r}_{i_{old}}^{\:\nu}+\vec{d}_1,\vec{r}_{i_{old}}^{\:m}+\vec{d}_2,...,\vec{r}_{i_{old}}^{\:M}+\vec{d}_2)$. 
The probability to accept the move becomes
  \begin{eqnarray}
  a\left(\left\lbrace R\right\rbrace_{new}\right) = \min \left(1,P_{tr}\right) \nonumber \\
  P_{tr}= A_1\cdot A_2
  \end{eqnarray}
where $A_1$ and $A_2$ are the probabilities to accept the translations of the two half polymers $S_1=(\vec{r}_{i_{old}}^{\:1},...,\vec{r}_{i_{old}}^{\:\nu})$ and 
$S_2=(\vec{r}_{i_{old}}^{\:m},...,\vec{r}_{i_{old}}^{\:M})$; namely:

\begin{eqnarray}
  V(R_{m_{new(old)}}) := \sum_{k\neq i} v(|\vec{r}_{i_{new(old)}}^{\:m}-\vec{r}_{k}^{\:m}|)\\
 A_1=\frac{\Psi_T(R_{1_{new}})e^{-\frac{\delta\tau}{2}V(R_{1_{new}})}e^{-\delta\tau V(R_{2_{new}})}...e^{-\frac{\delta\tau}{2}V(R_{{\nu}_{new}})}}
 {\Psi_T(R_{1_{old}})e^{-\frac{\delta\tau}{2}V(R_{1_{old}})}e^{-\delta\tau V(R_{2_{old}})}...e^{-\frac{\delta\tau}{2}V(R_{{\nu}_{old}})}}\\
 A_2 =\frac{e^{-\frac{\delta\tau}{2}V(R_{m_{new}})}e^{-\delta\tau V(R_{(m +1)_{new}})}...e^{-\frac{\delta\tau}{2}V(R_{M_{new}})}\Psi_T(R_{M_{new}})}
 {e^{-\frac{\delta\tau}{2}V(R_{m_{old}})}e^{-\delta\tau V(R_{(m+1)_{old}})}...e^{-\frac{\delta\tau}{2}V(R_{M_{old}})}\Psi_T(R_{M_{old}})} 
 \quad .
  \end{eqnarray}

In the PIMC case, a ring polymer $i$ with Worm extremities $\vec{r}_{i_{old}}^{\:\nu}$ and $\vec{r}_{i_{old}}^{\:m}$ becomes an open polymer. The translation in this case 
has only one parameter $\vec{d}$, so that if the polymer is defined by 
$S=(\vec{r}_{i_{old}}^{\:1},...,\vec{r}_{i_{old}}^{\:\nu},\vec{r}_{i_{old}}^{\:m},...,\vec{r}_{i_{old}}^{\:M}$, the new polymer will be 
$S^{new}=(\vec{r}_{i_{old}}^{\:1}+\vec{d},...,\vec{r}_{i_{old}}^{\:\nu}+\vec{d},\vec{r}_{i_{old}}^{\:m}+\vec{d},...,\vec{r}_{i_{old}}^{\:M} +\vec{d})$
The probability to accept this move is
  \begin{eqnarray}\label{pimctranslp}
  a\left(\left\lbrace R\right\rbrace_{new}\right) = \min \left(1,P_{tr}\right) \nonumber \\
  P_{tr}= \frac{e^{-\delta\tau V(R_{1_{new}})}...e^{-\frac{\delta\tau}{2}V(R_{\nu_{new}})}e^{-\frac{\delta\tau}{2}V(R_{m_{new}})}e^{-\delta\tau V(R_{(m+1)_{new}})}...e^{-\delta\tau V(R_{M_{new}})}}
 {e^{-\delta\tau V(R_{1_{old}})}...e^{-\frac{\delta\tau}{2}V(R_{\nu_{old}})}e^{-\frac{\delta\tau}{2}V(R_{m_{old}})}e^{-\delta\tau V(R_{(m+1)_{old}})}...e^{-\delta\tau V(R_{M_{old}})}}
  \quad .
  \end{eqnarray}

Particular care must be taken when applying the translation move to a PIMC configuration that has permutations. In this case, all the polymers that contribute to a permutation loop 
are translated by the same displacement vector $\vec{d}$; the probability to accept such a move is of the form of Eq.~\eqref{pimctranslp} but the correlation $V(R_m)$ 
must take into account not only the $i$--th polymer but also the other polymers involved in the translation: let the polymers in the permutation loop be $(i_1,i_2,...i_H)$ and 
the remaining polymers the elements of the set $W_{rem}$; then, considering that the translation does not change the correlations between these polymers, 
\begin{eqnarray}
  V(R_{m}) := \sum_{k \in W_{rem}} \sum_{l=1}^{H} v(|\vec{r}_{i_l}^{\:m}-\vec{r}_{k}^{\:m}|)
\quad .
\end{eqnarray}

\paragraph{Normalization of the One Body Density Matrix}

The Worm algorithm provides also a way to compute the correct normalization of the OBDM. We focus here on the PIMC case, 
the PIGS case is analogous; 
the OBDM is defined by Eq.~\eqref{obdmpimc}. This probability density is sampled in the $G$ sector. The normalization of 
Eq.~\eqref{obdmpimc} is the partition function $\mathcal{Z}$; $\mathcal{Z}$ corresponds to Eq.~\eqref{obdmpimc} with 
$\vec{r}=\vec{r}^{\:'}$; in the case of an homogeneous system, this is equivalent to the sampling of 
$\rho_1(r=|\vec{r}-\vec{r}^{\:'}|=0)$; within the Worm algorithm this quantity is related to the probability to switch 
from $Z$ to $G$. For an homogeneous system, the OBDM $\rho_1(r)$ is the histogram of the 
distance between the two worm extremities, $\vec{r}_{i}^{\:\nu}$ and $\vec{r}_{i}^{\:m}$, normalized as follows:
\begin{eqnarray}
\rho_1(r) = \frac{\left\langle\delta(r-\left|\vec{r}_{i}^{\:\nu}-\vec{r}_{i}^{\:m}\right|)\right\rangle}{V_{shell}(r)zC\rho}
\end{eqnarray}
where $C$ is the Worm parameter previously introduced, $\rho$ is the density of the system, $z$ is the number of 
Monte Carlo steps in the $Z$ sector. The Monte Carlo average $\langle ...\rangle$ in this context 
is the histogram of the distance $\left|\vec{r}_{i}^{\:\nu}-\vec{r}_{i}^{\:m}\right|$, each bin of the histogram 
has a width $dr$ and is divided by the volume of the spheric shell $V_{shell}(r)=V(r+dr)-V(r)$.

The introduction of a repulsive factor such as \eqref{frep1} does not preserve the Worm normalization. 
This happens because the repulsive factor interferes with the probability to switch from the $Z$ sector to the $G$ 
sector and {\it vice versa}. A workaround is to use a repulsive factor $f_w(r)$ that goes to unity for $r=0$: with this 
repulsive factor, in fact, when $\vec{r}=\vec{r}^{\:'}$ one obtains again the partition function $\mathcal{Z}$. 
A straightforward adaptation of the repulsive factor \eqref{frep1} that has been used in this work is the following
  \begin{eqnarray}\label{frep2}
  f_w(r)=\frac{1+A+C}{1+Ae^{-Br^2}+Ce^{-Dr}} \:\:.
  \end{eqnarray}

%% file: ch-appendicies/estimators.tex
\chapter{Estimators\label{ch:estimators}}

Here follows the derivation of some estimators with a fourth-order approximation of the propagator. \\
We consider the Pair Suzuki (PS) approximation:

\begin{eqnarray} \label{psuz1}
G\left(R_{m},R_{m+1},\tau\right)=\frac{1}{\left(4\pi\lambda\tau\right)^{Nd}}\int\:\prod_{i=1}^{N}d\vec{r}^{*}_{i}\exp^{-\frac{1}{4\lambda\tau}\sum_{i=1}^{N}\left[\left(\vec{r^{m}_{i}}-\vec{r^{*}_{i}}\right)^{2}+\left(\vec{r^{*}_{i}}-\vec{r^{m+1}_{i}}\right)^{2}\right]} \nonumber \\
\exp^{-\frac{\tau}{3}\sum_{i<j}^{N}\left[v_{e}\left(r^{m}_{ij}\right) + 4v_{c}\left(r^{*}_{ij}\right) + v_{e}\left(r^{m+1}_{ij}\right)\right]}
\end{eqnarray}
where $r_{ij}^{m}=|\vec{r}_{i}^{m}-\vec{r}_{j}^{m}|$ is the distance between the $i$-th bead and the $j$-th bead at an imaginary-time defined by the index $m$; $N$ is the particles number, $d$ is the dimensionality of the system, $2M$ is the effective\footnote{With the parameter $\alpha$ set to zero, the odd timeslices are those which describe the system, the even timeslices are the fictious beads used to express the fourth-order approximation of the Green's function} beads number, $\lambda=\frac{\hbar^{2}}{2m}$, $\tau=\frac{\beta}{2M}$ and
\begin{eqnarray}
v_{e}\left(r\right)= v\left(r\right) + \frac{2}{3}\alpha\tau^{2}\lambda\left(\frac{\partial v\left(r\right)}{\partial r}\right)^{2} \\
v_{c}\left(r\right)= v\left(r\right) + \frac{1}{3}\left(1-\alpha\right)\tau^{2}\lambda\left(\frac{\partial v\left(r\right)}{\partial r}\right)^{2}  
\end{eqnarray}
We remark that (\ref{psuz1}) is the Green's function that involves two adjacent \textit{real} timesliecs, thus the integration variables $\left\lbrace\vec{r}_{i}^{*}\right\rbrace$ represent the fictitious bead required by the PS approximation.
\subsection{Total Energy}
The thermodynamic estimator for the total energy is defined as follows
\begin{eqnarray}
E=-\frac{1}{2ZM}\frac{\partial Z}{\partial \tau}
\end{eqnarray}
where $Z$ is the partition function,
\begin{eqnarray} \label{partition}
Z=tr\left\lbrace\hat{\rho}\right\rbrace=\int dR^{1}...dR^{M} \:\prod_{m=1}^{M-1}G\left(R^{m},R^{m+1},\tau\right)
\end{eqnarray}
The imaginary-time derivative applied to the productory in (\ref{partition}) yelds a sum of M terms that may be viewed as the energy evaluated at an imaginary-time sector. The $\tau$ derivative applied to (\ref{psuz1}) gives three terms: the first comes from the normalization of the kinetic part, the second from the gaussian which expresses the kinetic propagator and the last one from the term involving the inter-polymer correlations. After some straightforward algebra, one finds
\begin{eqnarray} \label{totalen}
E^{m}=\left\langle\frac{Nd}{2\tau} - \frac{1}{4\lambda\tau^{2}}\sum_{i=1}^{N}\frac{\left(\vec{r}_{i}^{m}-\vec{r}_{i}^{*}\right)^2 +\left(\vec{r}_{i}^{*}-\vec{r}_{i}^{m+1}\right)^2 }{2} + \right.\nonumber \\
\left.\frac{\partial}{\partial \tau}\frac{\sum_{i<j}^{N} \frac{\tau}{3}\left[v_{e}\left(r_{ij}^{m}\right)+4v_{c}\left(r_{ij}^{*}\right)+v_{e}\left(r_{ij}^{m+1}\right)\right]}{2}\right\rangle
\end{eqnarray}
\subsection{Kinetic Energy}
The thermodynamic estimator for the kinetic energy is defined by the following formula
\begin{eqnarray}
K = \frac{m}{\beta Z}\frac{\partial Z}{\partial m} = -\frac{\lambda}{\beta Z}\frac{\partial Z}{\partial\lambda}
\end{eqnarray}
The arguments of the previous paragraph apply here too, resulting in the following expression
\begin{eqnarray}
K^{m}=\left\langle\frac{Nd}{2\tau} - \frac{1}{4\lambda\tau^{2}}\sum_{i=1}^{N}\frac{\left(\vec{r}_{i}^{m}-\vec{r}_{i}^{*}\right)^2 +\left(\vec{r}_{i}^{*}-\vec{r}_{i}^{m+1}\right)^2 }{2} + \right. \nonumber \\ 
\left.\frac{\lambda}{\tau}\frac{\partial}{\partial \lambda}\frac{\sum_{i<j}^{N} \frac{\tau}{3}\left[v_{e}\left(r_{ij}^{m}\right)+4v_{c}\left(r_{ij}^{*}\right)+v_{e}\left(r_{ij}^{m+1}\right)\right]}{2}\right\rangle
\end{eqnarray}

\subsection{Pressure}
The thermodynamic estimator for the pressure is obtained from a volume derivative of the partition function:
\begin{eqnarray}
P\left(N,V,\beta\right)=\frac{1}{\beta Z}\frac{\partial Z\left(N,V,\beta\right)}{\partial V}
\end{eqnarray}
In order to compute this volume derivative from (\ref{partition}) one has to perform the following change of variables
\begin{eqnarray}
\vec{r} =V^{\frac{1}{d}}\tilde{\vec{r}} \,\, \Rightarrow \,\, d\vec{r}=dxdydz=Vd\tilde{\vec{r}}
\end{eqnarray}
The estimator is composed of three terms, the first arises from a factor $V^{2NM}$ coming from the Jacobian transformation of the differentials {$d\vec{r}_{i}^{m}$}, the second is the derivative applied to the kinetic factor of the propagator and the last one comes from the inter-polymer correlations part. Shifting back to the former integration variables, one obtains
\begin{eqnarray}
P^{m}=\left\langle \frac{\rho}{\tau} - \frac{1}{2\lambda\tau^{2} Vd}\sum_{i=1}^{N}\frac{\left[\left(\vec{r}_{i}^{m}-\vec{r}_{i}^{*}\right)^{2}+\left(\vec{r}_{i}^{*}-\vec{r}_{i}^{m+1}\right)^{2}\right]}{2} \right. \nonumber \\
\left. -\frac{1}{6Vd}\sum_{i<j}r_{ij}\left[\left.\frac{\partial v_{e}}{\partial r}\right|_{r_{ij}^{m}}+\left.\frac{\partial 4v_{c}}{\partial r}\right|_{r_{ij}^{*}}+\left.\frac{\partial v_{e}}{\partial r}\right|_{r_{ij}^{m+1}}\right]\right\rangle
\end{eqnarray}
\subsection{T=0 limit}
Even though the previously introduced estimators are derived from a finite-temperature background, it can be shown that they are valid also in the zero temperature limit. Let's show this for the Hamiltonian operator
\begin{eqnarray}
\left\langle \Psi_{0}\hat{H}\Psi_{0}\right\rangle =- \lim_{\beta\rightarrow \infty} \frac{\partial}{\partial\beta} \log \int dR^{1} dR^{2M} \Psi_{T}\left(R^{1}\right)G\left(R^{1},R^{2M},\beta\right)\Psi_{T}\left(R^{2M}\right) = \nonumber \\
= -\lim_{\beta\rightarrow\infty}\frac{1}{\mathcal{N}}\int dR^{1} dR^{2M} \Psi_{T}\left(R^{1}\right)\frac{\partial G\left(R^{1},R^{2M},\beta\right)}{\partial\beta}\Psi_{T}\left(R^{2M}\right)
\end{eqnarray}
Because of the formal similarities between PIGS and PIMC, chosen a large enough imaginary-time $\beta$, this expression evaluated at the central timeslices expresses a zero-temperature quantum average.
\subsection{Virial Energy Estimator}
Eq. (\ref{totalen}), such as any estimator involving the Kinetic Energy, contains an higly fluctuating term which comes from the derivative applied to the kinetic part of the propagator. This results in a variance of the averages that increases with the beads number $M$. The virial estimator provides a way to get rid of these fluctuations. Let's derive it for the total energy estimator. Consider the quantity
\begin{eqnarray} \label{letot}
E_{1,L+1}=\left\langle \frac{NLd}{2\tau} - \frac{M}{2}\alpha + \frac{\partial \tilde{U}}{\partial\tau} \right\rangle
\end{eqnarray} 
where
\begin{eqnarray}
\alpha = \sum_{m=1}^{L} \frac{\sum_{i=1}^{N}\left[\left(\vec{r}_{i}^{m}-\vec{r}_{i}^{*}\right)^{2}+\left(\vec{r}_{i}^{*}-\vec{r}_{i}^{m+1}\right)^{2}\right]}{4\lambda\tau^{2} M} \\
\tilde{U} = \sum_{m=1}^{L}\sum_{i<j}^{N}\frac{\tau}{3}\left[v_{e}\left(r_{ij}^{m}\right) + 4v_{c}\left(r_{ij}^{*}\right) + v_{e}\left(r_{ij}^{m+1}\right)\right]
\end{eqnarray}
The quantity $E_{1,L+1}$ represents $L$ times the total energy of the system, where $L$ is a parameter arbitrarily chosen between 1 and $M$. For simplicity, let's rewrite the definitions of $\alpha$ and $U$ in a more treatable way:
\begin{eqnarray}
\alpha = \sum_{m=1}^{2L} \frac{\sum_{i=1}^{N}\left(\vec{r}_{i}^{m}-\vec{r}_{i}^{*}\right)^{2}}{4\lambda\tau^{2} M} \\
\tilde{U} = \sum_{m=1}^{2L}U\left(R^{m},R^{m+1}\right)
\end{eqnarray}
where now the index $m$ denotes every beads, both physical and fictitious, and
\begin{eqnarray}
U^{*}\left(r^{m},r^{m+1}\right)=v_{e}\left(r_{ij}^{m}\right)+2v_{c}\left(r_{ij}^{m+1}\right) \,\,\, m \quad \mbox{odd} \\
U^{*}\left(r^{m},r^{m+1}\right)=2v_{c}\left(r_{ij}^{m}\right)+v_{e}\left(r_{ij}^{m+1}\right) \,\,\, m \quad \mbox{even} \\
U\left(R^{m},R^{m+1}\right)=\sum_{i<j}^{N}U^{*}\left(r_{ij}^{m},r_{ij}^{m+1}\right)
\end{eqnarray}
Now define $dR^{m}=\prod_{i=1}^{N}d\vec{r}_{i}^{m}$, $\left(R^{m}-R^{n}\right)=\sum_{i=1}^{N}\left(\vec{r}_{i}^{m}-\vec{r}_{i}^{n}\right)$, $\frac{\partial}{\partial R^{m}} = \sum_{i=1}^{N}\frac{\partial}{\partial \vec{r}_{i}^{m}}$ and consider the quantity
\begin{eqnarray} \label{gfirst}
G=\frac{\int dR^{2}...dR^{2L}\sum_{m=2}^{2L}\left(R^{m}-R^{1}\right)\left(-\frac{1}{\beta}\right)\frac{\partial}{\partial R^{m}}\exp ^{-\beta g}}{\int dR^{2}...dR^{2L}\sum_{m=2}^{2L}\exp ^{-\beta g}} \\
\end{eqnarray}
with $g = \alpha + \frac{\tilde{U}}{\beta}$. If we make a change of integration variables, namely $\delta^{m}=R^{m}-R^{m-1}$, eq. (\ref{gfirst}) becomes
\begin{eqnarray} \label{gsec}
G=\frac{\int d\delta^{2}...d\delta^{2L}\sum_{m=2}^{2L}\delta^{m}\left(-\frac{1}{\beta}\right)\frac{\partial}{\partial \delta^{m}}\exp ^{-\beta g}}{\int dR^{2}...dR^{2L}\sum_{m=2}^{2L}\exp ^{-\beta g}} 
\end{eqnarray}
The integral at the numerator of (\ref{gsec}) can be computed by parts. The surface term vanishes if $\tau\lambda << V^{\frac{2}{d}}$, thus
\begin{eqnarray} \label{gdef1}
G = \frac{1}{\beta}\sum_{m=2}^{2L}\left\langle\frac{\partial \delta^{m}}{\partial \delta^{m}}\right\rangle = \frac{N\left(2L-1\right)d}{\beta}
\end{eqnarray}
The quantity in the RHS of (\ref{gfirst}) can be expressed by explicitely computing the derivative over the positions
\begin{eqnarray} \label{gdef2}
G= \left\langle \sum_{m=2}^{2L}\left(R^{m}-R^{1}\right)\frac{\partial \alpha}{\partial R^{m}}\right\rangle + 
\left\langle\frac{1}{\beta}\sum_{m=2}{2L}\left(R^{m}-R^{1}\right)\frac{\partial u}{\partial R^{m}}\right\rangle
\end{eqnarray}
The first term on the RHS, after some algebra, becomes
\begin{eqnarray} \label{gsum1}
\sum_{m=2}^{2L}\left(R^{m}-R^{1}\right)\frac{\partial \alpha}{\partial R^{m}}= 2\alpha + \frac{1}{4\lambda\tau^{2}M}\left(R^{2L}-R^{2L+1}\right)\left(R^{2L+1}-R^{1}\right)
\end{eqnarray}
Equating both (\ref{gdef1}) and (\ref{gdef2}) and using (\ref{gsum1}), the quantity $\alpha$ may be re-expressed as
\begin{eqnarray} \label{alpha1}
\alpha = \frac{N\left(2L-1\right)d}{2\beta} - \frac{1}{8\lambda\tau^{2}M}\left(R^{2L}-R^{2L+1}\right)\left(R^{2L+1}-R^{1}\right) \nonumber \\ - \frac{1}{2\beta}\sum_{m=2}^{2L}\left(R^{m}-R^{1}\right)\frac{\partial \tilde{U}}{\partial R^{m}}
\end{eqnarray}
Substituting $\alpha$ in eq. (\ref{letot}) we finally obtain the virial estimator for the total energy per particle:
\begin{eqnarray}
E_{virial}=\left\langle \frac{d}{2\tau} + \frac{1}{4\lambda\tau^{2}N}\sum_{i=1}^{N}\left(\vec{r}_{i}^{2L}-\vec{r}_{i}^{2L+1}\right)\left(\vec{r}_{i}^{2L+1}-\vec{r}_{i}^{1}\right) + \right. \nonumber \\
\left. + \frac{1}{2\tau N}\sum_{m=2}^{2L}\sum_{i<j1}^{N}\left(\vec{r}_{i}^{m}-\vec{r}_{1}^{m}\right)\cdot\frac{\partial U^{*}\left(r_{ij}^{m},r_{ij}^{m+1}\right)}{\partial \vec{r}_{i}^{m}} + \frac{2}{N}\sum_{m=1}^{2L}\frac{\partial U\left(R^{m},R^{m+1}\right)}{\partial \tau} \right\rangle
\end{eqnarray}
This estimator may be used also in the zero temperature limit if one performs the following substitutions: $R^{1} \rightarrow R^{\Gamma}$, $L \rightarrow \tilde{L}$ ,$\sum_{m=1}^{2L} \rightarrow \sum_{m=\Gamma}^{\Gamma+2\tilde{L}}$ and $\sum_{m=2}^{2L} \rightarrow \sum_{m=\Gamma +1}^{\Gamma+2\tilde{L}}$, where $\Gamma$ represents the index of the first time-sector that can be considered a ground-state description of the system and $\tilde{L}$ is an arbitrary number between 1 and the number of physical timeslices available for the evaluation of ground-state expectation values.

%% file: ch-appendicies/higord.tex
 \def\onlinecite{\cite}

 \chapter{Higher order approximations for the density matrix}\label{sechighorder}
  In this appendix we show approximations for the small imaginary--time density matrix that go beyond the 
  Primitive Approximation (PA) introduced in Eq. (\ref{pasimm}).
  We have already shown the ``Pair" Suzuki (PS) approximation in Sec. \ref{secpigs},  
  we will show now the Pair Product approximation (PPA) and the 
  Multi Product Expansion (MPE). The MPE and the PS, as well as the Primitive Approximation, have the advantage to be analytic and thus estimators can be
   derived exactly; the PPA on the other hand is numerical and only a restricted set of estimators, such as 
   those diagonal on the coordinate representation and the one body density matrix, can be simply derived. 
   The PPA, however, requires fewer imaginary--time projection than the other two and this feature could be 
   useful in some contexts.\\
   As mentioned before, given the Hamiltonian $\hat{H}=\hat{T}+\hat{V}$, one has to use a small imaginary--time 
   approximation of the propagator $e^{-\delta\tau\hat{H}}$ in order to obtain an analytic expression for 
   $G\left(R,R',\delta\tau\right)=\left\langle R\left|e^{-\delta\tau\hat{H}}\right|R'\right\rangle$. The simplest 
   approximation is the PA
   \begin{eqnarray} \label{propPA}
   G_2\left(R_i,R_j,\delta\tau\right) = e^{-\frac{\delta\tau}{2}\hat{V}_i}e^{-\delta\tau\hat{T}}e^{-\frac{\delta\tau}{2}\hat{V}_j}
   \end{eqnarray}
   which is correct up to second--order in $\delta\tau$; this approximation is obtained by ignoring the commutator $\left[\hat{T},\hat{V}\right]$ when factorizing the Hamiltonian.
   The effective potential  $U\left(R_m,R_{m+1},\delta\tau\right)$ for the beads represented with the PA is
   \begin{eqnarray}
   U\left(R_m,R_{m+1},\delta\tau\right) = \frac{\delta\tau}{2}\left[V\left(R_m\right)+V\left(R_{m+1}\right)\right]
   \end{eqnarray}
   so that the density matrix takes the form
   \begin{eqnarray}
   G\left(R_m,R_{m+1},\delta\tau\right) = G_0\left(R_m,R_{m+1},\delta\tau\right)e^{-U\left(R_m,R_{m+1},\delta\tau\right)}
   \end{eqnarray}
   where $G_0$ is the density matrix for free particles
   \begin{eqnarray} \label{freedm}
   G_0\left(R_m,R_{m+1},\delta\tau\right)=\left\langle R_m\left|e^{-\delta\tau\hat{T}}\right|R_{m+1}\right\rangle
   \end{eqnarray}

   \subsection{The Pair Product}
   Following Ref.~\onlinecite{pimc_app}, the PPA is a decomposition of the density matrix in which the effective potential is written as:
    \begin{eqnarray}\label{pprodform}
    U\left(R_m,R_{m+1},\delta\tau\right) = \sum_{i<j} u_2\left(\vec{r}_{m}^{\:i} - \vec{r}_{m}^{\:j},\vec{r}_{m+1}^{\:i} - \vec{r}_{m+1}^{\:j},\delta\tau\right)
    \end{eqnarray}
   where $\vec{r}_{m}^{\:i}$ is the position of the $i$--th particle at a timestep $\tau_m=m\delta\tau$. $u_2$ is the exact effective potential for 
   two atoms. This approximation states that, if the imaginary time is sufficiently small, the many--body propagator can be described as a product of two--body 
   propagators. 
    The density matrix for two particles can be obtained with different methods. For instance, 
   one can use the matrix--squaring method; this is shown in detail in Ref.~\onlinecite{pimc_app} and we limit here to state the final result for the two--body effective potential,
   \begin{eqnarray}
   u_2(\vec{r}_1,\vec{r}_2,\delta\tau) = \frac{u_0(\vec{r}_1,\delta\tau)+u_0(\vec{r}_2,\delta\tau)}{2} + \sum_{k=1}^{n}\sum_{j=0}^{k}u_{kj}(q,\delta\tau)z^{2j}s^{2(k-j)}
   \end{eqnarray}
   where $q=(|\vec{r}_1|+|\vec{r}_2|)/2$, $s=|\vec{r}_1-\vec{r}_2|$ and $z=|\vec{r}_1|-|\vec{r}_2|$. The first term is the effective potential of the PA and the 
   functions $u_{kj}$ are off--diagonal terms that are obtained from the partial wave expansion of the two--particles propagator. These off--diagonal terms  
   are usually obtained by numerical means and an analytic description of the pair density matrix is not available.

   \subsection{The Multi Product Expansion}
 From Eq. (\ref{propPA}), the $2n$--th order multi--product expansion is built with the following relation
 \begin{eqnarray} \label{mpgen}
 G_{2n}\left(\delta\tau\right) = \sum_{i=1}^{n}c_i G_2^{k_i}\left(\delta\tau/k_i\right) \\
 c_i = \prod_{j=1\left(\neq i\right)}^{n}\frac{k_i^2}{k_i^2-k_j^2} 
 \end{eqnarray}
Let's focus on the case $n=4$. Following ref \cite{c1rmulti}, the convenient choice for $\left\lbrace k_i\right\rbrace$ 
that yelds an eight order multi--product approximation suitable for PIGS calculations is 
$\left\lbrace k_i\right\rbrace = \left\lbrace 1,2,3,6\right\rbrace$. This, infact, produces elements $G_{2}^{k_i}$ 
with time--steps $\delta\tau/k_1$ that have a common divisor $\delta\tau/6$; these elements can thus be represented by a 
path integral with a time--step $\delta\tau/6$. 
This choice for $\left\lbrace k_i\right\rbrace$, combined with Eq.~\ref{mpgen}, gives the definition of the propagator used in Sec. \ref{sec:fluorographene}
\begin{eqnarray} 
G_8\left(\left\lbrace R_i\right\rbrace_{i=1}^{7},6\delta\tau\right) = G_0\left(1,2,\delta\tau\right)...G_0\left(6,7,\delta\tau\right)\times \nonumber \\
\left[\frac{54}{35}e^{-\frac{\delta\tau}{2}\hat{V}_{1}}e^{-\delta\tau\hat{V}_{2}}e^{-\delta\tau\hat{V}_{3}}e^{-\delta\tau\hat{V}_{4}} 
	e^{-\delta\tau\hat{V}_{5}}e^{-\delta\tau\hat{V}_{6}}e^{-\frac{\delta\tau}{2}\hat{V}_{7}} \right. \nonumber \\
-\frac{27}{40}e^{-\delta\tau\hat{V}_{1}}e^{-2\delta\tau\hat{V}_{3}}e^{-2\delta\tau\hat{V}_{5}}e^{-\delta\tau\hat{V}_{7}} \nonumber \\
+\frac{2}{15}e^{-\frac{3}{2}\delta\tau\hat{V}_{1}}e^{-3\delta\tau\hat{V}_{4}}e^{-\frac{3}{2}\delta\tau\hat{V}_{7}} \nonumber \\
- \left.\frac{1}{840}e^{-3\delta\tau\hat{V}_{1}}e^{-3\delta\tau\hat{V}_{7}}\right] 
\end{eqnarray}
where $G_0$ has been defined in Eq. (\ref{freedm}).